\documentclass[twocolumn]{article}

\usepackage{iftex}
\ifPDFTeX
  \usepackage[T1]{fontenc}
  \usepackage[utf8]{inputenc}
\else
  \usepackage{fontspec}
\fi

\usepackage{amsmath}
\usepackage{amssymb}
\usepackage{amsfonts}
\usepackage{amsthm}
\usepackage{bm}
\usepackage{dsfont}
\usepackage{booktabs}
\usepackage{nicefrac}
\usepackage{xcolor}
\usepackage{soul}
\usepackage{colortbl}
\usepackage{graphicx}
\usepackage{textcomp}
\usepackage{threeparttable}
\usepackage{tabularx}
\usepackage{multicol}
\usepackage{multirow}
\usepackage{adjustbox}
\usepackage{wrapfig}
\usepackage[normalem]{ulem}
\usepackage{arydshln}
\usepackage[linesnumbered,lined,vlined,ruled]{algorithm2e}
\usepackage{xpatch}
\usepackage{tikz}
\usetikzlibrary{arrows}
\usepackage{tikz-cd}
\usepackage{tcolorbox}
\usepackage{paralist}
\usepackage{enumitem}
\usepackage{array}
\usepackage{makecell}
\usepackage{float}
\usepackage{capt-of}
\usepackage{subfigure}
\usepackage{cite}
\PassOptionsToPackage{hyphens}{url}
\usepackage[hidelinks]{hyperref}
\usepackage[labelfont=bf, justification=justified]{caption}

\def\eqref#1{equation~\ref{#1}}

\def\1{\bm{1}}

\def\vx{{\bm{x}}}
\def\vy{{\bm{y}}}

\DeclareMathAlphabet{\mathsfit}{\encodingdefault}{\sfdefault}{m}{sl}
\SetMathAlphabet{\mathsfit}{bold}{\encodingdefault}{\sfdefault}{bx}{n}

\def\gA{{\mathcal{A}}}

\def\gF{{\mathcal{F}}}

\def\gI{{\mathcal{I}}}

\def\gS{{\mathcal{S}}}

\def\gV{{\mathcal{V}}}

\definecolor{darkblue}{rgb}{0.0, 0.0, 0.55}
\definecolor{mygray}{HTML}{5F5F5F}
\definecolor{myblue}{HTML}{0076BA}
\definecolor{myred}{HTML}{B51800}
\definecolor{mygreen}{HTML}{017100}
\definecolor{Gray}{gray}{0.9}

\hypersetup{
    colorlinks=true,
    citecolor=darkblue, 
    linkcolor=red, 
    urlcolor=magenta 
}

\theoremstyle{definition}

\newcommand{\myparagraph}[1]{%
  \par\smallskip\noindent\textbf{#1.}\enspace\ignorespaces
}

\newcommand{\myparagraphnp}[1]{%
  \par\smallskip\noindent\textbf{#1}\enspace\ignorespaces
}

\renewcommand{\sectionautorefname}{Section}
\providecommand{\figureautorefname}{Figure}
\providecommand{\tableautorefname}{Table}

\newcommand{\Loss}{\text{Loss}}
\newcommand{\Cost}{\text{Cost}}

\newcommand{\cc}{\cellcolor{Gray}}

\newcommand{\mybox}[2]{\colorbox{#1}{\strut #2}}
\definecolor{predcolor}{RGB}{212, 233, 199}
\definecolor{deccolor}{RGB}{250, 233, 178}

\begin{document}

\title{When Efficiency Backfires: Cascading LLMs Trigger Cascade Failure under Adversarial Attack}
\author{
  Zehan Sun\(^1\) \quad Dingfan Chen\(^2\) \quad Songze Li\(^1\) \\
  \(^1\)Southeast University \qquad
  \(^2\)Max Planck Institute for Intelligent Systems
}
\date{}
\maketitle

\begin{abstract}
Large Language Model (LLM) cascade systems are designed to balance efficiency and performance by processing queries with lightweight models while selectively escalating complex cases to more powerful ones. Such systems seek to reduces computational cost and latency while maintaining task performance, making it an appealing choice for large-scale deployment. However, the cascade design introduces new vulnerabilities through an expanded attack surface: the inclusion of lightweight front-end models and internal decision mechanisms introduces new weaknesses. In this work, we present the first study demonstrating that LLM cascade systems are susceptible to targeted adversarial manipulation, which disrupts both performance objectives and the intended cost advantages of the cascade design. We propose a novel attack framework that employs constrained sequential collaborative optimization of adversarial suffix under cascade dependencies, enabling simultaneous exploitation of lightweight models and decision mechanisms. This framework adapts to adversaries with varying capabilities, inducing controllable degradation in both cost-efficiency and accuracy. Unlike prior attacks targeting standalone models, our approach strategically leverages the cascade structure to achieve significantly stronger impact. Extensive experiments across diverse datasets and representative LLM cascade systems validate the practicality and severity of this attack. Our findings highlight the urgent need to rigorously scrutinize the security of LLM cascade systems and call for broader attention to the systemic risks inherent in such designs.
\end{abstract}

\vspace{0.5em}
\noindent\textbf{Keywords:} LLM Cascade Systems, Adversarial Attacks, Robustness, Cascade Vulnerabilities, Cost-Efficiency Degradation, Routing Manipulation
\vspace{0.5em}

\section{Introduction}
\label{sec:intro}

In recent years, large language models (LLMs) have gained widespread application across a broad spectrum of tasks, ranging from everyday applications to cutting-edge technological domains. 
The success of prominent models such as
the GPT series~\cite{achiam2023gpt}, the Claude family~\cite{anthropic2024claude3}, Gemini~\cite{team2023gemini}, DeepSeek~\cite{liu2024deepseek}, Qwen~\cite{bai2023qwen}, and Kimi~\cite{team2025kimi} exemplifies the growing influence of LLMs in fields such as natural language understanding, mathematical reasoning, scientific analysis, software development, and educational applications.

Despite these achievements, the development and deployment of LLMs remain expensive, as their large-scale architectures—often comprising more than billions of parameters—require substantial computational resources for training.
Moreover, the inference stage itself is resource intensive, often requiring substantial hardware, latency, and energy to process even a single query. The real-time nature of user-facing applications further amplifies the need for efficient inference. In contrast, simple and lightweight models provide much lower inference latency and computational overhead, though typically at the expense of accuracy or generality. 
This naturally raises the question of \textit{ whether it is possible to rely on inexpensive models whenever they suffice, and reserve the more powerful (but costly) models for only those queries that truly require them, thereby reducing overall budget or resource consumption}.

This motivates the idea of designing a cascade system of LLMs~\cite{li2021cascadebert,chen2023llmcascade}, where models are organized into a sequence of stages and executed in order from the first to the last. Lightweight but weaker models occupy the early stages, while stronger but more expensive models appear in later stages. Each stage is paired with a decision-making module that determines whether to terminate early at the current stage or escalate the query to the next one.  The goal of such a system is to reduce the overall inference cost while maintaining performance comparable to the strongest models.

Indeed, promising empirical results have been demonstrated: FrugalGPT~\cite{chen2023llmcascade} reports up to 98.3\% cost reduction compared to running a single strong LLM, with only minor performance degradation. 
Following this initial demonstration, a number of subsequent works have explored improvements across various dimensions~\cite{erbacher2024navigating, chen2024are, zhang2024llm, wang2024mixture, wang2025mixllm, chen2025improving, zhou2025adaptive, nie2024online, zhu2023optimal, zhang2023ecoassistant, gupta2024language, yue2024large, ramirez2024optimising}, such as enhanced decision heuristics, integration of intensively trained decision models, and hybrid architectures combining multiple weak and strong LLMs. These studies consistently demonstrate that carefully designed cascades can achieve appealing trade-offs between cost and performance, often approaching the accuracy of large proprietary models while notably reducing computational and monetary expenditure.

The practical deployment of cascade systems is also gaining traction. For example, Microsoft's Azure AI Foundry employs a language model as a decision-maker, selecting between fast, smaller models or routing queries to the more powerful GPT-5 model based on query complexity, cost, and latency requirements, reporting a up to 60\% reduction in inference costs while ensuring output quality~\cite{microsoft2025modelrouter,88hours2025gpt5}. 
Similarly, Lemony.ai has open-sourced Cascadeflow, which processes prompts with low-cost models, evaluates output quality, and escalates only when necessary, reducing development costs by up to 85\% with minimal latency overhead~\cite{lemony2025cascadeflow,siliconangle2025lemony}. Terminus Technology Group adopts a comparable approach in AIoT systems, deploying distilled edge models with decision mechanisms that selectively offload tasks to cloud-scale models, thereby reducing cost and latency while preserving performance~\cite{terminus2024edgecomputing,terminus2025aiotagent,terminus2025edgecloudllm}.
However, such gains should not be regarded as a  ``free lunch''. This raises a critical concern: is a cascade system reliable enough as a practical solution for reducing inference cost? From a systems perspective, cascades could be structurally vulnerable because prediction and decision modules are executed sequentially, and the output of each stage directly determines the input distribution of the next. This sequential dependency implies that errors or misjudgments made early in the pipeline can prematurely terminate or misdirect processing, effectively allowing early-stage decisions to dominate the system’s final outcome. 
Intuitively, a cascade behaves like a chain whose strength is \textit{limited by its weakest links: once a query is filtered out at an early stage, the later, more accurate models never have a chance to respond}. As a result, the system’s overall behavior could be highly sensitive to the correctness and robustness of these early predictions and routing decisions. 

More broadly, the cascade's layered architecture introduces additional decision boundaries and routing mechanisms, each of which creates new opportunities for adversarial manipulation. This naturally leads to the question of \textit{whether cascade architectures introduce failure modes and security vulnerabilities that are absent when a single strong model is used in isolation}. For instance, an adversary could exploit fragile decision logic to trigger unnecessary escalation, inflating inference cost or diverting queries to weaker components. Conversely, an adversary might suppress escalation, causing the system to rely on weaker models, thereby degrading task accuracy or increasing susceptibility to adversarial behaviors, which could otherwise be mitigated by stronger models. These scenarios suggest that the appealing cost-performance trade-offs reported by prior work may not be robust under adversarial conditions, especially when low-cost decision modules play a pivotal role in determining the execution path of the system.

In this work, we pioneer this direction and take a first step toward exploring whether cascade systems indeed exhibit such vulnerabilities. 
We begin by formulating a threat model that closely aligns with practical deployment settings, and we introduce a generic attack framework together with novel attack techniques. In particular, we note that existing attacks targeting a single LLM are sub-optimal when attacking cascade systems, as the coordination between prediction models and decision modules can often compensate for or correct errors introduced in any single component. As a result, attacks that consider only one module in isolation may be neutralized by other stages of the pipeline. We therefore explicitly exploit the cascade’s sequential dependency and design an alternating constrained optimization procedure, which enables new forms of manipulation that substantially strengthen attacks beyond what is possible when targeting a single LLM in isolation. We further provide analytical insights showing that, under typical conditions of practical cascade design, these vulnerabilities are inherent to the (vanilla) cascade architecture. 

Our experimental results demonstrate that, when targeting a single objective, our attack method can reduce prediction accuracy by up to 84.6\%, inflate normalized token cost by 148.9\%, and increase execution time by up
to 108.9\%. Furthermore, when jointly targeting both objectives, our method produces adversarial perturbations that simultaneously degrade task performance and system efficiency. These results consistently outperform existing baseline methods across multiple datasets, tasks, cascade architectures, and attack scenarios. Our findings suggest that vulnerability in cascade systems is not a corner-case artifact of specific implementations, but a structural consequence of how such systems are composed. This underscores an urgent need for new architectural principles, training objectives, and defense mechanisms that explicitly account for cross-module interactions in cascade systems, rather than treating them as loosely coupled collections of individually optimized components.

\myparagraph{Contributions}
We summarize the key contributions of this work as follows:
\begin{itemize}
\item \textbf{Formalizing a new security problem.}  
We identify inherent security vulnerabilities in LLM cascade architectures and introduce the first attack framework capturing diverse adversarial objectives.
    
\item \textbf{Novel attack techniques.} 
We propose a general adversarial attack framework that, unlike prior work focusing on a single model, jointly exploits the interactions between the weak prediction LLM and the decision module to coordinate cross-component attacks. This framework enables effective degradation of inference efficiency while supporting diverse adversarial objectives, revealing fundamental weaknesses of LLM cascade systems.

\item \textbf{Comprehensive empirical validation.} 
Through extensive experiments spanning 9 benchmark datasets, 4 language inference tasks, 12 LLM architectures, and 33 cascade configurations, we show that a single adversarial attack can simultaneously compromise both core objectives of LLM cascades. In particular, our attacks reduce prediction accuracy by up to 84.6\%, inflate normalized token cost by 148.9\%, and increase execution time by up to 108.9\%. For jailbreak scenarios, we also boost attack success rates by up to 81.4\%.
\end{itemize}

\section{Related work}
\label{sec:related_work}

\begin{figure*}[!t]
  \centering
    \includegraphics[width=\textwidth]{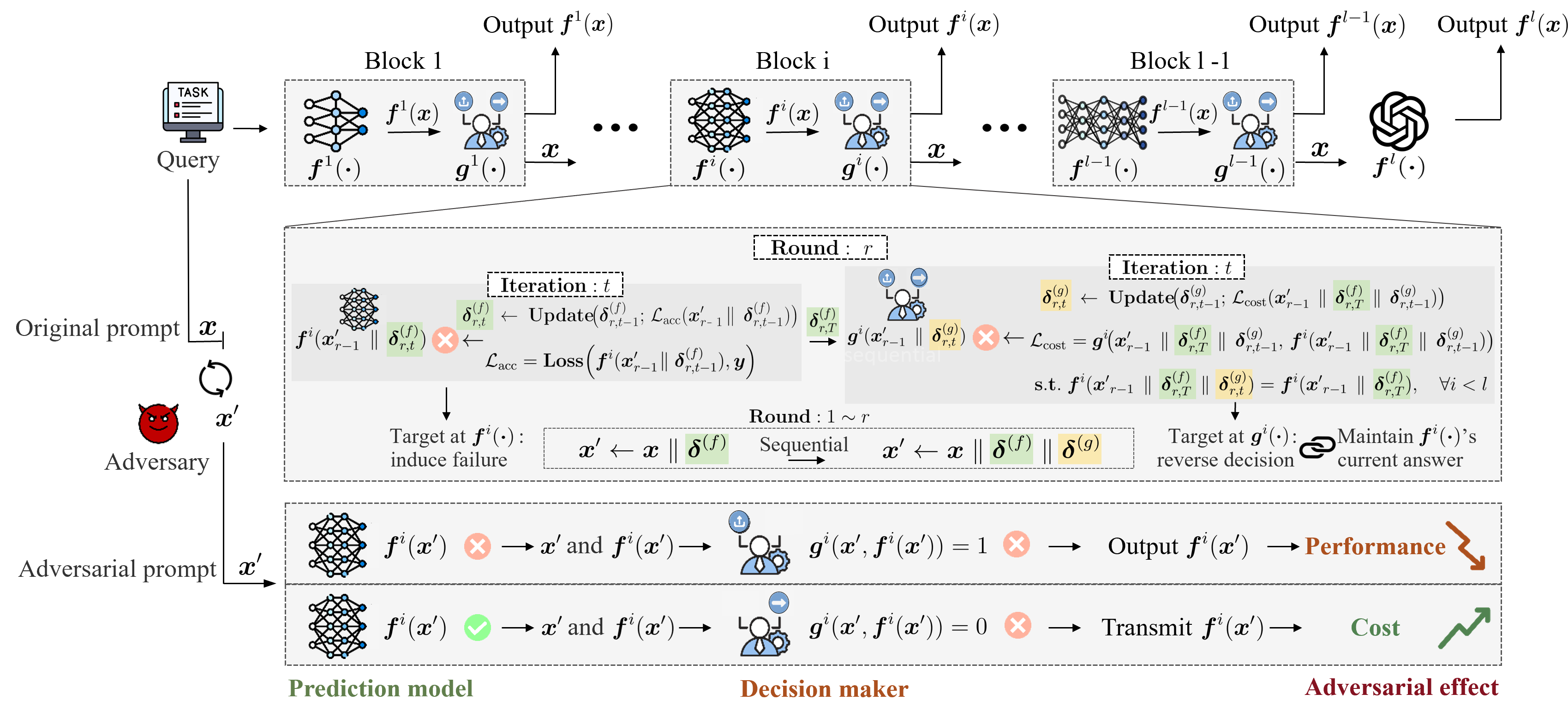}
    \vspace{-20pt}
  \caption{\textbf{Illustration of the LLM cascade and out joint-target attack (\sectionautorefname~\ref{subsec:alternating}).} \textit{``Top''}: Standard cascade inference. \textit{``Middle''}: Joint-target optimization, where the adversarial suffix is decomposed into \mybox{predcolor}{$\bm{\delta}^{(f)}$} (for prediction models) and \mybox{deccolor}{$\bm{\delta}^{(g)}$} (for decision modules).
  \textit{``Bottom''}: Attack-time inference with the optimized adversarial prompt.}
   \label{fig:teaser}
\end{figure*}

\subsection{LLM Cascade \& Routing}
With the growing demand for efficient LLM inference, \textit{cascading} and \textit{routing} strategies that orchestrate multiple models with different capabilities have received increasing attention as an option to balance computational cost and accuracy. While conceptually related, and sometimes treated interchangeably or combined into hybrid mixtures, we draw a clear distinction between cascades and routing mechanisms as discussed below.

\myparagraph{Cascade}
A cascade mechanism executes models \textit{\textbf{sequentially}} in a fixed order and allows early termination when an earlier stage is sufficiently confident, thereby avoiding the cost of running later, more expensive models. This idea can be traced back to early work such as CascadeBERT~\cite{li2021cascadebert}, which explored cascades of lightweight BERT-type variants~\cite{varshney2022model}.
More recently, this concept was first extended to LLMs by FrugalGPT~\cite{chen2023llmcascade}. FrugalGPT employs a lightweight language model as a decision module  that estimates the confidence of the current prediction and terminates early when this confidence is sufficiently high. 

Following FrugalGPT, numerous studies adhere to the scheme of adopting  \emph{separate lightweight language models} as decision-makers~\cite{erbacher2024navigating,chen2024are,zhang2024llm,wang2024mixture,wang2025mixllm,chen2025improving,zhou2025adaptive,nie2024online}. Another common strategy is to ask the \emph{prediction models themselves} to provide confidence scores alongside their answers, deferring to stronger models when the reported confidence is low~\cite{zhu2023optimal,zhang2023ecoassistant,gupta2024language}. Beyond these, several alternative heuristics have been explored. Examples include \emph{answer consistency} across multiple queries to a weak model~\cite{yue2024large} and \emph{margin sampling}, which derives confidence from the distribution over first-token predictions~\cite{ramirez2024optimising}. Despite these variations, using a lightweight language model as the decision module remains the mainstream approach, with FrugalGPT serving as the foundational architecture for most subsequent cascade designs.
In this paper, we focus on this mainstream class of cascade architectures for its representative nature and practical prevalence.  

\myparagraph{Routing} 
In contrast to cascade mechanisms, a routing paradigm typically places multiple models \textbf{\textit{in parallel}} and uses a dedicated router to determine which LLM to invoke based on the characteristics or requirements of the input. 

Routing strategies can be broadly grouped into \emph{static} and \emph{learned} methods. \textit{Static routing} uses fixed heuristics or comparison metrics to assess answer quality and model suitability. For instance, some approaches evaluate the output of a single model and escalate when a quality metric is not satisfied~\cite{aggarwal2024automix}, while others query multiple candidate LLMs and select the best response based on a comparison metric~\cite{wang2025mixllm,lee2024orchestrallm}. Although effective, such exhaustive evaluation incurs significant computational cost and latency, as all candidate models must be executed before selection.

\textit{Learned routing } methods aim to avoid exhaustive evaluation by training a router that predict the most suitable model~\cite{hari2023tryage,shnitzer2023large,hao2024fullanno,sakota2024fly}. 
Such routers are typically trained using supervised signals, reward functions, or other learned policies to guide model selection in advance, thereby reducing inference cost and latency. However, they require nontrivial offline training or calibration to characterize model behavior, and maintaining them becomes increasingly costly as the number of available models grows.

\subsection{Risks of LLM Systems}
While LLMs have demonstrated remarkable performance across diverse applications, their deployment introduces a broad range of security, privacy, fairness, and reliability risks. Recent studies highlight issues such as training-data leakage, prompt injection, bias propagation, and hallucinated content~\cite{yao2024survey,gallegos2024bias,das2025security,huang2025survey}. These findings underscore that LLM vulnerabilities are inherently multi-faceted, affecting model integrity, enabling misuse, and undermining trust. More recent discussions further argue that such risks are not confined to individual models but arise at the system level, where interactions among models, retrieval pipelines, and decision or routing components can introduce additional failure surfaces.

\myparagraph{Adversarial and Jailbreak Attacks}
Within this broader risk landscape, a substantial line of work investigates how adversarial perturbations and jailbreak prompts can deliberately elicit undesirable or unsafe behavior from language models. Early adversarial attacks~\cite{ebrahimi2018hotflip,jin2020bert,li2020bert} demonstrated that token-level or semantic perturbations can manipulate model predictions while preserving fluency. 
With the rise of LLMs, these approaches evolved into prompt-based attacks that circumvent alignment safeguards.
A prominent line of research focuses on \textit{adversarial prompt optimization}, including the Greedy Coordinate Gradient (GCG) attack~\cite{zou2023universal}, TAP~\cite{mehrotra2024tree}, and PAIR~\cite{chao2025jailbreaking}, showing that compact  adversarial suffixes can effectively induce harmful content across diverse LLMs. In parallel, \emph{mismatched generalization}, such as CodeAttack~\cite{jha2023codeattack}, FlipAttack~\cite{liu2024flipattack}, and ArtPrompt~\cite{jiang2024artprompt} use semantic or modality shifts to expose misalignment between training and inference distributions. \emph{Target-competition} attacks, including DAN~\cite{shen2024dany}, GUARD~\cite{jin2024guard}, and prefix/suffix injection~\cite{shi2024promptinjection}, embed high-authority or high-gradient competing instructions that override system prompts or safety policies. Collectively, these methods demonstrate that LLMs remain broadly vulnerable to a wide spectrum of prompt-level adversarial manipulations. 

These adversarial methods demonstrate that linguistic perturbations and structural manipulations can effectively compromise LLM behavior. However, despite this extensive progress, existing published evaluations consider vulnerabilities only for a \emph{single model in isolation}. When multiple models are orchestrated, as in LLM cascade systems, additional failure modes may arise from routing errors, miscalibrated confidence estimates, and adversarial manipulation of termination decisions. This gap motivates our central research question: \textit{Do LLM cascades face security risks beyond those inherent to their individual components, and how can such risks be quantitatively evaluated?}

\myparagraph{Robustness \& Inference Efficiency Attacks}
Parallel to adversarial research, a complementary body of work investigates \emph{robustness}, i.e., the stability of model predictions under input variation. 
Traditional robustness studies for simple language models have focused on substitution robustness, paraphrase invariance, and certified guarantees under norm-bounded perturbations~\cite{jia2019certified}. For modern LLMs, robustness analysis has shifted toward instruction-following and safety alignment, investigating how small lexical or syntactic changes to prompts affect factuality or compliance~\cite{ye-etal-2020-safer,huang2019achieving}. Formal tools such as Lipschitz bounds, input-to-state stability, and randomized smoothing have also been adapted for robustness verification~\cite{cohen2019certified,salman2020denoised,zhang2023certified,siska2024examining,howe2024exploring,ailem2024examining,beyer2025llmsafety}. 
 Despite these advances, robustness is overwhelmingly evaluated along semantic or behavioral axes, that is, whether the output content remains stable.

In contrast, our work examines a related but underexplored vulnerability: \emph{adversarial manipulation of inference cost and execution behavior}. While prior work in classical ML has shown that carefully crafted inputs can increase latency or energy consumption by disrupting activation sparsity~\cite{krithivasan2020sparsity,ilia2021spongeconference}, analogous attacks on LLM infrastructures (and especially on multi-stage or cascaded systems) remain largely unstudied. This gap is particularly consequential for cascaded LLM architectures, where lightweight models and decision modules conditionally invoke stronger (and more expensive) models. If these decision boundaries are fragile, subtle perturbations can redirect the cascade path, causing unnecessary escalation and inflating inference cost or latency without altering the semantic output. From a systems perspective, this exposes a form of robustness failure not only in \emph{prediction correctness} but also in \emph{execution stability}. 
Very recent work (much of it not yet published) has begun to expose security vulnerabilities in LLM routing systems~\cite{tang2026routeattack, shafran2025reroutingattack, lin2025lifecycle, yuan2025whoroutes}. These approaches primarily target a single routing component, aiming to manipulate the routing decision, i.e., forcing selection of a specific model. 
As such, these methods are instances of single-target attacks and largely reduce to the well-studied paradigm of adversarial attacks on a single model. Conceptually, they correspond to basic configurations of our framework (e.g., the single-target setting in Section~\ref{subsec:single-target}), differing primarily in that the attacked component is a (light-weight) router rather than a standalone (strong) LLM. In contrast, our work studies coordinated attacks on cascaded systems that simultaneously influence both prediction and routing behavior. This joint objective makes our setting fundamentally different and strictly more challenging, as it requires optimizing multiple interacting components rather than a single decision point in isolation.

\section{System and Threat Model}
\label{sec:threat_model}

\subsection{Threat Model}
We consider an adversary who manipulates \emph{only the input text} $\vx$, without altering any internal model parameters or routing policies. 
The adversary optimizes objectives defined over the system’s behavior, such as degrading prediction performance and/or inference efficiency. These objectives reflect practical adversarial incentives in cascade deployments. By shaping inputs, an attacker can force worst-case execution paths (e.g., repeatedly triggering expensive fallback models); at scale, this can resemble an application-layer denial-of-service, consuming shared resources and degrading service for other users. An attacker may also degrade reliability by inducing misrouting to weaker components, causing systematic errors that disrupt downstream use or erode trust. Additionally, they may increase policy violations by exploiting cross-stage inconsistencies, e.g., steering inputs toward weaker components that are easier to jailbreak, allowing restricted outputs that a stronger model would block.

We consider the modification within a bounded perturbation region (i.e., under the standard adversarial or robustness setting) so that the resulting perturbed input $\vx'$ lies in a neighborhood $N_\epsilon(\vx)$ around the original input. This constraint reflects realistic limits on how much an attacker can distort the input (e.g., via paraphrasing or prompt injection) while preserving semantic plausibility.
Notably, defining a threat model for cascade systems is more subtle than for single-model systems, as cascades involve multiple interacting components (e.g., routers, smaller models, and fallback large models), and a fully specified threat model could in principle require detailing adversarial access to each component. While this leads to a large and complex design space, we do not attempt to exhaustively enumerate all such configurations. Instead, we focus on the following primary threat scenarios that best reflect common real-world deployment settings.

\begin{itemize}[left=0pt]
    \item \textbf{White-box setting:} The adversary has full \textit{read access} to the internal components of the cascade system, including the architectures and parameters of the smaller models, the decision modules, and the overall cascade algorithm.  Such a setting corresponds to open-source or research environments where model checkpoints and the decision logic are publicly known. This enables gradient-based or structure-aware attacks that exploit specific decision boundaries or routing logic.

    \item \textbf{Black-box setting:} The adversary has no access to the cascade’s internal components. They can only interact with the overall system as a single black-box function, i.e., submitting input texts and observing final outputs, latency, or cost signals. This regime reflects public-facing systems such as chat-based LLM services (e.g., customer support bots or API endpoints), where attacks rely on input-output probing, response-time analysis, or transfer from externally trained surrogate models.
\end{itemize}

\subsection{System Model}
We consider an LLM cascade system designed to balance inference accuracy and computational efficiency. The system sequentially passes each input query through a series of language models with progressively increasing capability (and inference cost), optionally terminating early when a sufficiently confident response (determined by the decision modules) is obtained.
In normal operation, the cascade aims to minimize the average inference cost without significantly degrading task performance. By design, smaller models are responsible for handling easy queries, whereas larger models are reserved for harder or ambiguous cases. Decision modules control this process by determining whether to stop or escalate based on intermediate outputs and confidence signals (See \figureautorefname~\ref{fig:teaser} for a demonstration).

\subsubsection{LLM Cascade Formulation}
Let $\gF=\{f^i\}_{i=1}^l$ denote the ordered set of stage language models, typically arranged in ascending order of expressive power (and thus execution cost), and let $\mathcal{G}=\{g^i\}_{i=1}^{l-1}$ denote the corresponding decision modules.
We represent the $l$-stage cascade system as a function
$\Phi(\vx; \mathcal{F}, \mathcal{G}) $
parameterized by the stage models $\mathcal{F}$ and decision modules $\mathcal{G}$.
When clear from context, we write the cascade more simply as $\Phi(\vx)$.

For each input sample $\vx$, the cascade proceeds sequentially through stages $i=1,2,...,l$.
At stage $i$, the $i$-th language model $f_i$ produces an intermediate output $\vy^i = f^i(\vx)$. 
The corresponding decision module $g^i$ then receives both $\vx$ and $\vy^i$ and decides whether to terminate or proceed to the next stage. 
Formally, $g^i(\vx, \vy^i) \in \{0,1\}$ acts as a binary indicator, where $g^i=1$ denotes continuation and $g^i=0$ denotes stopping and returning the current output. 
The overall cascade can therefore be expressed as
\begin{equation}
\label{eq:llm_cascade}
\Phi(\vx; \mathcal{F}, \mathcal{G}) =
\begin{cases}
\vy^i, & \text{if } g^i(\vx, \vy^i) = 0 \text{ for some } i<l\\
\vy^l, & \text{if } g^i(\vx, \vy^i) = 1 \text{ for all } i<l
\end{cases}
\end{equation}
This formulation highlights the adaptive nature of the cascade, wherein progressively more capable (and costly) models are conditionally engaged based on intermediate predictions.

\myparagraph{Stopping Index and Inference Cost}
For each input sample $\vx$, let the cascade $\Phi(\vx)$ proceed sequentially through stages $i = 1, \dots, l$. 
We define the \emph{stopping index} $\tau$ as the stage at which the cascade terminates:
\begin{equation}
\label{eq:stop_index}
\tau =
\begin{cases}
\min\big\{\, i \,\big|\, g^i(\vx, f^i(\vx)) = 0 \,\big\} & \text{if such } i \text{ exists}\\[4pt]
l & \text{otherwise}
\end{cases}
\end{equation}
That is, $\tau$ indicates the index of the stage whose output is returned by the cascade for sample $\vx$ (with $\tau = l$ if all decision modules choose to continue).

Let $c_f^i(\vx)$ and $c_g^i(\vx, \vy^i)$ denote the  costs of executing the $i$-th model $f^i$ and decision module $g^i$, respectively, where $\vy^i = f^i(\vx)$. 
The total inference cost for sample $\vx$ is then given by
\begin{equation}
\label{eq:cascade_cost}
\Cost(\Phi(\vx))
=
\sum_{i=1}^{\tau} c_f^i(\vx)
\;+\;
\sum_{i=1}^{\tau} c_g^i(\vx, \vy^i)
\end{equation}
This captures the cumulative cost incurred up to termination.    
\section{Method}
\label{sec:method}

\subsection{General Adversarial Model}
As discussed in \sectionautorefname~\ref{sec:threat_model}, we consider an adversary that perturbs the input $\vx$ within a bounded neighborhood $N_\epsilon(\vx)$ to optimize an objective over the cascade system $\Phi$. In general, the attack can be formulated as
\begin{equation}
\label{eq:adv_general}
\max_{\vx' \in N_\epsilon(\vx)} \; F\big(\Phi(\vx'); \xi\big)
\end{equation}
where $F(\cdot)$ captures adversarial objectives and $\xi$ denotes optional auxiliary information.
For example, an accuracy-targeting attack maximizes task loss:
\begin{equation}
\label{eq:adv_acc}
\max_{\vx' \in N_\epsilon(\vx)} \; \Loss\big(\Phi(\vx'), \vy\big)
\end{equation}
where $\Loss(\cdot)$ is a task-specific (e.g., cross-entropy for classification), and $\vy$ is the ground-truth label for $\vx$.

In addition to degrading accuracy, an adversary may also aim to increase inference cost by forcing deeper execution:
\begin{equation}
\label{eq:adv_cost}
\max_{\vx' \in N_\epsilon(\vx)} \; \Cost\big(\Phi(\vx')\big)
\end{equation}
which encourages continuation decisions (i.e., $g^i = 1$) and thus engagement of more expensive stages.

To instantiate the neighborhood $N_\epsilon(\vx)$, we adopt the common \emph{suffix-based} perturbation strategy~\cite{zou2023universal}, in which a fixed-length suffix is appended to each query during optimization, following recent adversarial prompt studies~\cite{zhu2023promptrobust, chao2024jailbreakbench}. This induces a discrete neighborhood over token-level suffix modifications, serving as a simple and representative instantiation of our framework. Notably, our attack formulation is not tied to this specific choice: it is compatible with, possibly with minor modifications, any attack method that can be expressed as an optimization problem of the form in \equationautorefname~\ref{eq:adv_general}.

\subsection{Basic Single-target Attacks}
\label{subsec:single-target}
We first consider adversaries that target only \textit{one class of early-stage components} in the cascade, i.e., either the prediction models or the decision modules. 
This setting mirrors the standard single-model attack paradigm, allowing us to assess whether such attacks already pose a threat in cascaded systems. In particular, the inclusion of weaker early-stage models inherently enlarges the attack surface, suggesting that cascades may be more vulnerable than relying solely on a strong model. From an analytical standpoint, this restriction isolates the impact of early-stage components and provides a baseline to quantify how much performance the cascade may lose relative to using only the final, strongest model $f
^l$. 

Concretely, an accuracy-oriented attack can be instantiated by maximizing the prediction loss over selected early-stage models. Let $\gI \subseteq \{1, \cdots, l-1\}$ denote the targeted stages:
\begin{equation}
\label{eq:objective_acc}
\max_{\vx' \in N_\epsilon(\vx)} \; \sum_{i \in \gI} \Loss\big(f^i(\vx'), \vy\big)
\end{equation}
Similarly, an adversary aiming to increase inference cost may encourage continuation at these stages:
\begin{equation}
\label{eq:objective_cost}
\max_{\vx' \in N_\epsilon(\vx)} \; \sum_{i \in \gI} g^i\big(\vx', f^i(\vx')\big)
\end{equation}
This formulation restricts the adversary to manipulating a single class of components in isolation. While these single-target attacks are partially effective, they are generally suboptimal: targeting only prediction models cannot guarantee accuracy degradation, as errors may be corrected by later stronger models. Conversely, targeting only decision modules may be insufficient, as confident early-stage prediction outputs can still lead to premature termination.

\subsection{Joint-target Attacks}
\label{subsec:alternating}
To successfully attack a cascaded system, adversarial effects must propagate through sequentially dependent components, making single-target attacks inherently suboptimal. Optimizing only the prediction models or only the decision modules overlooks their coupling and fails to account for how errors are shaped and amplified across the pipeline. This motivates a joint-target optimization strategy that explicitly accounts for both components.
Specifically, we decompose the adversarial suffix into two parts: $\bm{\delta}^{(f)}$, targeting the \textit{prediction} models, and $\bm{\delta}^{(g)}$, targeting the \textit{decision} modules.

The optimization proceeds in multiple \emph{rounds} $r$. In each round, we append a new pair of suffixes $(\bm{\delta}^{(f)}_r, \bm{\delta}^{(g)}_r)$ to the input. After $r$ rounds, the perturbed input takes the form
\[
\vx'_r
= \vx \,\|\, \bm{\delta}^{(f)}_1 \,\|\, 
\bm{\delta}^{(g)}_1 \,\|\, \bm{\delta}^{(f)}_2 \,\|\, \bm{\delta}^{(g)}_2 \,\|\, \cdots \,\|\, \bm{\delta}^{(f)}_r \,\|\, \bm{\delta}^{(g)}_r
\]
where $\Vert$ denotes the concatenation (i.e., appending) operation.
Each round consists of two phases: one for prediction degradation and one for decision escalation, each internally solved by a few optimization \emph{iterations} $t \in \{1,\cdots,T\}$.
In the first phase of round \(r\), the prediction-targeted component is updated to increase the prediction loss:
\begin{equation}
\bm{\delta}^{(f)}_{r,t}
\;\leftarrow\;
\mathrm{Update} \,\big(\bm{\delta}^{(f)}_{r,t-1};\, \mathcal{L}_{\mathrm{acc}}(\vx'_{r-1} \,\|\, \bm{\delta}^{(f)}_{r,t-1})\big)
\end{equation}
where
\[
\mathcal{L}_{\mathrm{acc}}
= \sum_{i\in \gI} \Loss\;\big(f^i(\vx'_{r-1} \,\|\, \bm{\delta}^{(f)}_{r,t-1}),\,\vy\big)
\]

In the second phase, the decision-targeted component is adjusted to increase the escalation tendency:
\begin{equation}
\bm{\delta}^{(g)}_{r,t}
\;\leftarrow\;
\mathrm{Update}\,\big(\bm{\delta}^{(g)}_{r,t-1};\, \mathcal{L}_{\mathrm{cost}}(\vx'_{r-1} \,\|\, \bm{\delta}^{(f)}_{r,T} \,\|\,\bm{\delta}^{(g)}_{r,t-1})\big)
\end{equation}
where
\[
\mathcal{L}_{\mathrm{cost}}
= \sum_{i\in \gI} g^i\!\big(\vx'_{r-1} \,\|\, \bm{\delta}^{(f)}_{r,T} \,\|\,\bm{\delta}^{(g)}_{r,t-1},\, f^i(\vx'_{r-1} \,\|\, \bm{\delta}^{(f)}_{r,T} \,\|\,\bm{\delta}^{(g)}_{r,t-1})\big)
\]
Each \(\mathrm{Update}(\cdot)\) denotes an approximate improvement operator that performs inner iterations towards maximizing the corresponding objective within the perturbation boundary.  
This alternating procedure resembles block coordinate optimization: in each round, one subset of variables ($\bm{\delta}^{(f)}$ or $\bm{\delta}^{(g)}$) is updated to improve its associated objective, while treating the other as temporarily fixed.

\subsection{Coordinated joint optimization}
\label{subsubsec:constraint_preserve}
While the two target components can be optimized via alternation as described above, their updates are not independent. Because decision modules receive both the input and intermediate predictions, a decision-oriented update can inadvertently \emph{repair} prediction errors introduced by the preceding prediction-oriented phase: for example, by appending tokens that revert a previously incorrect early-stage output back to the correct label. To avoid this counterproductive effect, we require that \textit{decision-oriented updates} (i.e., updating $\bm{\delta}^{(g)}$) preserve the prediction errors already induced by the attack.

Let the perturbed input after the prediction phase of round \(r\) be
\(\vx'_r\),
and let \(\hat{\vy}^i_r = f^i(\vx'_r)\) denote the corresponding intermediate predictions at layer \(i\).
During the decision-oriented phase of round \(r\), we perform \(T\) inner \emph{iterations}.  
At each iteration \(t = 1,\dots,T\), a small discrete candidate set of feasible local modifications is generated around the current decision suffix:
$\mathcal{C}_{r,t} \subset N_\epsilon(\bm{\delta}^{(g)}_{r,t-1})$.
We then select the next update as
\[
\bm{\delta}^{(g)}_{r,t} \in 
\arg\max_{\bm{\delta} \in \mathcal{C}_{r,t}}
\mathcal{L}_{\mathrm{cost}}(\vx'_{r-1} \,\|\, \bm{\delta}^{(f)}_{r,T} \,\|\,\bm{\delta})
\]
subject to the following priority constraints:
\begin{enumerate}
\item[(i)] Prefer candidates that keep the previous stage predictions unchanged:
$$f^i(\vx'_{r-1} \,\|\, \bm{\delta}^{(f)}_{r,T} \,\|\,\bm{\delta}) = \hat{\vy}^i_r,\quad \forall i<l$$
\item[(ii)] If none satisfy (1), prefer candidates that maintain previously incorrect predictions as still incorrect (i.e., not necessarily identical):
\[
f^i(\vx'_{r-1} \,\|\, \bm{\delta}^{(f)}_{r,T} \,\|\,\bm{\delta}) \neq \vy,\quad \text{whenever } \;\hat{\vy}^i_r \neq \vy
\]
\item[(iii)] If neither (i) nor (ii) can be satisfied, drop the constraint and maximize \(\mathcal{L}_{\mathrm{cost}}\) freely.
\end{enumerate}
This prioritized selection ensures that decision-oriented updates \textit{coordinate with} the degradation already achieved by the prediction-oriented phase. In other words, the decision phase builds on (rather than overrides) the previous phase, maintaining the cumulative effect of the alternating attack process.

\subsection{Adaptive Pass-Rate Control}
\label{subsubsec:pass_rate}
There is an inherent trade-off between competing attack objectives in cascade systems:forcing inputs to proceed to later stages increases inference cost, while deeper models may counteract or alter the desired adversarial effects. To allow practical control over this trade-off, we propose a simple stochastic targeting rule that enforces escalation only for a controlled fraction of samples.

Let \( p \in [0,1] \) denote the target pass-rate for samples that already exhibit the desired adversarial effect after the prediction-oriented phase. Let \( \hat{\vy}_r \) denote the intermediate predictions at round \( r \). Samples that do not yet satisfy the attack objective are always escalated to later stages, while those that already meet the objective are escalated with probability \( p \) and allowed to terminate with probability \( 1 - p \). This stochastic control enables the attacker to balance competing objectives by tuning \( p \), thereby regulating overall attack behavior.

\section{Analytical Insights}
\label{sec:analysis}

We provide insights showing that a cascade system is generally more vulnerable than its strongest component (typically the most powerful LLM placed at the final stage) under adversarial conditions. 
Importantly, this vulnerability gap is not solely caused by adversarial perturbations. It already arises in the standard (non-adversarial) setting due to the routing structure of the cascade, and adversarial inputs further amplify this effect.
We illustrate this phenomenon using the accuracy degradation setting introduced above. 

Recall that the event that the adversarial text $\vx'$ triggers an incorrect output at stage $i$ is
\begin{equation}
    \gA_i(\vx') = \{f^i(\vx')\neq y \,\wedge\, g^i(\vx', f^i(\vx'))=0 \}
\end{equation}
The event that the cascade outputs an incorrect final prediction is
\begin{equation}
\begin{aligned}
    \gA_{\mathrm{cas}}(\vx')
    =
    \bigcup_{i=1}^l
    \Bigg(
    \bigcap_{j=1}^{i-1}\{g^j(\vx',f^j(\vx'))=1\}
    \cap
    \gA_i(\vx')
    \Bigg)
\end{aligned}
\end{equation}
For each stage $i$, define the routing event
\begin{align*}
\gS_i(\vx') = \biggl( & \bigcap_{j=1}^{i-1} \{g^j(\vx',f^j(\vx'))=1\} \\
& \cap \{g^i(\vx',f^i(\vx'))=0\} \biggr)
\end{align*}
We assume $g^l(\vx',f^l(\vx'))=0$, i.e., the cascade always terminates at the final stage. Then $\{\gS_i(\vx')\}_{i=1}^l$ forms a partition of the sample space. 
Using this partition, the cascade error probability decomposes as
\begin{equation}
\Pr[\gA_{\mathrm{cas}}(\vx')]
=
\sum_{i=1}^l
\Pr\!\left[
\gS_i(\vx') \cap \{f^i(\vx')\neq \vy\}
\right]
\end{equation}
where all probabilities are taken over $(\vx,\vy)\sim\mathcal D$ and any randomness in the perturbation $\bm{\delta}$.
Similarly, the error of the final-stage model admits
\begin{equation}
\Pr[f^l(\vx')\neq \vy]
=
\sum_{i=1}^l
\Pr\!\left[
\gS_i(\vx') \cap \{f^l(\vx')\neq \vy\}
\right]
\end{equation}
Therefore,
\begin{align}
\label{eq:gap_adv}
&\Pr[\gA_{\mathrm{cas}}(\vx')]
-
\Pr[f^l(\vx')\neq \vy]
= \sum_{i=1}^{l-1}
\Pr[\gS_i(\vx')] \cdot  \nonumber\\
&
\Big(
\Pr[f^i(\vx')\neq \vy \mid \gS_i(\vx')]
-
\Pr[f^l(\vx')\neq \vy \mid \gS_i(\vx')]
\Big)
\end{align}
This shows that both quantities are mixtures over the same routing-induced partition, but with different stage-wise error contributions. In particular, for each stage $i<l$, the cascade uses $f^i$ on inputs in $\gS_i(\vx')$, whereas the standalone system would instead apply $f^l$ on the same subset.
Consequently, if there exists some $i<l$ such that $
\Pr[\gS_i(\vx')] > 0$  and 
\begin{equation*}
\Pr[f^i(\vx')\neq \vy \mid \gS_i(\vx')]
>
\Pr[f^l(\vx')\neq \vy \mid \gS_i(\vx')]    
\end{equation*}
and for all other $k<l$,
\[
\Pr[f^k(\vx')\neq \vy \mid \gS_k(\vx')]
\ge
\Pr[f^l(\vx')\neq \vy \mid \gS_k(\vx')]
\]
then
\begin{equation}
\Pr[\gA_{\mathrm{cas}}(\vx')]
>
\Pr[f^l(\vx')\neq \vy]    
\end{equation}
This assumption requires that the final-stage model is at least as accurate as earlier stages on the routed subsets, with strict improvement on at least one subset. This condition is mild and aligns with practical cascade design, where later models are typically more powerful and intended to handle harder inputs. Under this condition, the cascade incurs strictly higher error than the strongest standalone model. Notably, this conclusion holds even in the absence of adversarial perturbations (i.e., when replacing $\vx'$ with $\vx$), indicating that the degradation arises inherently from the routing structure and the presence of weaker intermediate stages.

Adversarial perturbations further amplify this gap by jointly affecting both routing and prediction. To make this precise, define the gap on clean inputs
\[
\Delta(\vx)
=
\Pr[\gA_{\mathrm{cas}}(\vx)]
-
\Pr[f^l(\vx)\neq \vy],
\]
and under perturbation
\[
\Delta(\vx')
=
\Pr[\gA_{\mathrm{cas}}(\vx')]
-
\Pr[f^l(\vx')\neq \vy].
\]
Given \equationautorefname~\ref{eq:gap_adv}, adversarial perturbations increase the gap $\Delta$ whenever
\begin{equation}
\small
\begin{aligned}
&\sum_{i=1}^{l-1} \Pr(\gS_i(\vx')) \,\Pr(f^i(\vx')\neq \vy \mid \gS_i(\vx')) \\
&\qquad - \sum_{i=1}^{l-1} \Pr(\gS_i(\vx')) \,\Pr(f^l(\vx')\neq \vy \mid \gS_i(\vx')) \\
&> 
\sum_{i=1}^{l-1} \Pr(\gS_i(\vx)) \,\Pr(f^i(\vx)\neq \vy \mid \gS_i(\vx)) \\
&\qquad - \sum_{i=1}^{l-1} \Pr(\gS_i(\vx)) \,\Pr(f^l(\vx)\neq \vy \mid \gS_i(\vx))
\end{aligned} \nonumber
\end{equation}
This inequality can arise through two mechanisms: 
\begin{itemize}[left=1pt]
\item \textbf{Routing shift:} Adversarial perturbations can increase the probability of early termination at weaker stages, i.e.,
\[
\Pr[\gS_i(\vx')] > \Pr[\gS_i(\vx)]
\quad \text{for some } i<l,
\]
while the corresponding conditional gaps remain nonnegative. In this case, more probability mass is assigned to stages where $f^i$ is weaker than $f^l$, directly increasing $\Delta(\vx')$.

\item \textbf{Conditional gap amplification:} Even when routing probabilities remain unchanged, perturbations can increase the conditional error gap on routed subsets. Specifically, for some $i<l$,
\begin{align*}
&\Pr[f^i(\vx')\neq \vy \mid \gS_i(\vx')]
-
\Pr[f^l(\vx')\neq \vy \mid \gS_i(\vx')] \\
&>
\Pr[f^i(\vx)\neq \vy \mid \gS_i(\vx)]
-
\Pr[f^l(\vx)\neq \vy \mid \gS_i(\vx)]
\end{align*}
This occurs because $\gS_i(\vx')$ defines a different conditional distribution than $\gS_i(\vx)$, and adversarial perturbations may shift the routed subset toward inputs on which the weaker model $f^i$ degrades more than the stronger model $f^l$.
\end{itemize}
Taken together, these results highlight that adversarial perturbations can systematically exploit and amplify an intrinsic vulnerability of cascade systems induced by their routing structure. In particular, a cascade can be strictly more vulnerable than using a single strong (well-aligned) LLM alone. This suggests that the efficiency goal of cascaded designs can indeed ``backfire'', in that it may come with a non-negligible robustness cost, and such trade-offs should not be overlooked in practice.

\begin{figure*}[t]
  \centering
    \subfigure[Web-question (passrate)]{
    \includegraphics[width=0.56\columnwidth]{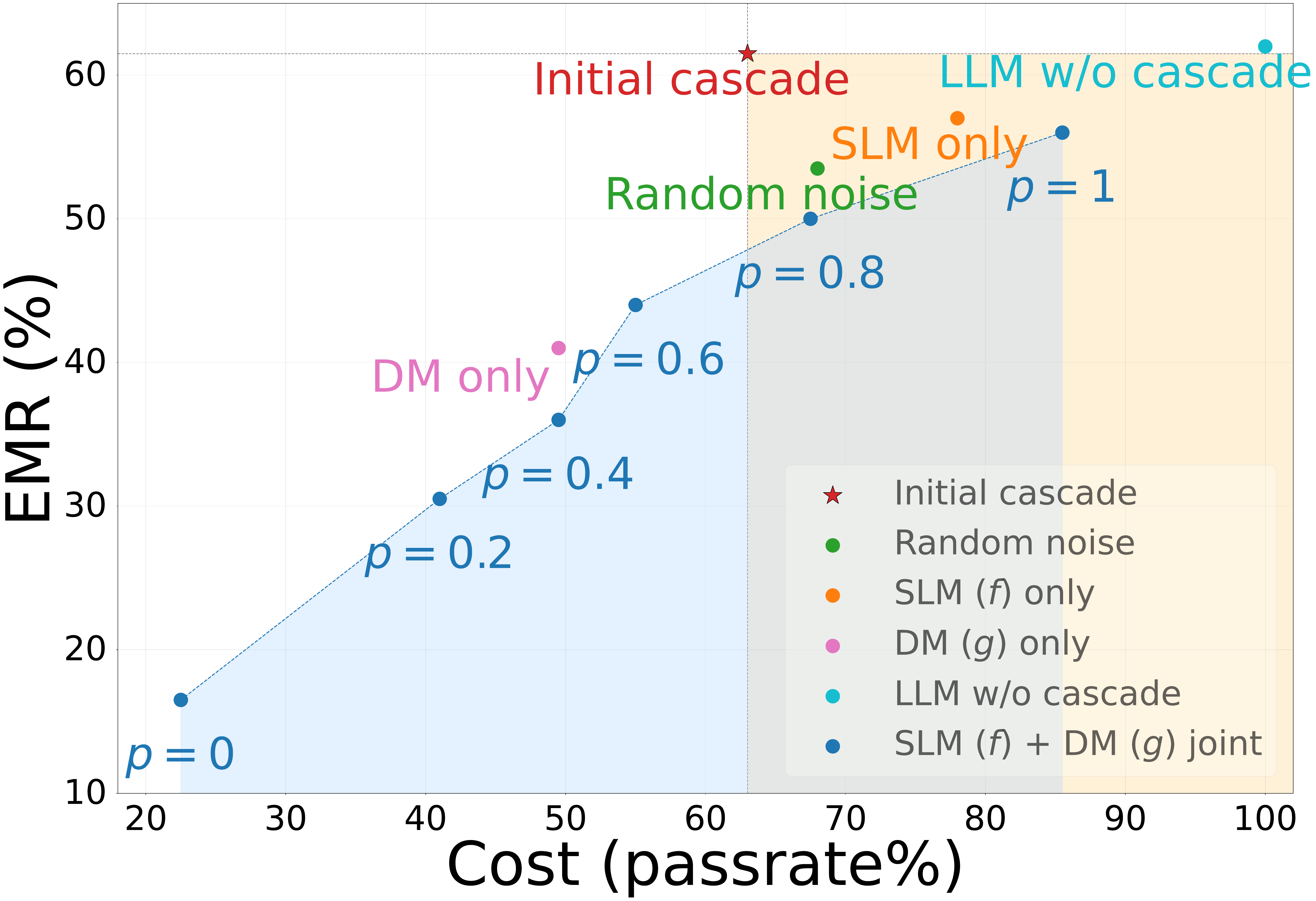}
  }
        \subfigure[SQuAD2.0 (passrate)]{
    \includegraphics[width=0.56\columnwidth]{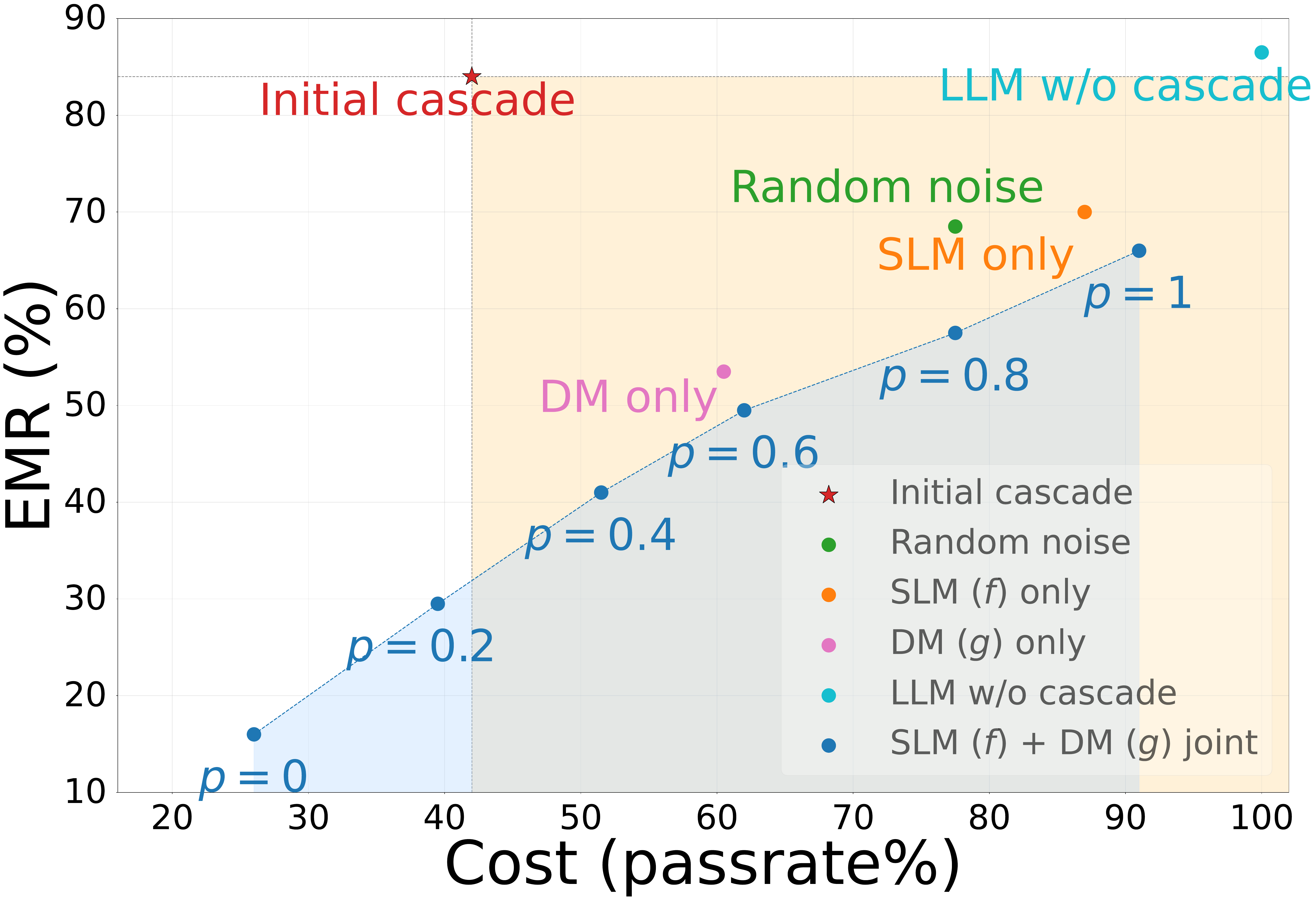}
  }
  \subfigure[WildJailbreak (passrate)]{
    \includegraphics[width=0.56\columnwidth]{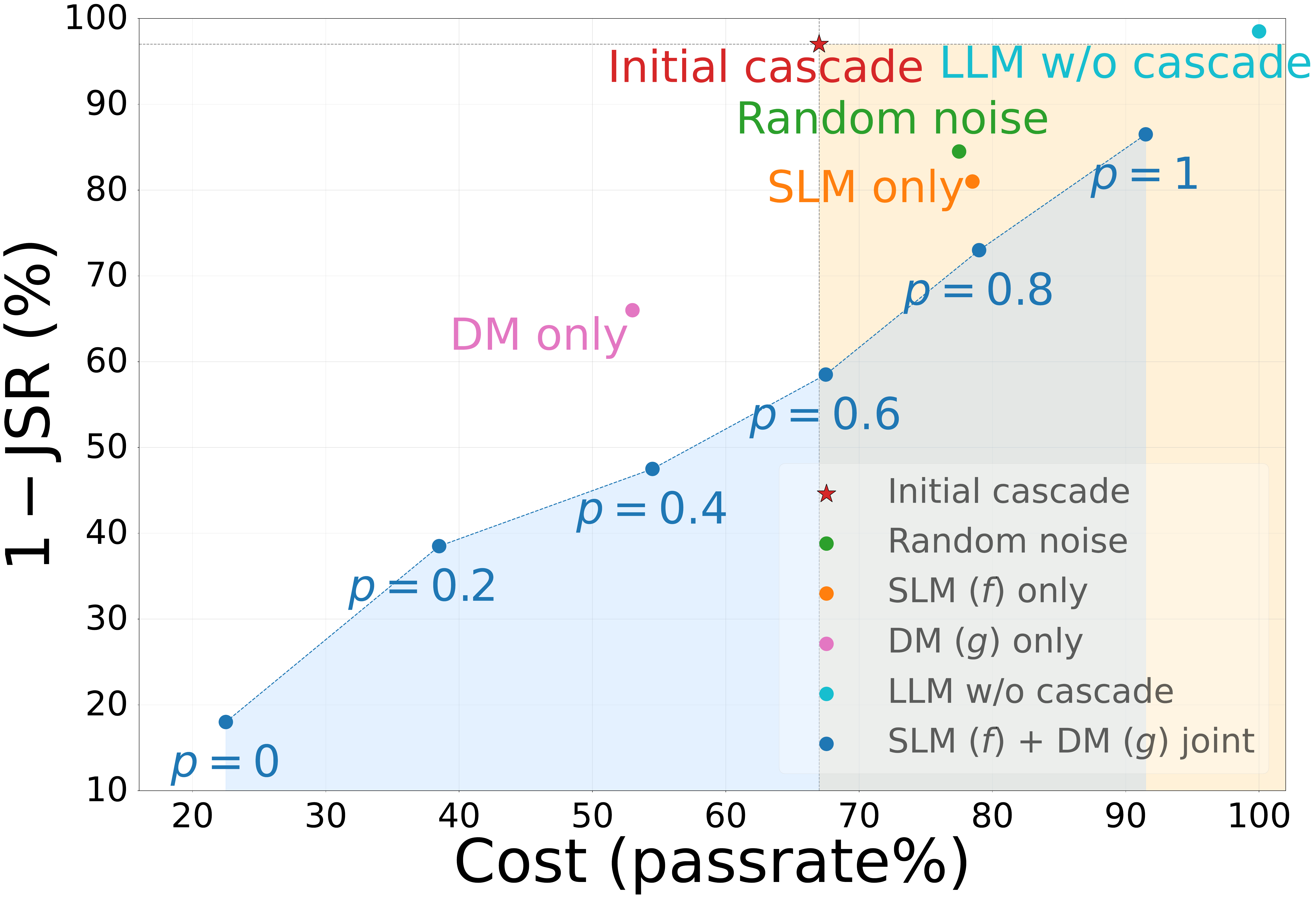}
  }
  \subfigure[Web-question (normalized tokens)]{
    \includegraphics[width=0.56\columnwidth]{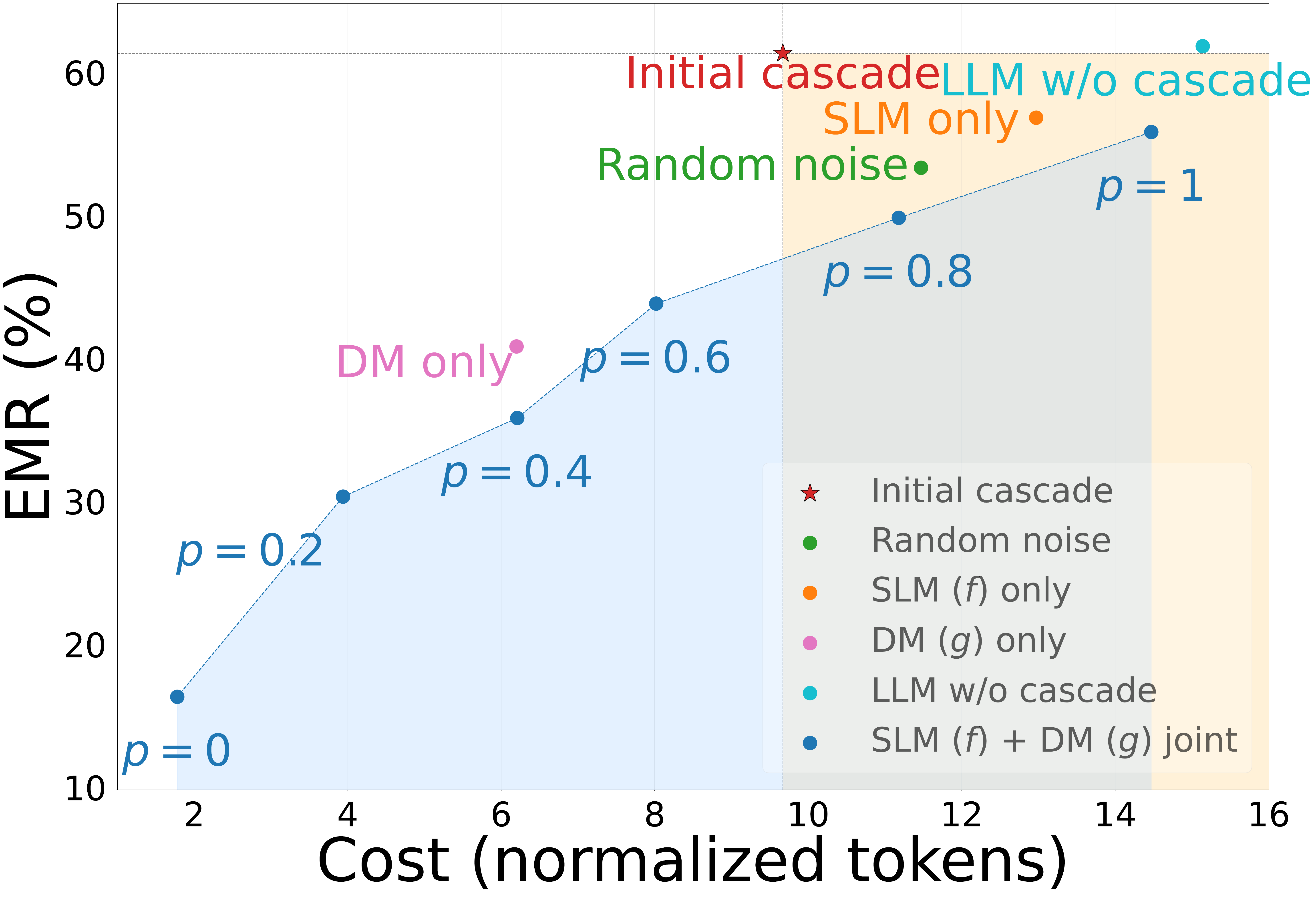}
  }
   \subfigure[SQuAD2.0 (normalized tokens)]{
    \includegraphics[width=0.56\columnwidth]{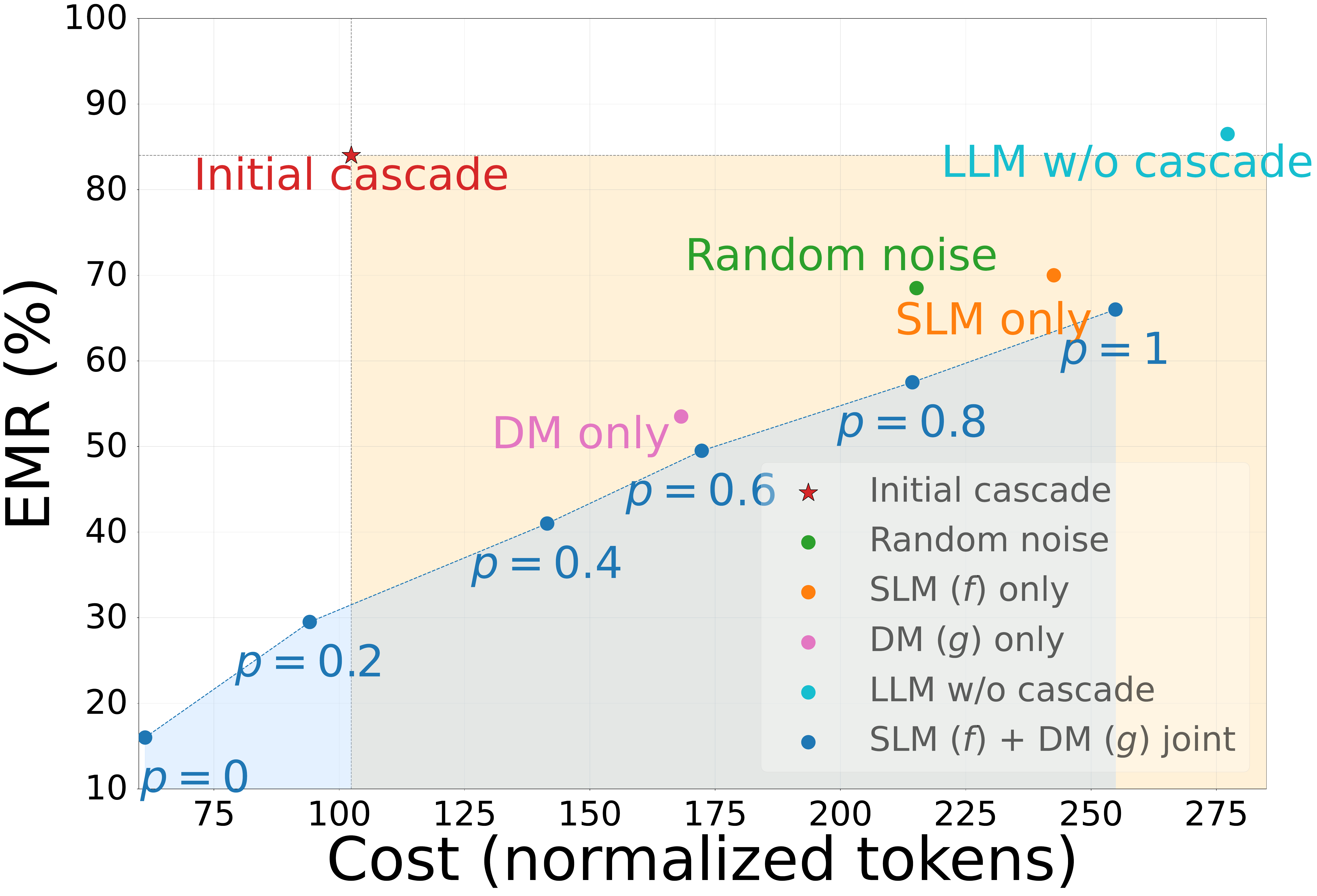}
  }
      \subfigure[WildJailbreak (normalized tokens)]{
    \includegraphics[width=0.56\columnwidth]{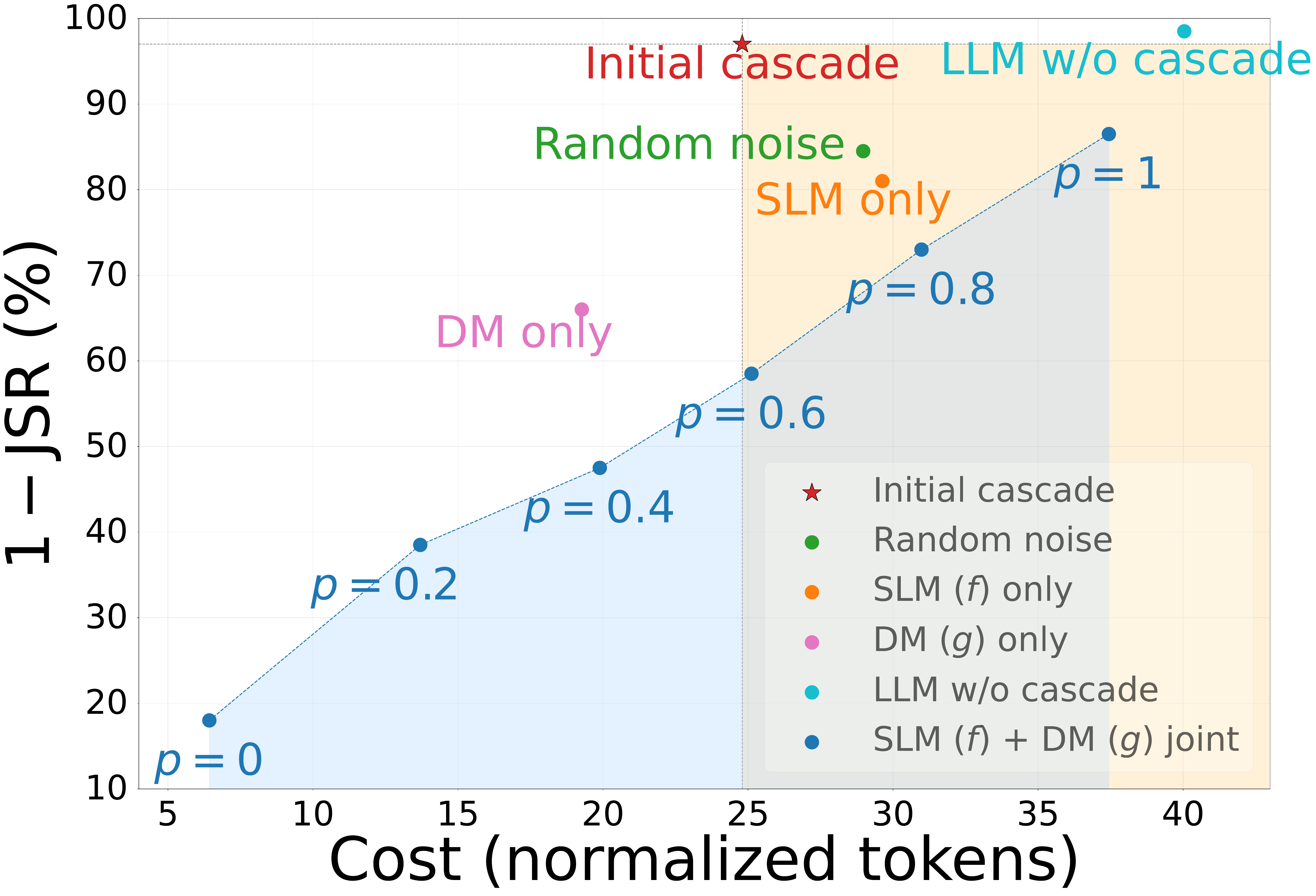}
  }
  \subfigure[Web-question (time)]{
    \includegraphics[width=0.56\columnwidth]{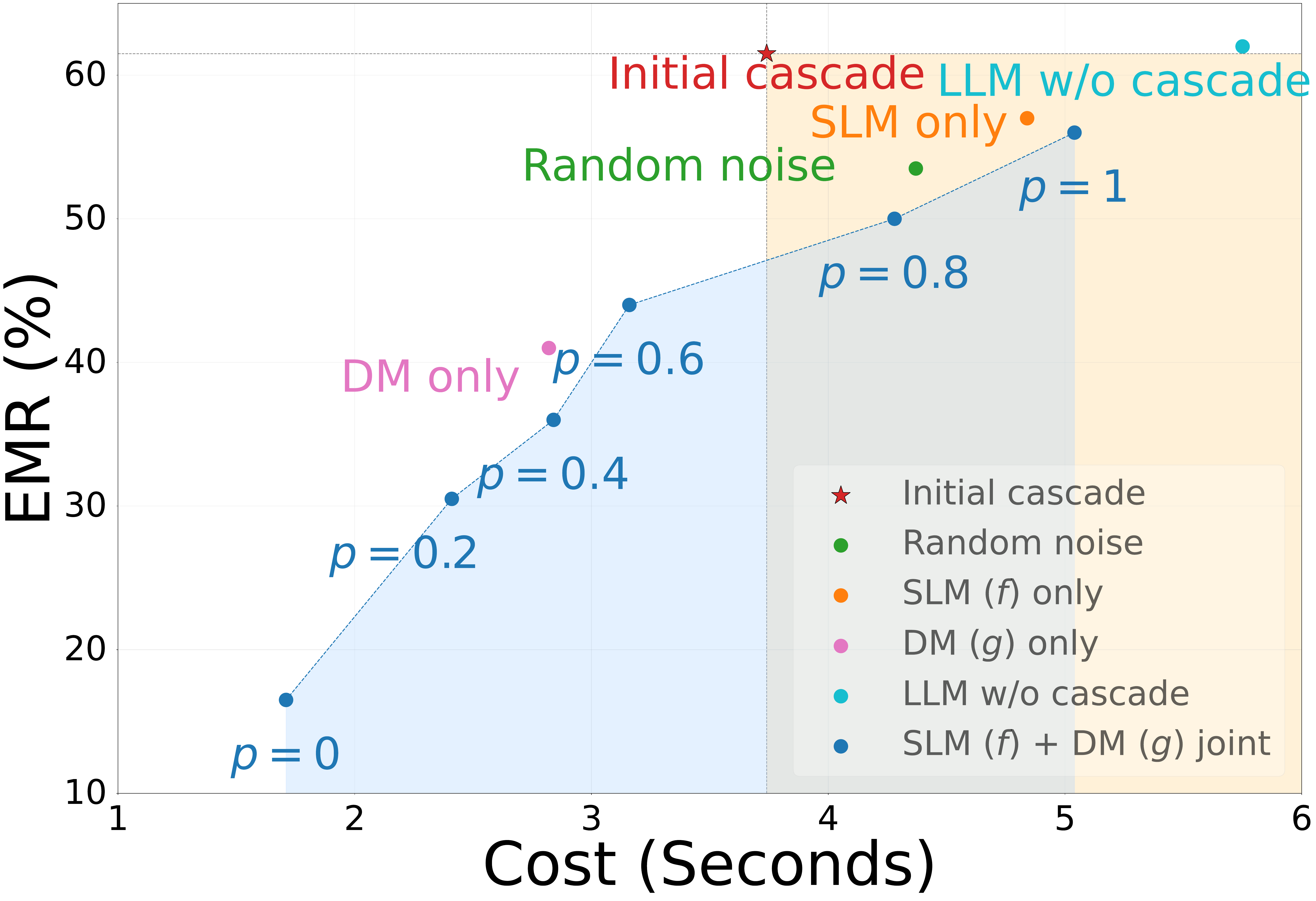}
  }
      \subfigure[SQuAD2.0 (time)]{
    \includegraphics[width=0.56\columnwidth]{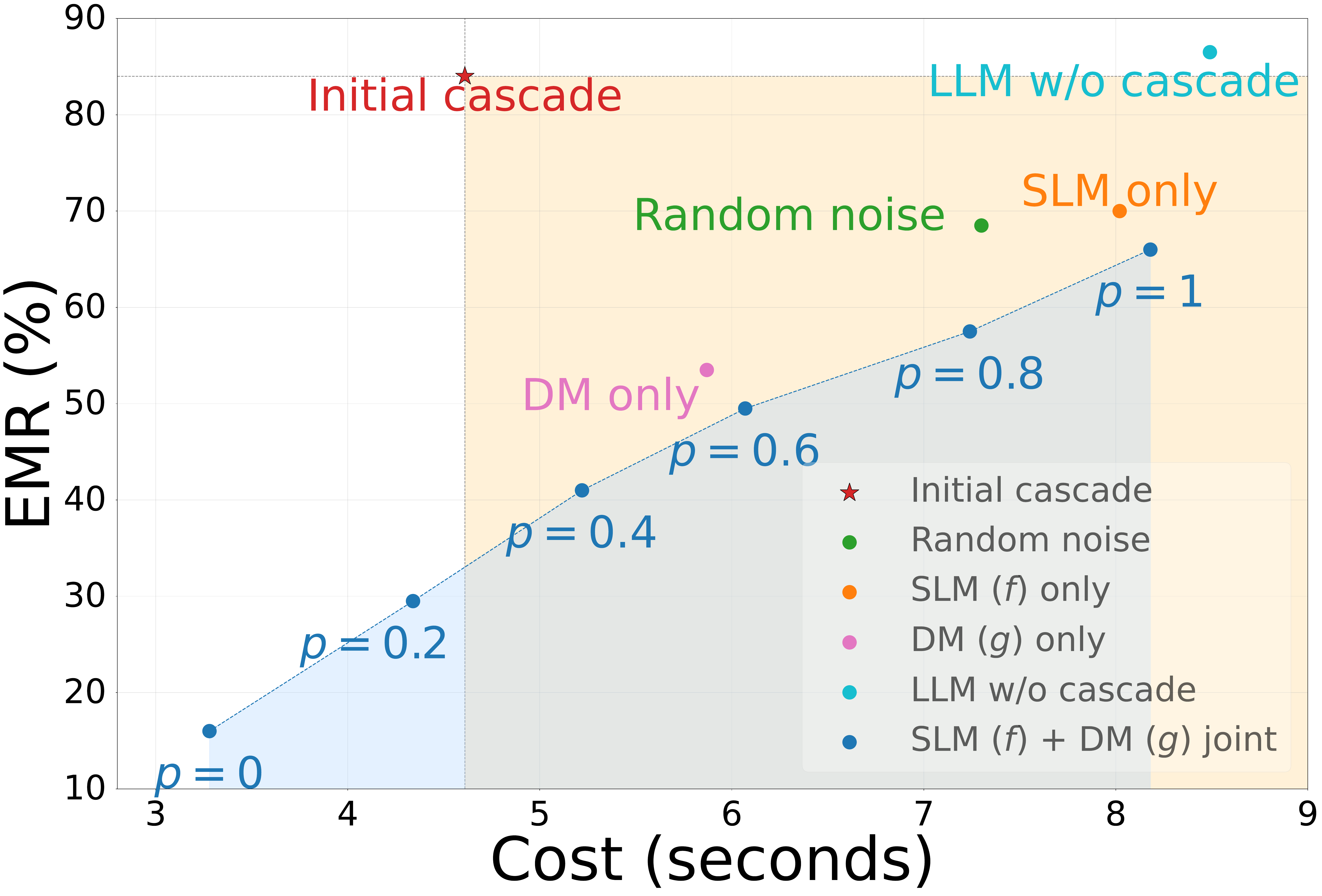}
  }
  \subfigure[WildJailbreak (time)]{
    \includegraphics[width=0.56\columnwidth]{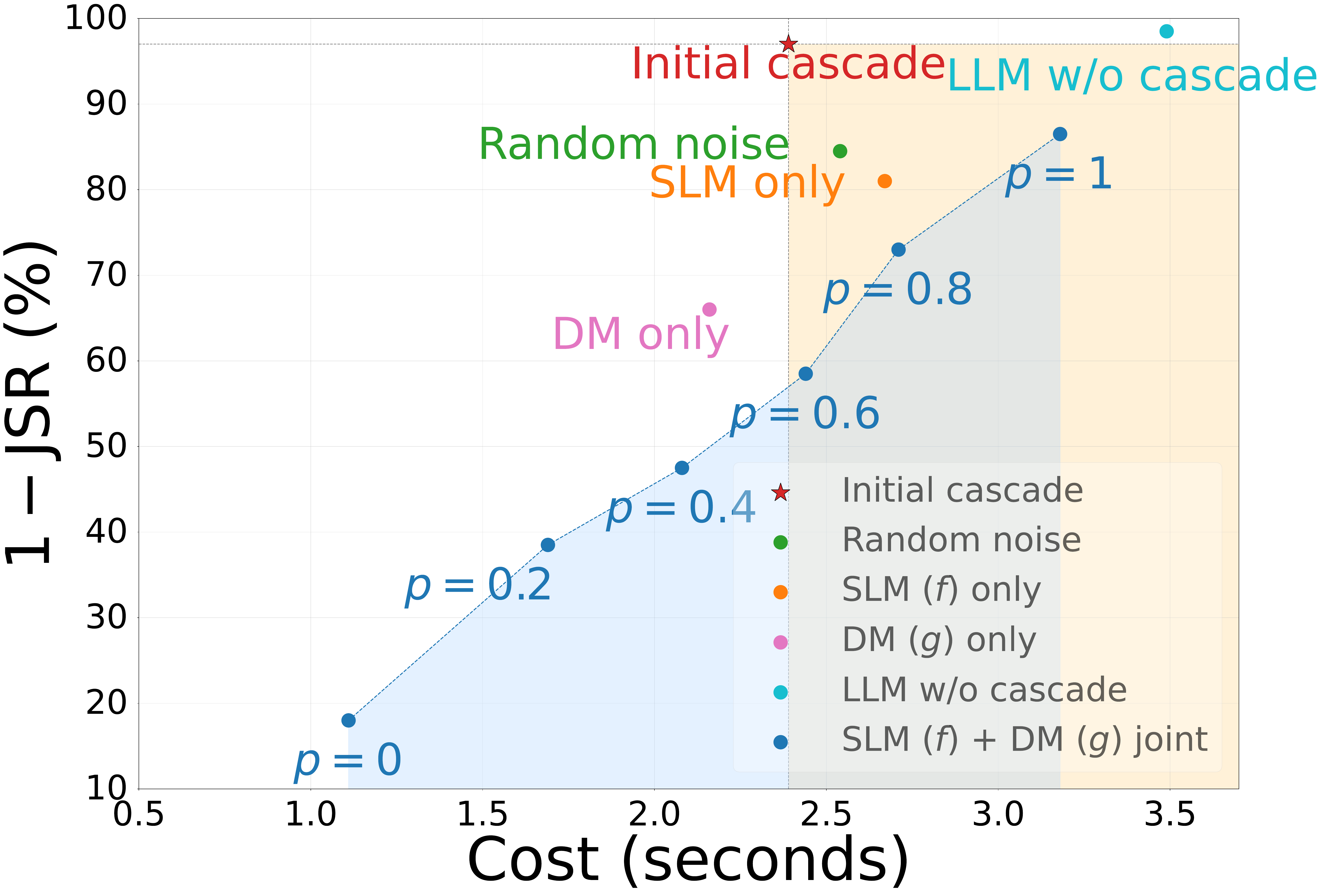}
  }
  \vspace{-6pt}
  \caption{
\textbf{Comparison of attack methods on a two-layer LLM cascade across the WebQuestions, SQuAD2.0, and WildJailbreak datasets, using cascade configurations~(1), (1), and (8), respectively (see Table~\ref{tab:cascade-configs-1}).}
The cost–accuracy trade-off achieved by our pass-rate control method (Section~\ref{subsubsec:pass_rate}) is illustrated by the {\color[HTML]{1F77B4}blue} dashed line. Points toward the lower-right represent more effective attacks.}
  \label{fig:2layer_results_1}
  \vspace{1pt}
\end{figure*}

\begin{figure*}[t]
  \centering
  \subfigure[AGnews (normalized tokens)]{
    \includegraphics[width=0.52\columnwidth]{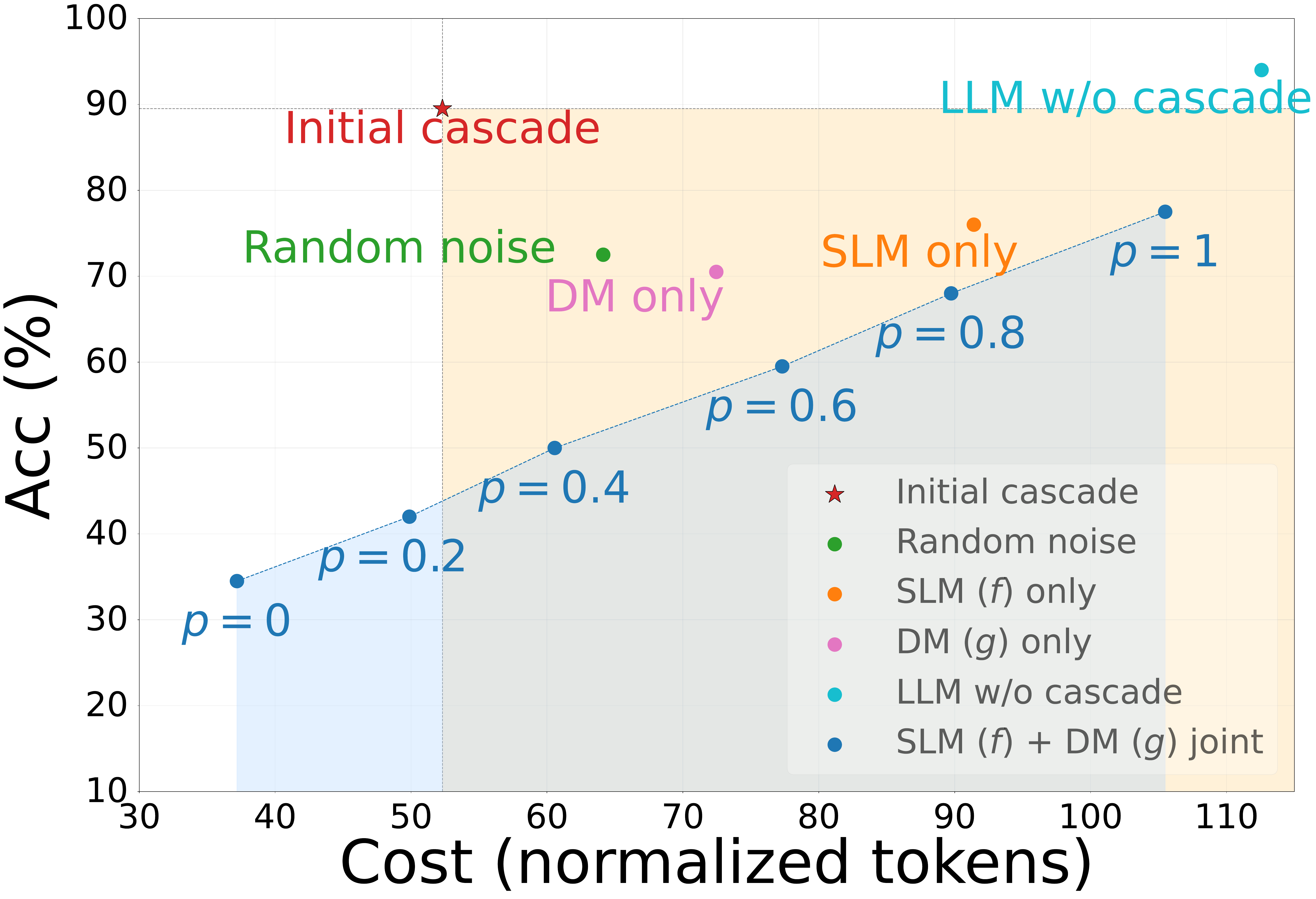}
  }
   \subfigure[SQuAD2.0 (normalized tokens)]{
    \includegraphics[width=0.52\columnwidth]{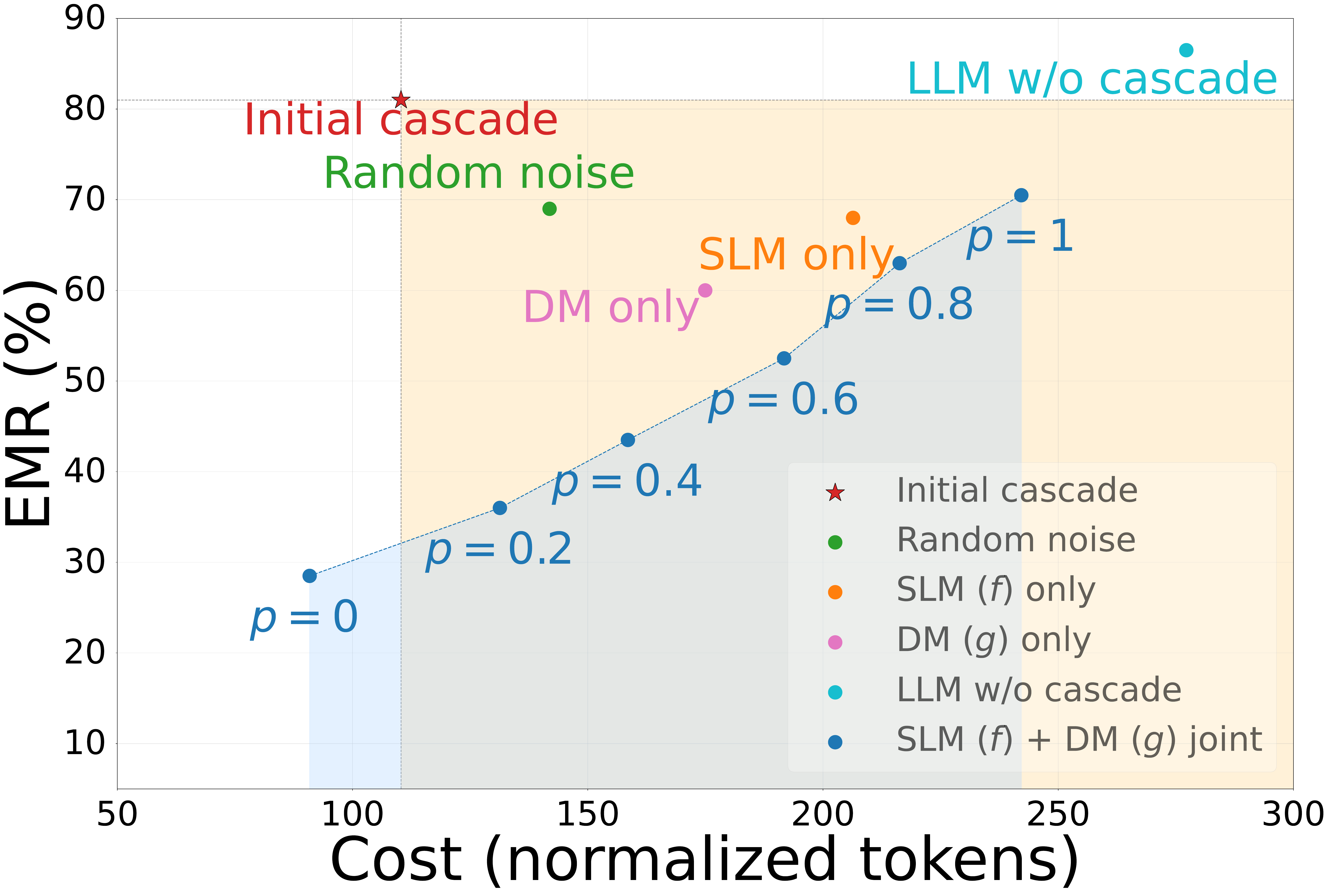}
  }
      \subfigure[WildJailbreak (normalized tokens)]{
    \includegraphics[width=0.52\columnwidth]{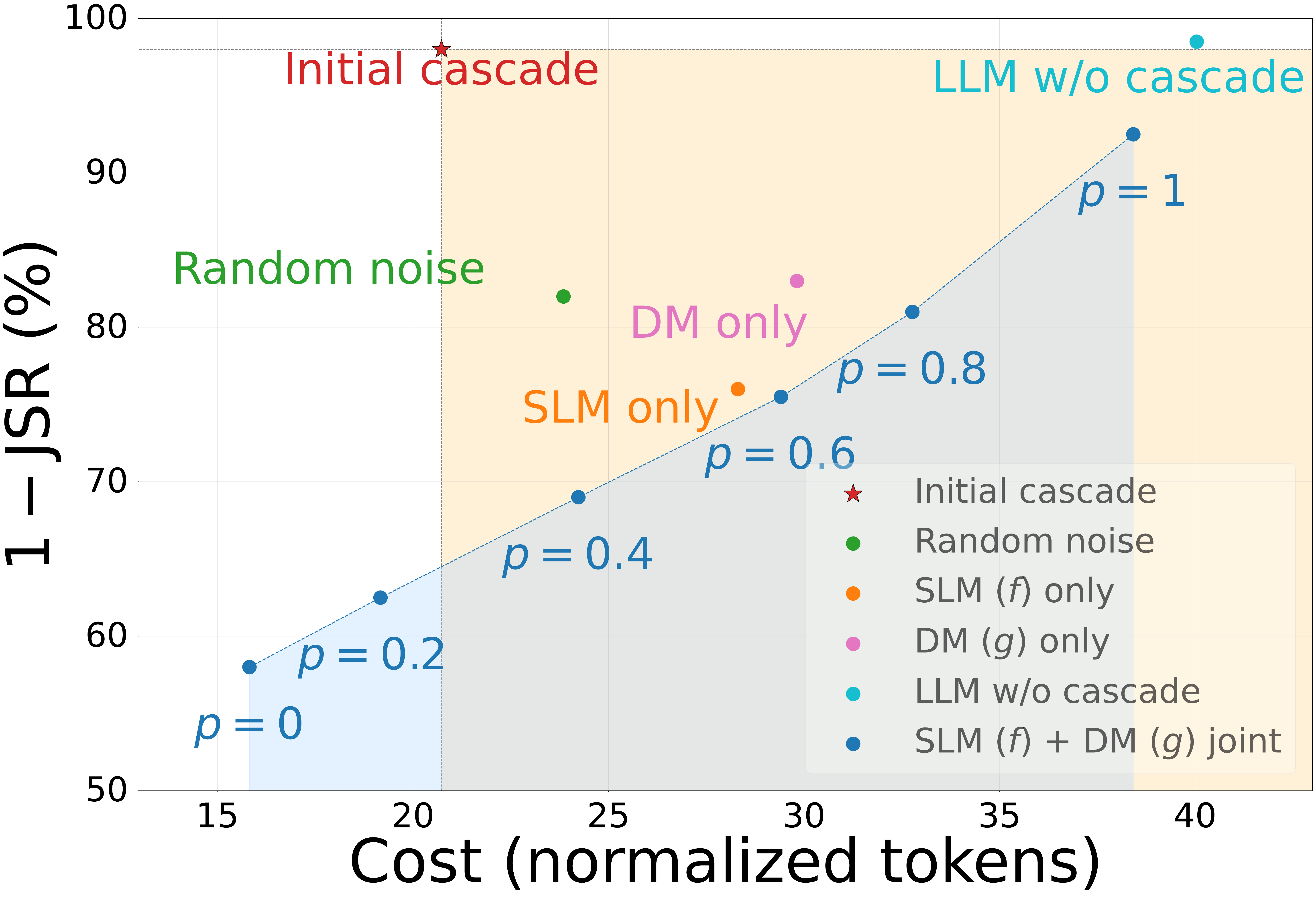}
  }
  \subfigure[AGnews (time)]{
    \includegraphics[width=0.52\columnwidth]{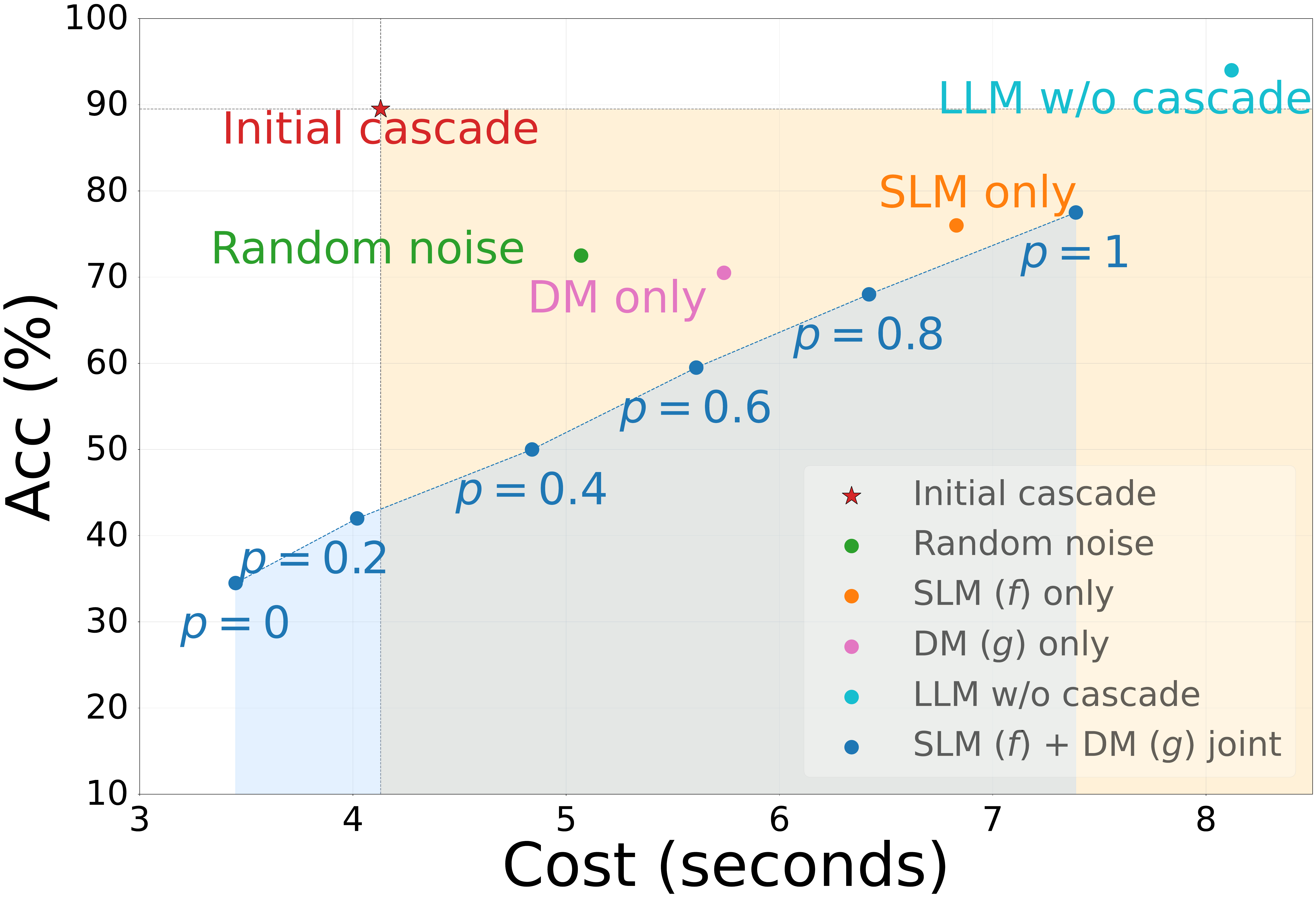}
  }
      \subfigure[SQuAD2.0 (time)]{
    \includegraphics[width=0.55\columnwidth]{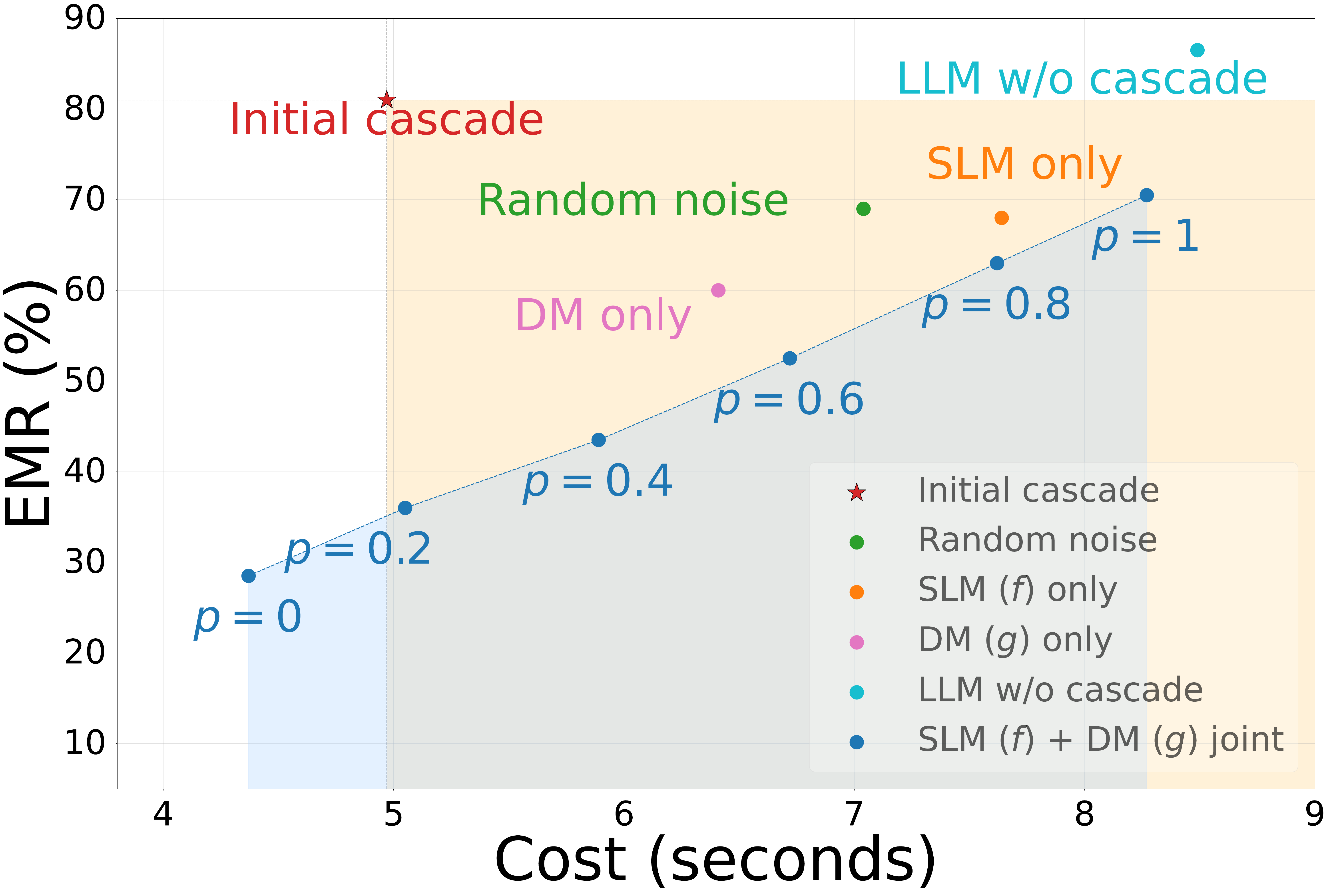}
  }
  \subfigure[WildJailbreak (time)]{
    \includegraphics[width=0.52\columnwidth]{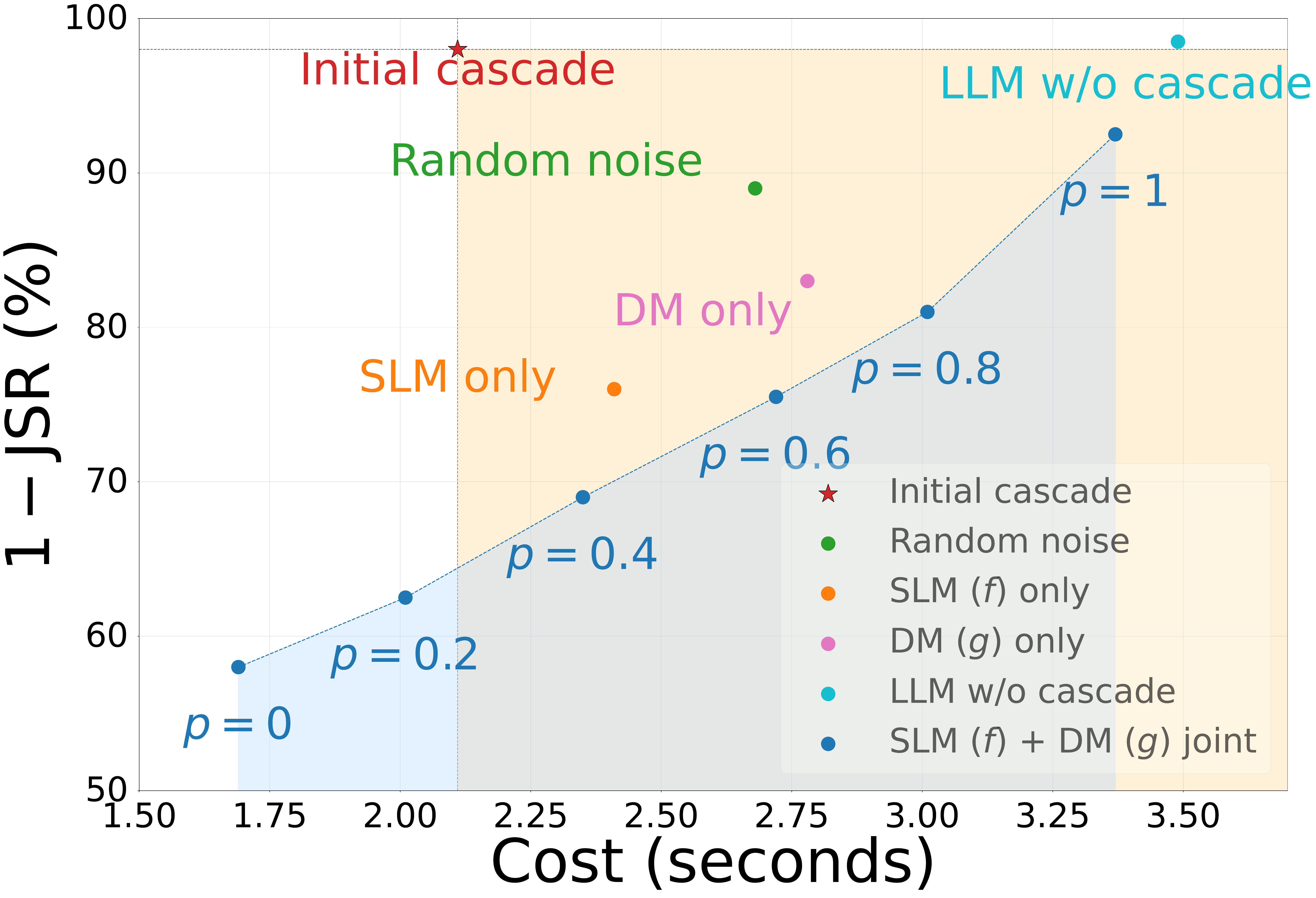}
  }
  \vspace{-6pt}
  \caption{\textbf{Comparison of attack methods across the AGnews, SQuAD2.0, and WildJailbreak datasets on a \textit{three-layer} LLM cascade target the\textit{ second stage} (configureation (24) in \tableautorefname~\ref{tab:cascade-configs-2}) .}
The cost–accuracy trade-off achieved by our pass-rate control method (introduced in \sectionautorefname~\ref{subsubsec:pass_rate}) is shown as the {\color[HTML]{1F77B4}blue} dashed line. Points toward the lower-right represent more effective attacks.}
  \label{fig:3layer_results_block2}
  \vspace{1pt}
\end{figure*}

\section{Experiment}

\subsection{Setup}
\label{sec:exp:setup}

\myparagraph{Datasets}
We conduct extensive evaluation by investigating nine widely used benchmark datasets across five diverse tasks: text classification (\texttt{Headlines}~\cite{sinha2021impact}, \texttt{Overruling}~\cite{zheng2021does}, \texttt{AG News}~\cite{zhang2015character},  \texttt{IMDB}~\cite{maas-EtAl:2011:ACL-HLT2011}), question answering (\texttt{WebQuestions}~\cite{berant-etal-2013-semantic}, \texttt{SQuAD-2.0}~\cite{SQuAD2,SQuAD1}), constrained generation (\texttt{CommonGen}~\cite{CommonGen-lin-EtAl:2020:FindingsEMNLP2020}), mathematical reasoning (\texttt{SVAMP}~\cite{svamp-kadlcik-EtAl:2023:EMNLP}), and safety evaluation/jailbreaking (\texttt{WildJailbreak}~\cite{wildteaming2024}). See Appendix~\ref{appendix:dataset}  for details.

\myparagraph{Target Cascade Architectures}
Our study focuses on cascade systems following the FrugalGPT paradigm~\cite{chen2023llmcascade}, one of the earliest and most widely adopted frameworks for LLM cascades. This design has become a canonical backbone for cascade-based inference and is broadly consistent with subsequent works~\cite{yue2024large,ramirez2024optimising,zhu2023optimal,zhang2023ecoassistant,gupta2024language,chen2025improving}. We consider both a standard two-stage cascade ($f_1\!\rightarrow\!f_2$) and an extended three-stage cascade ($f_1 \!\rightarrow\! f_2 \!\rightarrow\! f_3$), with decision modules between stages. In practice, such 2–3 stage designs are most commonly used, as deeper cascades typically offer diminishing returns while introducing additional complexity. Across all settings, we employ lightweight language models as the decision modules, a standard design choice that enables efficient routing with minimal overhead~\cite{erbacher2024navigating,chen2024are,zhang2024llm,wang2024mixture,wang2025mixllm,chen2025improving,zhou2025adaptive,nie2024online}. This design is also widely employed in real-world systems such as CascadeFlow and Microsoft Azure AI. We evaluate diverse combinations of LLMs and decision modules spanning a wide range of model sizes, architectures, capabilities, and cost–performance trade-offs. Full configurations are provided in \tableautorefname~\ref{tab:cascade-configs-1}-\ref{tab:cascade-configs-2} in Appendix~\ref{appendix:architectures}.

\myparagraph{Evaluation metrics}
We evaluate attack effectiveness along two complementary dimensions: \textbf{\emph{task-level objectives}} and \textbf{\emph{inference-cost objectives}}. For task performance, we adopt standard success-based metrics tailored to each task type, including \textit{classification accuracy}, \textit{exact-match style metrics} for question answering, \textit{coverage-based metrics} for constrained generation, \textit{solve rate} for mathematical reasoning, and \textit{jailbreak success/defense rates} for jailbreaking. For efficiency, we evaluate three complementary dimensions of inference cost: \textit{normalized token cost} (a proxy for computational and monetary expenditure, reflecting real-world usage), \textit{average execution time} (latency), and \textit{pass rate} (capturing routing behavior across cascade stages). Detailed definitions and implementation specifics are provided in Appendix~\ref{appendix:metrics}.

\myparagraph{Baselines and reference methods}
To our knowledge, this work is the first to study the vulnerability of LLM cascade systems. As this setting has not been explored in prior work, there are no directly comparable methods tailored to it.
Our approach follows the general paradigm of adversarial prompt optimization but is not tied to a specific instantiation. Instead, the framework is modular and allows methods from the adversarial LLM literature to be incorporated as plug-in components for the token update step. For concreteness and consistency with prior work, we adopt the discrete gradient-based update from the GCG method~\cite{zou2023universal} as the default optimization procedure (see Appendix~\ref{appendix:gray_box} for results with alternative optimization methods).

To contextualize our results, we consider several reference settings that reflect key aspects of the cascade architecture. We first include a \texttt{Random-Noise} setting, where the adversarial suffix is randomly sampled without optimization, representing an uninformed adversary and providing a lower-bound reference. 
We further consider two \emph{single-target} attack settings (Section~\ref{subsec:single-target}) that isolate different components of the cascade. The \texttt{SLM-only} attack optimizes the suffix solely against the prediction language model using the task performance objective, capturing the effect of perturbing the prediction model without influencing the cascade decision process. The \texttt{DM-only} attack targets only the decision mechanism: it encourages escalation when the intermediate prediction is correct (inflating cost) and early termination when it is incorrect (preventing correction by downstream models).

\subsection{White-box Setting}

We begin with the white-box setting, providing a detailed comparison of different attack configurations and examining the controlled accuracy--cost trade-off enabled by our pass-rate mechanism (Section~\ref{subsubsec:pass_rate}). The results are shown in \figureautorefname~\ref{fig:2layer_results_1}--\ref{fig:3layer_results_block2} (with additional results in \figureautorefname~\ref{fig:2layer_results_2}--\ref{fig:3layer_results_block1} and \tableautorefname~\ref{tab:gpt2_bert_phi3_results}-\ref{tab:qwen_mistral_phi35_block2_results}).

A key observation is that even a naive \texttt{Random-Noise} suffix leads to noticeable  degradation in both task performance and inference cost across all datasets and configurations. \textbf{This indicates that the cascade architecture is inherently vulnerable, even to an uninformative adversary.} Moreover, this setting provides only a \emph{lower bound} on vulnerability, i.e., more sophisticated attacks can cause significantly greater damage. This highlights the urgent need to account for cascade-level fragility in future system design.

\begin{table}[htbp]
\resizebox{\columnwidth}{!}{
\begin{tabular}{cccccccc}
\toprule
\multirow{2}{*}{Setting} 
& \multicolumn{2}{c}{Cascade ($\Phi$)} 
& \multicolumn{2}{c}{SLM ($f^1$)} 
& \multicolumn{2}{c}{LLM ($f^2$)} 
& DM ($g^1$)\\
& Acc & Cost 
& Acc & Cost 
& Acc & Cost 
& Acc  \\ 
\midrule
Initial cascade (1)
& 78.0\% & 10.37 
& 48.0\% & 0.51
& 76.0\% & 25.19
& 86.5\% 
\\

Attack $p=0$ (1)$\rightarrow$ (1) 
& 21.0\% & 9.30
& 16.0\% & 0.63 
& 63.0\% & 25.47 
& 18.0\% 
\\

\cc $\Delta$ & \cc -73.1\%		
& \cc -10.3\%	
& \cc -66.7\%
& \cc +23.5\%	
& \cc -17.1\%	
& \cc +1.1\%			
& \cc -79.2\%
\\
\hdashline

Initial cascade (8)      & 85.0\%  & 41.2  & 71.0\%  & 3.4   & 91.0\%   & 104.1 & 91.5\%     \\
Attack $p=0$ (8) $\rightarrow$(8) & 52.5\%  & 18.4 & 50.0\% & 3.4    & 84.0\% & 104.5    & 19.0\%  \\
\cc $\Delta$  & \cc -41.0\% & \cc -55.3\% & \cc -29.6\% & \cc - & \cc  -7.7\% & \cc +0.4\% & \cc -79.2\%  \\
\hdashline
\\[-0.85em]

Initial casacde(15)        & 83.5\%  & 37.5  & 71.0\%  & 3.4  & 85.0\%  & 51.8  & 87.0\%  \\
Attack $p=0$  (8)$\rightarrow$(15)  & 51.0\%  & 29.1  & 50.0\% & 3.4   & 74.0\%  & 53.2 & 59.0\%  \\
\cc $\Delta$  & \cc -38.9\% & \cc -22.4\% & \cc -29.6\% & \cc - & \cc -12.9\% & \cc +2.7\% & \cc -32.2\% \\
\midrule

Initial casacde(16)        & 82.5\%  & 18.9  & 72.5\%  & 4.3   & 84.0\%  & 49.8  & 88.0\%  \\
Attack $p=0$  (1)$\rightarrow$(16) & 48.0\%  & 11.1  & 42.5\%  & 4.7   & 69.0\%  & 55.5  & 49.0\%  \\
\cc $\Delta$  & \cc -41.8\% & \cc -41.3\% & \cc -41.4\% & \cc +9.3\% & \cc -17.9\% & \cc +11.4\% & \cc -44.3\% \\
\midrule

Initial casacde(16)        & 82.5\%  & 18.9  & 72.5\%  & 4.3   & 84.0\%  & 49.8  & 88.0\%  \\
Attack $p=0$  (8)$\rightarrow$(16)  & 53.0\%  & 14.4  & 49.5\%  & 4.5   & 74.0\%  & 53.2 & 61.0\%  \\
\cc $\Delta$  & \cc -35.8\% & \cc -23.8\% & \cc -31.7\% & \cc +4.7\% & \cc -11.9\% & \cc +6.8\% & \cc -30.7\% \\
\midrule

Initial casacde(17)        & 82.5\%  & 19.4  & 67.0\%  & 0.9   & 84.0\%  & 49.8  & 82.5\%  \\
Attack $p=0$  (1)$\rightarrow$(17)  & 43.5\%  & 15.6  & 29.0\%  & 1.0   & 69.0\%  & 55.5 & 52.0\%  \\
\cc $\Delta$   & \cc -47.3\% & \cc -19.6\% & \cc -56.7\% & \cc +11.1\% & \cc -17.9\% & \cc +11.4\% & \cc -37.0\% \\
\midrule

Initial casacde(17)        & 82.5\%  & 19.4  & 67.0\%  & 0.9   & 84.0\%  & 49.8  & 82.5\%  \\
Attack $p=0$  (8)$\rightarrow$(17)  & 49.5\%  &16.7   & 36.0\%  & 1.1   & 74.0\%  & 53.2 & 67.0\%  \\
\cc $\Delta$   & \cc -40.0\% & \cc -13.9\% & \cc -46.3\% & \cc +22.2\% & \cc -11.9\% & \cc +6.8\% & \cc -18.8\% \\
\midrule

Initial casacde(18)        & 82.0\%  & 33.4  & 39.0\%  & 0.7   & 84.0\%  & 49.8  & 89.5\%  \\
Attack $p=0$  (1)$\rightarrow$(18) & 37.0\%  & 14.4  & 35.5\%  & 0.8   & 69.0\%  & 55.5 & 36.0\%  \\
\cc $\Delta$   & \cc -54.9\% & \cc -56.9\% & \cc -9.0\%  & \cc +14.3\% & \cc -17.9\% & \cc +11.4\% & \cc -59.8\% \\
\midrule
Initial casacde(19)        & 84.5\%  & 59.1  & 54.5\%  & 0.6   & 85.0\%  & 102.0 & 85.0\%  \\
Attack $p=0$  (1)$\rightarrow$(19) & 45.5\%  & 40.8  & 34.5\%  & 0.7   & 51.0\%  & 113.4 & 59.5\%  \\
\cc $\Delta$  & \cc -46.2\% & \cc -31.0\% & \cc -36.7\% & \cc +16.7\% & \cc -40.0\% & \cc +11.2\% & \cc -30.0\% \\
\midrule
Initial casacde(20)        & 86.0\%  & 62.4  & 54.5\%  & 0.6   & 86.5\%  & 109.3 & 91.0\%  \\
Attack $p=0$  (1)$\rightarrow$(20) & 52.5\%  & 41.8  & 34.5\%  & 0.7   & 60.0\%  & 119.4 & 54.0\%  \\
\cc $\Delta$  & \cc -39.0\% & \cc -33.0\% & \cc -36.7\% & \cc +16.7\% & \cc -30.6\% & \cc +9.2\% & \cc -40.7\% \\
\bottomrule
\end{tabular}}
\vspace{1pt}
\caption{\textbf{Transferability of attack performance across two-layer LLM cascades where at least two components are altered from the baseline (\textit{open-source models}).} Cost is measured as the \textit{average normalized token cost.}}

\label{tab:transfer_2layer_opensource}
\end{table}

Second, the \texttt{SLM-only} attack also achieves strong degradation across both objectives, indicating that attacking the weakest component in the chain can significantly disrupt the entire cascade. In contrast, the \texttt{DM-only} attack is generally more effective at degrading accuracy but less successful at inflating cost. This is expected: by design, our DM baseline \emph{flips} the intended decision behavior, i.e., escalating when the first-stage prediction is correct and terminating when it is incorrect, rather than forcing all queries to be passed forward. The latter behavior is naturally captured in our joint attack when the pass-rate is set to $p=1$.

\textbf{In comparison, our default joint \texttt{SLM}+\texttt{DM} attack enables explicit control over both objectives and consistently outperforms other attack configurations.} Using the two-layer cascade as an example, at the extremes, setting \(p = 0\) drives accuracy to very low levels, resulting in performance drops of up to 84.6\%. Conversely, setting \(p = 1\) substantially inflates inference cost, increasing normalized token cost by 148.9\%  and execution time by up to 108.9\%.

Beyond these extremes, our framework is flexible and readily adapts to task-specific objectives. Notably, on jailbreak evaluation benchmarks, our attack achieves \textbf{substantially higher success rates (79\% increase in the jailbreak success rate)  than attacking a standalone strong LLM, indicating that cascade structures can amplify vulnerability rather than mitigate it}.
Moreover, the attack remains effective across a wide range of cascade configurations, including varying numbers of stages and different model compositions, and can target arbitrary subsets of components, demonstrating its generality (e.g., \figureautorefname~\ref{fig:3layer_results_block2} and Appendix~\ref{appendix:white-box}). 
This generality also extends to targeting intermediate stages in multi-stage cascades. For instance, in the 3-layer setting, attacking the second stage still leads to a substantial impact on overall cascade behavior, accompanied by a clear increase in computational overhead. Compared with targeting the first stage, the effect on task performance is more moderate, while the increase in cost is more pronounced. This demonstrates that our method \textbf{can effectively target non-initial stages while still achieving the intended impact.}

Collectively, these results show that LLM cascade systems are highly vulnerable to adversarial manipulation: \textbf{a single carefully crafted suffix can simultaneously degrade predictive accuracy, significantly increase inference cost, and achieve high-success evasion or jailbreak behavior, effectively undermining the intended efficiency–performance trade-off of the cascade design.}

\begin{table}[htbp]
\resizebox{\columnwidth}{!}{
\begin{tabular}{cccccccc}
\toprule
\multirow{2}{*}{Setting} 
& \multicolumn{2}{c}{Cascade ($\Phi$)} 
& \multicolumn{2}{c}{SLM ($f^1$)} 
& \multicolumn{2}{c}{LLM ($f^2$)} 
& DM ($g^1$)\\
& Acc & Cost 
& Acc & Cost 
& Acc & Cost 
& Acc  \\ 
\midrule

Initial casacde(21)        & 93.5\%  & \$ 0.52  & 82.5\%  & \$ 0.27    & 94.5\%  & \$ 1.37  & 98.5\%  \\
Attack $p=0$  (1)$\rightarrow$(21) & 76.0\%  & \$ 0.44 & 75.0\%  & \$ 0.29   & 87.0\%  & \$ 1.42   & 73.0\%  \\
\cc $\Delta$  &\cc -18.7\% &\cc -15.4\% &\cc -9.1\% &\cc +7.4\% &\cc -7.9\% &\cc +3.6\% &\cc -25.9\% \\
\midrule

Initial casacde(21)        & 93.5\%  & \$ 0.52  & 82.5\%  & \$ 0.27    & 94.5\%  & \$ 1.37  & 98.5\%  \\
Attack $p=0$  (8)$\rightarrow$(21) & 74.0\%  & \$ 0.42 & 71.0\%  & \$ 0.28   & 89.0\%  & \$ 1.41   & 75.0\%  \\
\cc $\Delta$  &\cc -20.9\% &\cc -19.2\% &\cc -13.9\% &\cc +3.7\% &\cc -5.8\% &\cc +2.9\% &\cc -23.9\% \\
\midrule
Initial casacde(22)        & 94.0\%  & \$ 0.38 & 81.5\%  & \$  
0.087   & 94.0\%  & \$ 0.93 & 96.0\%  \\
Attack $p=0$ (1)$\rightarrow$(22) & 74.5\%  & \$ 0.21  & 71.0\%  & \$ 0.092  & 88.0\%  & \$ 0.95 & 74.0\%  \\
\cc $\Delta$            & \cc -20.7\% & \cc -44.7\% & \cc -12.9\% & \cc +5.7\% & \cc -6.4\% & \cc +2.2\% & \cc -22.9\% \\
\midrule
Initial casacde(23)        & 91.0\%  & \$ 0.19 & 80.0\%  & \$ 0.092   & 92.0\%  & \$ 0.57 & 91.0\%  \\
Attack $p=0$ (1)$\rightarrow$(23) & 75.5\%  & \$ 0.14  & 72.0\%  & \$ 0.097  & 89.0\%  &\$ 0.60 & 71.5\%  \\
\cc $\Delta$            & \cc -17.0\% & \cc -26.3\% & \cc -10.0\% & \cc +5.4\% & \cc -3.3\% & \cc +5.3\% & \cc -21.4\% \\

\bottomrule
\vspace{1pt}
\end{tabular}}
\caption{\textbf{Transferability of attack performance across different two-layer LLM cascades (with \textit{API}-based models).} Cost is measured as the \textit{average monetary cost} (in \$).} 
\label{tab:transfer_2layer_API}
\end{table}

\begin{figure*}[t]
  \centering
  \includegraphics[width=0.9\textwidth]{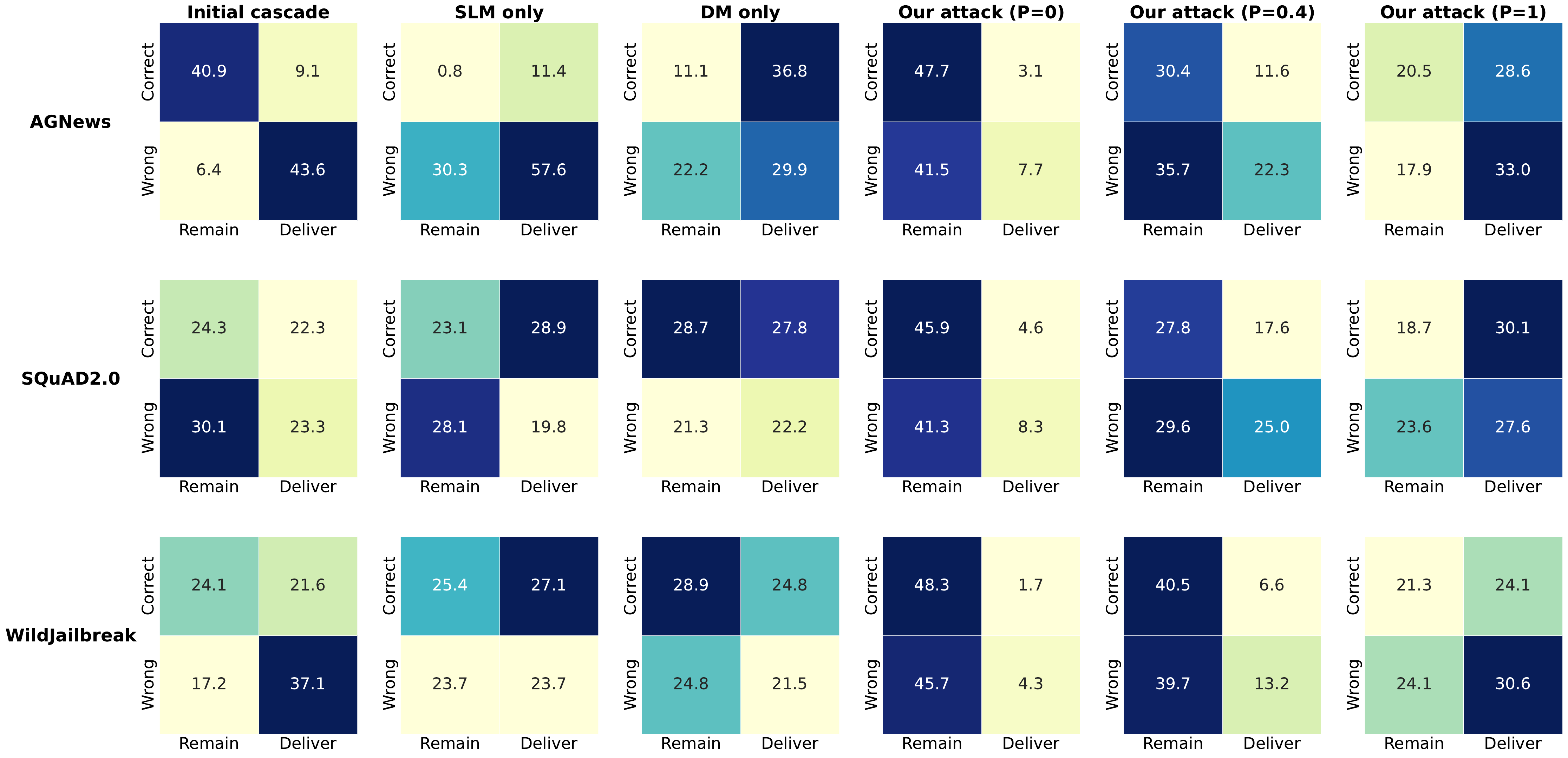}
  \caption{\textbf{
  Confusion matrices of the decision module (\texttt{DM$_2$}) when the attack is targeting the second stage of a three-layer LLM cascade with configuration~(24) in \tableautorefname~\ref{tab:cascade-configs-2}.
}}
  \label{fig:confusion_matrix_3layers_block2}
\end{figure*}

\begin{figure*}[t]
  \centering
  \subfigure[Headline]{
    \includegraphics[width=0.47\columnwidth]{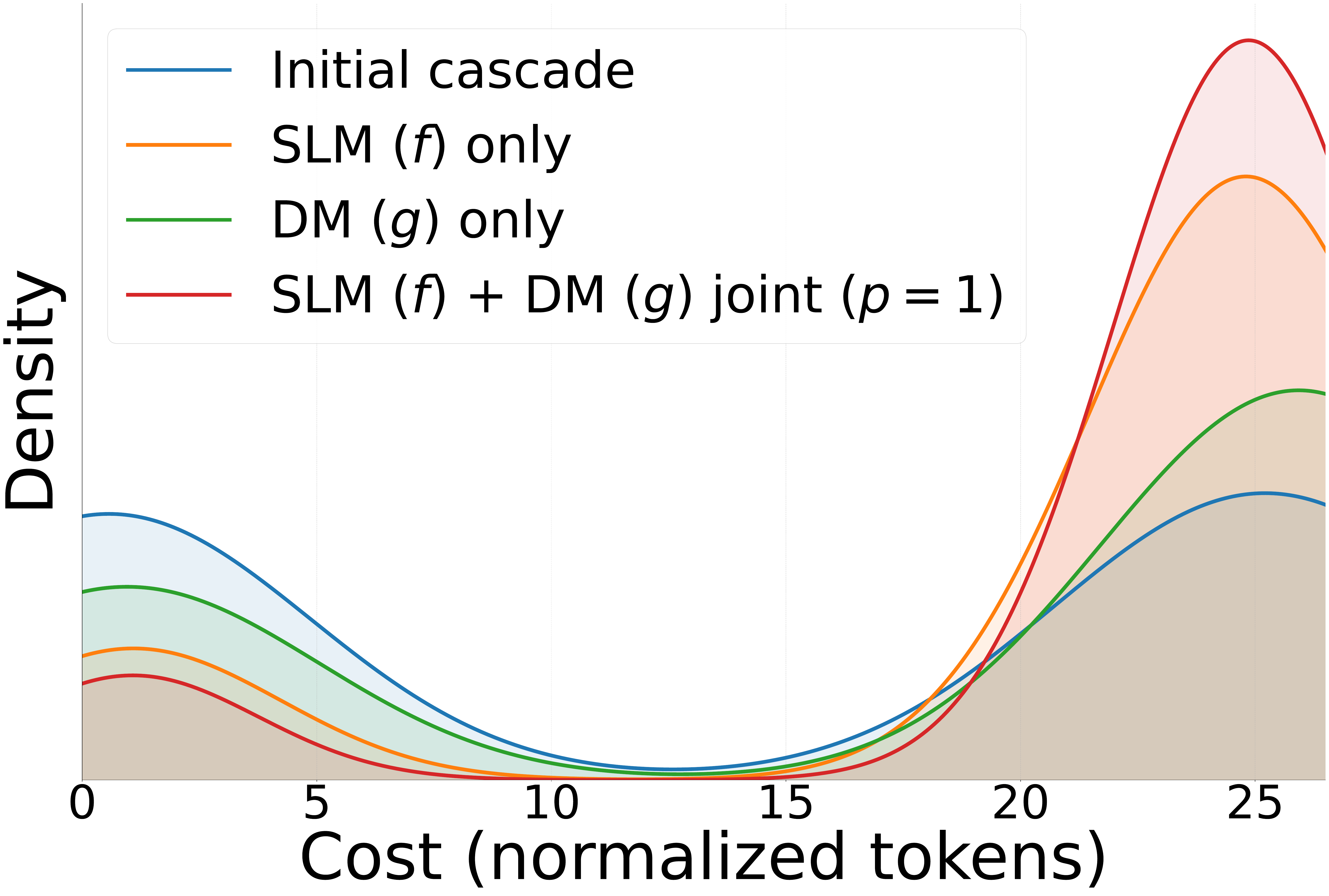}
    \label{fig:headline}
  }
  \subfigure[IMDB]{
    \includegraphics[width=0.47\columnwidth]{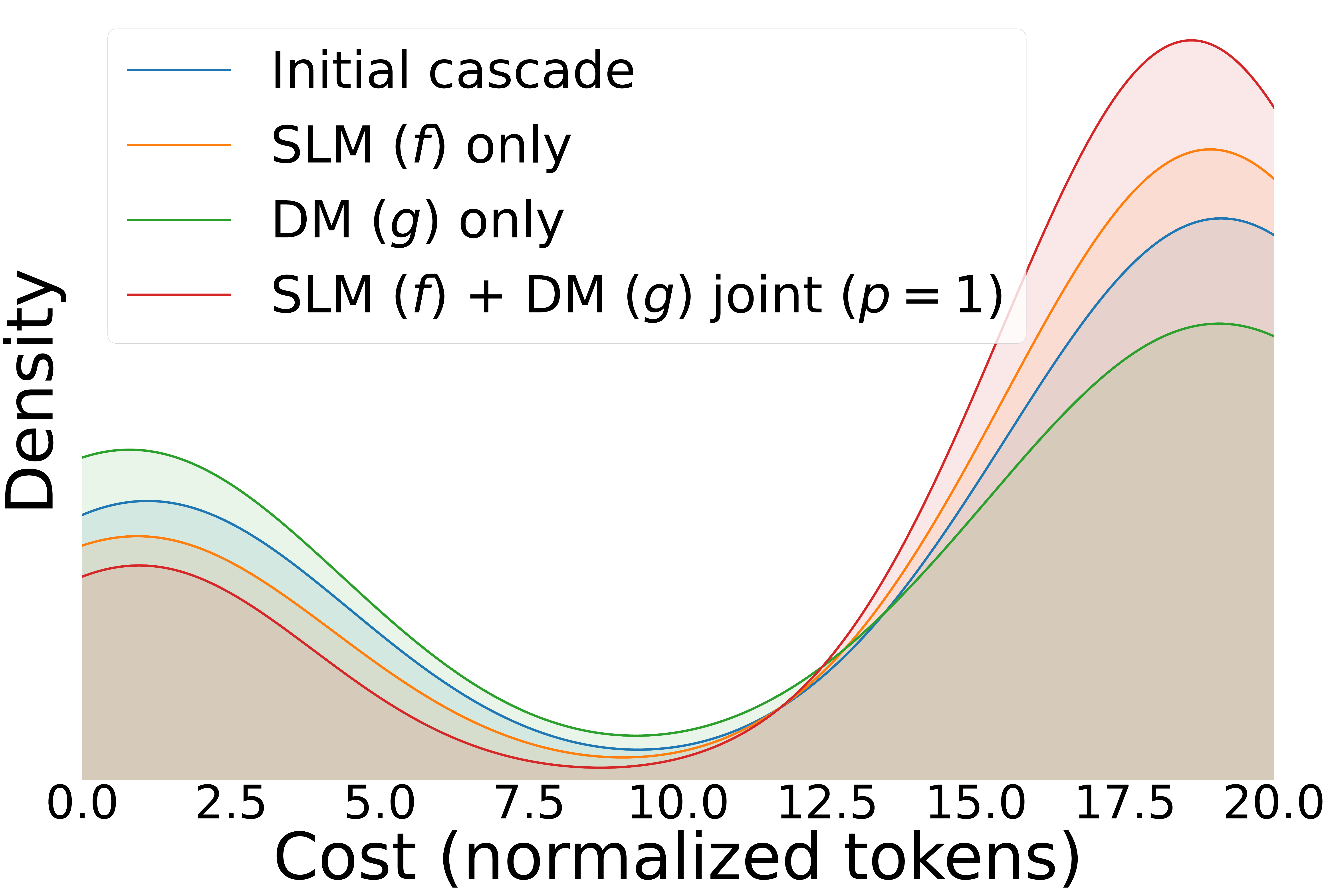}
    \label{fig:overruling}
  }
  \subfigure[AGnews]{
    \includegraphics[width=0.47\columnwidth]{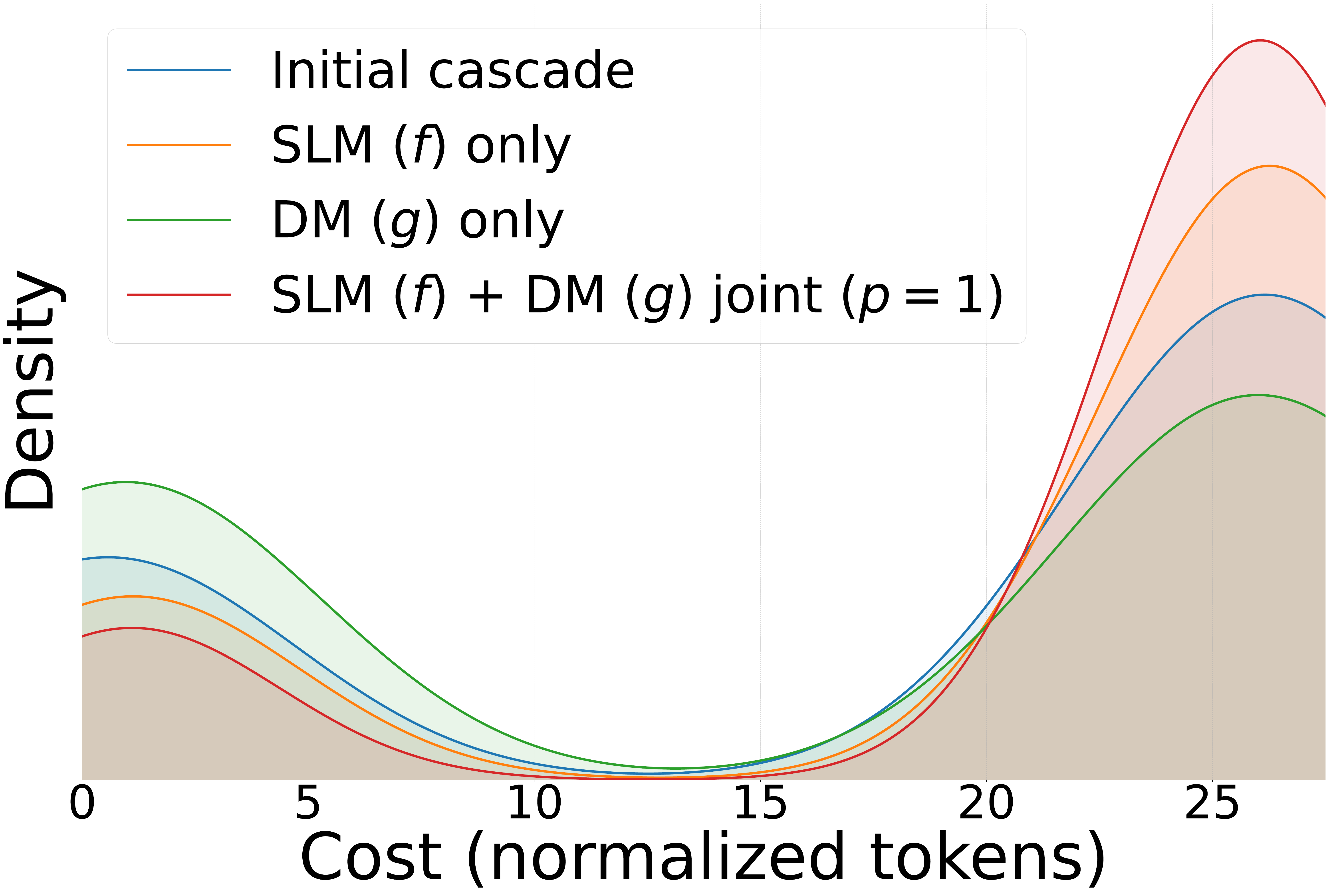}
    \label{fig:AGnews}
  }
  \subfigure[Web-question]{
    \includegraphics[width=0.47\columnwidth]{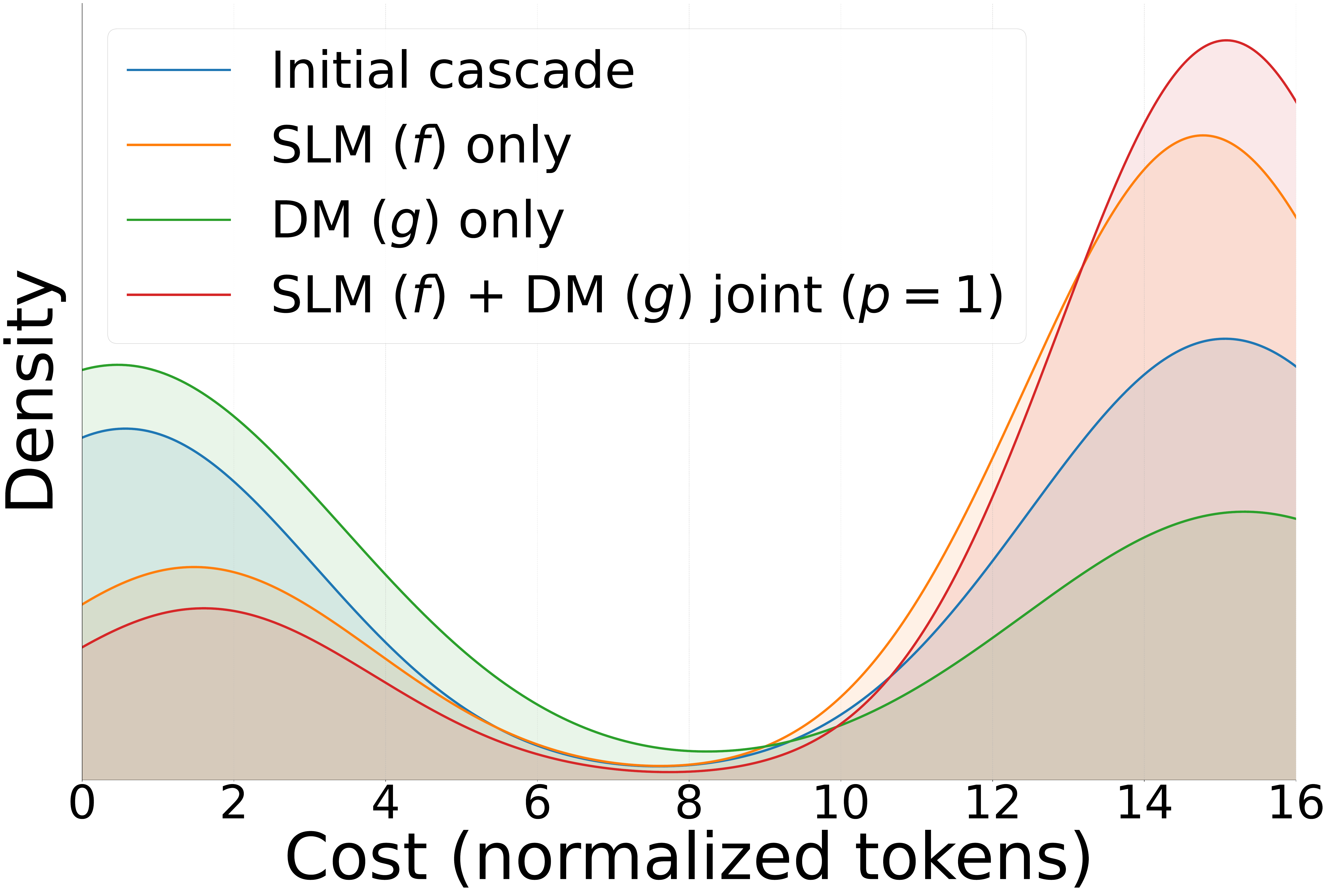}
    \label{fig:Web_Q}
  }
  
  \vspace{-1pt}
  \caption{\textbf{Distributions of cost quantified by normalized tokens on a two-layer LLM cascade across the Headline, IMDB, AGnews, WebQuestions datasets, using cascade configurations~(1) (see Table~\ref{tab:cascade-configs-1})}.}
  \label{fig:cost_distribution}
  \vspace{1pt}
\end{figure*}
\subsection{Black-box Setting}
We present the results of black-box attack in Table~\ref{tab:transfer_2layer_opensource}-\ref{tab:transfer_2layer_API} (and \tableautorefname~\ref{tab:transfer_dm}-\ref{tab:transfer_slm} in Appendix)), where we optimize an adversarial suffix on one cascade configuration and then apply it directly, (i.e., without any re-optimization nor adaptation) to other cascade architectures. The underlying model configurations for all cascades are summarized in Table~\ref{tab:cascade-configs-1}-\ref{tab:cascade-configs-2}. 

Across all evaluations, we observe that adversarial suffixes remain highly transferable: \textbf{even when applied to cascades built from different models than those used during optimization, they consistently degrade both task performance and inference cost.}  While the transferred attacks are naturally weaker than the fully white-box setting—where the attacker knows the exact models deployed——their impact remains substantial, revealing non-trivial vulnerability even under black-box conditions. This reflects realistic deployment scenarios in which adversaries may infer or guess the general cascade structure but lack precise knowledge of the underlying model checkpoints. Our findings demonstrate that in such settings, attackers can still significantly compromise the cascade, highlighting yet another layer of systemic fragility introduced by multi-stage architectures. This underscores the need for rigorous robustness assessments and stronger safeguards when deploying LLM cascades in practical applications.

Our approach also remains certain effect in attacking cascade API scenarios (\tableautorefname~\ref{tab:transfer_2layer_API}), where intermediate model components are completely inaccessible for manipulation nor inspection. Despite this limited access, the method is able to substantially degrade the overall system performance by increasing the computational cost while simultaneously reducing accuracy. This disruption results in a significantly deteriorated trade-off between accuracy and inference cost, forcing the cascade system to either incur higher latency or suffer from weakened predictive quality. Importantly, this vulnerability cannot be mitigated simply by strengthening individual LLM components—a common practice in current deployments aimed at improving benchmark performance. In cascade architectures, weaker or less robust components can still be exposed and exploited through routing and interaction effects, allowing adversarial inputs to bypass stronger models or amplify inefficiencies. \textbf{Consequently, our attack reveals a fundamental weakness in cascade design: even without white-box access, attackers can leverage heterogeneous components to impair both efficiency and effectiveness.}

\subsection{Qualitative Results}

\myparagraph{Confusion Matrices of the Decision Modules} 
We visualize the confusion matrices of the decision module \texttt{DM} under different attack settings across all datasets in \figureautorefname~\ref{fig:confusion_matrix_3layers_block2}. In a well-functioning cascade, the decision module is expected to follow a simple and intuitive rule: correctly predicted queries should trigger early termination, whereas incorrectly predicted ones should be passed to the next stage for correction. This corresponds to strong diagonal dominance in the confusion matrix.

As shown, the \texttt{DM} in the initial (non-attacked) cascade behaves reasonably and aligns well with the intended routing logic. However, under adversarial manipulation—particularly the \texttt{DM Only} attack and our full \texttt{SLM+DM (p=0)} joint attack—the decision boundaries become severely distorted. These attacks effectively fool the decision module into reversing or randomizing its decisions, causing many incorrect queries to be terminated prematurely or forcing unnecessary escalation for correct ones. Interestingly, while the \texttt{SLM Only} attack does not directly target the decision module, it substantially increases the portion of mispredicted queries produced by the first-stage model. This indirectly shifts the decision module's input distribution toward harder or incorrect samples, leading to widespread error propagation and degraded system accuracy—even without explicitly altering \texttt{DM} decision boundaries.
In summary, \textbf{these results highlight that both direct and indirect adversarial influences can destabilize decision behavior in LLM cascades}. The decision module, despite being lightweight, plays a pivotal role in the system pipeline and thus represents a critical—and highly vulnerable—attack surface.

\myparagraph{Execution Time Distributions}
We further analyze the distribution of execution times for adversarial samples generated under different attack configurations across all benchmarks, which complements the reported average cost statistics by providing a more detailed view of runtime behavior. As shown in \figureautorefname~\ref{fig:cost_distribution}, the distribution is bimodal, reflecting the two possible termination points of the cascade: early exit at the first stage or escalation to the second.
Both the \texttt{``SLM only''} and \texttt{``SLM + DM (p=1)''} attacks—which attempt to defer all samples to the next stage—shift most mass toward the right mode, corresponding to longer runtime. The default \texttt{``joint attack''} configuration produces an even more pronounced shift, confirming its stronger cost-inflation effect.

In contrast, the \texttt{``DM only''} attack (recall its flipped optimization target: escalate when the prediction is correct and terminate when incorrect) has a much weaker impact on runtime. This is expected: a large fraction of samples are mispredicted by the weak \texttt{SLM} at stage 1, so attacking only the decision module tends to force termination rather than escalation. This behavior is beneficial for accuracy degradation (as incorrect outputs are finalized early) but is counterproductive for increasing inference cost, since most samples are never escalated to the expensive second-stage model.

\begin{table}[htbp]
\centering
\resizebox{0.49\textwidth}{!}{
\begin{tabular}{lcccc}
\toprule
\multirow{2}{*}{Setting} & \multicolumn{2}{c}{Original Attack Performance (Headline)} & \multicolumn{2}{c}{After Random Smooth (Headline)} \\
\cmidrule(lr){2-3} \cmidrule(lr){4-5}
                        & Acc & Cost    & Acc & Cost \\
\midrule
Initial cascade         & 78.00\% & 13.43   & -    & - \\
Attack $p=0$            & 23.00\% & 7.26    & 29.00\% & 7.83 \\
Attack $p=0.2$          & 41.00\% & 14.62   & 44.00\% & 14.83 \\
Attack $p=0.4$          & 43.50\% & 15.99   & 45.50\% & 16.54 \\
Attack $p=0.6$          & 50.00\% & 18.34   & 50.50\% & 18.67 \\
Attack $p=0.8$          & 56.50\% & 20.09   & 57.00\% & 20.79 \\
Attack $p=1$            & 63.00\% & 22.45   & 64.50\% & 22.93 \\
\midrule
\multirow{2}{*}{Setting} & \multicolumn{2}{c}{Original Attack Performance (Overruling)} & \multicolumn{2}{c}{After Random Smooth (Overruling)} \\
\cmidrule(lr){2-3} \cmidrule(lr){4-5}
                        & Acc & Cost    & Acc & Cost \\
\midrule
Initial cascade         & 71.00\% & 14.11   & -    & - \\
Attack $p=0$            & 21.00\% & 4.78    & 24.00\% & 5.01 \\
Attack $p=0.2$          & 29.50\% & 7.94    & 31.50\% & 9.21 \\
Attack $p=0.4$          & 45.50\% & 12.57   & 40.50\% & 11.58 \\
Attack $p=0.6$          & 57.00\% & 14.66   & 45.50\% & 12.14 \\
Attack $p=0.8$          & 61.50\% & 16.94   & 58.00\% & 15.09 \\
Attack $p=1$            & 67.50\% & 19.46   & 65.50\% & 18.51 \\
\midrule
\multirow{2}{*}{Setting} & \multicolumn{2}{c}{Original Attack Performance (AGnews)} & \multicolumn{2}{c}{After Random Smooth (AGnews)} \\
\cmidrule(lr){2-3} \cmidrule(lr){4-5}
                        & Acc & Cost    & Acc & Cost \\
\midrule
Initial cascade         & 78.00\% & 17.38   & -    & - \\
Attack $p=0$            & 19.50\% & 5.78    & 22.50\% & 7.36 \\
Attack $p=0.2$          & 37.50\% & 9.64    & 39.50\% & 10.28 \\
Attack $p=0.4$          & 44.50\% & 13.85   & 46.00\% & 14.01 \\
Attack $p=0.6$          & 53.50\% & 17.24   & 52.00\% & 16.69 \\
Attack $p=0.8$          & 61.50\% & 20.97   & 61.00\% & 20.05 \\
Attack $p=1$            & 70.50\% & 24.15   & 67.00\% & 22.65 \\
\bottomrule
\end{tabular}
}
\vspace{1pt}
\caption{\textbf{Attack performance under \textit{randomized smoothing} defense}. Cost is measured as the \textit{average token cost.}}
\label{tab:under_defense}
\end{table}

\begin{table}[htbp]
\centering
\resizebox{0.49\textwidth}{!}{
\begin{tabular}{lcccc}
\toprule
\multirow{2}{*}{Setting} & \multicolumn{2}{c}{Original Attack Performance (Headline)} & \multicolumn{2}{c}{After \textbf{PPL} Filter (Headline)} \\
\cmidrule(lr){2-3} \cmidrule(lr){4-5}
                        & Acc & Cost    & Acc & Cost \\
\midrule
Initial cascade         & 78.00\% & 13.43   & -    & - \\
Attack $p=0$            & 23.00\% & 7.26    & 42.00\% & 9.04 \\
Attack $p=0.2$          & 41.00\% & 14.62   & 63.00\% & 16.49 \\
Attack $p=0.4$          & 43.00\% & 15.99   & 61.00\% & 16.49 \\
Attack $p=0.6$          & 50.00\% & 18.34   & 55.00\% & 15.31 \\
Attack $p=0.8$          & 56.50\% & 20.09   & 62.50\% & 16.75 \\
Attack $p=1$            & 63.00\% & 22.45   & 70.50\% & 18.48 \\
\midrule
\multirow{2}{*}{Setting} & \multicolumn{2}{c}{Original Attack Performance (Headline)} & \multicolumn{2}{c}{After \textbf{Regex} Filter (Headline)} \\
\cmidrule(lr){2-3} \cmidrule(lr){4-5}
                        & Acc & Cost    & Acc & Cost \\
\midrule
Initial cascade         & 78.00\% & 13.43   & -    & - \\
Attack $p=0$            & 23.00\% & 7.26    & 50.00\% & 8.98 \\
Attack $p=0.2$          & 41.00\% & 14.62   & 66.00\% & 16.13 \\
Attack $p=0.4$          & 43.00\% & 15.99   & 66.50\% & 16.16 \\
Attack $p=0.6$          & 50.00\% & 18.34   & 63.50\% & 14.98 \\
Attack $p=0.8$          & 56.50\% & 20.09   & 67.00\% & 15.86 \\
Attack $p=1$            & 63.00\% & 22.45   & 73.50\% & 17.37 \\
\midrule
\multirow{2}{*}{Setting} & \multicolumn{2}{c}{Original Attack Performance (Headline)} & \multicolumn{2}{c}{After \textbf{CPT} Filter (Headline)} \\
\cmidrule(lr){2-3} \cmidrule(lr){4-5}
                        & Acc & Cost    & Acc & Cost \\
\midrule
Initial cascade         & 78.00\% & 13.43   & -    & - \\
Attack $p=0$            & 23.00\% & 7.26    & 41.50\% & 8.82 \\
Attack $p=0.2$          & 41.00\% & 14.62   & 59.00\% & 17.68 \\
Attack $p=0.4$          & 43.00\% & 15.99   & 57.50\% & 17.75 \\
Attack $p=0.6$          & 50.00\% & 18.34   & 52.00\% & 15.89 \\
Attack $p=0.8$          & 56.50\% & 20.09   & 60.50\% & 17.52 \\
Attack $p=1$            & 63.00\% & 22.45   & 65.50\% & 20.38 \\
\bottomrule
\end{tabular}
}
\vspace{1pt}
\caption{\textbf{Attack performance of cascade configurations~(1) (see Table~\ref{tab:cascade-configs-1}) under \textit{filtering defenses} on HEADLINE.} Cost is measured as the \textit{average token cost}.}
\label{tab:headline_under_defense_3method}
\end{table}

\begin{table}[htbp]
\centering
\resizebox{\columnwidth}{!}{
\begin{tabular}{lccccccc}
\toprule
\multicolumn{8}{c}{\textbf{Perplexity (PPL) Filter}} \\
\cmidrule(lr){1-8}
\multirow{2}{*}{Parameter} & \multicolumn{6}{c}{Defense Success Rate (\%)} & \multirow{2}{*}{Over-defense Rate (\%)} \\
\cmidrule(lr){2-7}
& $p=0$ & $p=0.2$ & $p=0.4$ & $p=0.6$ & $p=0.8$ & $p=1.0$ & \\
\midrule
$q=0.990$, trim=4  & 14.50 & 17.00 & 14.50 & 17.00 & 14.00 & 17.00 & 6.64 \\
$q=0.955$, trim=6  & 28.50 & 29.00 & 29.00 & 29.50 & 28.00 & 28.50 & 13.85 \\
$q=0.920$, trim=8  & 34.50 & 34.50 & 34.50 & 33.50 & 34.00 & 32.00 & 17.05 \\
$q=0.885$, trim=10 & 47.50 & 37.00 & 36.00 & 38.00 & 39.00 & 38.50 & 19.97 \\
$q=0.850$, trim=12 & 62.50 & 61.00 & 59.00 & 61.50 & 62.50 & 62.00 & 24.82 \\
\midrule
\multicolumn{8}{c}{\textbf{Regex Filter}} \\
\cmidrule(lr){1-8}
\multirow{2}{*}{Parameter} & \multicolumn{6}{c}{Defense Success Rate (\%)} & \multirow{2}{*}{Over-defense Rate (\%)} \\
\cmidrule(lr){2-7}
& $p=0$ & $p=0.2$ & $p=0.4$ & $p=0.6$ & $p=0.8$ & $p=1.0$ & \\
\midrule
ratio=0.05000, trim=4  & 3.50 & 6.50 & 9.00 & 9.50 & 8.50 & 16.00 & 2.00 \\
ratio=0.03775, trim=6  & 7.50 & 8.50 & 10.00 & 9.50 & 8.50 & 11.00 & 3.50 \\
ratio=0.02550, trim=8  & 13.50 & 16.50 & 17.00 & 19.50 & 19.00 & 16.50 & 4.00 \\
ratio=0.01325, trim=10 & 15.50 & 19.00 & 22.00 & 22.00 & 21.50 & 20.50 & 5.00 \\
ratio=0.00100, trim=12 & 25.50 & 26.50 & 24.00 & 22.00 & 22.50 & 24.50 & 6.50 \\
\midrule
\multicolumn{8}{c}{\textbf{Characters Per Token (CPT) Filter}} \\
\cmidrule(lr){1-8}
\multirow{2}{*}{Parameter} & \multicolumn{6}{c}{Defense Success Rate (\%)} & \multirow{2}{*}{Over-defense Rate (\%)} \\
\cmidrule(lr){2-7}
& $p=0$ & $p=0.2$ & $p=0.4$ & $p=0.6$ & $p=0.8$ & $p=1.0$ & \\
\midrule
$\tau=2.3$, trim=4  & 32.00 & 31.50 & 33.50 & 32.00 & 30.50 & 31.50 & 14.74 \\
$\tau=2.5$, trim=6  & 46.50 & 41.00 & 44.50 & 44.50 & 43.50 & 44.00 & 21.79 \\
$\tau=2.7$, trim=8  & 49.50 & 47.00 & 49.00 & 48.50 & 47.00 & 48.00 & 26.92 \\
$\tau=2.9$, trim=10 & 64.00 & 60.50 & 64.00 & 61.50 & 63.00 & 60.00 & 33.97 \\
$\tau=3.1$, trim=12 & 79.00 & 76.50 & 79.50 & 81.00 & 80.50 & 78.50 & 45.51 \\
\bottomrule
\end{tabular}
}
\vspace{1pt}
\caption{\textbf{Defense performance with varying strength on HEADLINE when attack 2-layer cascade with configuration (1) (see Table~\ref{tab:cascade-configs-1})}.}
\label{tab:defense_strength_compact}
\end{table}

\subsection{Attack Performance under Defense} 
For completeness, we evaluate the robustness of our attack under several representative defenses. We first apply \texttt{randomized smoothing}~\cite{cohen2019certified,zhang2024random}, a widely used noise-based defense, by injecting Gaussian noise (\(\sigma=0.35\)) into the embeddings of adversarial suffixes. Despite this perturbation, the attack remains effective (Table~\ref{tab:under_defense}), indicating that general noise-based defenses are insufficient to eliminate the induced vulnerabilities.

We further consider three lightweight input-filtering defenses: (i) \texttt{PPL-based filtering}~\cite{alon2023detectingPPL}, which removes suspicious suffixes based on perplexity; (ii) \texttt{Regex-based filtering}~\cite{khachaturov2025adversarialregx}, which trims hand-crafted abnormal patterns; and (iii) \texttt{CPT-based filtering}~\cite{zychlinski2025brokentokenCPT}, which detects anomalous character-to-token ratios. Across all methods, we observe only partial recovery under attack (Table~\ref{tab:headline_under_defense_3method}), with improvements varying by defense type and attack strength. A consistent trade-off emerges when sweeping defense strength (Table~\ref{tab:defense_strength_compact}): stronger defenses improve attack mitigation but significantly increase over-defense on benign inputs. For instance, while CPT achieves the highest filtering effectiveness at strong settings, it also incurs the largest degradation on clean samples; conversely, lighter defenses such as Regex introduce less collateral damage but offer limited protection.

\textbf{Overall, no evaluated defense simultaneously achieves strong mitigation and low collateral impact.} These results indicate that existing generic or lightweight defenses are either ineffective against coordinated cascade attacks or overly aggressive on benign inputs, highlighting the need for more targeted, structure-aware defense mechanisms for LLM cascade systems.

\section{Discussion}

\myparagraph{Generalization Across Tasks and Configurations}
While we aim to cover a diverse range of cascade architectures, tasks, and data domains, our evaluation is inherently non-exhaustive. Nevertheless, the core mechanisms underlying cascade systems, i.e., the sequential model invocation and intermediate routing, are largely task-agnostic. Any setting in which an early, weaker model produces an intermediate prediction and a decision module determines whether to escalate can be formulated in a similar manner.
More broadly, any system that introduces intermediate decision boundaries and exposes routing logic is susceptible to similar forms of adversarial manipulation. We therefore expect the fragility demonstrated here to generalize to a wide class of multi-model pipelines, task-oriented LLM agents, and hybrid tool-use systems, highlighting an important direction for future research.

\myparagraph{Practical Guidance for Designing Robust Cascades}
Our findings highlight that, while standalone LLMs have become increasingly robust and resistant to naive adversarial attempts, introducing a cascade architecture can inadvertently create new points of failure. Even fully black-box adversaries—those with no knowledge of model parameters can manipulate routing decisions or destabilize early-stage predictions with simple adversarial suffixes. This does not imply that cascades should be abandoned, as they remain a powerful and cost-effective strategy in many real-world deployments. Rather, our results emphasize the need for careful design, particularly around the decision modules that govern escalation. Practitioners should avoid overly lightweight or uncalibrated decision mechanisms, incorporate monitoring for abnormal escalation patterns, and evaluate cascades under adversarial stress rather than relying solely on clean-data performance.

\myparagraph{Risk of Multi-LLM Collaborative Systems}

To the best of our knowledge, the security risks of multi-LLM collaborative systems—such as cascades and routing-based architectures—remain largely underexplored. Several recent and concurrent efforts (many not yet published) have begun to examine adversarial vulnerabilities in such systems~\cite{tang2026routeattack, shafran2025reroutingattack, lin2025lifecycle, yuan2025whoroutes}. These works all focus on attacking individual components, i.e., a single routing module, for example by using prefix-based perturbations or search-based methods to influence routing decisions.

While this work reveals important vulnerabilities in routing-based systems, its analysis focuses on isolated routing components and can be regarded as the basic case (\sectionautorefname~\ref{subsec:single-target}) of our framework. In contrast, we consider collaborative multi-LLM systems as a whole, where multiple models interact and decisions emerge from their joint behavior. More broadly, these findings point to a systemic and largely unaddressed risk in multi-LLM collaborative architectures, suggesting that the security of such systems cannot be understood by analyzing individual components in isolation and that substantially more investigation is needed.

\section{Conclusion}
This work presents the first systematic investigation of adversarial vulnerabilities in LLM cascade systems—an increasingly popular paradigm for cost-efficient deployment at scale. While prior work highlights their favorable performance–cost trade-offs, we show that the same architectural features enabling these gains also introduce new attack surfaces absent in standalone LLMs. Specifically, we formalize a previously unexplored threat model for LLM cascades 
and develop adversarial techniques that exploit the interaction between prediction and routing components, enabling coordinated manipulation across stages. Our analysis shows that these vulnerabilities stem from structural properties of typical cascade designs, making them inherently difficult to eliminate without architectural changes.
Extensive experiments across diverse datasets, model combinations, and transfer settings demonstrate that cascades can be severely compromised: a single adversarial suffix can simultaneously reduce accuracy, jailbreak cascade system, increase inference cost, and remain effective under transfer to unseen configurations. These results suggest that cascade systems can be more fragile than the strong LLMs they aim to approximate.
Overall, our findings motivate a re-examination of cascade design from a robustness perspective and emphasize the need for more reliable decision modules, adversarial stress testing during development, and more principled routing mechanisms that are less susceptible to manipulation.


\section*{Ethical Considerations}
\vspace{-0.5em}
This work examines the security and robustness of LLM cascade systems under adversarial settings. While we develop techniques capable of degrading accuracy and inflating inference cost, our objective is not to facilitate misuse but to reveal structural vulnerabilities that may arise in widely adopted cost-efficient LLM architectures. All experiments are conducted exclusively on publicly available datasets and open-source models; no proprietary, personal, or sensitive data is used. We intentionally refrain from releasing harmful prompts, jailbreak strings, or attack payloads that could directly enable malicious exploitation. Instead, we focus on methodological insights, high-level attack principles, and aggregated quantitative results.

The attacks studied in this paper are intended solely for defensive and diagnostic purposes—to help practitioners understand potential failure modes, strengthen decision modules, harden routing mechanisms, and design more reliable multi-model inference systems. All experiments are carried out in a controlled research environment isolated from production systems, ensuring that no deployed services or commercial APIs are impacted. Our work does not target or disadvantage any specific individuals, organizations, or model vendors. Rather, we aim to contribute to the community’s awareness of emerging risks and to motivate the development of safeguards, monitoring tools, and countermeasures that protect against adversarial manipulation, cost exploitation, and performance degradation in LLM cascades.

\section*{Open Science}
All artifacts necessary to evaluate and reproduce the results of this paper rely entirely on publicly accessible resources. The cascade system itself is built upon existing open-source implementations, for which we provide the relevant repository links and configuration details to ensure faithful reconstruction. All models used in our cascades are open-source and can be obtained directly from their official repositories, and we specify the exact checkpoints used whenever necessary to ensure unambiguous reproducibility. The datasets used in our evaluation (e.g., Poem-Sentiment, Overruling, Headlines) are likewise publicly available, and we provide precise references to facilitate identical reproduction. The full implementation of our attacks, cascade evaluation pipeline, and experiment scripts will be released upon publication, accompanied by instructions for environment setup, running experiments, and reproducing all reported results. Together, these materials ensure that the review committees can fully assess the validity and reproducibility of our contribution.

\section*{LLM Usage Considerations}
The authors affirm that in the whole process of this research, the authors utilize LLMs solely for grammar checking and language refinement when writing. All LLM-assisted text was reviewed and edited to ensure compliance with ethical guidelines and to avoid plagiarism or scientific misconduct. The authors take full responsibility for this declaration and for all content presented in the manuscript.

\bibliographystyle{unsrt}
\bibliography{reference}

\appendix
\section{Additional Experimental Setup Details}
\subsection{Cascade Architectures}
\label{appendix:architectures}
We present in \tableautorefname~\ref{tab:cascade-configs-1}-\ref{tab:cascade-configs-2} the summary of investigated cascade architectures in this work.
\begin{table}[h!]
\centering
\resizebox{\columnwidth}{!}{%
\begin{tabular}{llll}
\toprule
 & $f_1$ (small model) & $f_2$ (large model) & $g_1$ (decision module) \\
\midrule
(1) & GPT2             & Phi3-medium      & BERT       \\
(2) & GPT2             & Phi3-medium      & DistilBERT \\
(3) & GPT2             & Phi3-medium      & GPT2       \\
(4) & GPT2             & Phi3-medium      & AlBERT     \\
(5) & BLOOM-0.556B     & Phi3-medium      & BERT       \\
(6) & Qwen3-0.6B       & Phi3-medium      & BERT       \\
(7) & TinyLlama-1.1B   & Phi3-medium      & BERT       \\
(8) & Qwen2.5          & Phi3.5-MOE      & DistilBERT       \\
(9) & Qwen2.5             & Phi3.5-MOE      & BERT \\
(10) & Qwen2.5             & Phi3.5-MOE      & GPT2       \\
(11) & Qwen2.5             & Phi3.5-MOE      & AlBERT     \\
(12) & BLOOM-0.556B     & Phi3.5-MOE      & BERT       \\
(13) & Qwen3-0.6B       & Phi3.5-MOE      & BERT       \\
(14) & TinyLlama-1.1B   & Phi3.5-MOE      & BERT       \\
(15) & Qwen2.5         & Qwen3-32B      & BERT   \\
(16) & Phi2-2.7B        & Qwen3-32B        & GPT2       \\
(17) & Qwen3-0.6B       & Qwen3-32B        & DistilBERT \\
(18) & BLOOM-0.556B     & Qwen3-32B        & BERT       \\
(19) & TinyLlama-1.1B   & DeepSeek-R1-70B  & DistilBERT \\
(20) & TinyLlama-1.1B    & Llama-3.1-70B     & GPT2 \\
(21) & GPT-3.5-Turbo    & Chatgpt-4o-latest & DistilBERT \\
(22) & Grok-3-mini    & Claude-opus-4-5        & BERT \\
(23) & Claude-3-haiku    & Gemini-3.1-pro-preview  & GPT-2 \\

\bottomrule
\end{tabular}%
}
\vspace{1pt}
\caption{\normalfont\bfseries Investigated 2-layer cascade system configurations.}
\label{tab:cascade-configs-1}
\end{table}

\begin{table}[h!]
\centering
\resizebox{\columnwidth}{!}{%
\begin{tabular}{llllll}
\toprule
 & $f_1$ (small) & $f_2$ (medium) & $f_3$ (large) & $g_1$   & $g_2$\\
\midrule
(24) & Qwen2.5    & Mistral-7B       & Phi3.5-MOE       & BERT   & BERT \\
(25)& Qwen2.5    & Mistral-7B       & Phi3.5-MOE   & GPT2   & GPT2 \\
(26) & GPT2        &TinyLlama-1.1B        & Phi3-medium      & BERT   & BERT  \\
(27) & GPT2        &TinyLlama-1.1B        & Phi3-medium      & GPT2   & GPT2  \\
(28) & BLOOM-0.556B        &Phi2-2.7B        & Phi3-medium      & BERT   & BERT  \\
(29) & BLOOM-0.556B        &Phi2-2.7B       & Phi3-medium      & GPT2   & GPT2  \\
(30)& GPT-3.5 Turbo        & Gemini-2.5-pro   & Chatgpt-4o-latest     & BERT   & BERT  \\
(31) & GPT-3.5 Turbo        & Gemini-2.5-pro      &Chatgpt-4o-latest     & GPT2   & GPT2 \\
(32) & Claude-3-haiku       & Kimi-K2   & GPT-5.1    & BERT   & BERT  \\
(33) & Claude-3-haiku       & Kimi-K2     & GPT-5.1     & GPT2   & GPT2 \\
\bottomrule
\end{tabular}%
}
\vspace{1pt}
\caption{\normalfont\bfseries Investigated 3-layer cascade system configurations.}
\label{tab:cascade-configs-2}
\end{table}

\subsection{Dataset}
\label{appendix:dataset}

\begin{table}[thbp]
  \centering
  \resizebox{\columnwidth}{!}{ 
  \begin{tabular}{lllc}
    \toprule
    \textbf{Main Task} & \textbf{Subcategory} & \textbf{Dataset} & \textbf{Train/Val} \\
    \midrule
    \multirow{4}{*}{Classification}
    & \multirow{2}{*}{Binary Classification} & Overruling & 2000/200 \\
    & & IMDB & 3000/200 \\
    \cmidrule(lr){2-4}
    & \multirow{2}{*}{Multi-class Classification} & Headlines & 3000/200 \\
    & & AG News & 3000/200 \\
    \midrule
    \multirow{6}{*}{Generation}
    & \multirow{2}{*}{Question Answering} & WebQuestions & 3000/200 \\
    & & SQuAD-2.0 & 3000/200 \\
    \cmidrule(lr){2-4}
    & Constrained Generation & CommonGen & 3000/200 \\
    \cmidrule(lr){2-4}
    & Mathematical Reasoning & SVAMP & 800/200 \\
    \cmidrule(lr){2-4}
    & Safety assessment & WildJailbreak & 3000/200 \\
    \bottomrule
  \end{tabular}
  }
  \vspace{1pt}
    \caption{\normalfont\bfseries Summary of datasets.}
  \label{tab:datasets}
\end{table}

\myparagraphnp{Headlines}~\cite{sinha2021impact} utilizes gold-related financial news headlines to conduct four-category predictions  (``up'', ``down'', ``neutral'', or ``none'') on the future trend of gold price fluctuations. It contains 11,412 manually annotated entries. Our work randomly samples 3,000 instances for training and 200 instances for validation.

\myparagraphnp{Overruling}~\cite{zheng2021does} is a binary classification benchmark dataset used to determine whether a sentence in U.S. legal text overrules a previous judgment. It contains 2,400 English legal sentences manually annotated by lawyers. Our work randomly samples 2,000 instances for training and 200 instances for validation. 

\myparagraphnp{AG News}~\cite{zhang2015character} is a classic benchmark dataset for news topic classification, encompassing four major categories: ``world'', ``sports'', ``business'', and ``technology''. It provides approximately 120,000 training samples and 7,600 test samples for training and evaluating. Our work randomly samples 3,000 instances for training and 200 instances for validation.

\myparagraphnp{IMDB}~\cite{maas-EtAl:2011:ACL-HLT2011} is used for sentiment classification through movie reviews and descriptions. It typically contains 50,000 movie reviews labeled as ``positive'' or ``negative''. Our work randomly samples 3,000 instances for training and 200 instances for validation.

\myparagraphnp{WebQuestions}~\cite{berant-etal-2013-semantic} is constructed for Knowledge Base Question Answering (KBQA) tasks. It requires the model to answer factual questions based on a structured knowledge base It contains 5,810 natural language questions based on Freebase Q\&A. Our work randomly samples 3,000 instances for training and 200 instances for validation.

\myparagraphnp{SQuAD-2.0}~\cite{SQuAD2,SQuAD1} is a extractive machine reading comprehension dataset. It contains over 100,000 questions from Wikipedia, and innovatively introduces nearly 50,000 adversarial questions that are similar but cannot find answers in the original text to test the depth of model reading comprehension. Our work randomly samples 3,000 instances for training and 200 instances for validation.

\myparagraphnp{CommonGen}~\cite{CommonGen-lin-EtAl:2020:FindingsEMNLP2020} is a constrained text generation dataset containing approximately 35,000 unique concept sets as the generation tasks. It aims to make the model generate a coherent sentence using the given concept. Our work randomly samples 3,000 instances for training and 200 instances for validation.

\myparagraphnp{SVAMP}~\cite{svamp-kadlcik-EtAl:2023:EMNLP} consists of simple arithmetic word problems, with each problem generating multiple variants through slight modifications (such as changing the order of numbers, asking a different question, or adjusting irrelevant details). It contains 1000 arithmetic word problems, which aims to rigorously test the model's true understanding of the quantitative relationships and structures in the problems. Our work randomly samples 800 instances for training and 200 instances for validation.

\myparagraphnp{WildJailbreak}~\cite{wildteaming2024}: This dataset is an open-source synthetic dataset designed to improve the safety alignment of large language models (LLMs). and aims to test whether the LLM refuses to answer harmful queries or gets jailbroken. It contains two main types of data: Vanilla Harmful (direct malicious requests, e.g., “How to make a bomb?”) and Adversarial Harmful (complex jailbreak prompts that attempt to bypass model safety guardrails). The dataset covers 13 risk categories (e.g., violence, hate speech, illegal acts). We focus on the Vanilla Harmful subset. These malicious queries are directly generated by GPT-4. We randomly sample 3,000 instances for training and 200 instances for validation/testing to evaluate the jailbreak robustness of cascade systems.

\subsection{Evaluation Metrics}
\label{appendix:metrics}

\subsubsection{Task-specific performance} We first present the details of the adopted task-specific performance evaluation metrics.

\myparagraph{Classification tasks}
For both binary and multi-class classification, we adopt the standard \textit{\textbf{classification accuracy}} as the evaluation metric. For a sample with ground-truth label \(\vy_i \in \{1, 2, \dots, C\}\), where \(C\) denotes the number of classes, and predicted label \(\hat{\vy}_i\), the accuracy is defined as
\[
\mathrm{Acc} = \frac{1}{N} \sum_{i=1}^{N} \mathds{1}\!\left[ \hat{\vy}_i = \vy_i \right],
\]
where \(\mathds{1}[\cdot]\) is the indicator function and \(N\) is the size of the evaluation dataset.

\myparagraph{Open-ended question answering (free-form generation)}
This task requires the model to generate a natural-language answer in an open-ended manner. The output is a token sequence \(\hat{\vy}_i \in \gV^*\), where \(\gV\) denotes the vocabulary and \(\gV^*\) the set of all finite-length token sequences. The ground-truth answer is denoted by \(\vy_i \in \gV^*\).
For each test instance, we evaluate whether the predicted sequence \(\hat{\vy}_i\) matches the reference \(\vy_i\) after standard normalization (e.g., lowercasing, punctuation removal):
\[
m_i = \mathds{1}\bigl[ \text{normalize}(\hat{\vy}_i) = \text{normalize}(\vy_i) \bigr].
\]
An instance is considered a \emph{success} if and only if \(m_i = 1\).
We define the task-specific metric, \textbf{Exact Match Rate (EMR)}, as
\[
\mathrm{EMR} = \frac{1}{N} \sum_{i=1}^{N} m_i,
\]
where \(N\) is the total number of test instances. EMR measures the proportion of instances for which the normalized predicted answer exactly matches the normalized ground-truth answer.

\myparagraph{Constrained generation (keyword coverage)}
This task requires the model to generate a sentence that explicitly incorporates all given concept keywords. The output is a token sequence \(\hat{\vy}_i \in \gV^*\), where \(\gV\) denotes the vocabulary and \(\gV^*\) the set of all finite-length token sequences. The ground-truth requirement is a concept set \(\vy_i = \gS_i\), where \(\gS_i \subseteq \gV\).
For each test instance, we evaluate the proportion of required concepts that appear in the generated sequence \(\hat{\vy}_i\):
\[
\mathrm{coverage}_i = \frac{\left| \{ s \in \gS_i : s \text{ appears in } \hat{\vy}_i \} \right|}{|\gS_i|}.
\]
An instance is considered a \emph{success} if and only if all concepts are covered, i.e., \(\mathrm{coverage}_i = 1\).
We then define the task-specific metric, \textbf{Full Coverage Rate (FCR)}, as
\[
\mathrm{FCR} = \frac{1}{N} \sum_{i=1}^{N} \mathds{1}\!\left[ \mathrm{coverage}_i = 1 \right],
\]
where \(N\) is the total number of test instances. FCR measures the proportion of generated sequences that successfully incorporate every required concept.

\myparagraph{Mathematical reasoning (numerical answers)}
This task requires the model to solve an arithmetic problem and produce the final answer. The output is a token sequence \(\hat{\vy}_i \in \gV^*\), where \(\gV\) denotes the vocabulary. The ground-truth answer is denoted by \(\vy_i \in \gV^*\).
For each test instance, we compare the numerical values extracted from the predicted sequence \(\hat{\vy}_i\) and the ground-truth sequence \(\vy_i\):
\[
m_i = \mathds{1}\!\left[ \text{value}(\hat{\vy}_i) = \text{value}(\vy_i) \right],
\]
where \(\text{value}(\cdot)\) extracts and normalizes the numerical value from a sequence. An instance is considered a \emph{success} if and only if \(m_i = 1\).
We define the task-specific metric, \textbf{Solve Rate (SR)}, as
\[
\mathrm{SR} = \frac{1}{N} \sum_{i=1}^{N} m_i,
\]
where \(N\) is the total number of test instances. SR measures the proportion of problems for which the predicted answer is numerically equivalent to the ground-truth answer.

\myparagraph{Safety assessment (jailbreak robustness)}
This task evaluates whether the model can safely refuse harmful or adversarial requests under jailbreak attempts. Given an input query \(\vy_i \in \gV^*\), the model produces an output sequence \(\hat{\vy}_i \in \gV^*\), where \(\gV\) denotes the vocabulary.
For each test instance, we employ an external judge \(J(\cdot,\cdot)\) (e.g., a strong LLM API) that takes \((\vy_i, \hat{\vy}_i)\) as input and outputs a binary verdict:
\[
m_i = J(\vy_i, \hat{\vy}_i) \in \{0,1\},
\]
where \(m_i = 1\) indicates a successful jailbreak (i.e., the model produces unsafe content), and \(m_i = 0\) otherwise.
We define the \textbf{Jailbreak Success Rate (JSR)} as
\[
\mathrm{JSR} = \frac{1}{N} \sum_{i=1}^{N} m_i.
\]
In addition, we report the complementary metric \(1 - \mathrm{JSR}\), referred to as the \textbf{Defense Success Rate (DSR)}, which measures the proportion of jailbreak attempts that are successfully blocked. Higher \(\mathrm{DSR}\) indicates better safety alignment.

\subsubsection{Inference cost}
For inference cost, we evaluate three complementary metrics widely used in LLM efficiency studies: (i) the \textit{average normalized token cost}, which approximates model-dependent computational expenditure; (ii) the \textit{average execution time} per test sample (latency); and (iii) the \textit{escalation rate}, which captures routing behavior in cascade systems. For API-based models, we additionally report the monetary cost in \$.

\myparagraph{Normalized token cost}
This metric approximates model-dependent computational expenditure by weighting input and output tokens using fixed per-token coefficients (0.01 for input tokens and 0.03 for output tokens), following the pricing structure of common LLM API services (e.g., GPT-4 Turbo~\cite{wiggers2023openaigpt4turbo}) as a standardized proxy. The weighted token count is further scaled by the size of each model component (in billions of parameters). For each test sample, the resulting cost is normalized by the length of the input query to ensure comparability across examples. The final reported value is averaged over all test samples.

\myparagraph{Average execution time}
We measure the end-to-end inference time per test sample (in seconds), averaged over the test set. This metric reflects the practical latency of the system under a given hardware setup, where smaller models typically yield faster responses and larger models incur higher latency due to increased computational demands.

\myparagraph{Pass rate}
In a cascaded system, the cost of a query increases as it is forwarded to deeper, more expensive models. For an \(n\)-layer cascade, let \(E\) denote the number of test samples escalated from a given layer to the next stage, and let \(N\) be the total number of test samples. The escalation rate is defined as \(\rho = E / N\). A lower \(\rho\) indicates that more queries are resolved by earlier, less expensive components, reflecting more efficient routing. Unlike token cost or execution time, the escalation rate is largely hardware-agnostic and captures the intrinsic decision-making behavior of the cascade.

\myparagraph{Experimental Setup}
Unless otherwise specified, we use the attack procedure in Algorithm~\ref{alg:attack_framework}, jointly targeting the first-stage prediction model $f^1$ (the small language model, \texttt{SLM}) and the decision module $g^1$ (\texttt{DM}), denoted as \texttt{SLM+DM}. All techniques introduced in Sections~\ref{subsec:alternating}--\ref{subsubsec:pass_rate} are enabled by default.

For the GPT-2-based cascade system, the budget is set to 20, the temperature is set to 0, and the maximum generation length is 250 tokens. For the Qwen2.5-based cascade system, the budget is set to 100, the temperature is set to 0, and the maximum generation length is 300 tokens. For all models using top-$k$ decoding, we set $k=1$ to ensure deterministic decoding. The numbers of training and test instances for each dataset follow the descriptions in the main text.

During inference, we use few-shot in-context learning with a task-specific prefix. For classification tasks, the prefix contains one demonstration for each answer class. For generation tasks, it contains two question-answer demonstrations. For each dataset, the demonstrations are fixed throughout inference and are randomly sampled from correctly labeled training examples.

\myparagraph{Attack Hyper-parameters.}
For all datasets, we use a unified interleaved attack setting with a total suffix length of 8 tokens. Specifically, we adopt a 2+2+2+2 schedule, corresponding to two rounds of alternating optimization, where each round optimizes a 2-token suffix for the SLM objective and a 2-token suffix for the DM objective.

For adversarial suffix optimization on functional language models for question-answering tasks, the objective is defined to maximize the deviation of the generated response from the ground-truth answer. We use a loss threshold of $-40$ for early stopping, with a maximum of 300 optimization iterations. In each iteration, 512 candidate suffixes are evaluated, and two suffix positions are replaced. For DM attacks, we use the same search budget, namely 300 iterations and 512 candidate suffixes per iteration, but do not apply the loss-threshold-based early stopping criterion.

For DM suffix optimization, we further impose a constraint on candidate selection: the preferred suffix must preserve the response of the preceding language model with respect to the original cascade input, while keeping the final cascaded output incorrect. This constraint takes precedence over the loss value. If no candidate satisfies this constraint after the current search is exhausted, we restart the search from a different initialization. At most three restarts are allowed. If no feasible suffix is found after all three restarts, we use the suffix with the minimum loss value as the final result.

\myparagraph{Evaluation and Hardware Setup.}
Unless otherwise specified, all reported results are averaged over three independent runs, and sample standard deviations are also reported. Computational overhead is measured using end-to-end runtime.

All experiments are conducted under the same software and hardware environment for fair comparison. Specifically, experiments are run on a shared server with Ubuntu 22.04.5 LTS, two Intel Xeon Platinum 8378C CPUs, approximately 1.0~TiB RAM, and eight NVIDIA GeForce RTX 4090 GPUs (24~GB each), using NVIDIA driver 535.288.01 and CUDA 12.2. Because the server is shared by multiple users, absolute runtime may vary with system load. Therefore, runtime comparisons are made only under the same execution environment.

\subsection{Defense Setup}
For PPL filtering, the tuned strength hyperparameter is the perplexity quantile threshold $q$ (smaller $q$ indicates stricter filtering).
For Regex filtering, the tuned strength hyperparameter is the allowed special-character ratio \texttt{ratio} (smaller \texttt{ratio} indicates stricter filtering).For CPT filtering, the tuned strength hyperparameter is the CPT threshold $\tau$ (larger $\tau$ indicates stricter filtering), where CPT denotes the character-per-token statistic.In all methods, \texttt{trim} denotes the number of tail words removed after triggering.

\section{Additional Results}
\subsection{White-box setting}
\label{appendix:white-box}

\begin{figure*}[thbp]
  \centering
    \subfigure[Headline (passrate)]{
    \includegraphics[width=0.6\columnwidth]{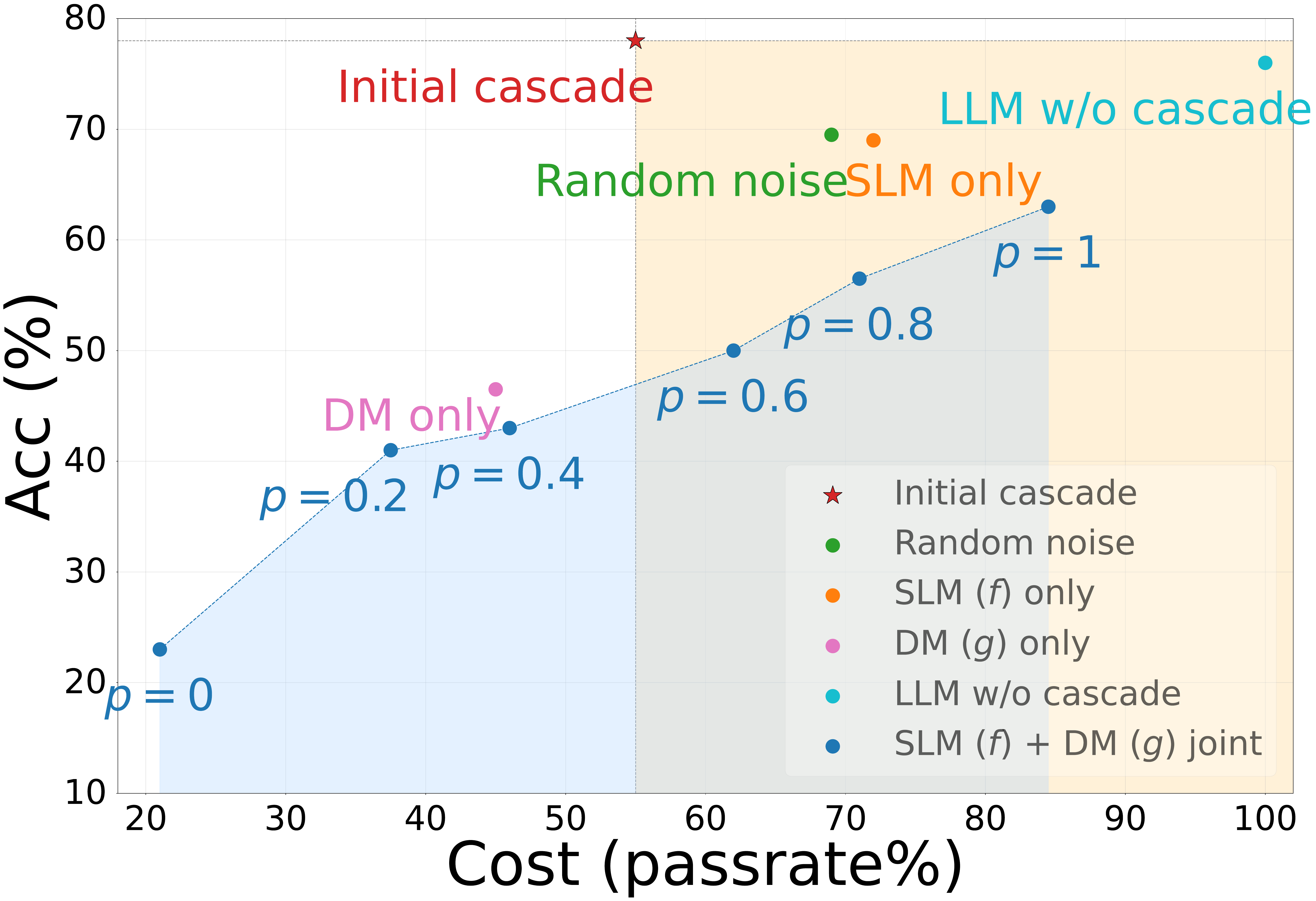}
  }
        \subfigure[IMDB (passrate)]{
    \includegraphics[width=0.6\columnwidth]{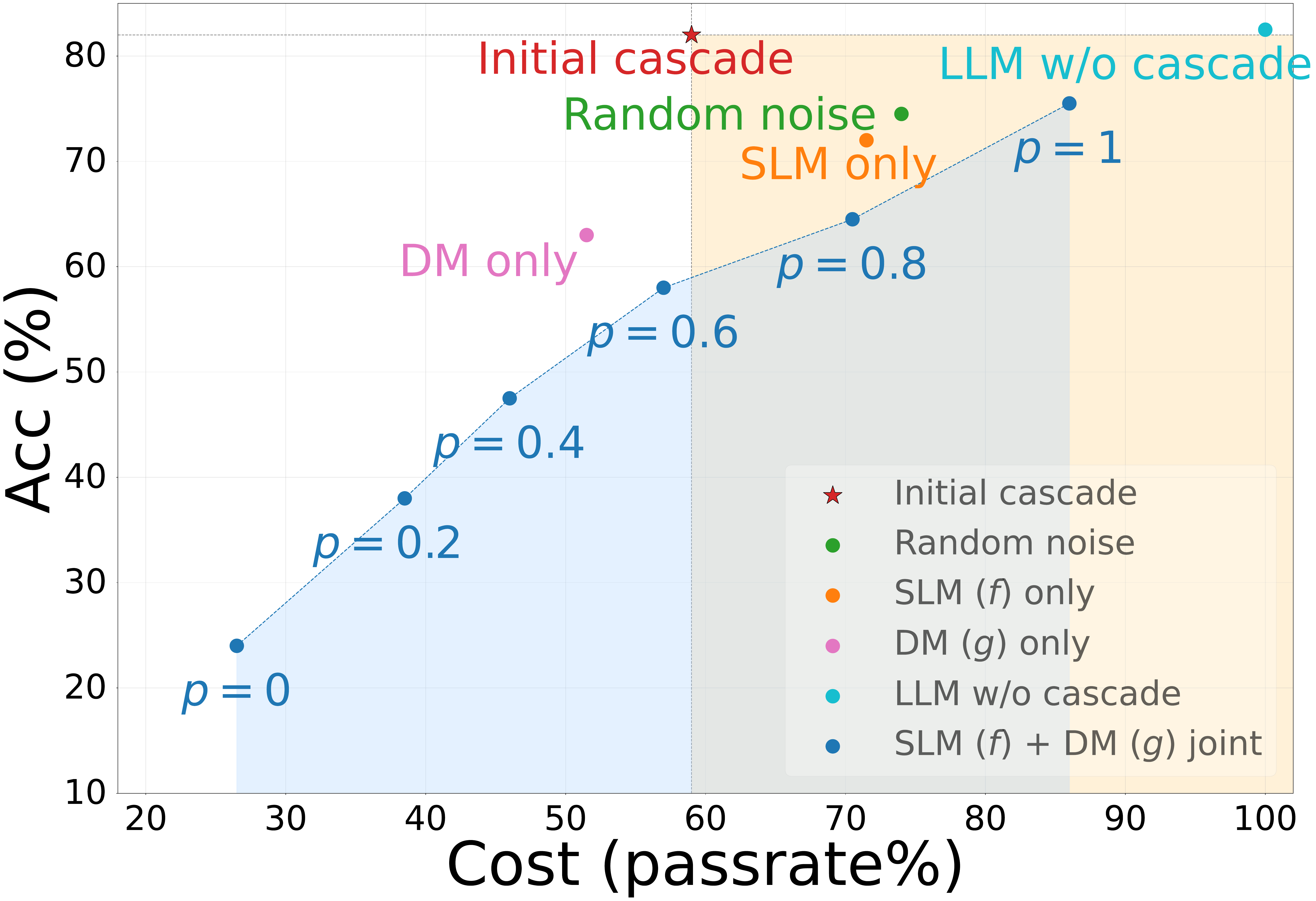}
  }
  \subfigure[AGnews (passrate)]{
    \includegraphics[width=0.6\columnwidth]{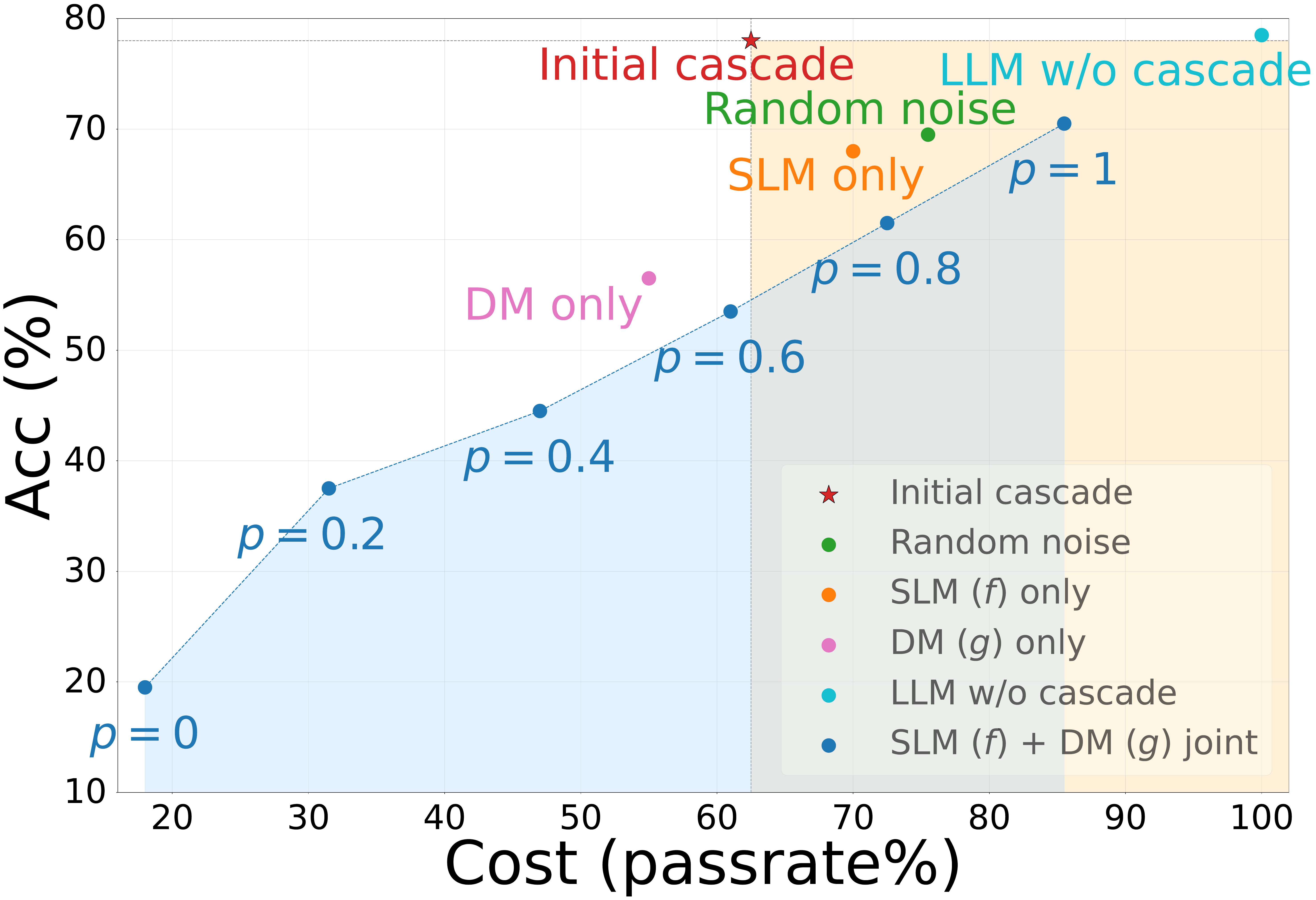}
  }
  \subfigure[Headline (normalized tokens)]{
    \includegraphics[width=0.6\columnwidth]{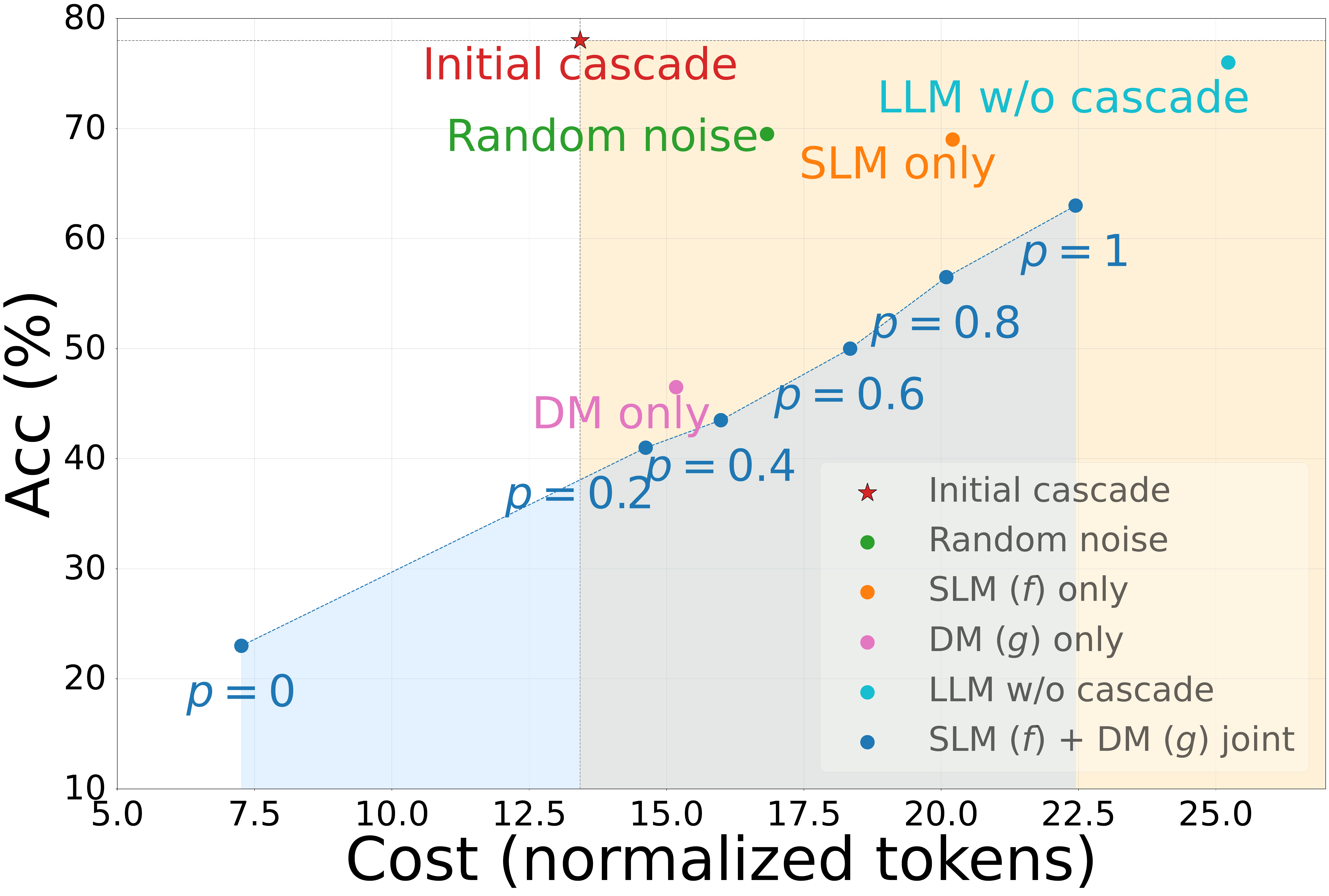}
  }
   \subfigure[IMDB (normalized tokens)]{
    \includegraphics[width=0.6\columnwidth]{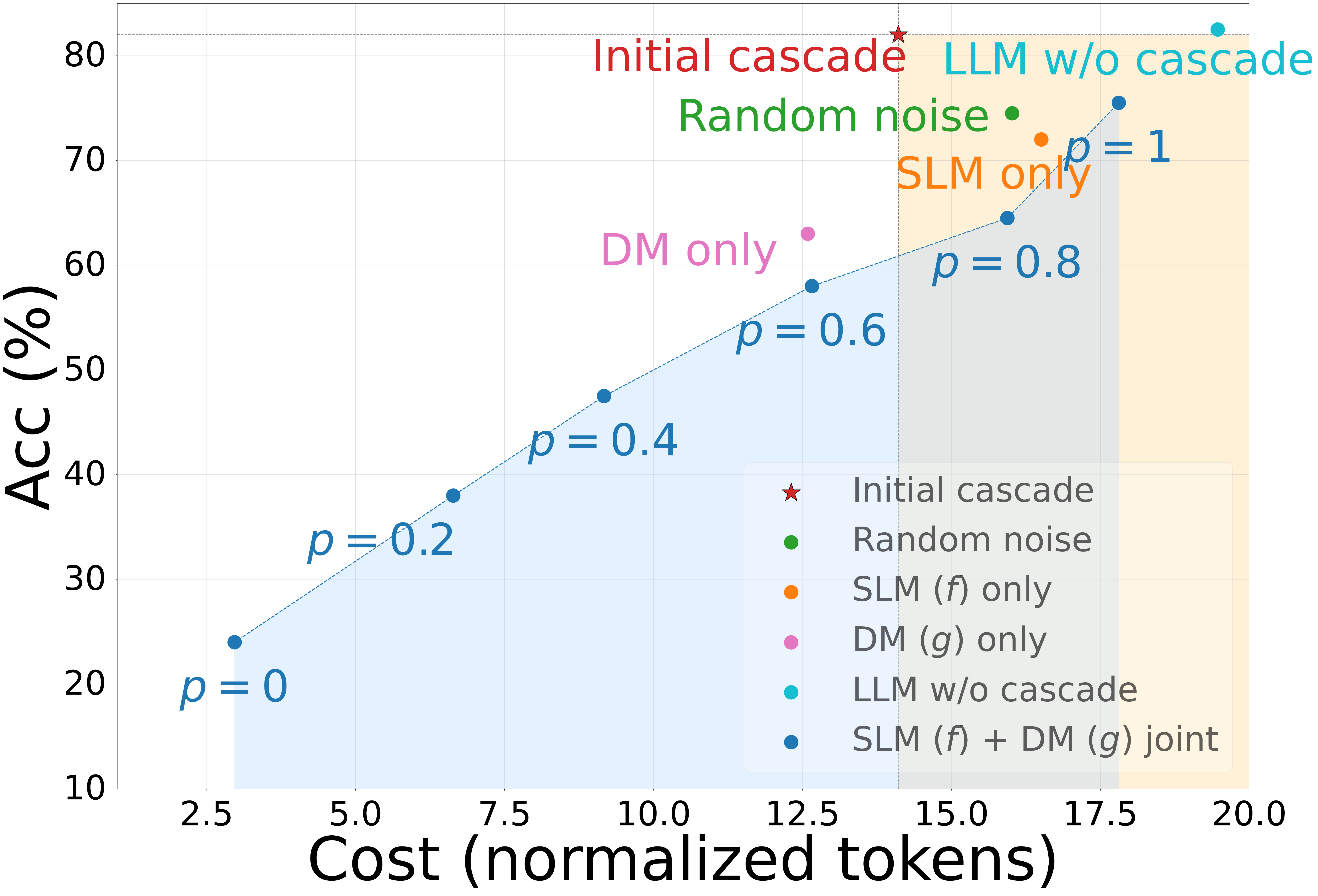}
  }
      \subfigure[AGnews (normalized tokens)]{
    \includegraphics[width=0.6\columnwidth]{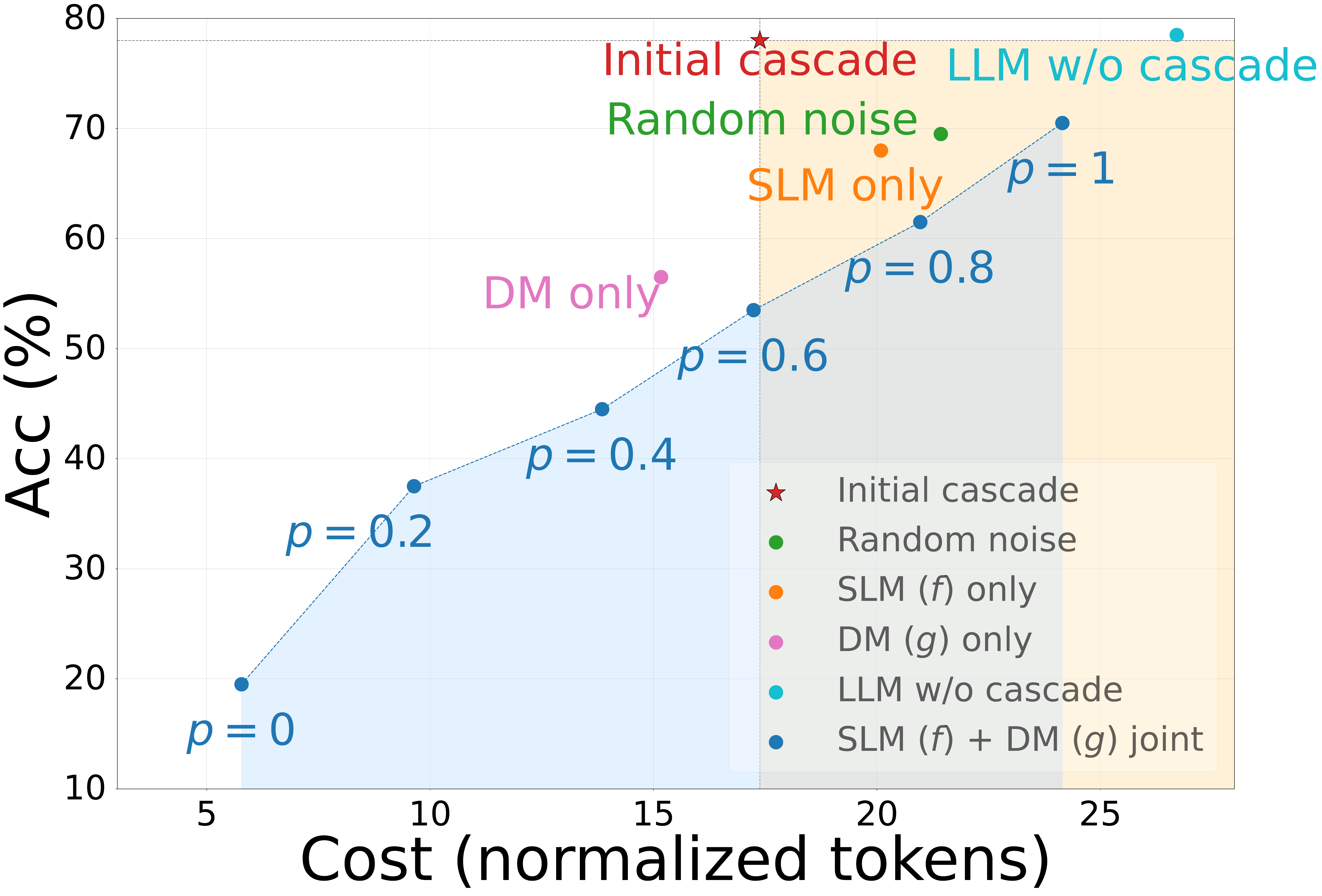}
  }
  \subfigure[Headline (time)]{
    \includegraphics[width=0.6\columnwidth]{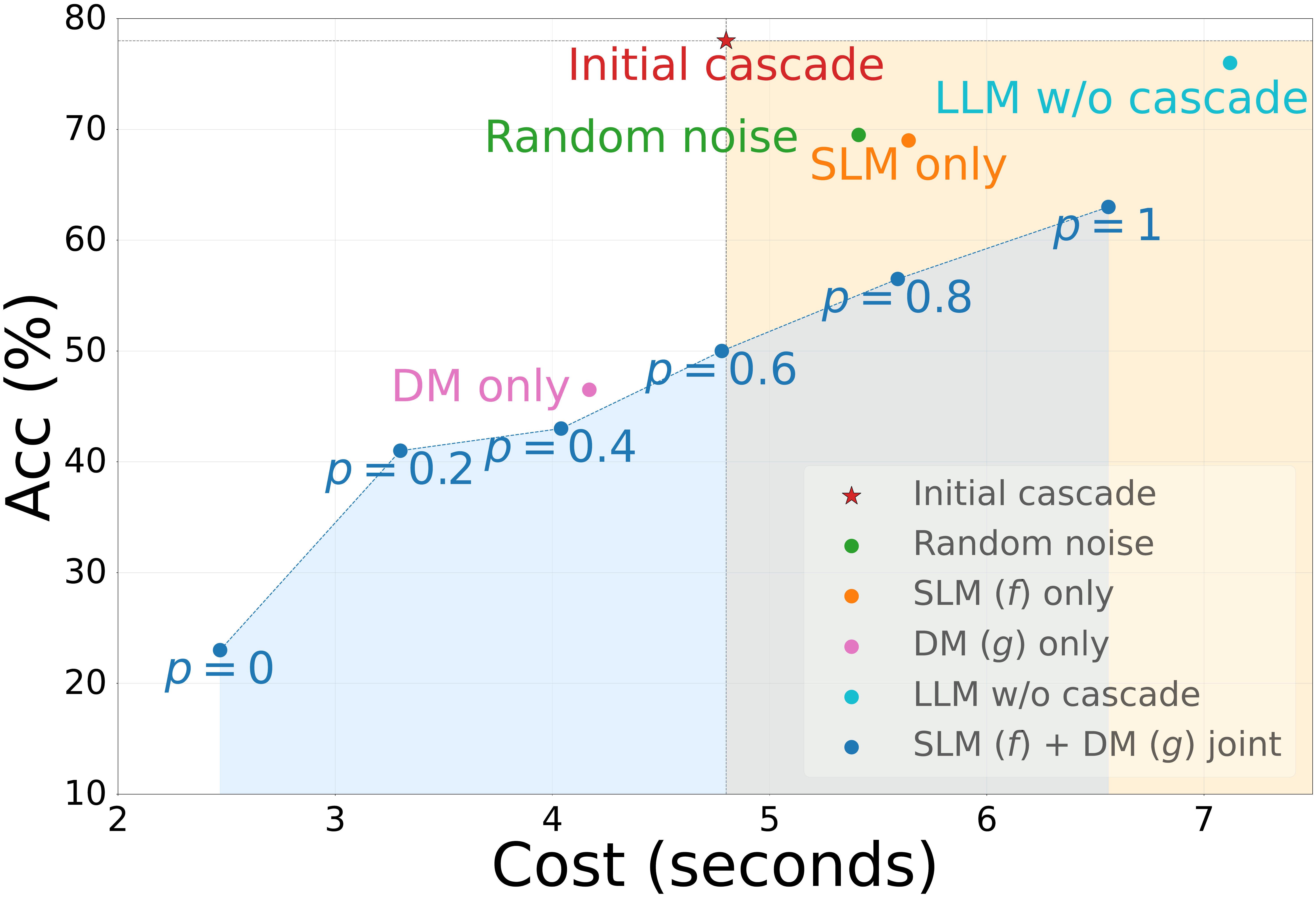}
  }
      \subfigure[IMDB (time)]{
    \includegraphics[width=0.6\columnwidth]{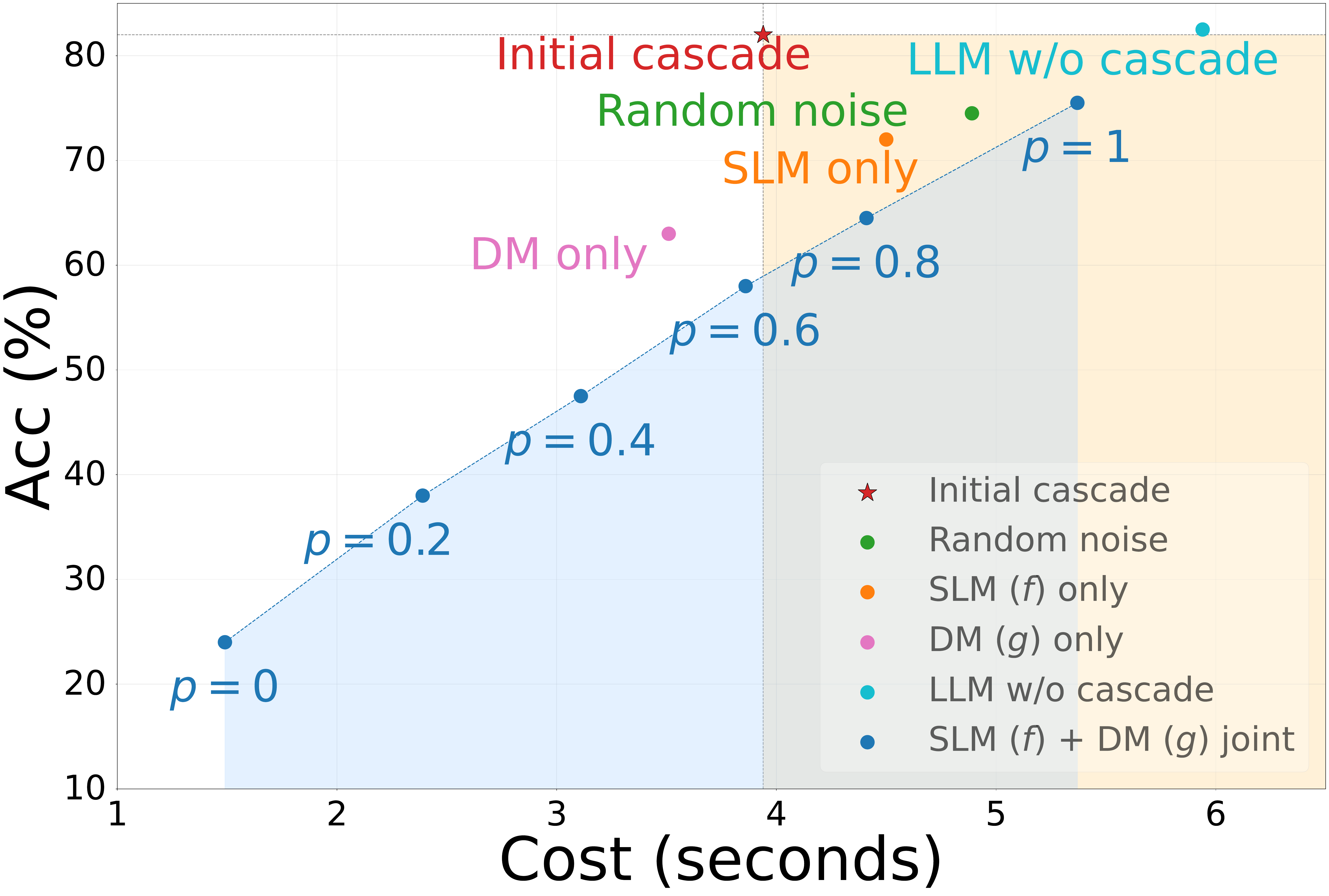}
  }
  \subfigure[AGnews (time)]{
    \includegraphics[width=0.6\columnwidth]{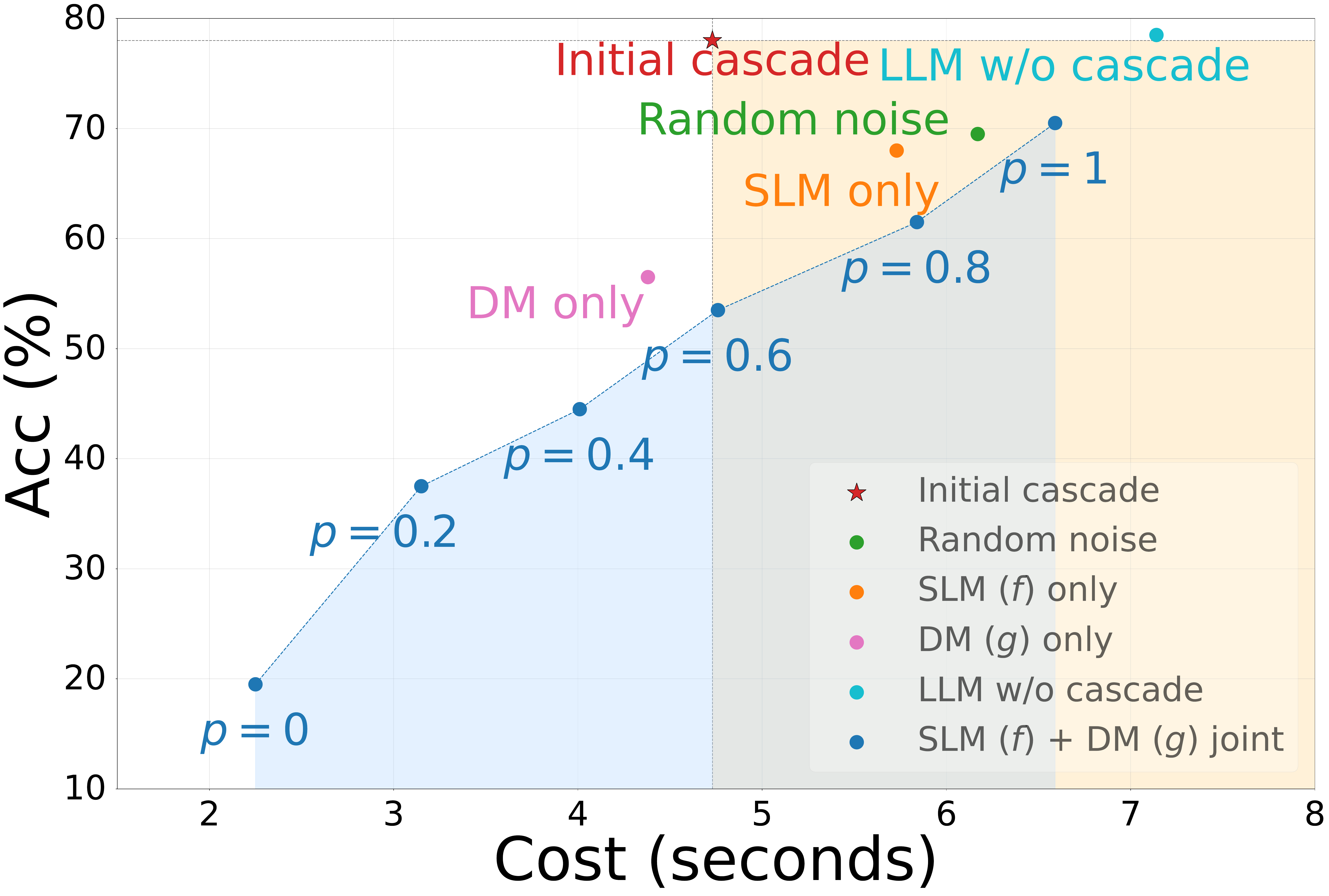}
  }
  \vspace{-6pt}
  \caption{\textbf{Comparison of attack methods across the Headline, IMDB, and AGNews datasets on a \textit{two-layer} LLM cascade (Configuration~(1) in \tableautorefname~\ref{tab:cascade-configs-1}) . The cost–accuracy trade-off achieved by our pass-rate control method(Section~\ref{subsubsec:pass_rate}) is illustrated by the {\color[HTML]{1F77B4}blue} dashed line. Points toward the lower-right represent more effective attacks.}}
  \label{fig:2layer_results_2}
  \vspace{6pt}
\end{figure*}

\begin{figure*}[thbp]
  \centering
    \subfigure[Overruling (passrate)]{
    \includegraphics[width=0.6\columnwidth]{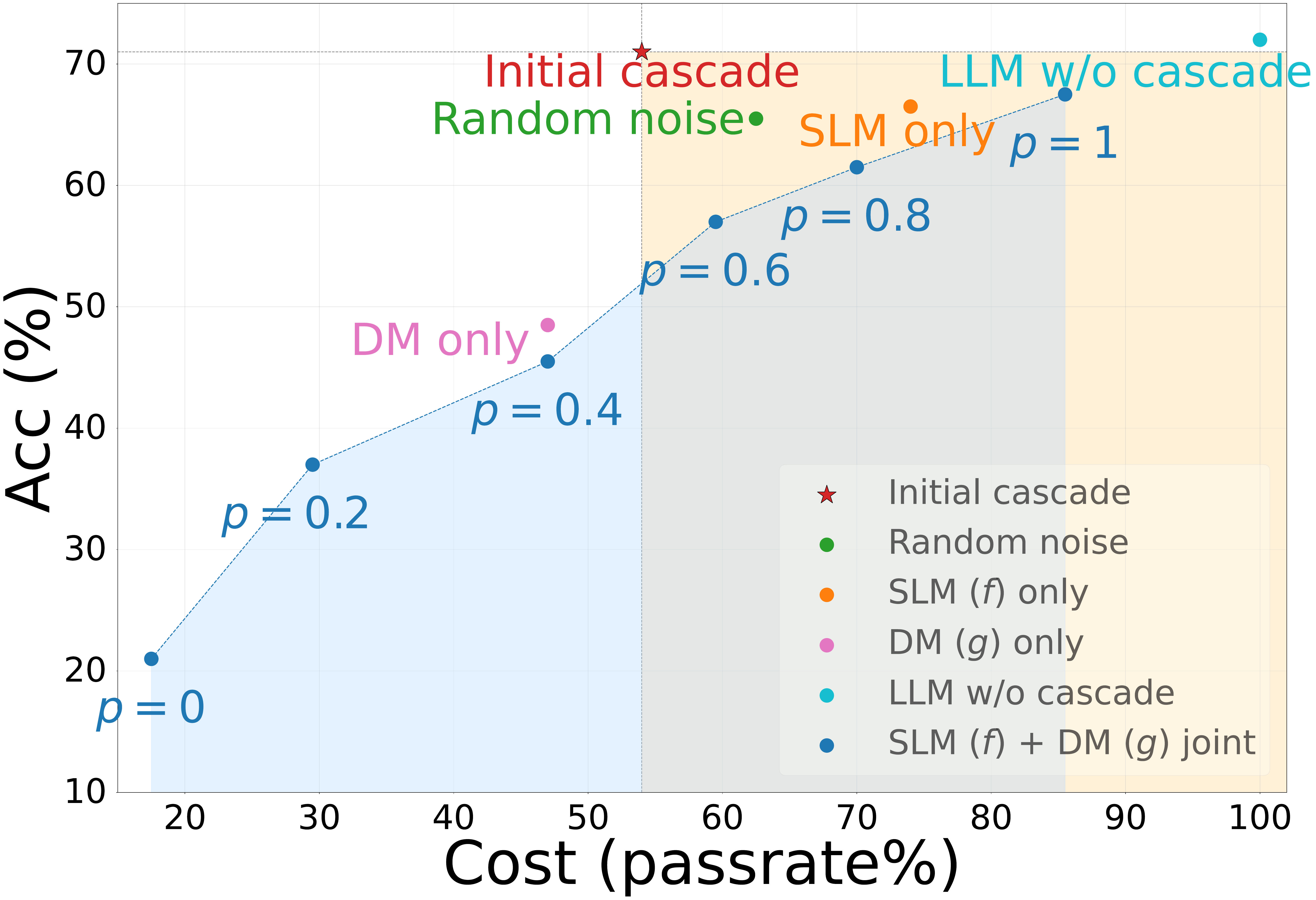}
  }
        \subfigure[CommonGen (passrate)]{
    \includegraphics[width=0.6\columnwidth]{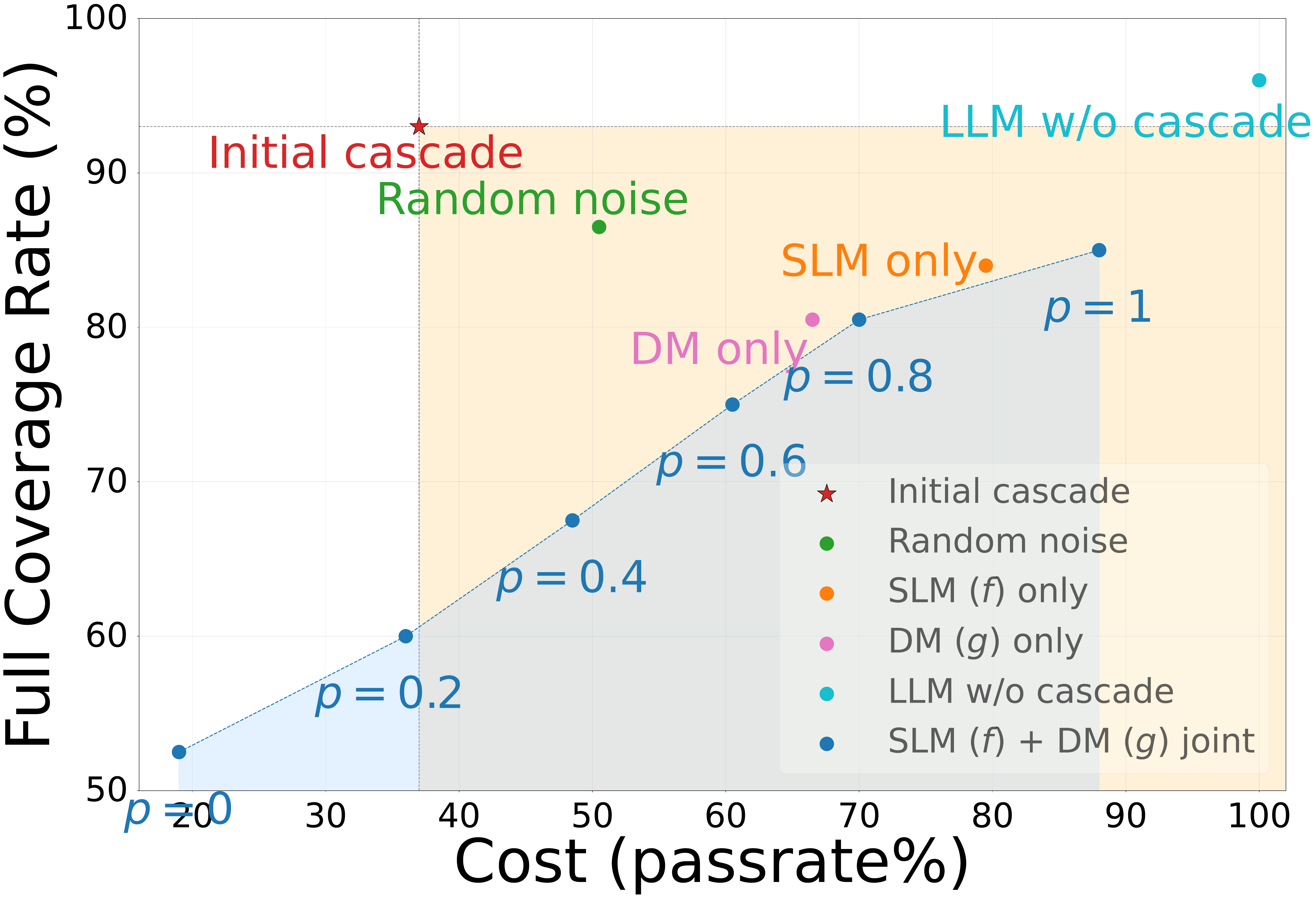}
  }
  \subfigure[SVAMP (passrate)]{
    \includegraphics[width=0.6\columnwidth]{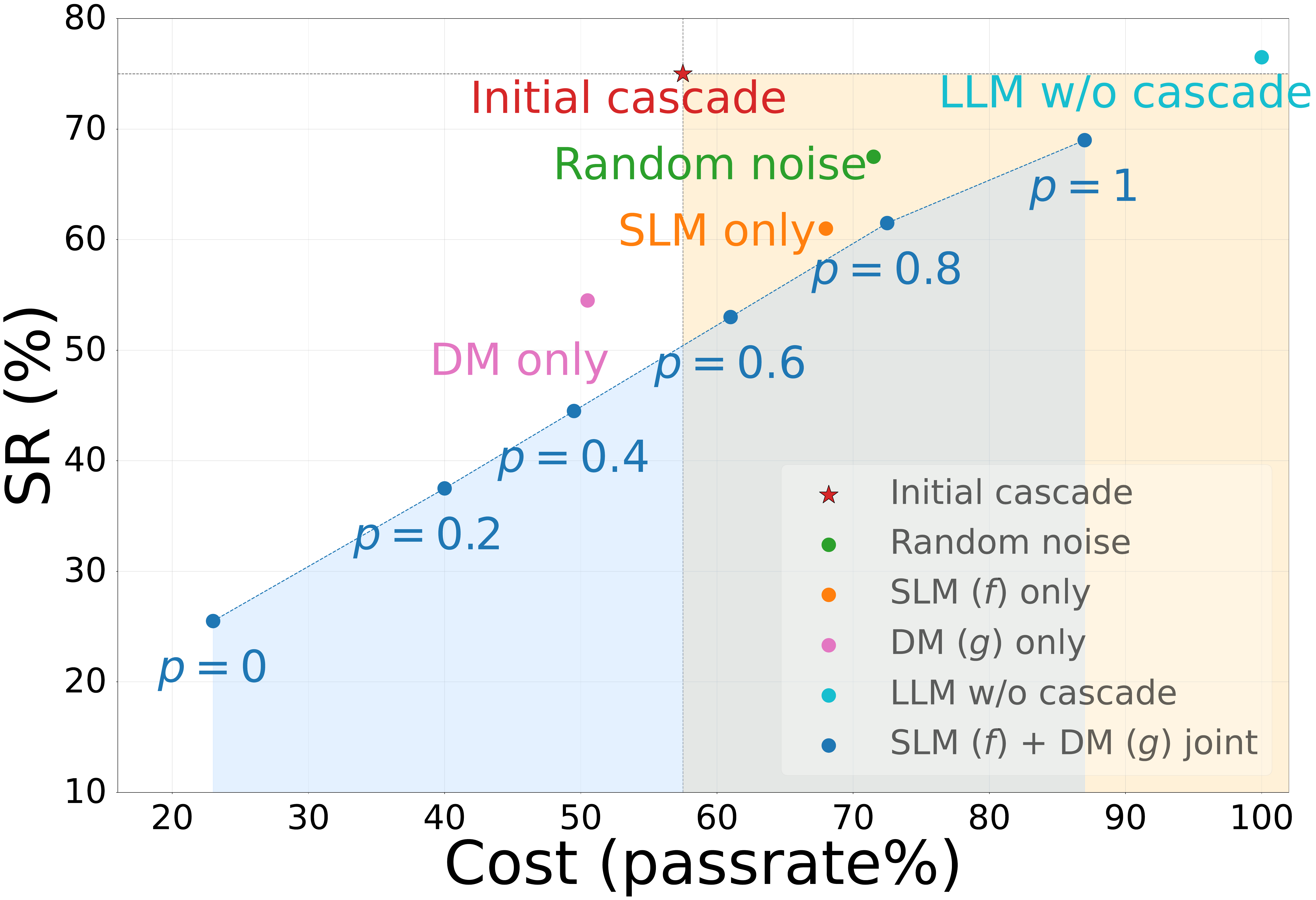}
  }
  \subfigure[Overruling (normalized tokens)]{
    \includegraphics[width=0.6\columnwidth]{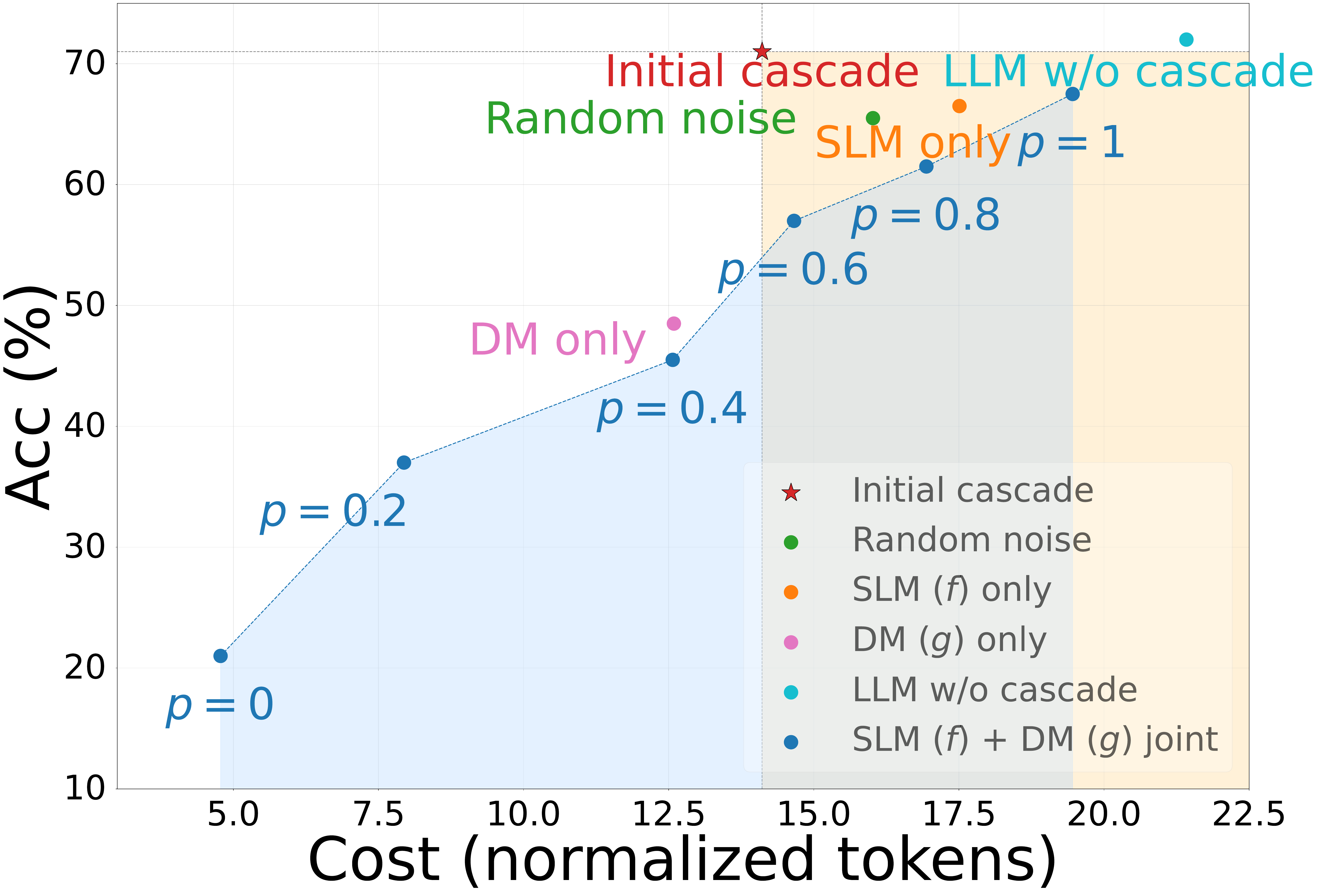}
  }
   \subfigure[CommonGen (normalized tokens)]{
    \includegraphics[width=0.6\columnwidth]{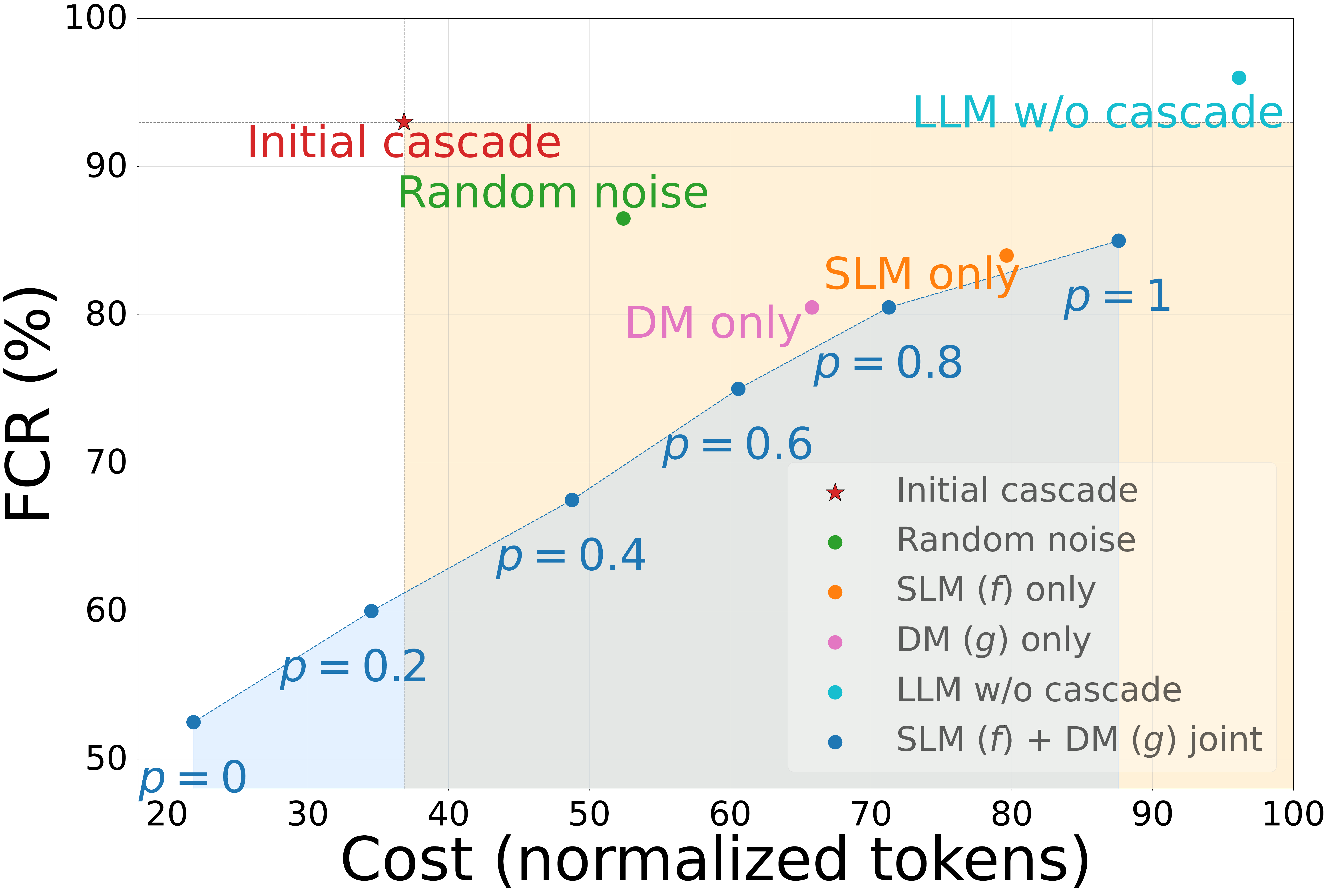}
  }
      \subfigure[SVAMP (normalized tokens)]{
    \includegraphics[width=0.6\columnwidth]{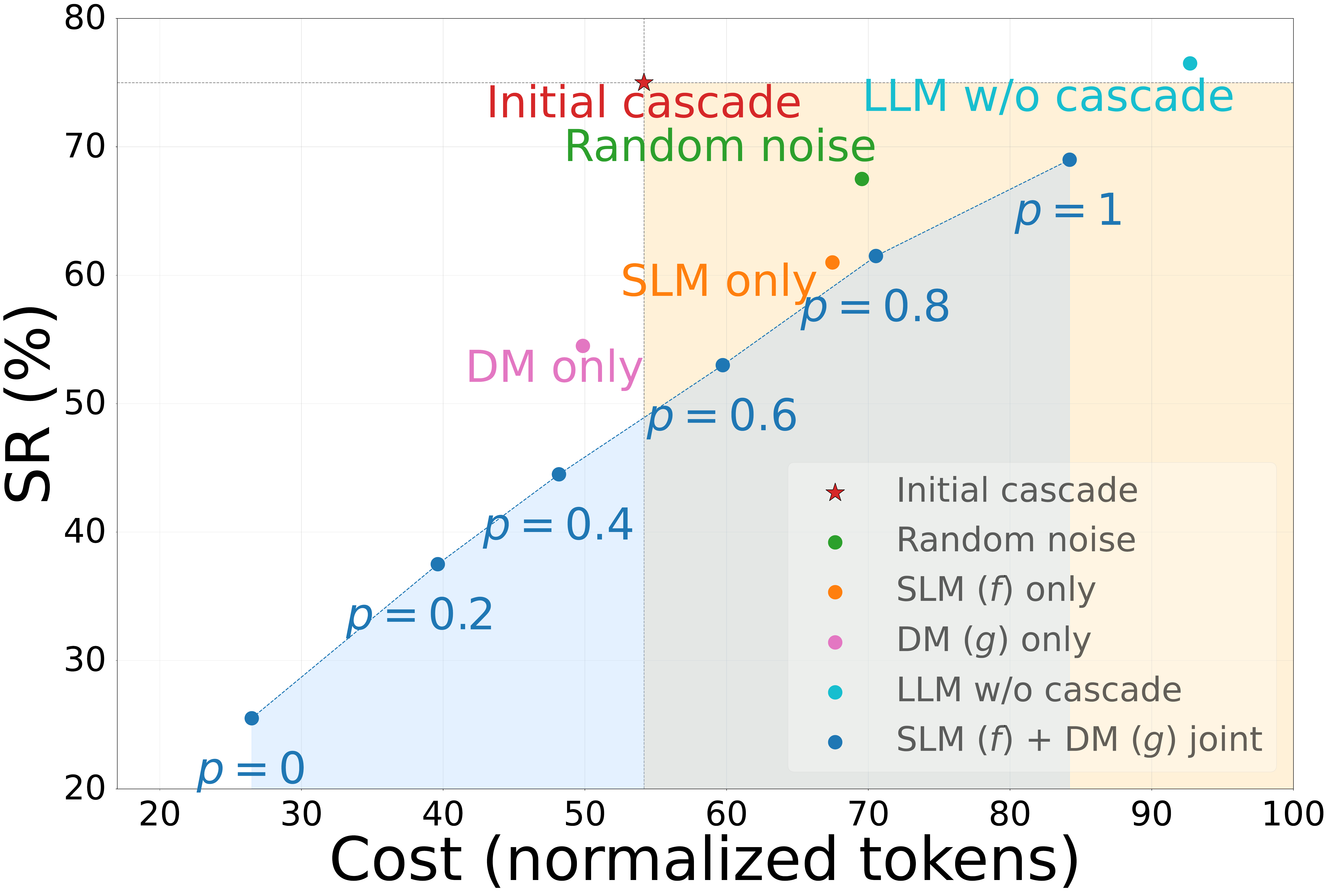}
  }
  \subfigure[Overruling (time)]{
    \includegraphics[width=0.6\columnwidth]{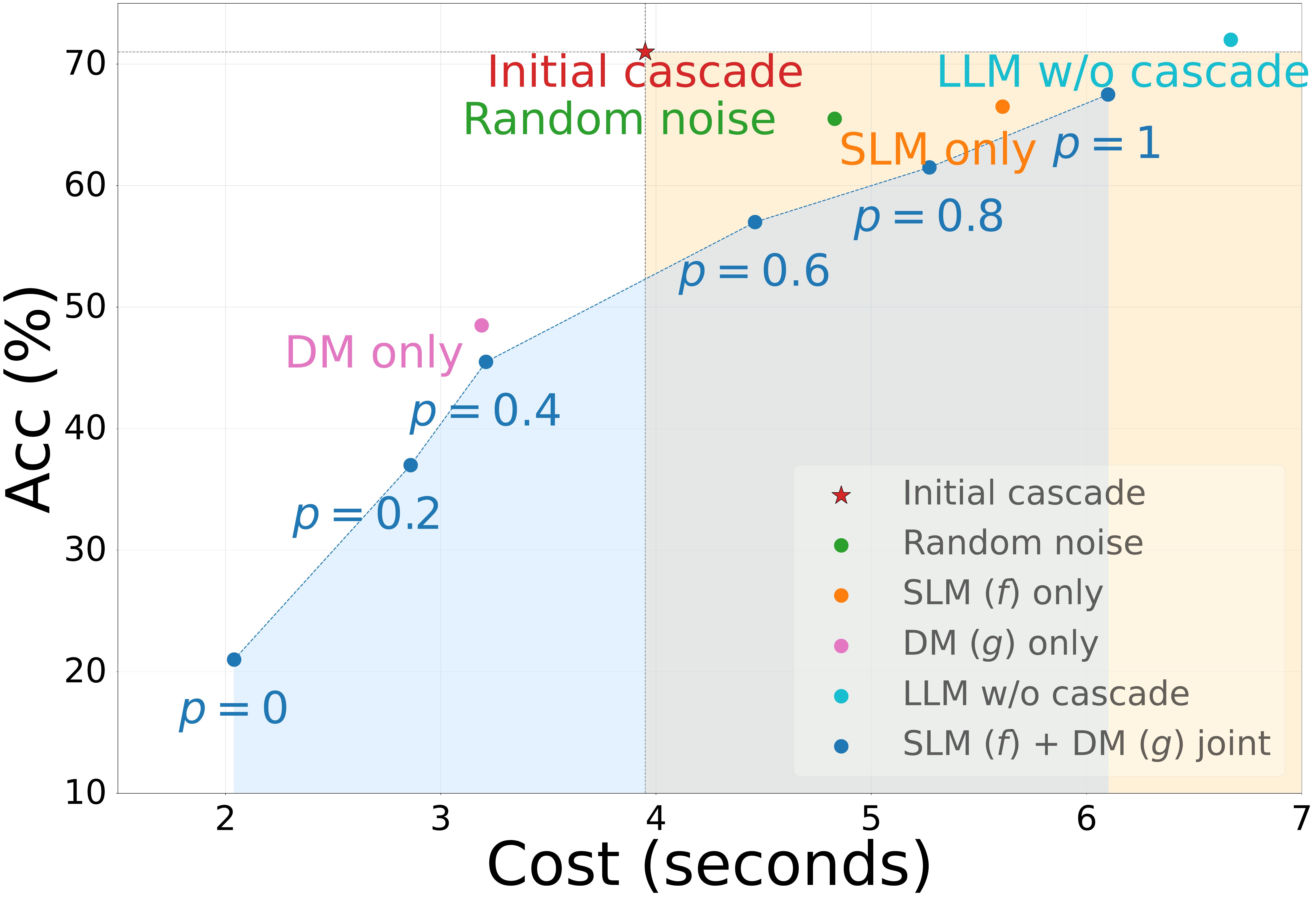}
  }
      \subfigure[CommonGen (time)]{
    \includegraphics[width=0.6\columnwidth]{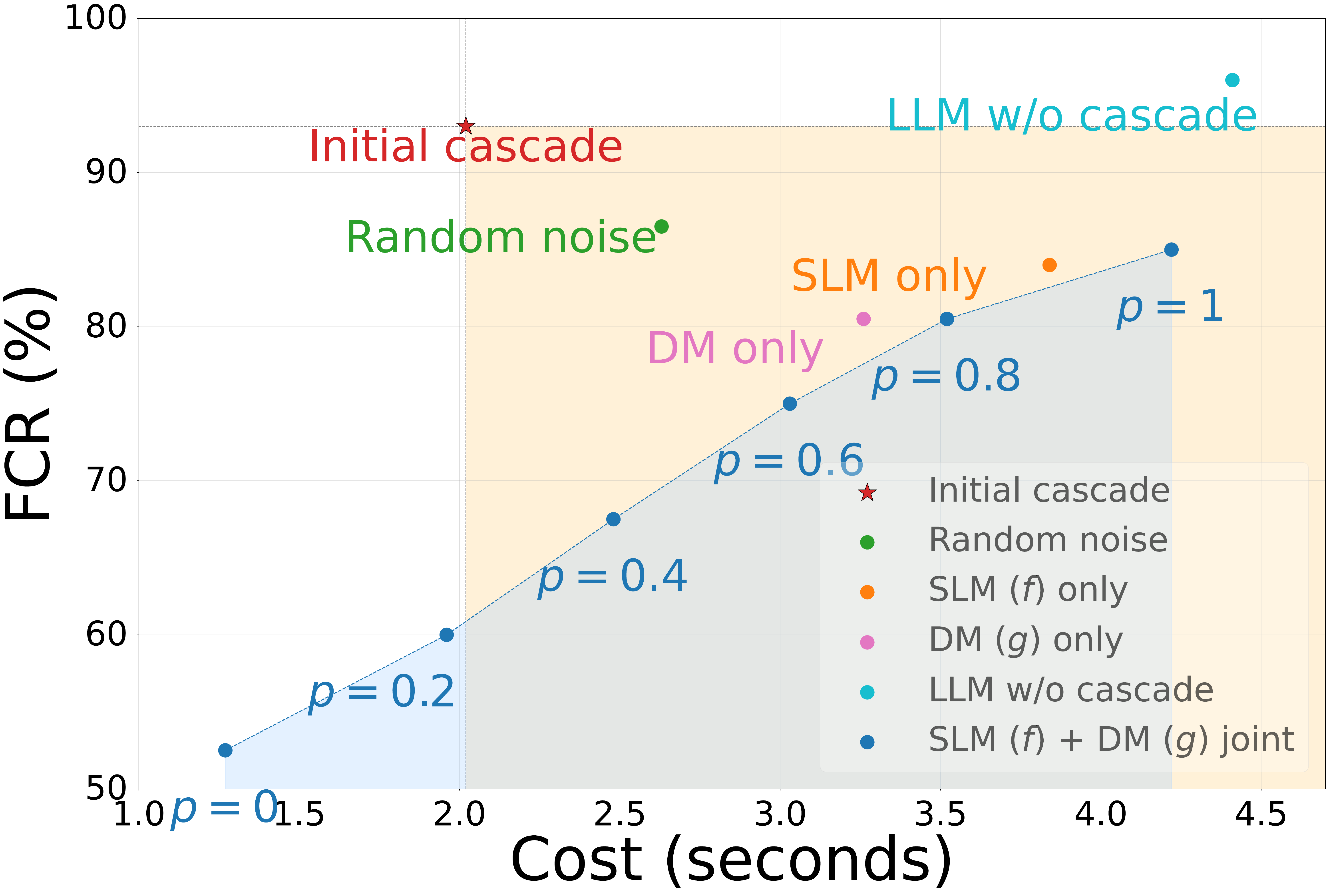}
  }
  \subfigure[SVAMP (time)]{
    \includegraphics[width=0.6\columnwidth]{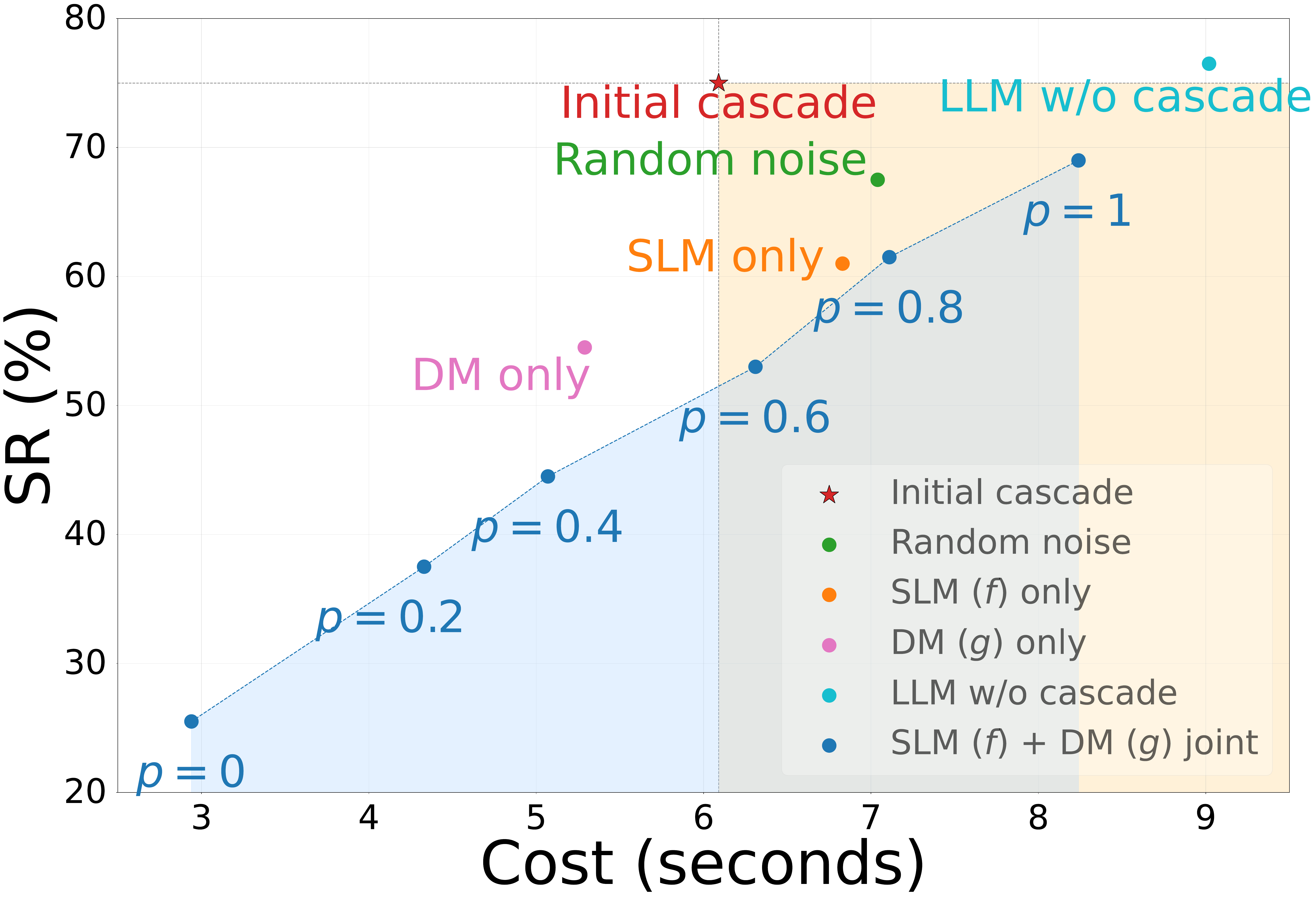}
  }
  \vspace{-6pt}
  \caption{\normalfont\bfseries Comparison of attack methods on a two-layer LLM cascade across the across the Overrulling, CommonGen, and SVAMP datasets using cascade configurations~(1) (see Table~\ref{tab:cascade-configs-1}). The cost–accuracy trade-off achieved by our pass-rate control method(Section~\ref{subsubsec:pass_rate}) is illustrated by the {\color[HTML]{1F77B4}blue} dashed line. Points toward the lower-right represent more effective attacks.}
  
  \label{fig:2layer_results_3}
  \vspace{6pt}
\end{figure*}

\begin{figure*}[thbp]
  \centering
  \subfigure[AGnews (normalized tokens)]{
    \includegraphics[width=0.6\columnwidth]{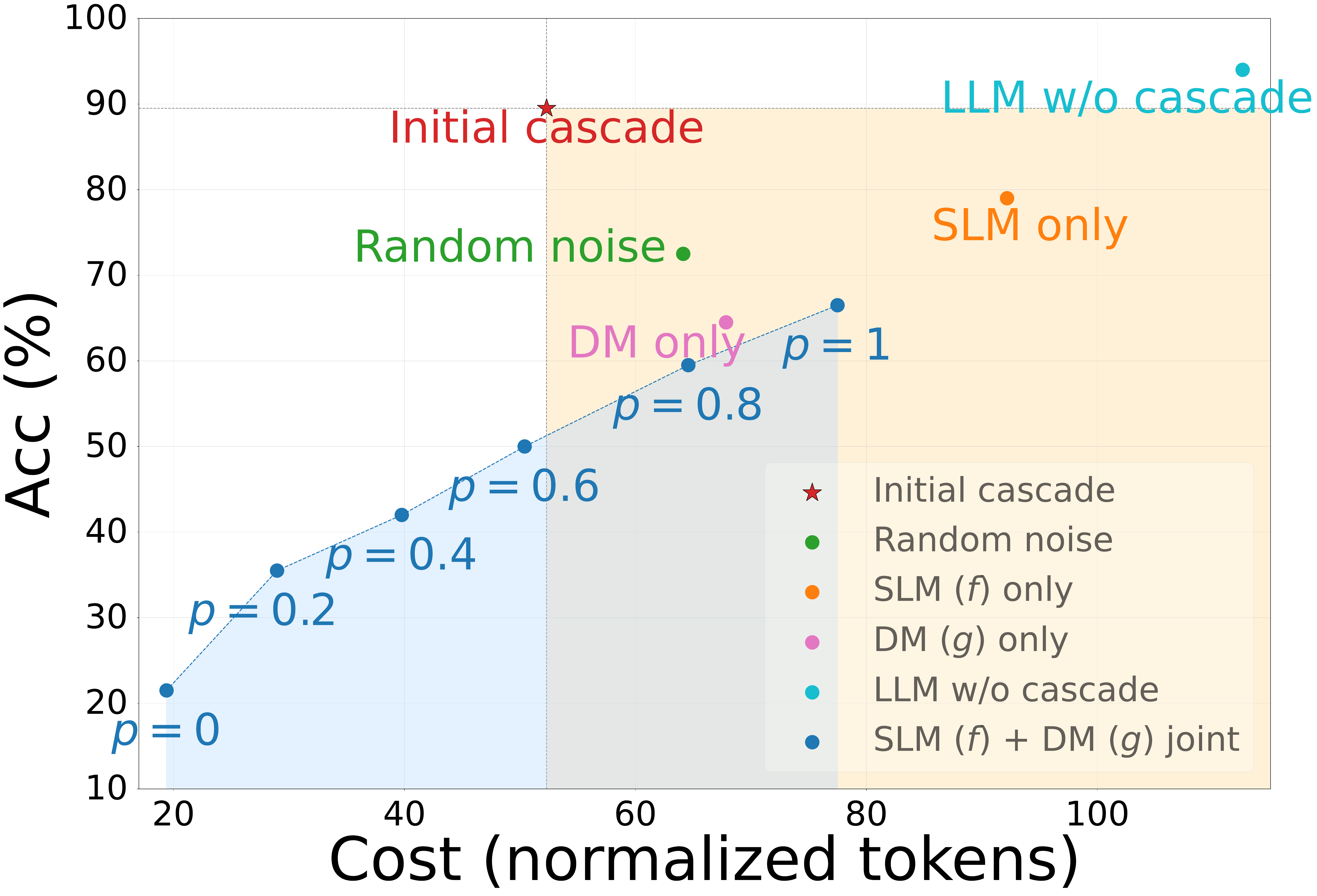}
  }
   \subfigure[SQuAD2.0 (normalized tokens)]{
    \includegraphics[width=0.6\columnwidth]{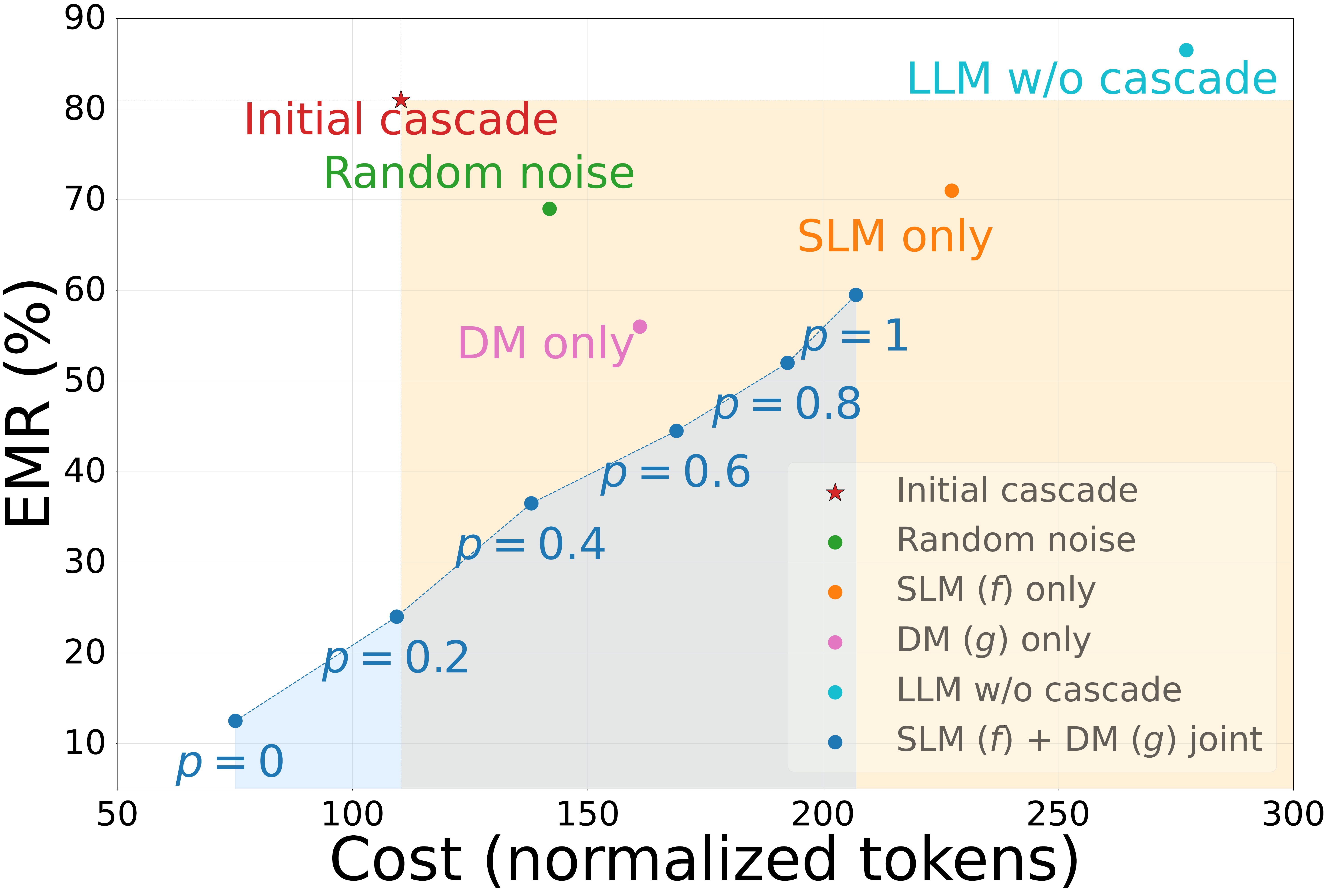}
  }
      \subfigure[WildJailbreak (normalized tokens)]{
    \includegraphics[width=0.6\columnwidth]{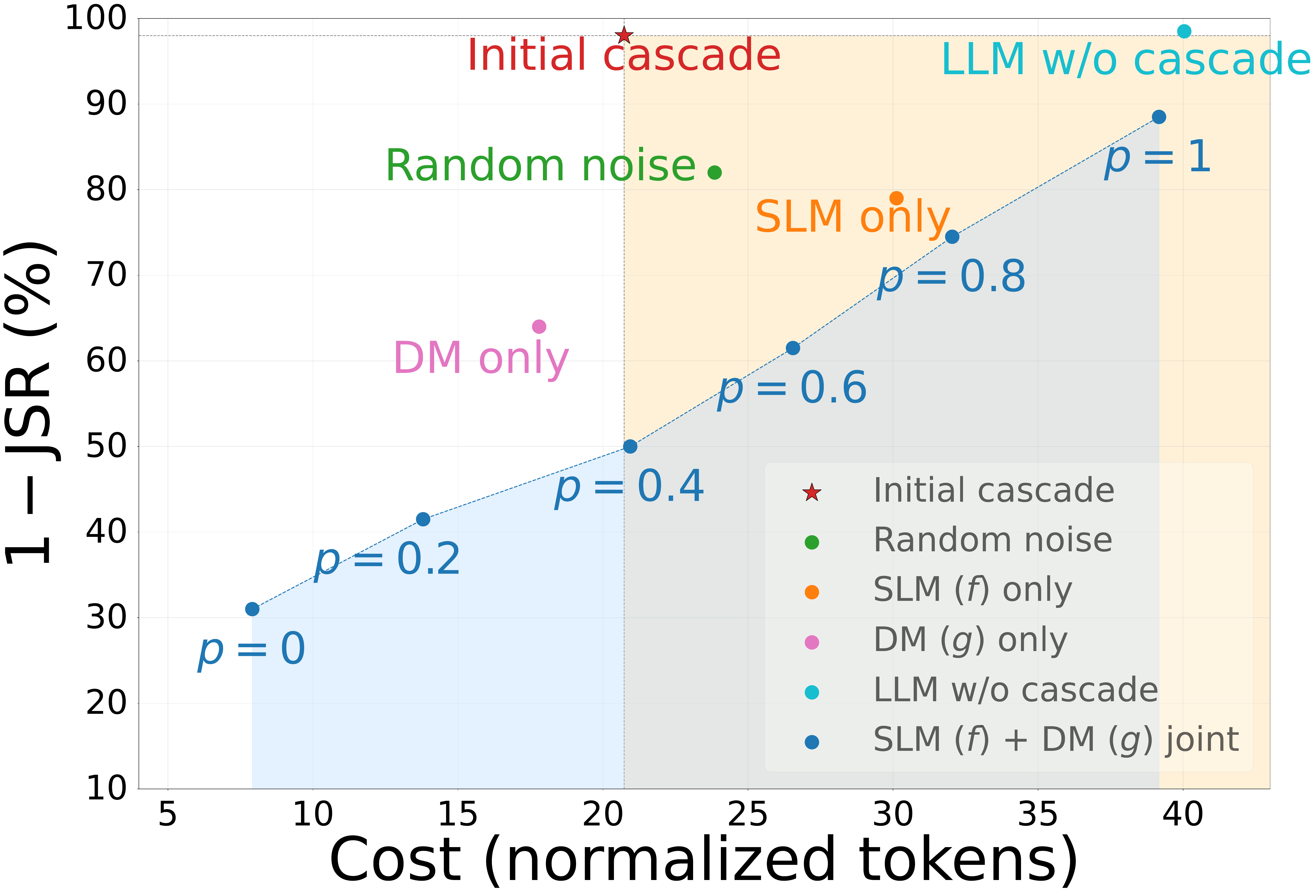}
  }
  \subfigure[AGnews (time)]{
    \includegraphics[width=0.6\columnwidth]{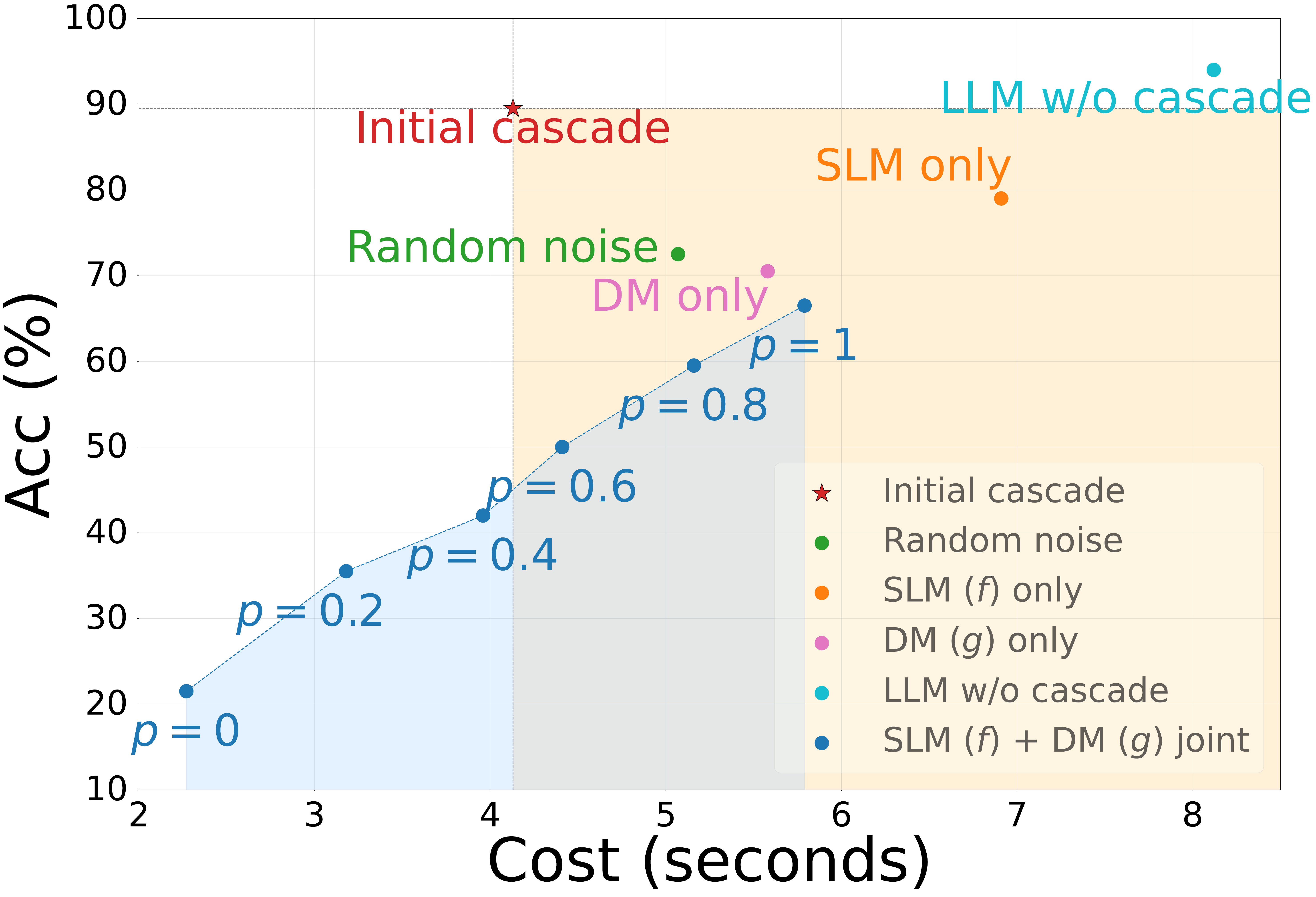}
  }
      \subfigure[SQuAD2.0 (time)]{
    \includegraphics[width=0.6\columnwidth]{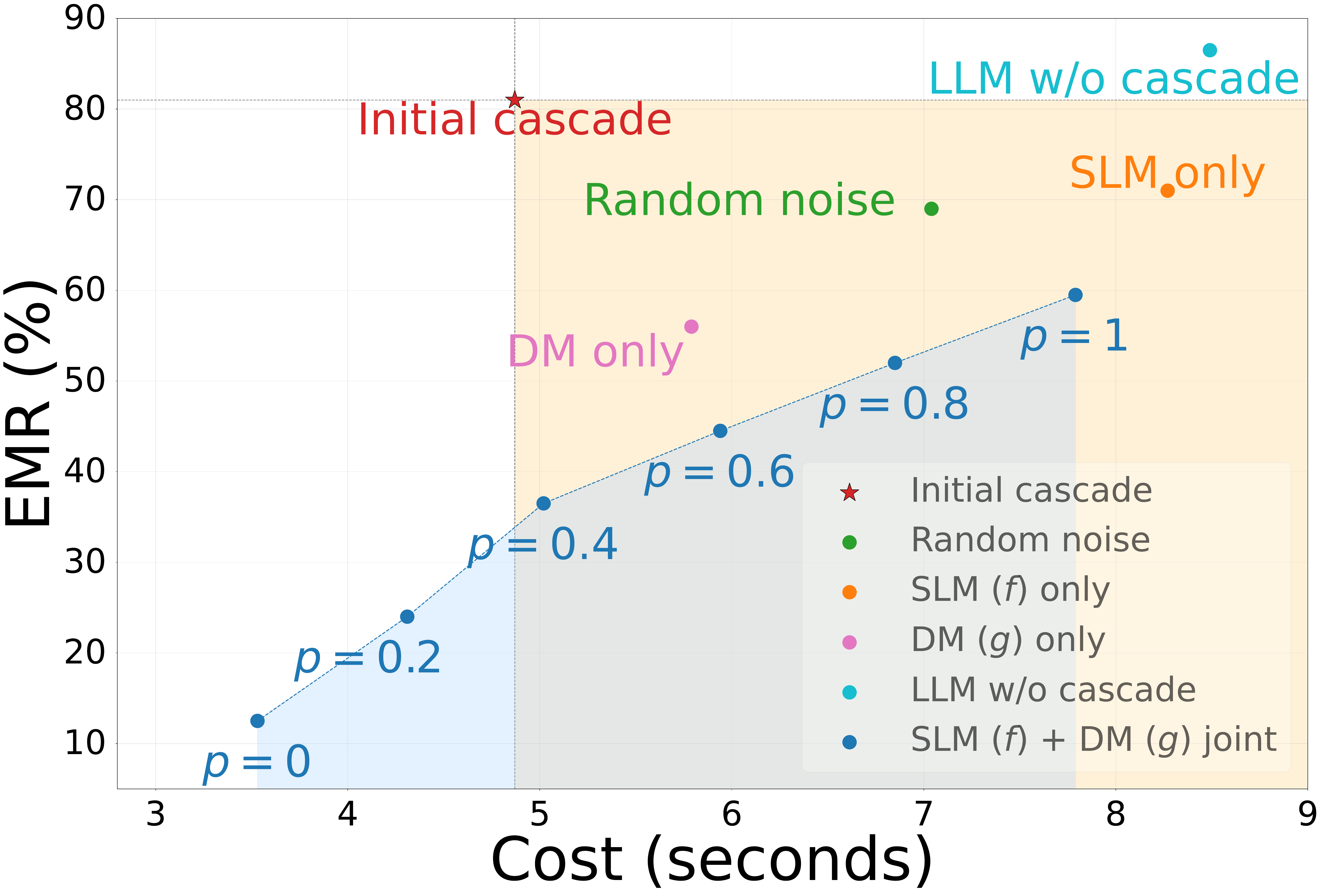}
  }
  \subfigure[WildJailbreak (time)]{
    \includegraphics[width=0.6\columnwidth]{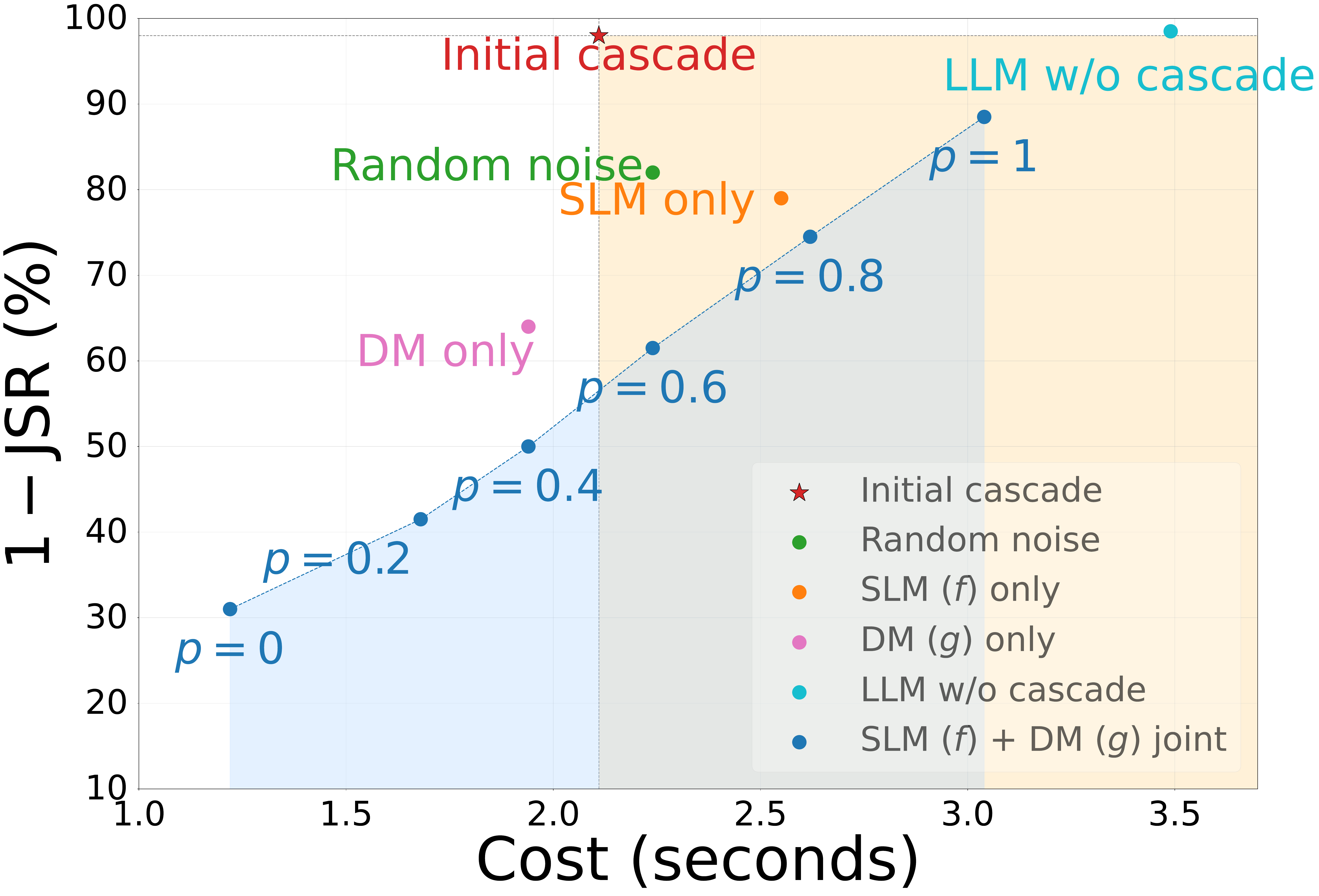}
  }
  \vspace{-6pt}
  \caption{\normalfont\bfseries Comparison of attack methods across the AGNews, SQuAD2.0, and WildJailbreak datasets on a \textit{three-layer} LLM cascade targeting the \textit{first stage} (Configuration~(24) in \tableautorefname~\ref{tab:cascade-configs-2}) . The cost–accuracy trade-off achieved by our pass-rate control method (introduced in \sectionautorefname~\ref{subsubsec:pass_rate}) is shown as the {\color[HTML]{1F77B4}blue} dashed line. Points toward the lower-right represent more effective attacks}
  \label{fig:3layer_results_block1}
  \vspace{6pt}
\end{figure*}

\begin{figure*}[thbp]  
  \centering         
  \includegraphics[width=1\textwidth]{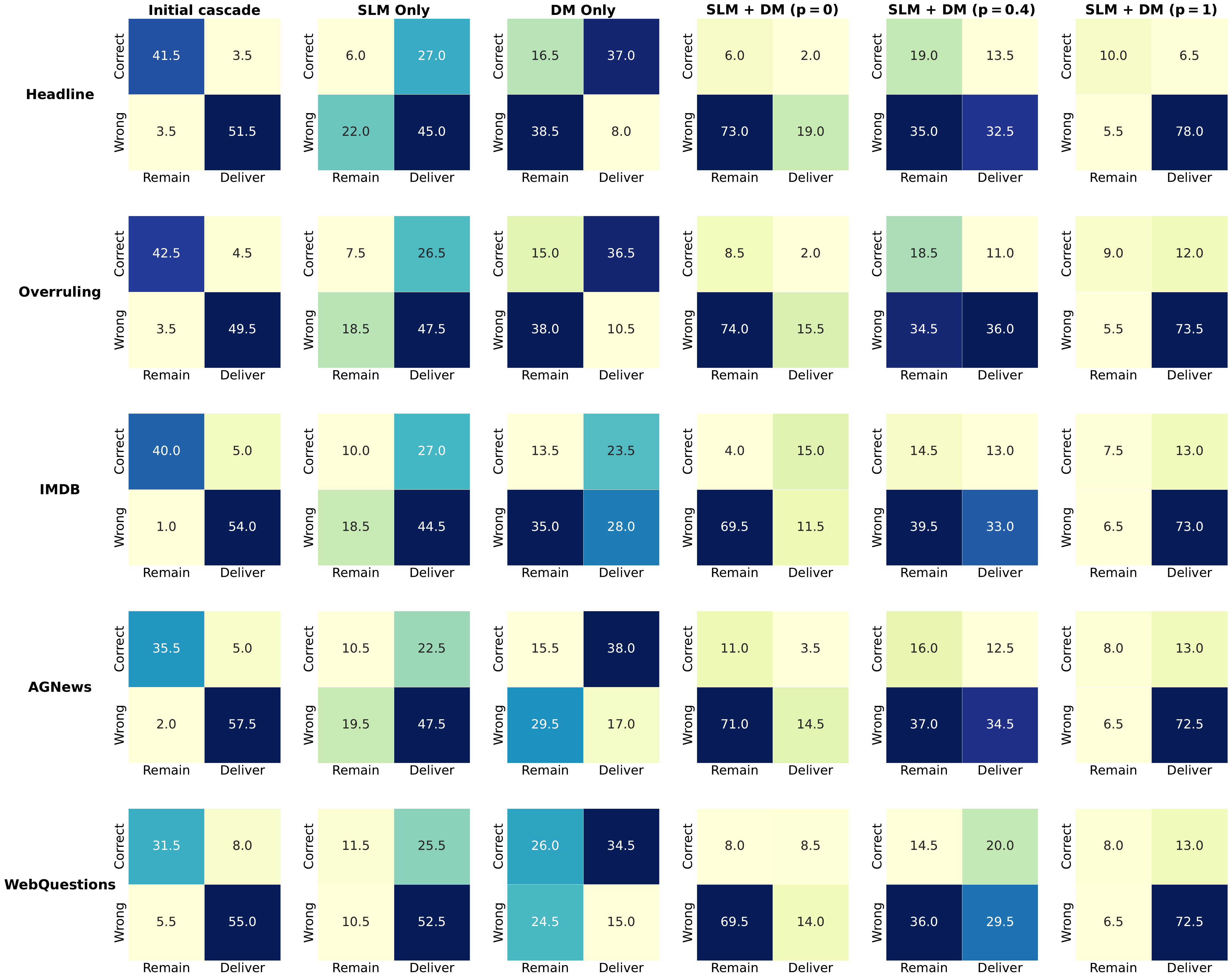}
  \caption{\textbf{Confusion matrices of the decision module (\texttt{DM})} for the GPT2-medium + BERT +Phi3-medium under the initial cascade and different attack configurations. }  
  \label{fig:confusion_matrix_gpt2}  
\end{figure*}

\begin{figure*}[thbp]  
  \centering          
  \includegraphics[width=1\textwidth]{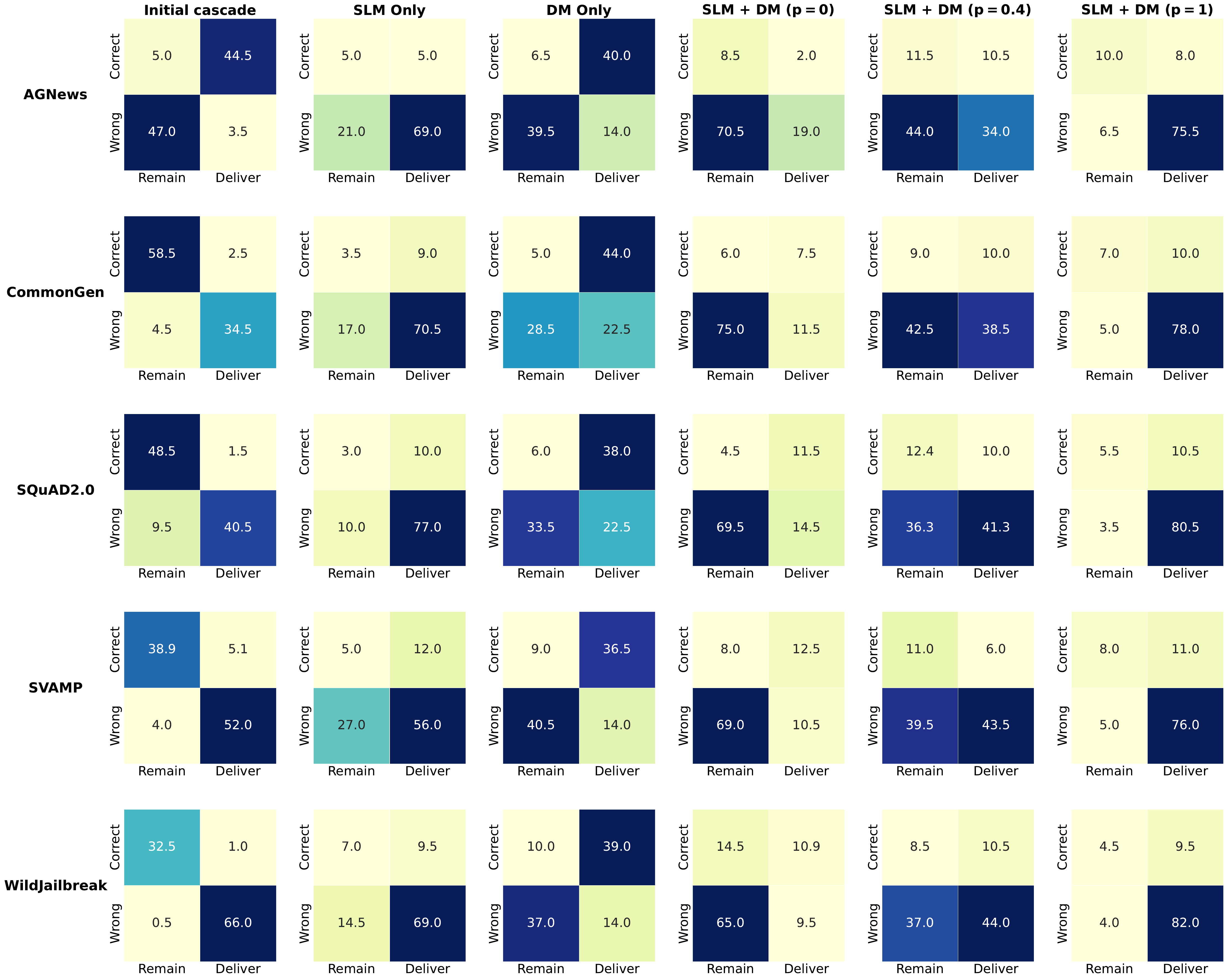}
  \caption{\textbf{Confusion matrices of the decision module (\texttt{DM})} for the wen2.5+DistilBERT+Phi-3.5-MoE under the initial cascade and different attack configurations. }  
  \label{fig:confusion_matrix_qwen}  
\end{figure*}

\begin{figure*}[thbp]
  \centering
  \includegraphics[width=1\textwidth]{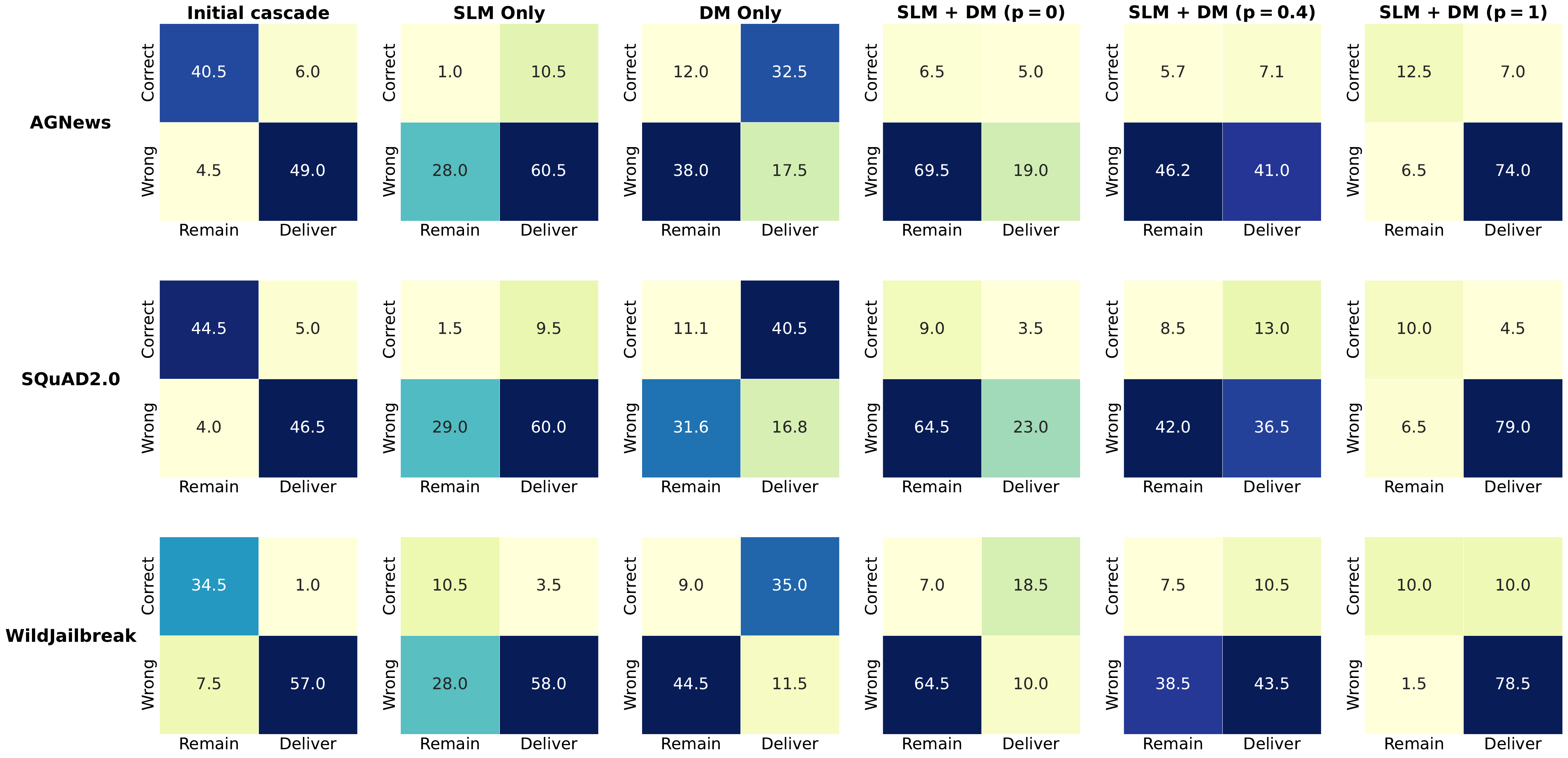}
  \caption{\textbf{When suffix optimization target is (\texttt{DM\_1}), the confusion matrices of the decision module (\texttt{DM\_1}) for the Qwen2.5 + BERT-base + Mistral-7B + BERT-base + Phi-3.5-MoE cascade system under the initial cascade and different attack configurations.}}
  \label{fig:appendix_confusion_matrix_3layers_block1}
  \vspace{30pt}
\end{figure*}

We present additional white-box results for two-layer cascades in Figures~\ref{fig:2layer_results_2} and Figures~\ref{fig:2layer_results_3}, with the heat map for the specific query transmission situation of the decision maker \tableautorefname{~\ref{fig:confusion_matrix_gpt2}} and \tableautorefname{~\ref{fig:confusion_matrix_qwen}}. The corresponding complete experimental results of two-layer cascades reported in Tables~\ref{tab:gpt2_bert_phi3_results} and \ref{tab:qwen_distilbert_phi35_results} respectively. We further provide additional white-box results for three-layer cascades in Figure~\ref{fig:3layer_results_block1}, with the heat map for the specific query transmission situation of the decision maker \tableautorefname{~\ref{fig:appendix_confusion_matrix_3layers_block1}} and \tableautorefname{~\ref{fig:confusion_matrix_3layers_block2}}. The corresponding complete experimental results of three-layer cascades reported in Tables~\ref{tab:qwen_mistral_phi35_block1_results} and \ref{tab:qwen_mistral_phi35_block2_results}. These results complement Figures~\ref{fig:2layer_results_1} and \ref{fig:3layer_results_block2} in the main paper.

These results are consistent with the main findings: all attack configurations remain effective across different datasets and cascade setups. Notably, our joint attack can drive the cascade system to extremely poor performance, often substantially worse than using a single LLM alone. This further highlights the vulnerability of cascaded architectures under adversarial manipulation.

\subsection{Black-box setting}
\label{appendix:black_box}
We present additional results of black-box attack in following Table~\ref{tab:transfer_dm}-\ref{tab:transfer_slm}. The suffix is optimized on the initial cascade with gradient and the target of optimization is making weak LLM directly output the answer for every query. We observe that our attack architecture for generating adversarial suffixes exhibits good transferability in a two-layer cascade model. For replacing only the Decision-maker, it reduced Acc by up to 64.2\%; for replacing only the Weak LLM, it reduced Acc by up to 52.2\%; and for replacing both the Weak LLM and the Decision-maker, it reduced Acc by up to 54.9\%. Furthermore, we tested its attack performance transferability in a three-layer cascade model. As a structural extension of the two-layer model, the three-layer cascade reduced Acc by up to 83.0\%, even achieving better attack results than attacking any two-layer model. This indicates that as the number of layers increases, the cascade structure introduces more vulnerabilities and attack surfaces.
\begin{table}[!h]
\centering
\aboverulesep=0ex
\belowrulesep=0ex
\resizebox{\columnwidth}{!}{
\begin{tabular}{cccccccc}
\toprule
\multirow{2}{*}{Setting} 
& \multicolumn{2}{c}{Cascade ($\Phi$)} 
& \multicolumn{2}{c}{SLM ($f^1$)} 
& \multicolumn{2}{c}{LLM ($f^2$)} 
& DM ($g^1$)\\
& Acc & Cost 
& Acc & Cost 
& Acc & Cost 
& Acc  \\ 
\midrule
Initial cascade (1)
& 78.0\% & 10.37 
& 48.0\% & 0.51
& 76.0\% & 25.19
& 86.5\% 
\\

Attack $p=0$ (1)$\rightarrow$ (1) 
& 21.0\% & 9.30
& 16.0\% & 0.63 
& 63.0\% & 25.47 
& 18.0\% 
\\

\cc $\Delta$ & \cc -73.1\%		
& \cc -10.3\%	
& \cc -66.7\%
& \cc +23.5\%	
& \cc -17.1\%	
& \cc +1.1\%			
& \cc -79.2\%
\\
\hdashline
Initial cascade (8)      & 85.0\%  & 41.2  & 71.0\%  & 3.4   & 91.0\%   & 104.1 & 91.5\%     \\
Attack $p=0$ (8) $\rightarrow$(8) & 52.5\%  & 18.4 & 50.0\% & 3.4    & 84.0\% & 104.5    & 19.0\%  \\
\cc $\Delta$  & \cc -41.0\% & \cc -55.3\% & \cc -29.6\% & \cc - & \cc  -7.7\% & \cc +0.4\% & \cc -79.2\%  \\
\hdashline
\\[-0.85em]

Initial cascade (2)
& 79.5\% & 13.5 
& 48.0\% & 0.6
& 76.0\% & 25.2 
& 88.0\% \\

Attack $p=0$ (1) $\rightarrow$ (2) 
& 28.5\% & 8.6
& 16.0\% & 0.6  
& 63.0\% & 28.6 
& 32.5\% \\

\cc $\Delta$ & \cc -64.2\% & \cc -36.3\% & \cc -66.7\% & \cc - & \cc -17.1\% & \cc +13.5\% & \cc -63.1\% \\

\midrule

Initial cascade (3)
& 76.0\% & 10.8 
& 48.0\% & 0.6
& 76.0\% & 25.2
& 84.0\% \\

Attack $p=0$ (1) $\rightarrow$ (3)
& 43.0\% & 13.7 
& 16.0\% & 0.6  
& 63.0\% & 28.6 
& 55.0\% \\

\cc $\Delta$ & \cc -43.4\% & \cc +26.9\% & \cc -66.7\% & \cc - & \cc -17.1\% & \cc +13.5\% & \cc -34.5\% \\

\midrule

Initial cascade (4)
& 75.0\% & 12.1 
& 48.0\% & 0.6 
& 76.0\% & 25.2 
& 85.5\% \\

Attack $p=0$  (1) $\rightarrow$ (4)
& 52.0\% & 19.2 
& 16.0\% & 0.6  
& 63.0\% & 28.6 
& 67.0\% \\

\cc $\Delta$ & \cc -30.7\% & \cc +58.7\% & \cc -66.7\% & \cc - & \cc -17.1\% & \cc +13.5\% & \cc -21.6\% \\

\midrule

Initial cascade (9)
& 86.0\% & 42.1 
& 71.0\%  & 3.4   & 91.0\%   & 104.1
& 91.0\% \\

Attack $p=0$ (8) $\rightarrow$ (9)
& 59.0\% & 28.2 
& 50.0\% & 3.4    & 84.0\% & 104.5 
& 46.0\% \\

\cc $\Delta$ & \cc -31.4\% & \cc -33.0\% & \cc -29.6\% & \cc - & \cc -7.7\% & \cc +0.4\% & \cc -49.5\% \\

\midrule

Initial cascade (10)
& 81.0\% & 40.8 
& 71.0\%  & 3.4   & 91.0\%   & 104.1
& 85.5\% \\

Attack $p=0$ (8) $\rightarrow$ (10)
& 61.0\% & 32.9 
& 50.0\% & 3.4    & 84.0\% & 104.5 
& 50.5\% \\

\cc $\Delta$ & \cc -24.7\% & \cc -19.4\% & \cc -29.6\% & \cc - & \cc -7.7\% & \cc +0.4\% & \cc -40.9\% \\

\midrule

Initial cascade (11)
& 86.5\% & 41.9 
& 71.0\%  & 3.4   & 91.0\%   & 104.1
& 89.0\% \\

Attack $p=0$ (8) $\rightarrow$ (11)
& 67.0\% & 37.3
& 50.0\% & 3.4    & 84.0\% & 104.5  
& 64.0\% \\

\cc $\Delta$ & \cc -22.5\% & \cc -11.0\% & \cc -29.6\% & \cc - & \cc -7.7\% & \cc +0.4\% & \cc -28.1\% \\
\bottomrule
\end{tabular}%
 }
 \vspace{1pt}
\caption{Transferability of attack performance across different decision modules \texttt{DM}. Cost is measured as the \textit{average normalized token cost}} 
\label{tab:transfer_dm}
\end{table}

\begin{table}[!h]
\resizebox{\columnwidth}{!}{
\begin{tabular}{cccccccc}
\toprule
\multirow{2}{*}{Setting} 
& \multicolumn{2}{c}{Cascade ($\Phi$)} 
& \multicolumn{2}{c}{SLM ($f^1$)} 
& \multicolumn{2}{c}{LLM ($f^2$)} 
& DM ($g^1$)\\
& Acc & Cost 
& Acc & Cost 
& Acc & Cost 
& Acc  \\ 
\midrule
Initial cascade (1)
& 78.0\% & 10.37 
& 48.0\% & 0.51
& 76.0\% & 25.19
& 86.5\% 
\\

Attack $p=0$ (1)$\rightarrow$ (1) 
& 21.0\% & 9.30
& 16.0\% & 0.63 
& 63.0\% & 25.47 
& 18.0\% 
\\

\cc $\Delta$ & \cc -73.1\%		
& \cc -10.3\%	
& \cc -66.7\%
& \cc +23.5\%	
& \cc -17.1\%	
& \cc +1.1\%			
& \cc -79.2\%
\\
\hdashline
Initial cascade (8)      & 85.0\%  & 41.2  & 71.0\%  & 3.4   & 91.0\%   & 104.1 & 91.5\%     \\
Attack $p=0$ (8) $\rightarrow$(8) & 52.5\%  & 18.4 & 50.0\% & 3.4    & 84.0\% & 104.5    & 19.0\%  \\
\cc $\Delta$  & \cc -41.0\% & \cc -55.3\% & \cc -29.6\% & \cc - & \cc  -7.7\% & \cc +0.4\% & \cc -79.2\%  \\
\hdashline
\\[-0.85em]

Initial cascade (5)      & 76.5\%  & 17.1  & 39.0\%  & 0.7   & 76.0\%  & 25.2 & 88.0\%  \\
Attack $p=0$ (1) $\rightarrow$(5) & 54.5\%  & 17.5  & 35.5\%  & 0.8   & 63.0\%  & 28.6 & 58.5\%  \\
\cc $\Delta$  & \cc -28.8\% &  \cc +2.3\% & \cc -9.0\%  & \cc +14.3\% & \cc -17.1\% & \cc +13.5\% & \cc -33.5\% \\
\midrule
Initial cascade (6)       & 76.0\%  & 11.5   & 67.0\%  & 0.9   & 76.0\%  & 25.2 & 79.0\%  \\
Attack $p=0$ (1)$\rightarrow$(6) & 43.0\%  & 13.0    & 29.0\%  & 1.0   & 63.0\%  & 28.6 & 51.5\%  \\
\cc $\Delta$  & \cc -43.4\% & \cc +13.0\% & \cc -56.7\% & \cc +11.1\% & \cc -17.1\% & \cc +13.5\% & \cc -34.8\% \\
\midrule
Initial cascade (7)       & 79.5\%  & 14.5   & 54.5\%  & 0.7   & 76.0\%  & 25.2 & 89.5\%  \\
Attack $p=0$ (1)$\rightarrow$(7) & 38.0\%  & 4.5    & 34.5\%  & 0.8    & 63.0\%  & 28.6 & 42.5\%  \\
\cc $\Delta$  & \cc -52.2\% & \cc -69.0\% & \cc -36.7\% & \cc +14.3\% & \cc -17.1\% & \cc +13.5\% & \cc -52.5\% \\
\midrule
Initial cascade (12)       & 86.0\%  & 40.5   & 39.0\%  & 0.7   & 91.0\%   & 104.1 & 86.0\%  \\
Attack $p=0$ (8)$\rightarrow$(12) & 60.0\%  & 27.4   & 35.5\%  & 0.8   & 84.0\% & 104.5  & 69.5\%  \\
\cc $\Delta$  & \cc -30.2\% & \cc -32.3\% & \cc -9.0\% & \cc +14.3\% & \cc -7.7\% & \cc +0.4\% & \cc -19.2\% \\
\midrule
Initial cascade (13)       & 84.0\%  & 37.2   & 67.0\%  & 0.9   & 91.0\%   & 104.1 & 92.0\%  \\
Attack $p=0$ (8)$\rightarrow$(13) & 58.0\%  & 22.6    & 36.0\%  & 1.1   & 84.0\% & 104.5  & 55.5\%  \\
\cc $\Delta$  & \cc -31.0\% & \cc -39.2\% & \cc -46.3\% & \cc +22.2\% & \cc -7.7\% & \cc +0.4\% & \cc -39.7\% \\
\midrule
Initial cascade (14)       & 83.0\%  & 39.8   & 54.5\%  & 0.6   & 91.0\%   & 104.1 & 88.0\%  \\
Attack $p=0$ (8)$\rightarrow$(14) & 56.5\%  & 21.2    & 34.5\%  & 0.7   & 84.0\% & 104.5  & 53.5\%  \\
\cc $\Delta$  & \cc -31.9\% & \cc -46.7\% & \cc -36.7\% & \cc +16.7\% & \cc -7.7\% & \cc +0.4\% & \cc -39.2\% \\

\bottomrule
\end{tabular}}
\vspace{1pt}
\caption{Transferability of attack performance across different small language models \texttt{SLM}. Cost is measured as the \textit{average normalized token cost}} 
\label{tab:transfer_slm}
\end{table}

\begin{table}[!h]
\resizebox{\columnwidth}{!}{
\begin{tabular}{ccccccccccc}
\toprule
\multirow{2}{*}{Setting} 
& \multicolumn{2}{c}{Cascade ($\Phi$)} 
& \multicolumn{2}{c}{SLM ($f^1$)} 
& \multicolumn{2}{c}{SLM ($f^2$)} 
& \multicolumn{2}{c}{LLM ($f^3$)} 
& DM ($g^1$)
& DM ($g^2$)\\

& Acc & Cost 
& Acc & Cost 
& Acc & Cost 
& Acc & Cost 
& Acc
& Acc  \\ 
\midrule
Initial cascade (1)
& 78.0\% & 10.37 
& 48.0\% & 0.51
& N/A & N/A
& 76.0\% & 25.19
& 86.5\% 
& N/A\\

Attack $p=0$ (1)$\rightarrow$ (1) 
& 21.0\% & 9.30
& 16.0\% & 0.63 
& N/A & N/A
& 63.0\% & 25.47 
& 18.0\% 
& N/A\\

\cc $\Delta$ & \cc -73.1\%		
& \cc -10.3\%	
& \cc -66.7\%
& \cc +23.5\%	
& \cc N/A 
& \cc N/A
& \cc -17.1\%	
& \cc +1.1\%			
& \cc -79.2\%
& \cc N/A 

\\
\hdashline
Initial cascade (8)      & 89.0\%  & 41.2  & 71.0\%  & 3.4   & N/A  & N/A  & 91.0\%   & 104.1 & 91.5\%  & N/A   \\
Attack $p=0$ (8) $\rightarrow$(8) & 52.5\%  & 18.4 & 50.0\% & 3.4    & N/A  & N/A & 84.0\% & 104.5    & 19.0\% & N/A  \\
\cc $\Delta$  & \cc -41.0\% & \cc -55.3\% & \cc -29.6\% & \cc - & \cc N/A & \cc N/A & \cc -7.7\% & \cc +0.4\% & \cc -79.2\% & \cc N/A \\
\hdashline
\\[-0.85em]
Initial cascade (24)      & 87.5\%  & 21.4  & 71.0\%  & 3.4   & 84.5\%  & 13.7  & 91.0\%   & 104.1  & 88.0\%  & 90.0\%   \\
Attack $p=0$ (8) $\rightarrow$(24) & 67.5\%  & 12.6 & 50.0\% & 3.4    & 78.5\%  & 13.8 & 84.0\% & 104.5    & 22.5\% & 65.0\%  \\
\cc $\Delta$  & \cc -22.9\% & \cc -41.1\% & \cc -29.6\% & \cc - & \cc -7.1\% & \cc +0.7\% & \cc -7.7\% & \cc +0.4\% & \cc -74.4\% & \cc -27.8\% \\
\midrule
Initial cascade (25)      & 88.0\%  & 26.8  & 71.0\%  & 3.4   & 84.5\%  & 13.7  & 91.0\%   & 104.1  & 87.0\%  & 84.5\%   \\
Attack $p=0$ (8) $\rightarrow$(25) & 69.5\%  & 14.9 & 50.0\% & 3.4    & 78.5\%  & 13.8 & 84.0\% & 104.7    & 28.0\% & 63.0\%  \\
\cc $\Delta$  & \cc -21.0\% & \cc -44.4\% & \cc -29.6\% & \cc - & \cc -7.1\% & \cc +0.7\% & \cc -7.7\% & \cc +0.6\% & \cc -67.8\% & \cc -25.4\% \\
\midrule
Initial cascade (26)      & 77.0\%  & 12.5  & 48\%  & 0.6   & 54.5\%  & 2.0   & 76.0\%  & 25.2  & 81.0\%  & 83.5\%   \\
Attack $p=0$ (1) $\rightarrow$(26) & 30.5\%  & 8.0   & 16.0\% & 0.6   & 22.5\%  & 2.1   & 63.0\%  & 25.6  & 18.0\%  & 43.0\%  \\
\cc $\Delta$  & \cc -60.4\% & \cc -36.0\% & \cc -66.7\% & \cc - & \cc -58.7\% & \cc +5.0\% & \cc -17.1\% & \cc +1.6\% & \cc -77.8\% & \cc -48.5\% \\
\midrule
Initial cascade (27)      & 76.5\%  & 12.0  & 48\%  & 0.6   & 54.5\%  & 2.0   & 73.0\%  & 25.2  & 82.0\%  & 85.5\%  \\
Attack $p=0$ (1)$\rightarrow$(27) & 13.0\%  & 4.8   & 16.0\% & 0.6   & 22.5\%  & 2.1   & 63.0\%  & 25.6  & 13.5\%  & 27.5\% \\
\cc $\Delta$  & \cc -83.0\% & \cc -60.0\% & \cc -66.7\% & \cc - & \cc -58.7\% & \cc +5.0\% & \cc -13.7\% & \cc +1.6\% & \cc -83.5\% & \cc -67.8\% \\
\midrule
Initial cascade (28)      & 78.0\%  & 11.4  & 39.0\% & 0.7   & 72.5\%  & 4.3   & 84.0\%  & 49.8  & 76.5\%  & 83.5\%  \\
Attack $p=0$ (1)$\rightarrow$(28) & 61.0\%  & 9.0   & 34.0\% & 0.8   & 61.5\%  & 4.4   & 73.0\%  & 50.7  & 48.5\%  & 67.0\%  \\
\cc $\Delta$  & \cc -21.8\% & \cc -21.1\% & \cc -12.8\% & \cc +14.3\% & \cc -15.2\% & \cc +2.3\% & \cc -13.1\% & \cc +1.8\% & \cc -36.6\% & \cc -19.8\% \\
\midrule
Initial cascade (29)      & 77.0\%  & 11.4  & 39.0\% & 0.7   & 72.5\%  & 4.3   & 84.0\%  & 49.8  & 80.5\%  & 87.5\% \\
Attack $p=0$ (1)$\rightarrow$(29) & 59.5\%  & 10.1  & 34.0\% & 0.8   & 61.5\%  & 4.4   & 73.0\%  & 50.7  & 44.5\%  & 61.0\%  \\
\cc $\Delta$  & \cc -22.7\% & \cc -11.4\% & \cc -12.8\% & \cc +14.3\% & \cc -15.2\% & \cc +2.3\% & \cc -13.1\% & \cc +1.8\% & \cc -44.7\% & \cc -30.3\% \\

\bottomrule
\end{tabular}
}
\vspace{1pt}
\caption{Transferability of attack performance across 3-layers cascade with different open-source language models. Cost is measured as the \textit{average normalized token cost}} 
\label{tab:transfer_slm}
\vspace{1em}
\end{table}

\begin{table}[!h]
\resizebox{\columnwidth}{!}{
\begin{tabular}{ccccccccccc}
\toprule
\multirow{2}{*}{Setting} 
& \multicolumn{2}{c}{Cascade ($\Phi$)} 
& \multicolumn{2}{c}{SLM ($f^1$)} 
& \multicolumn{2}{c}{SLM ($f^2$)} 
& \multicolumn{2}{c}{LLM ($f^3$)} 
& DM ($g^1$)
& DM ($g^2$)\\

& Acc & Cost 
& Acc & Cost 
& Acc & Cost 
& Acc & Cost 
& Acc
& Acc  \\ 
\midrule
Initial cascade (1)
& 78.0\% & 10.37 
& 48.0\% & 0.51
& N/A & N/A
& 76.0\% & 25.19
& 86.5\% 
& N/A\\

Attack $p=0$ (1)$\rightarrow$ (1) 
& 21.0\% & 9.30
& 16.0\% & 0.63 
& N/A & N/A
& 63.0\% & 25.47 
& 18.0\% 
& N/A\\

\cc $\Delta$ & \cc -73.1\%		
& \cc -10.3\%	
& \cc -66.7\%
& \cc +23.5\%	
& \cc N/A 
& \cc N/A
& \cc -17.1\%	
& \cc +1.1\%			
& \cc -79.2\%
& \cc N/A 
\\
\hdashline
Initial cascade (8)      & 89.0\%  & 41.2  & 71.0\%  & 3.4   & N/A  & N/A  & 91.0\%   & 104.1 & 91.5\%  & N/A   \\
Attack $p=0$ (8) $\rightarrow$(8) & 52.5\%  & 18.4 & 50.0\% & 3.4    & N/A  & N/A & 84.0\% & 104.5    & 19.0\% & N/A  \\
\cc $\Delta$  & \cc -41.0\% & \cc -55.3\% & \cc -29.6\% & \cc - & \cc N/A & \cc N/A & \cc -7.7\% & \cc +0.4\% & \cc -79.2\% & \cc N/A \\
\hdashline
\\[-0.85em]
Initial cascade (30)       & 94.0\%  & \$ 0.49  & 82.5\%  & \$ 0.27 & 87.5\%  & \$ 0.30  & 94.5\%  & \$ 1.37 & 96.5\% & 98.0\% \\
Attack $p=0$ (8)$\rightarrow$(30) & 76.0\%  & \$ 0.44    & 71.0\%  & \$ 0.28 & 78.0\%  & \$ 0.31  & 89.0\%  & \$ 1.41  & 68.0\%  &81.0\%  \\
\\[-0.85em]
\cc $\Delta$  & \cc -19.1\% & \cc -10.2\% & \cc -13.9\% & \cc +3.7\% & \cc -10.9\% & \cc +3.3\% & \cc -5.8\% & \cc +2.9\% & \cc -29.5\% & \cc -17.3\% \\
\midrule
Initial cascade (31)       & 93.0\%  & \$ 0.49   & 82.5\%  & \$ 0.27 & 87.5\%  & \$ 0.30  & 94.5\%  & \$ 1.37 & 95.0\% & 94.0\% \\
Attack $p=0$ (8)$\rightarrow$(31) & 72.0\%  & \$ 0.40    & 71.0\%  & \$ 0.28 & 78.0\%  & \$ 0.31  & 89.0\%  & \$ 1.41  & 62.0\%  &85.5\%  \\
\cc $\Delta$  & \cc -22.6\% & \cc -18.4\% & \cc -13.9\% & \cc +3.7\% & \cc -10.9\% & \cc +3.3\% & \cc -5.8\% & \cc +2.9\% & \cc -34.7\% & \cc -9.0\% \\
\midrule
Initial cascade (30)       & 94.0\%  & \$ 0.49  & 82.5\%  & \$ 0.27 & 87.5\%  & \$ 0.30  & 94.5\%  & \$ 1.37 & 96.5\% & 98.0\% \\
Attack $p=0$ (1)$\rightarrow$(30) & 77.5\%  & \$ 0.42    & 75.0\%  & \$ 0.29 & 79.5\%  & \$ 0.32  & 87.0\%  & \$ 1.42  & 71.0\%  &76.0\%  \\
\cc $\Delta$  & \cc -17.5\% & \cc -14.3\%   &\cc -9.1\% &\cc +7.4\% &\cc -9.1\% &\cc +6.7\% &\cc -7.9\% &\cc +3.6\% &\cc -26.4\% &\cc -22.4\% \\
\midrule
Initial cascade (31)       & 93.0\%  & \$ 0.49   & 82.5\%  & \$ 0.27 & 87.5\%  & \$ 0.30  & 94.5\%  & \$ 1.37 & 95.0\% & 94.0\% \\
Attack $p=0$ (1)$\rightarrow$(31) & 72.0\%  & \$ 0.40    & 71.0\%  & \$ 0.28 & 78.0\%  & \$ 0.31  & 89.0\%  & \$ 1.41  & 62.0\%  &85.5\%  \\
\cc $\Delta$  & \cc -20.4\% & \cc -16.3\%   &\cc -9.1\% &\cc +7.4\% &\cc -9.1\% &\cc +6.7\% &\cc -7.9\% &\cc +3.6\% &\cc -27.4\% &\cc -17.6\% \\
\midrule
Initial cascade (32)       & 92.0\%  & \$ 0.24   & 80.0\%  & \$ 0.092   & 89.0\%  & \$ 0.19  & 95.0\%  & \$ 0.46 & 92.0\% & 89.5\% \\
Attack $p=0$ (1)$\rightarrow$(32) &  72.5\%  & \$ 0.18    & 72.0\%  & \$ 0.097  & 77.5\%  & \$ 0.20  & 90.0\%  & \$ 0.50  & 75.5\%  & 80.0\%  \\
\cc $\Delta$  & \cc -21.2\% & \cc -25.0\%   &\cc -10.0\% &\cc +5.4\% &\cc -12.9\% &\cc +5.3\% &\cc -5.3\% &\cc -8.7\% &\cc -17.9\% &\cc -10.6\% \\
\midrule
Initial cascade (33)     & 93.5\%  & \$ 0.31   & 80.0\%  & \$ 0.092   & 89.0\%  & \$ 0.19  & 95.0\%  & \$ 0.46 & 94.0\% & 95.5\% \\
Attack $p=0$ (1)$\rightarrow$(33) &  70.5\%  & \$ 0.17    & 72.0\%  & \$ 0.097  & 77.5\%  & \$ 0.20  & 90.0\%  & \$ 0.50  & 71.0\%  & 74.0\%  \\
\cc $\Delta$  & \cc -24.6\% & \cc -45.2\%   &\cc -10.0\% &\cc +5.4\% &\cc -12.9\% &\cc +5.3\% &\cc -5.3\% &\cc -8.7\% &\cc -24.5\% &\cc -22.5\% \\
\bottomrule
\end{tabular}
}
\vspace{12pt}
\caption{\textbf{Attack performance of 3-layers cascade when transferred across different language models by API. For (1) and (16), the cost is calculated by averaging the number of tokens per query. For (30) to (33), the cost is directly calculated based on the money spent on using the API}.} 
\label{tab:transfer_slm}
\end{table}

\begin{table}[!h]
\resizebox{\columnwidth}{!}{
\begin{tabular}{ccccccccccc}
\toprule
\multirow{2}{*}{Setting} 
& \multicolumn{2}{c}{Cascade ($\Phi$)} 
& \multicolumn{2}{c}{SLM ($f^1$)} 
& \multicolumn{2}{c}{SLM ($f^2$)} 
& \multicolumn{2}{c}{LLM ($f^3$)} 
& DM ($g^1$)
& DM ($g^2$)\\

& Acc & Cost 
& Acc & Cost 
& Acc & Cost 
& Acc & Cost 
& Acc
& Acc  \\ 
\midrule
Initial cascade (1)
& 78.0\% & 2771.9 
& 48.0\% & 114.0
& N/A & N/A
& 76.0\% & 5037.3
& 86.5\% 
& N/A\\

Attack $p=0$ (1)$\rightarrow$ (1) 
& 21.0\% & 2006.2
& 16.0\% & 125.5 
& N/A & N/A
& 63.0\% & 5728.5  
& 18.0\% 
& N/A\\

\cc $\Delta$ & \cc -73.1\%		
& \cc +27.6\%	
& \cc -66.7\%
& \cc +10.0\%	
& \cc N/A 
& \cc N/A
& \cc -17.1\%	
& \cc +13.7\%			
& \cc -79.2\%
& \cc N/A 

\\
\hdashline
\\[-0.85em]
Initial cascade (16)      & 77.0\%  & 2490.4  & 48\%  & 114.0   & 54.5\%  & 402.2  &76.0\%   & 5037.3  & 81.0\%  & 83.5\%   \\
Attack $p=0$ (1) $\rightarrow$(16) & 30.5\%  & 1608.3 & 16.0\% & 125.5    & 22.5\%  & 417.7 & 63.0\% & 5728.5    & 18.0\% & 43.0\%  \\
\cc $\Delta$  & \cc -60.4\% &  \cc -35.4\%     & \cc -66.7\%  &\cc +10.0\%	  &\cc -58.7\% & \cc +3.9\%  &\cc -17.1\% & \cc +13.7\%   &\cc -76.5\% &\cc -48.5\% \\
\midrule
Initial cascade (17)       & 76.5\%  & 2398.6  & 48\%  & 114.0   & 54.5\%  & 402.2  & 73.0\% & 5037.3    & 82.0\%  & 85.5\%  \\
Attack $p=0$ (1)$\rightarrow$(17) & 13.0\%  & 962.6  & 16.0\% & 125.5    & 22.5\%  & 417.7 & 63.0\%  & 5728.5  & 13.5\% & 27.5\% \\
\cc $\Delta$  & \cc -83.0\% &\cc -59.9\%    & \cc -66.7\%  &\cc +10.0\%	  &\cc -58.7\% & \cc +3.9\%  &\cc -17.1\% & \cc +13.7\% &\cc -83.5\% &\cc -58.0\% \\
\midrule
Initial cascade (18)       & 78.0\%  & 2272.8   & 39.0\%  & 148.2 & 72.5\%  & 856.5  & 84.0\%  & 9967.4 & 76.5\% & 83.5\%  \\
Attack $p=0$ (1)$\rightarrow$(18) & 61.0\%  & 1794.3    & 34.0\%  & 154.6 & 61.5\%  & 878.7  & 73.0\%  & 10942.7 & 48.5\%  &67.0\%  \\
\cc $\Delta$  & \cc -21.8\% & \cc -21.1\%   &\cc -13.2\% &\cc +4.3\% &\cc -15.2\% &\cc +2.6\% &\cc -13.1\% &\cc +9.8\% &\cc -36.6\% &\cc -19.8\% \\
\midrule
Initial cascade (19)       & 77.0\%  & 2281.4   & 39.0\%  & 148.2 & 72.5\%  & 856.5  & 84.0\%  & 9967.4 & 80.5\% & 87.5\% \\
Attack $p=0$ (1)$\rightarrow$(19) &  59.5\%  & 2017.5    & 34.0\%  & 154.6 & 61.5\%  & 878.7  & 73.0\%  & 10942.7 & 44.5\%  &61.0\%  \\
\cc $\Delta$  & \cc -22.7\% & \cc -11.6\%   &\cc -13.2\% &\cc +4.3\% &\cc -15.2\% &\cc +2.6\% &\cc -13.1\% &\cc +9.8\% &\cc -44.7\% &\cc -30.3\% \\

\bottomrule
\end{tabular}
}
\vspace{12pt}
\caption{\textbf{Attack performance of 3-layers cascade when transferred across different open-source language models. Cost is measured as the \textit{average normalized token cost}}} 
\label{tab:transfer_slm}
\vspace{1em}
\end{table}

\begin{table}[!h]
\resizebox{\columnwidth}{!}{
\begin{tabular}{ccccccccccc}
\toprule
\multirow{2}{*}{Setting} 
& \multicolumn{2}{c}{Cascade ($\Phi$)} 
& \multicolumn{2}{c}{SLM ($f^1$)} 
& \multicolumn{2}{c}{SLM ($f^2$)} 
& \multicolumn{2}{c}{LLM ($f^3$)} 
& DM ($g^1$)
& DM ($g^2$)\\

& Acc & Cost 
& Acc & Cost 
& Acc & Cost 
& Acc & Cost 
& Acc
& Acc  \\ 
\midrule

Initial cascade (1)
& 78.0\% & 2771.9 
& 48.0\% & 114.0
& N/A & N/A
& 76.0\% & 5037.3
& 86.5\% 
& N/A\\

Attack $p=0$ (1)$\rightarrow$ (1) 
& 21.0\% & 2006.2
& 16.0\% & 125.5 
& N/A & N/A
& 63.0\% & 5728.5  
& 18.0\% 
& N/A\\

\cc $\Delta$ & \cc -73.1\%		
& \cc +27.6\%	
& \cc -66.7\%
& \cc +10.0\%	
& \cc N/A
& \cc N/A
& \cc -17.1\%	
& \cc +13.7\%			
& \cc -79.2\%
& \cc N/A 

\\
\hdashline
\\[-0.85em]
Initial cascade (20)       & 94.0\%  & \$ 0.49  & 82.5\%  & \$ 0.27 & 87.5\%  & \$ 0.30  & 94.5\%  & \$ 1.37 & 96.5\% & 98.0\% \\
Attack $p=0$ (1)$\rightarrow$(20) & 77.5\%  & \$ 0.42    & 75.0\%  & \$ 0.29 & 79.5\%  & \$ 0.32  & 87.0\%  & \$ 1.42  & 71.0\%  &76.0\%  \\
\cc $\Delta$  & \cc -17.5\% & \cc -14.3\%   &\cc -9.1\% &\cc +7.4\% &\cc -9.1\% &\cc +6.7\% &\cc -7.9\% &\cc +3.6\% &\cc -26.4\% &\cc -22.4\% \\
\midrule

Initial cascade (21)       & 93.0\%  & \$ 0.49   & 82.5\%  & \$ 0.27 & 87.5\%  & \$ 0.30  & 94.5\%  & \$ 1.37 & 95.0\% & 94.0\% \\
Attack $p=0$ (1)$\rightarrow$(21) & 74.0\%  & \$ 0.41    & 75.0\%  & \$ 0.29 & 79.5\%  & \$ 0.32  & 87.0\%  & \$ 1.42  & 69.0\%  &77.5\%  \\
\cc $\Delta$  & \cc -20.4\% & \cc -16.3\%   &\cc -9.1\% &\cc +7.4\% &\cc -9.1\% &\cc +6.7\% &\cc -7.9\% &\cc +3.6\% &\cc -27.4\% &\cc -17.6\% \\
\midrule

Initial cascade (22)       & 92.0\%  & \$ 0.24   & 80.0\%  & \$ 0.092   & 89.0\%  & \$ 0.19  & 95.0\%  & \$ 0.46 & 92.0\% & 89.5\% \\
Attack $p=0$ (1)$\rightarrow$(22) &  72.5\%  & \$ 0.18    & 72.0\%  & \$ 0.097  & 77.5\%  & \$ 0.20  & 90.0\%  & \$ 0.50  & 75.5\%  & 80.0\%  \\
\cc $\Delta$  & \cc -21.2\% & \cc -25.0\%   &\cc -10.0\% &\cc +5.4\% &\cc -12.9\% &\cc +5.3\% &\cc -5.3\% &\cc +8.7\% &\cc -17.9\% &\cc -10.6\% \\
\midrule
Initial cascade (23)     & 93.5\%  & \$ 0.31   & 80.0\%  & \$ 0.092   & 89.0\%  & \$ 0.19  & 95.0\%  & \$ 0.46 & 94.0\% & 95.5\% \\
Attack $p=0$ (1)$\rightarrow$(23) &  70.5\%  & \$ 0.17    & 72.0\%  & \$ 0.097  & 77.5\%  & \$ 0.20  & 90.0\%  & \$ 0.50  & 71.0\%  & 74.0\%  \\
\cc $\Delta$  & \cc -24.6\% & \cc -45.2\%   &\cc -10.0\% &\cc +5.4\% &\cc -12.9\% &\cc +5.3\% &\cc -5.3\% &\cc +8.7\% &\cc -24.5\% &\cc -22.5\% \\
\bottomrule
\end{tabular}
}
\vspace{12pt}
\caption{\textbf{Attack performance of 3-layers cascade when transferred across different language models by API }.} 
\label{tab:transfer_slm}
\end{table}

In the black box scenario, we further replaced all language models except for the decision maker with commercial closed source models, called them in API form, and constructed the  new cascade structure with better task performance to test our attack performance in generating adversarial suffixes. The task cost is measured by the dollars spent directly using API calls.
We conducted migration testing using an adversarial suffix of P=0 testing for three kinds of decision makers BERT, GPT-2, DistilBERT and found that the overall Acc of the cascade system decreased by up to 20.7\% under a two-layer structure. The performance of SLM alone decreased by up to 12.9\%, and the decision accuracy of DM decreased by up to 25.9\%. It can be observed that the adversarial suffix we constructed has slightly lower attack performance in closed source models compared to open source models, but still has significant effects, indicating that the inherent vulnerability of the cascade structure will not be resolved by the improvement of model components

\subsection{Alternative Optimization backbone}
\label{appendix:gray_box}
We additionally consider alternative optimization backbones beyond the full gradient-based GCG method, focusing on a more realistic \textit{gray-box} setting where the adversary has limited internal knowledge. In this scenario, the attacker may know the general cascade structure or routing algorithm, but only has black-box query access to individual models (e.g., via APIs) or partial routing outputs. This reflects practical deployments such as model-as-a-service systems, where an attacker can probe the system through crafted queries and observe outputs or routing signals without access to model parameters.

Within this setting, we instantiate our attack framework using a genetic algorithm to perform gradient-free adversarial suffix optimization. This approach enables query-based search guided by observed system responses. Experimental results on the AGNews, Overruling, and Headline datasets show that the attack remains effective, although its performance is slightly weaker than gradient-based methods. Nevertheless, the genetic algorithm provides a robust and practical alternative when only interactive input–output access is available.

\begin{figure*}[htbp]
  \centering
  \subfigure[Headline (passrate)]{
    \includegraphics[width=0.6\columnwidth]{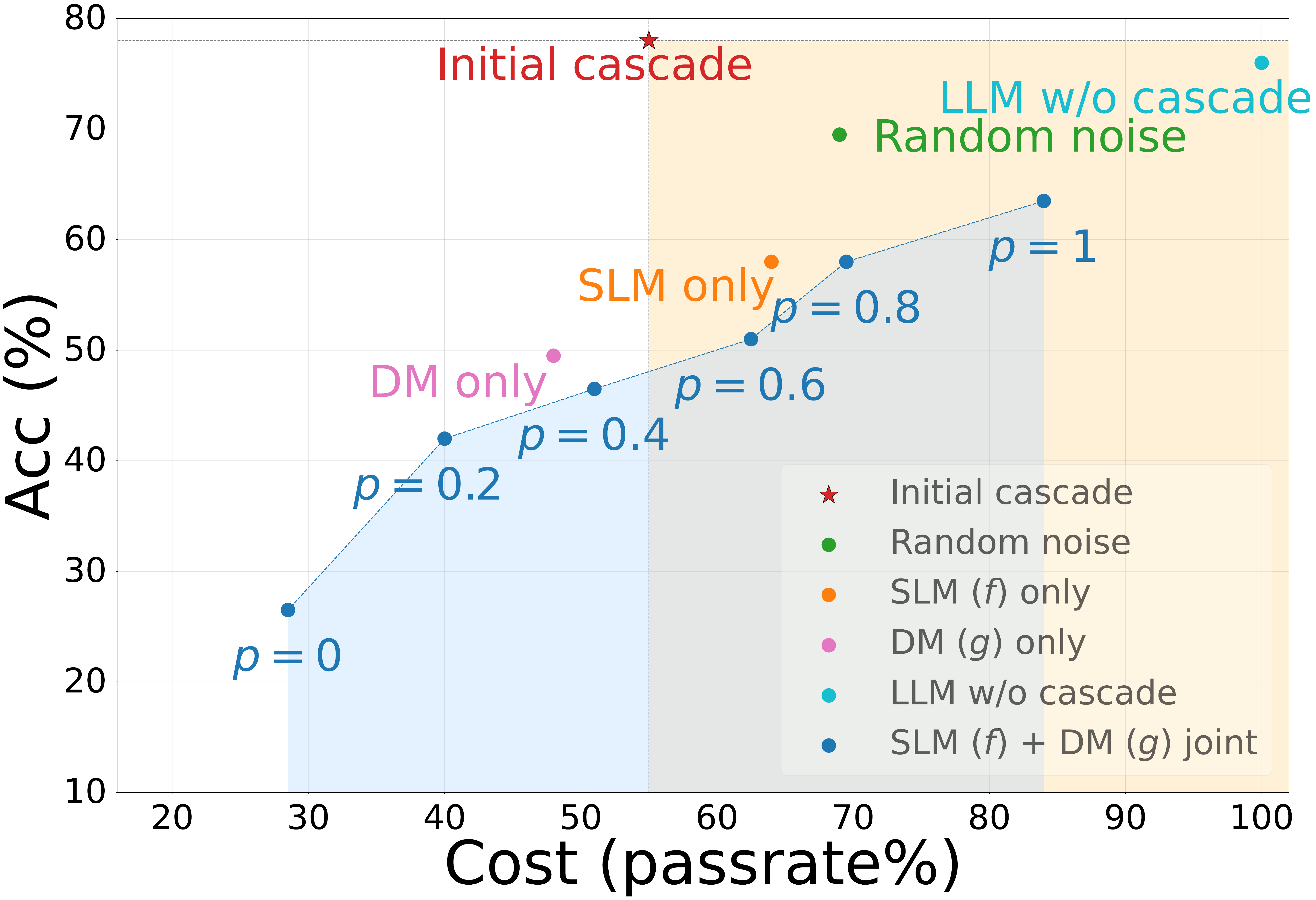}
    \label{fig:}
  }
    \subfigure[IMDB (passrate)]{
    \includegraphics[width=0.6\columnwidth]{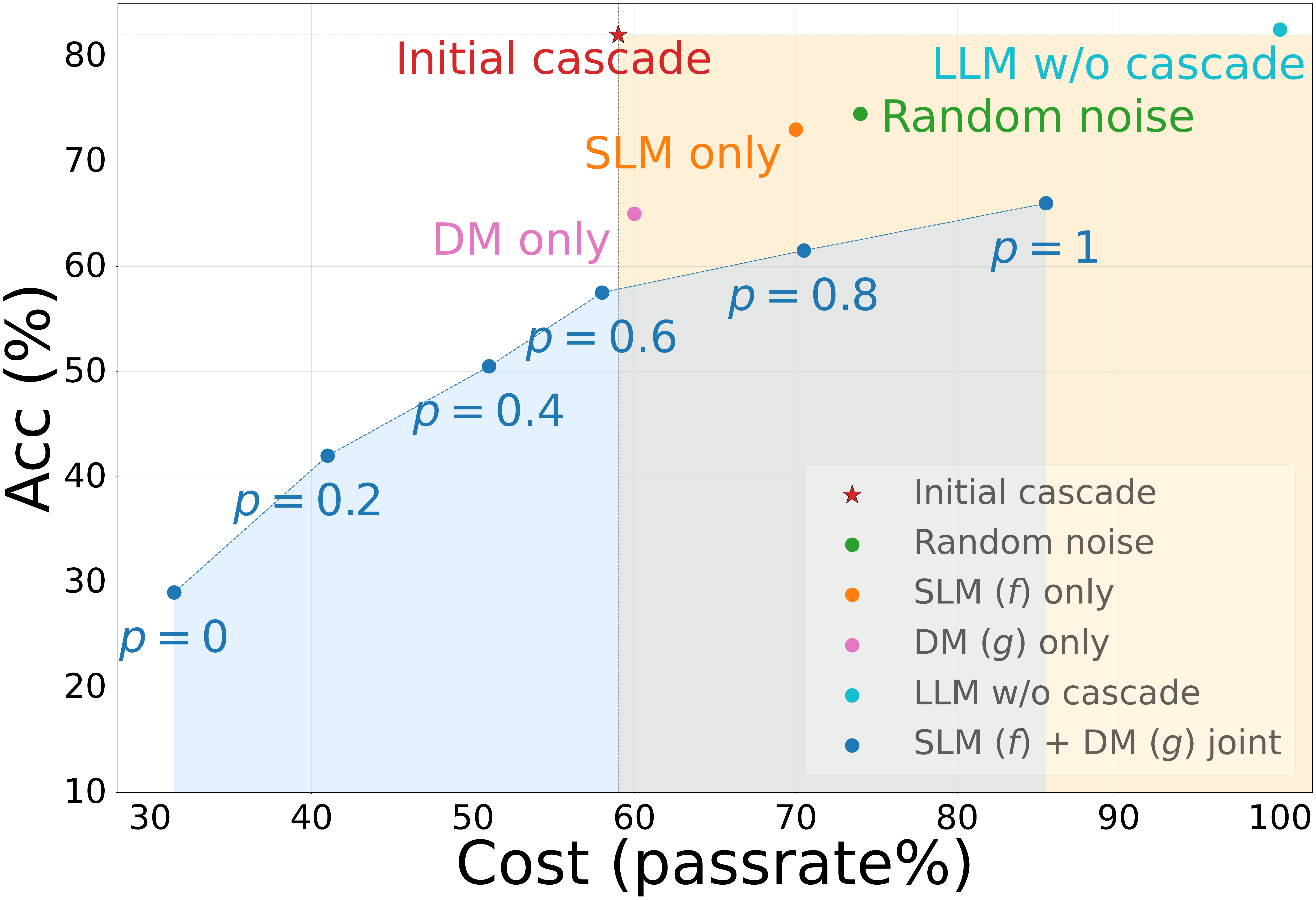}
    \label{fig:}
  }
      \subfigure[Web-question (passrate)]{
    \includegraphics[width=0.6\columnwidth]{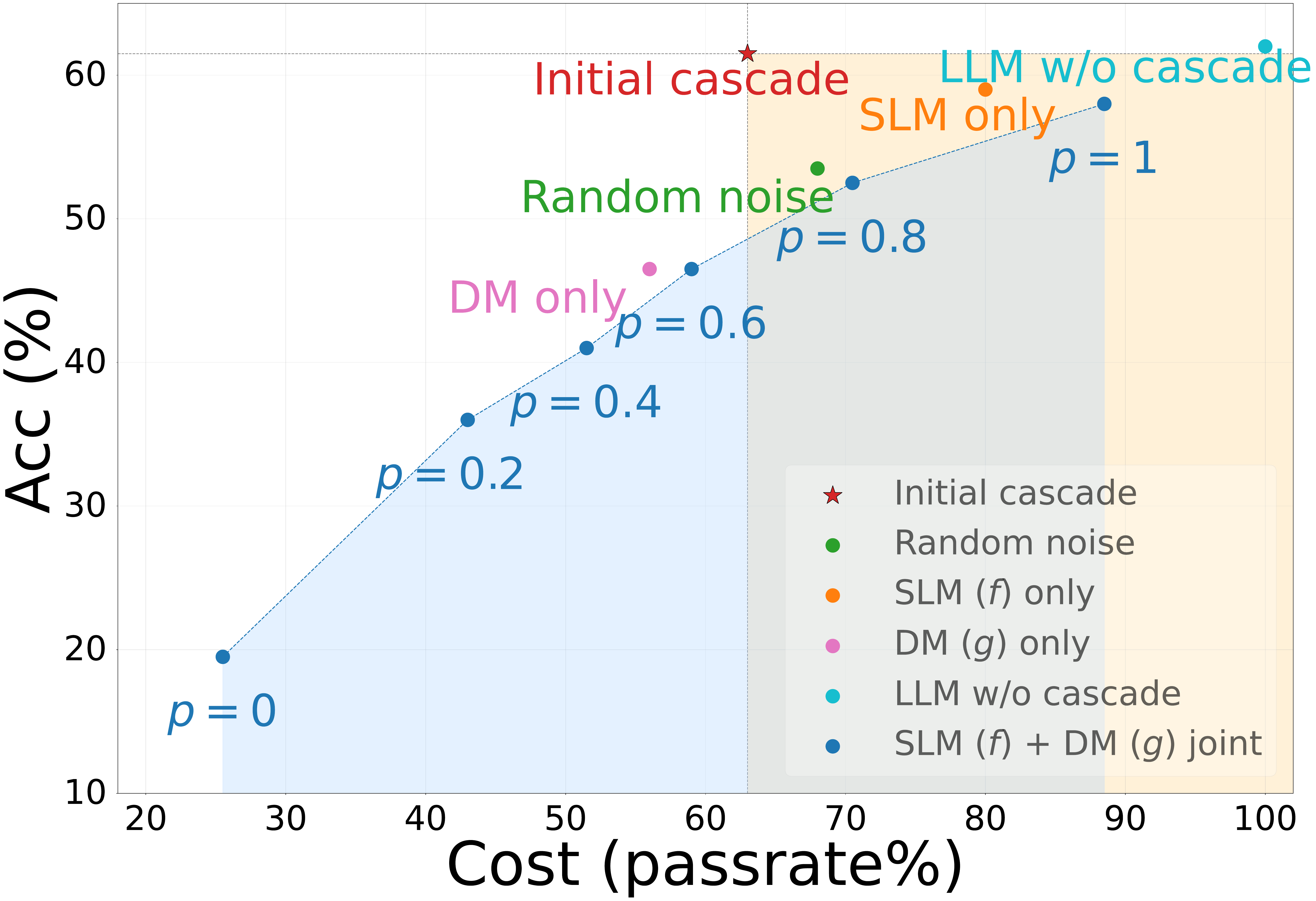}
    \label{fig:}
  }
  \subfigure[Headline (normalized tokens)]{
    \includegraphics[width=0.6\columnwidth]{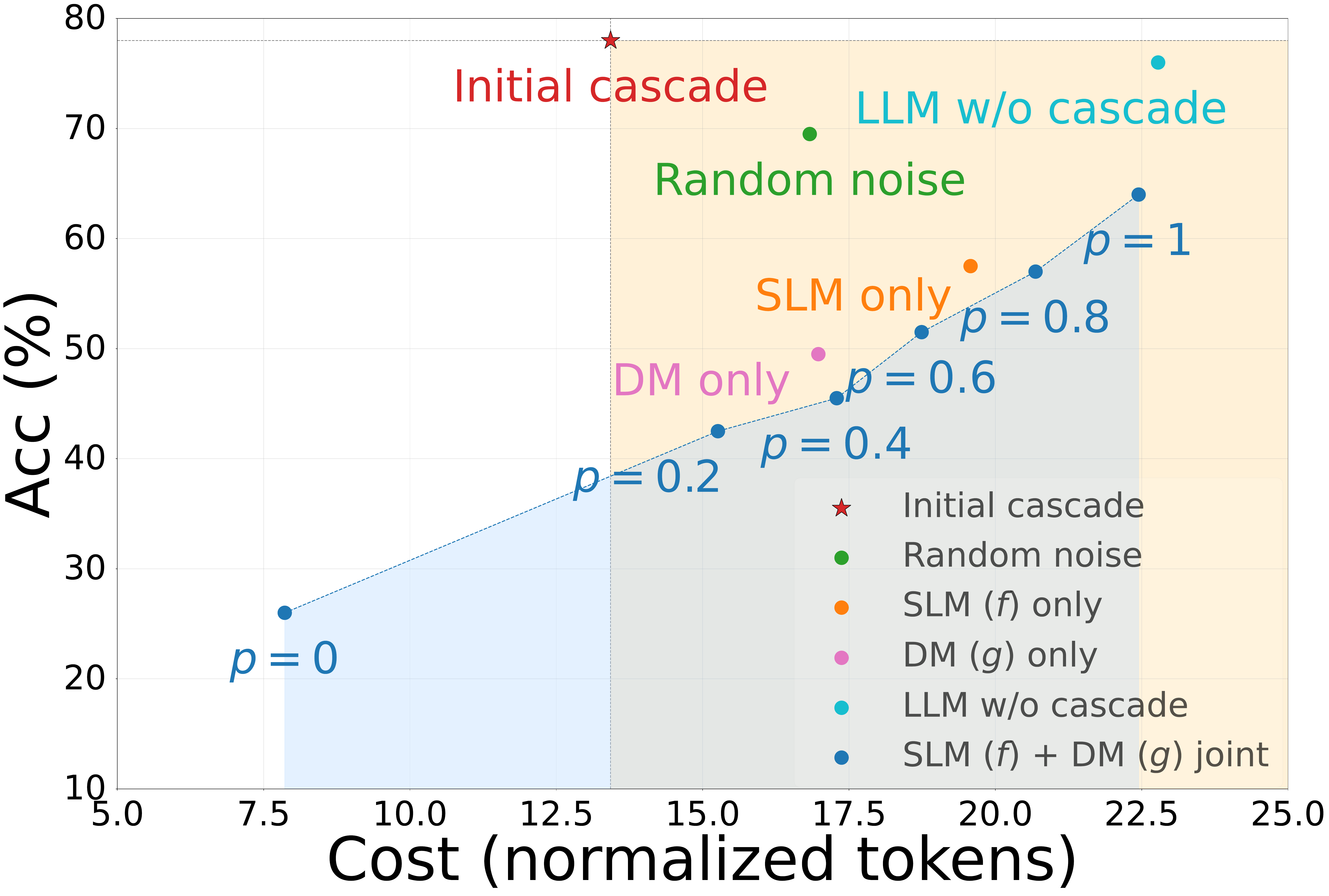}
    \label{fig:}
  }
  \subfigure[IMDB (normalized tokens)]{
    \includegraphics[width=0.6\columnwidth]{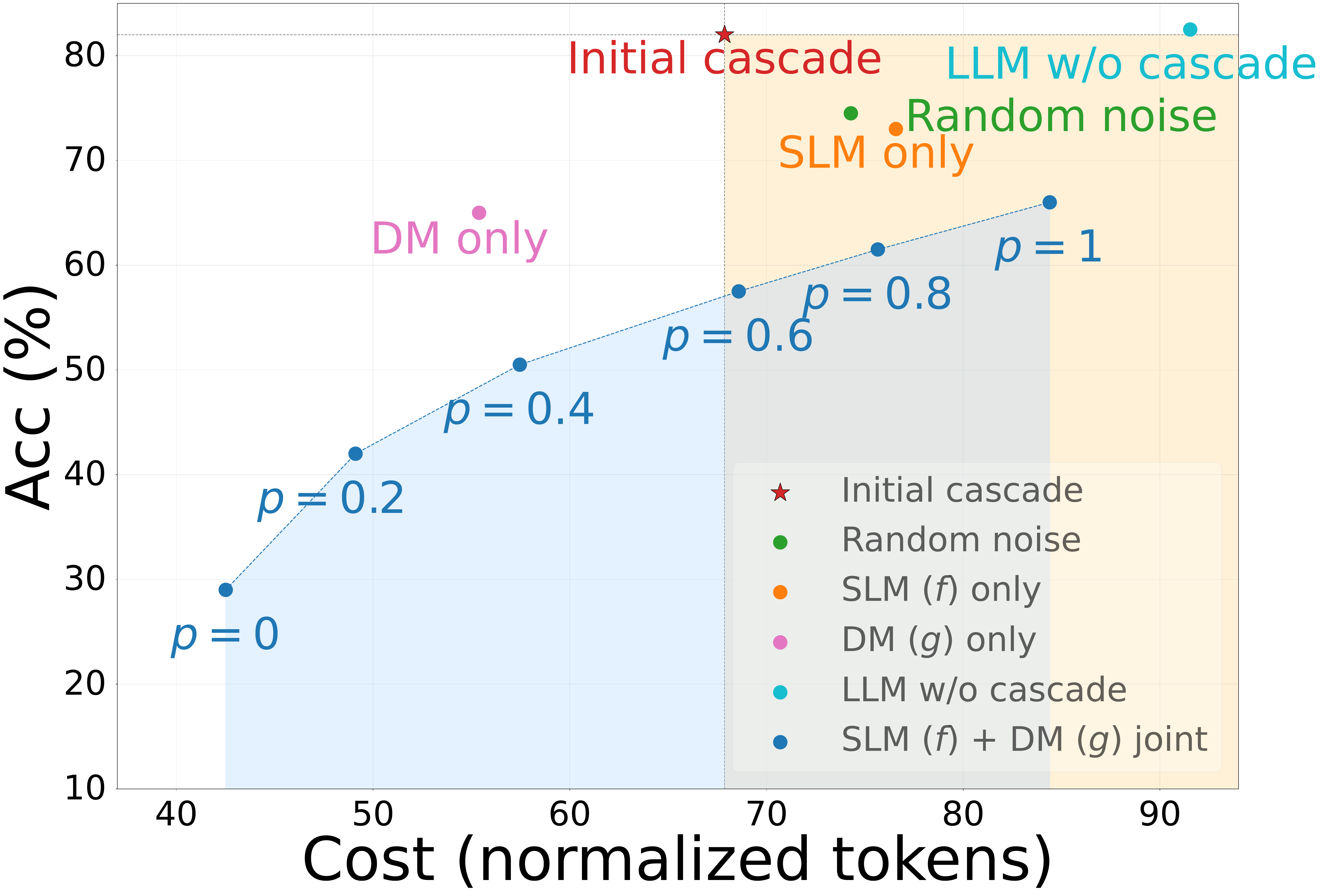}
    \label{fig:}
  }
  \subfigure[Web-question (normalized tokens)]{
    \includegraphics[width=0.6\columnwidth]{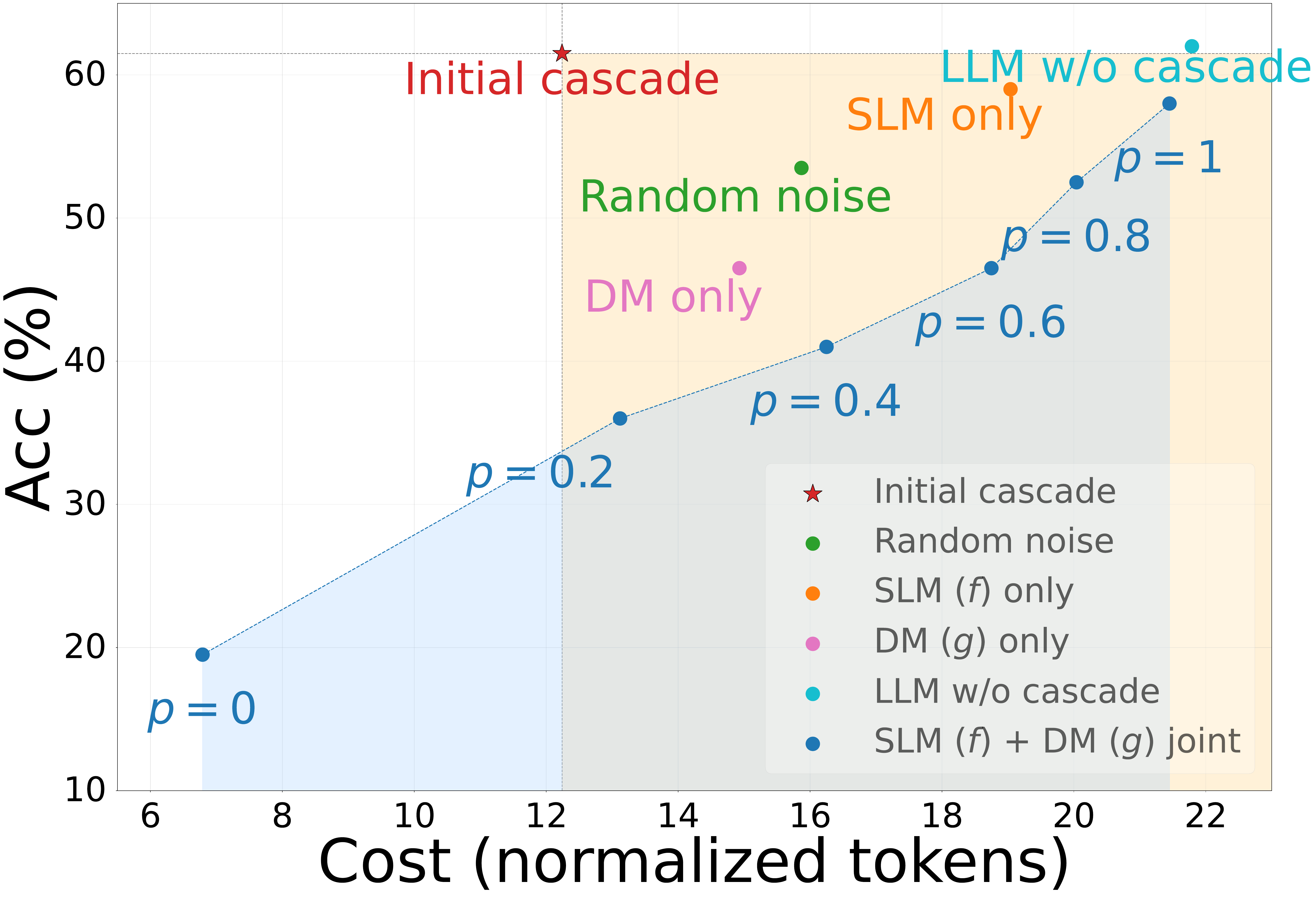}
    \label{fig:}
  }
   \subfigure[Headline (time)]{
    \includegraphics[width=0.6\columnwidth]{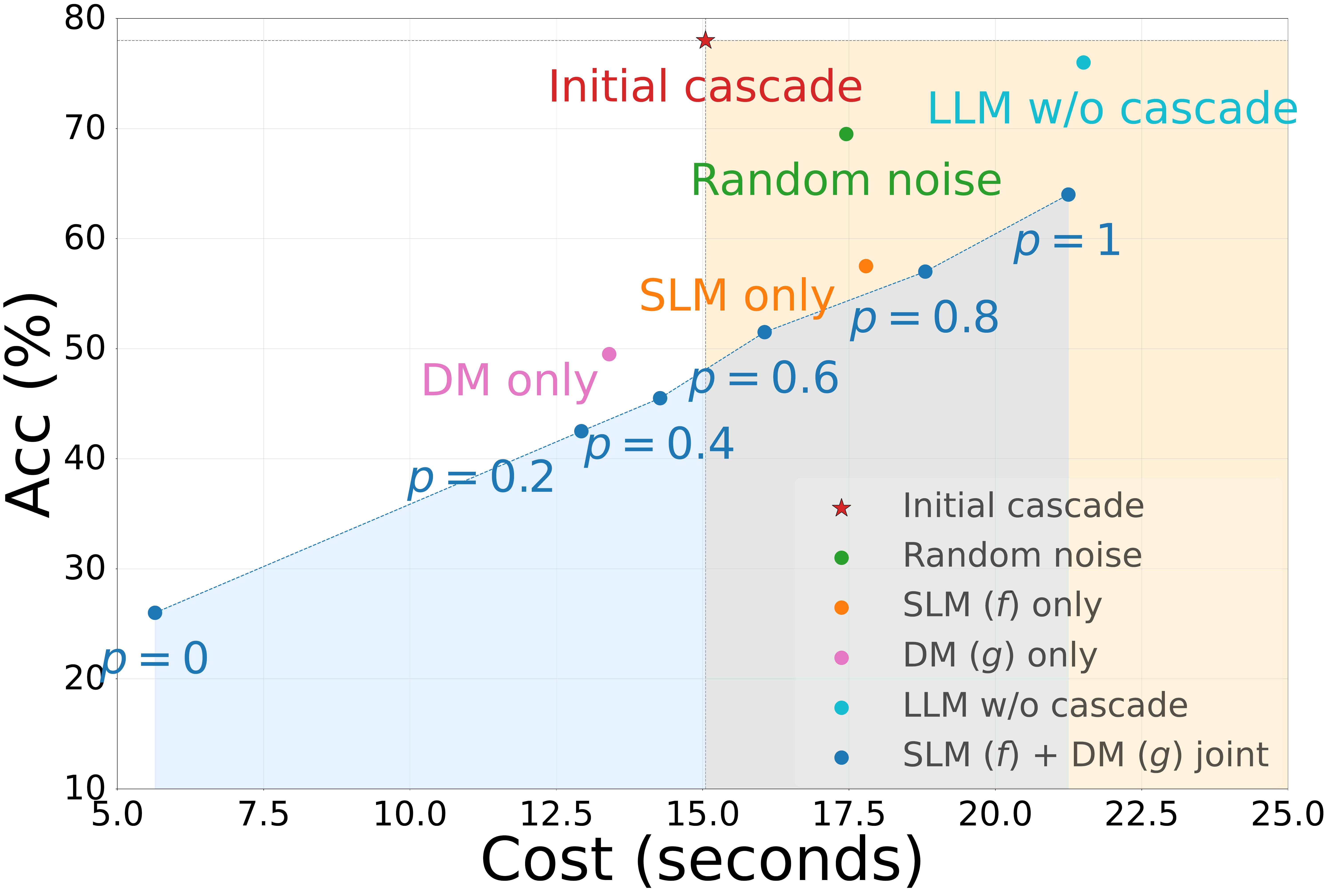}
    \label{fig:}
  }
  \subfigure[IMDB (time))]{
    \includegraphics[width=0.6\columnwidth]{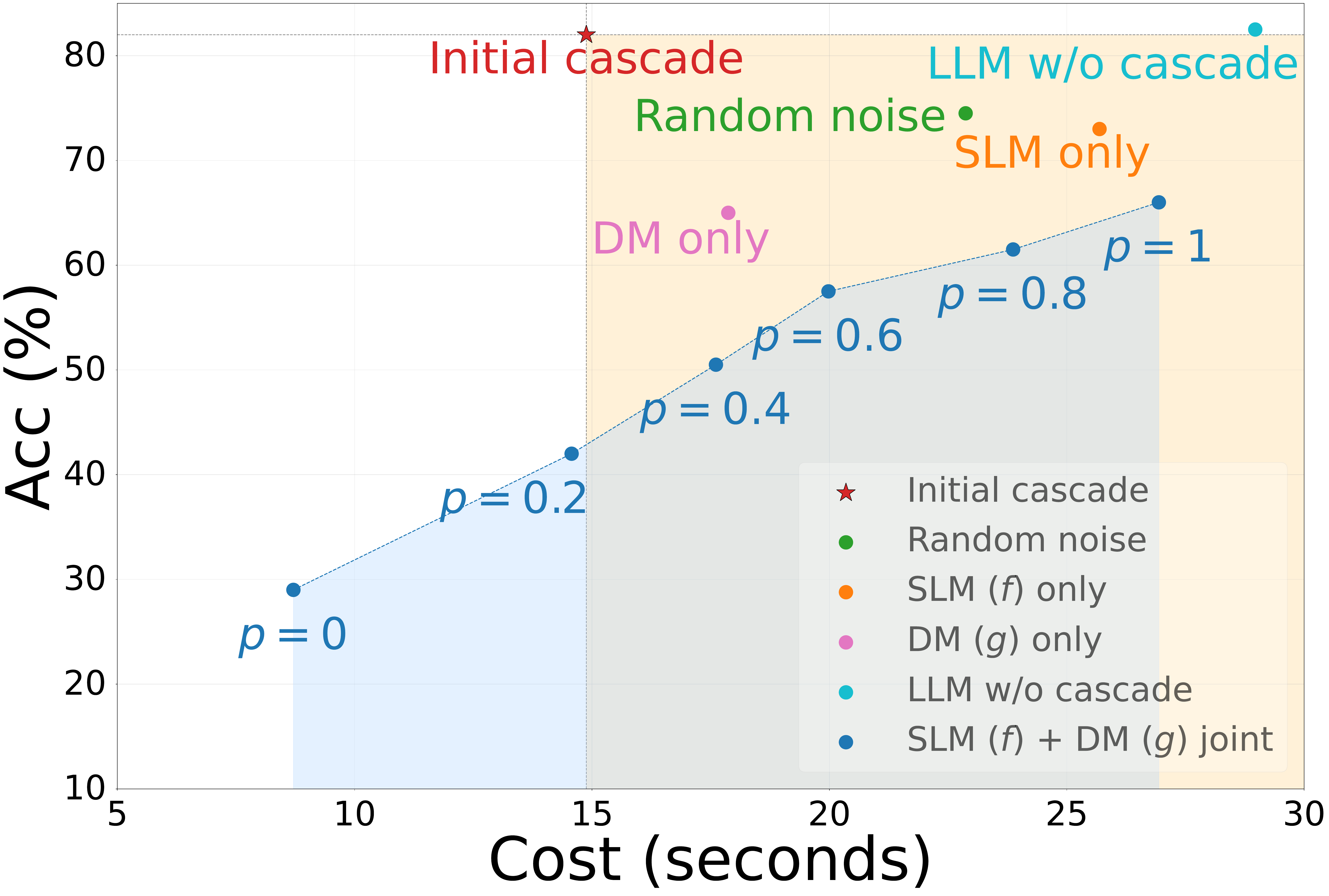}
    \label{fig:}
  }
  \subfigure[Web-question (time)]{
    \includegraphics[width=0.6\columnwidth]{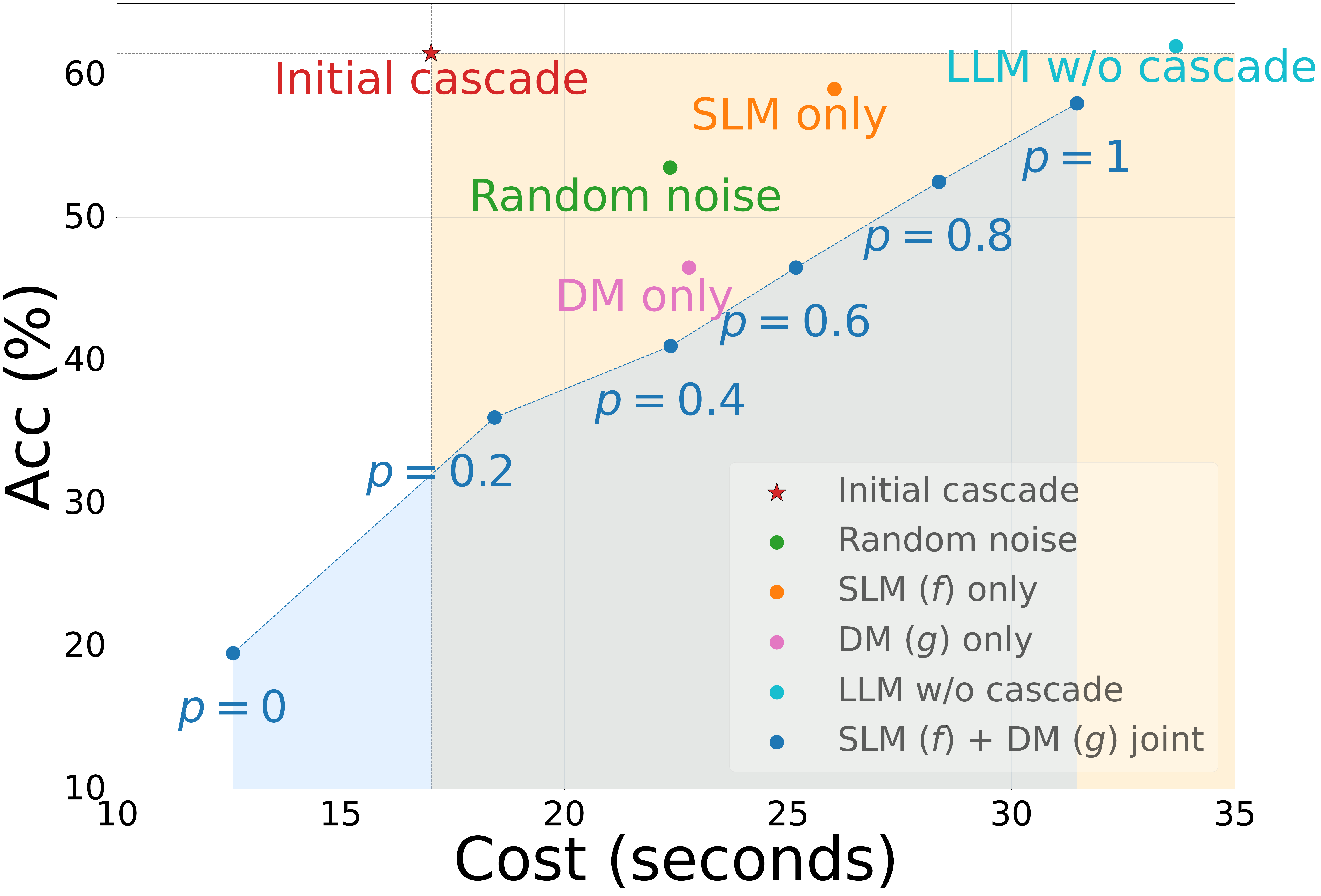}
    \label{fig:}
  }
  \vspace{-6pt}
  \caption{\textbf{Comparison of different attack configurations.}
The cost–accuracy trade-off achieved by our pass-rate control method in gray box setting without the information of gradient(introduced in \sectionautorefname~\ref{subsubsec:pass_rate}) is shown as the {\color[HTML]{1F77B4}blue} dashed line.}
  \label{fig:tradeoff_comparison1}
  \vspace{6pt}
\end{figure*}

\begin{figure*}[htbp]
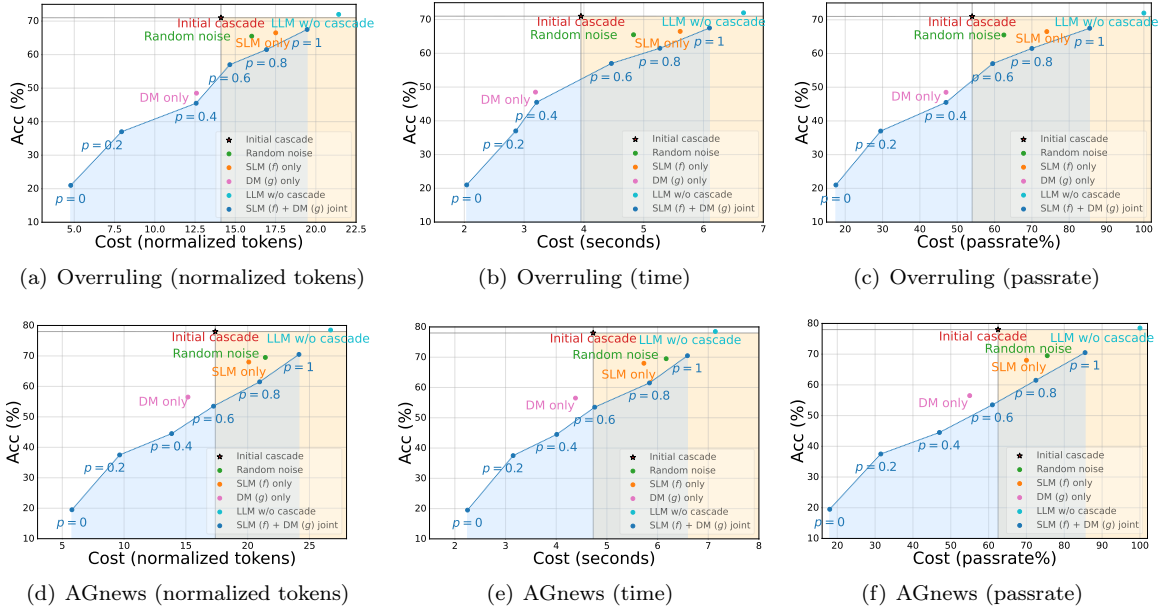

  \centering
    \subfigure[Overruling (normalized tokens)]{
    \includegraphics[width=0.6\columnwidth]{fig/Tradeoff_Overruling.pdf}
    \label{fig:}
  }
  \subfigure[Overruling (time)]{
    \includegraphics[width=0.6\columnwidth]{fig/Tradeoff_Overruling_time.pdf}
    \label{fig:}
  }
  \subfigure[Overruling (passrate)]{
    \includegraphics[width=0.6\columnwidth]{fig/Tradeoff_Overruling_passrate.pdf}
    \label{fig:}
  }

  \subfigure[AGnews (normalized tokens)]{
    \includegraphics[width=0.6\columnwidth]{fig/Tradeoff_AGnews.pdf}
    \label{fig:}
  }
  \subfigure[AGnews (time)]{
    \includegraphics[width=0.6\columnwidth]{fig/Tradeoff_AGnews_time.pdf}
    \label{fig:}
  }
  \subfigure[AGnews (passrate)]{
    \includegraphics[width=0.6\columnwidth]{fig/Tradeoff_AGnews_passrate.pdf}
    \label{fig:}
  }
  \vspace{-6pt}
  \caption{\textbf{Comparison of different attack configurations on extra datasets.}
The cost–accuracy trade-off on dataset IMDB ~\cite{berant-etal-2013-semantic} and AGnews~\cite{maas-EtAl:2011:ACL-HLT2011} achieved by our pass-rate control method in white box setting using the information of gradient  (introduced in \sectionautorefname~\ref{subsubsec:pass_rate}) is shown as the {\color[HTML]{1F77B4}blue} dashed line.}
  \label{fig:tradeoff_comparison2}
  \vspace{6pt}
\end{figure*}

The results of such attacks are presented in Figures~\ref{fig:tradeoff_comparison1}--\ref{fig:tradeoff_comparison2}. Overall, these findings demonstrate that our attack framework generalizes beyond white-box assumptions and remains effective under limited-knowledge conditions.

\subsection{Qualitative Examples}
We present several qualitative examples in Table~\ref{table:example}, showcasing how adversarial suffixes affect both components of the cascade. For each input query, we report the initial predictions made by the small language model (\texttt{SLM}) and the decision module (\texttt{DM}), followed by the predictions after appending the adversarial suffix. These examples illustrate how an attack can (i) flip the \texttt{SLM}'s prediction, (ii) manipulate the \texttt{DM} into escalating or prematurely terminating a query, or (iii) achieve both effects simultaneously. The cases highlight typical failure modes we observe across datasets: even short adversarial suffixes can reliably push the \texttt{DM} to misroute queries and induce substantial errors in the \texttt{SLM}, demonstrating how coordinated manipulation of both components can degrade cascade behavior in practice.

\begin{table*}[!thbp]
\centering
\aboverulesep=0ex
\belowrulesep=0ex
\hspace{-5pt}
\begin{tabular}{
p{3cm} 
|>{\centering\arraybackslash}p{0.9cm}
|>{\centering\arraybackslash}p{1.5cm}
|>{\centering\arraybackslash}p{1.7cm}
|p{4.5cm}
|>{\centering\arraybackslash}p{1.5cm}
|>{\centering\arraybackslash}p{1.7cm}
}
\toprule
\multicolumn{1}{>{\centering\arraybackslash}p{3cm}|}{\textbf{Query $\vx$}} &
\textbf{Ground-truth $\vy$ } &
\textbf{SLM Answer $f^1(\vx)$} &
\textbf{DM Prediction $g^1 (\vx)$} &
\multicolumn{1}{>{\centering\arraybackslash}m{4.5cm}|}{\textbf{Adversarial Suffix $\delta$}} &
\textbf{SLM (After Attack) $f^1(\vx  \,\|\, \delta)$} &
\textbf{DM (After Attack)  $g^1(\vx  \,\|\, \delta)$} \\
\midrule
Q: Gold falls on low demand, global cues &
"Down" &
"Down" \textcolor{green}{$\checkmark$} &
"Terminate" \textcolor{green}{$\checkmark$}&
A WanMale*/( Females climb whilst 670 rose 930 &
"Down" \textcolor{green}{$\checkmark$} &
"Pass" \textcolor{red}{$\times$} \\
\hline

Q: Gold prices lower in Asia with China markets shut, Fed minutes eyed &
"Down" &
"Down" \textcolor{green}{$\checkmark$}&
"Terminate" \textcolor{green}{$\checkmark$}&
Pagesclaimer771 Allaah amplification climb rises lbs gain 670 &
"Down" \textcolor{green}{$\checkmark$}&
"Pass" \textcolor{red}{$\times$}\\
\hline

Q: Gold rises 1\% on renewed safe-haven buying &
"Up" &
"Up" \textcolor{green}{$\checkmark$}&
"Terminate" \textcolor{green}{$\checkmark$} &
Manufacturer*: be financially and sts vaudeville newfound leaps &
"None" \textcolor{red}{$\times$} &
"Pass" \textcolor{green}{$\checkmark$}\\
\hline

Q: Gold tumbles below Rs 30,000 mark as intense selloff continues &
"Down" &
"Down" \textcolor{green}{$\checkmark$}&
"Terminate" \textcolor{green}{$\checkmark$} &
hardships Webster wors BO Abilities climb bust championed &
"Up" \textcolor{red}{$\times$} &
"Pass" \textcolor{green}{$\checkmark$}\\
\hline

Q: June gold soars after jobs report, up \$21.80 (1.7\%) to \$1,294/oz &
"Up" &
"Up" \textcolor{green}{$\checkmark$}&
"Terminate" \textcolor{green}{$\checkmark$} &
GENERAL Cosponsors Darkannon Bolton appellatestand passes lifts wrath &
"Neutral" \textcolor{red}{$\times$} &
"Terminate" \textcolor{green}{$\checkmark$}\\
\hline

Q: Gold, silver crash on shift to \$, cash &
"Down" &
"Neutral"\textcolor{red}{$\times$} &
"Pass" \textcolor{green}{$\checkmark$}&
AuthorTeXoine Components ILCS attained slices wrote novella accolades &
"Up" \textcolor{red}{$\times$} &
"Terminate" \textcolor{red}{$\times$} \\
\hline

Q: Gold prices slip on stronger dollar; bullion seen re-testing low &
"Down" &
"Down" \textcolor{green}{$\checkmark$}&
"Terminate" \textcolor{green}{$\checkmark$} &
fungi Career qualified Requirements credential quarterfinalsdown climb surpassing &
"Down" \textcolor{green}{$\checkmark$} &
"Terminate" \textcolor{green}{$\checkmark$}\\
\hline

Q: Gold to platinum ratio hits all-time highs &
"None" &
"Up"  \textcolor{red}{$\times$}  &
"Pass" \textcolor{green}{$\checkmark$} &
gran plunged Yenwal rallies individually 740 advancing builds &
"None" \textcolor{green}{$\checkmark$} &
"Terminate" \textcolor{green}{$\checkmark$}\\
\bottomrule
\end{tabular}
\vspace{12pt}
\caption{\textbf{Headline dataset examples illustrating attack effects on both the \texttt{SLM} and \texttt{DM} components.}}
\label{table:example}
\end{table*}

\subsection{Ablation Study}
\label{sec:exp:attack_techniques}
\begin{figure*}[thbp]
  \centering
  
  \subfigure[Accuracy of the small prediction model $f^1$]{
    \includegraphics[width=0.95\columnwidth]{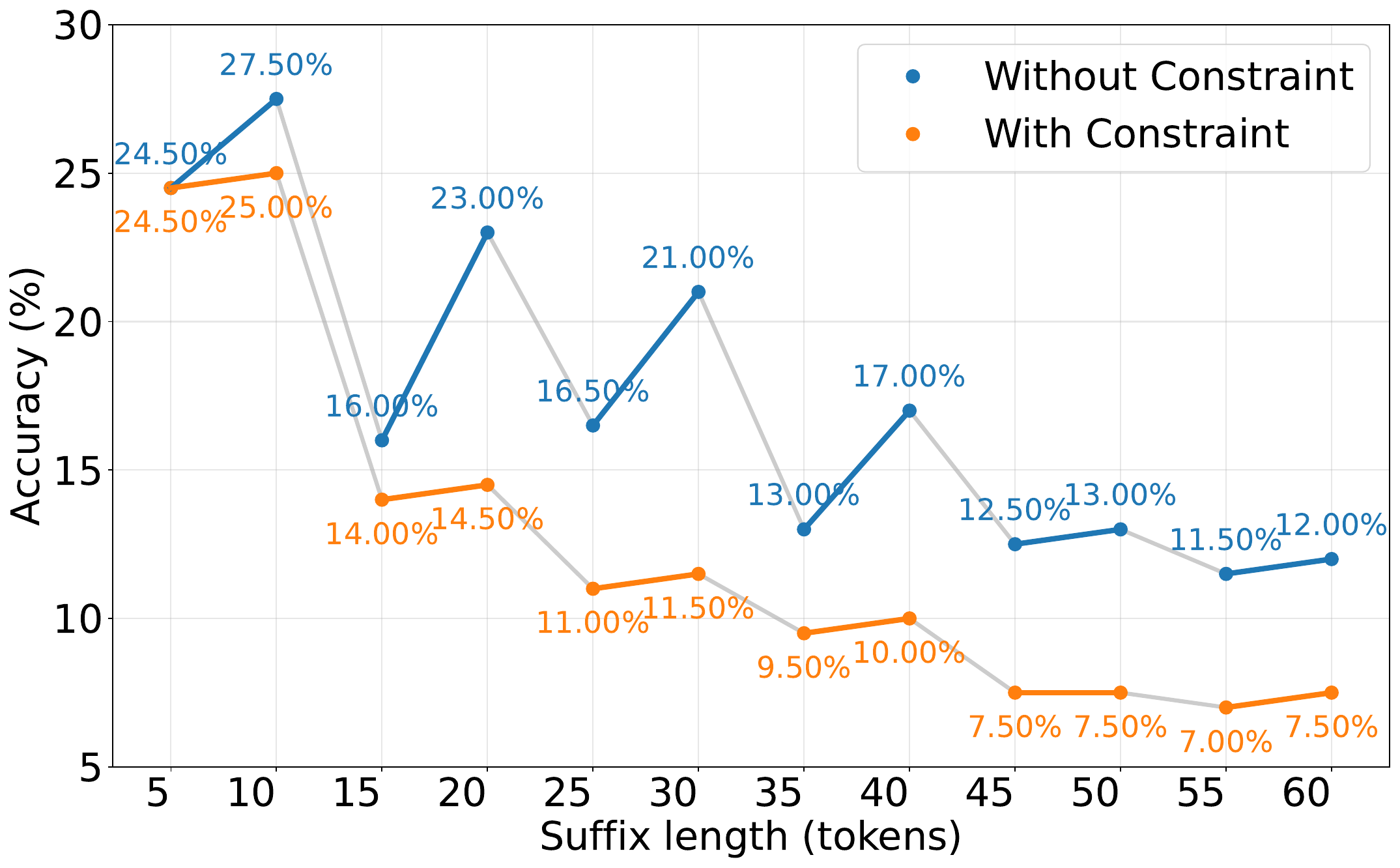}
    \label{fig:constraint_comparison_SLMACC}
  }
  \subfigure[Accuracy of the overall cascade system $\Phi$]{
    \includegraphics[width=0.95\columnwidth]{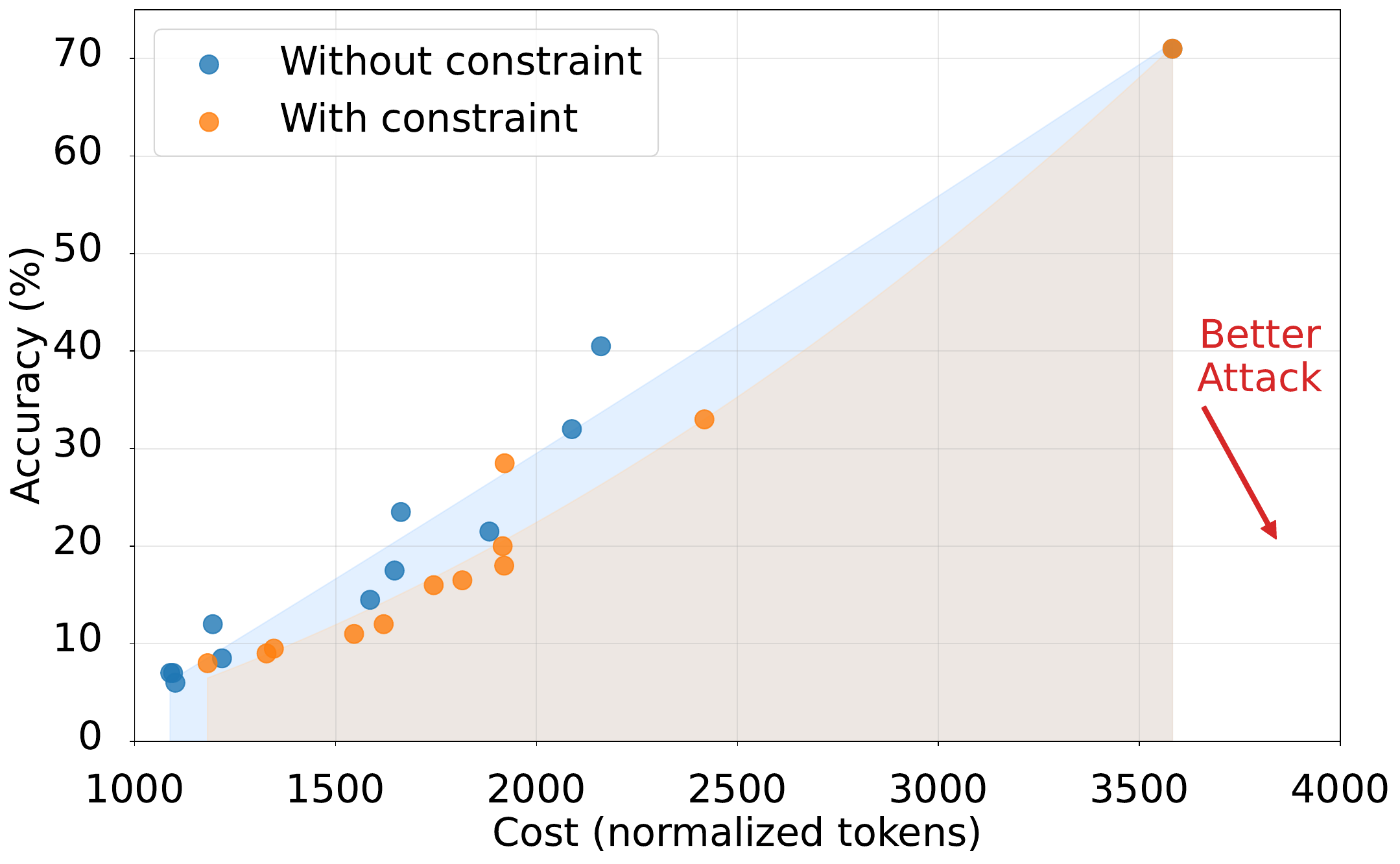}
    \label{fig:constraint_comparison_other}
  }
  \vspace{-6pt}
  \caption{\textbf{Comparison of ``with'' vs. ``without'' constraint setting } (see \sectionautorefname~\ref{subsubsec:constraint_preserve}) of our method.}
  \label{fig:constraint_comparisons}
  \vspace{10pt}
\end{figure*}

We evaluate the effectiveness of the proposed attack techniques described in \sectionautorefname~\ref{subsec:alternating}–\ref{subsubsec:pass_rate}. In \figureautorefname~\ref{fig:constraint_comparisons}, we compare attack performance with and without the constraint-preserving mechanism introduced in \sectionautorefname~\ref{subsubsec:constraint_preserve}.
 Overall, the cascade suffers greater degradation (as shown in the right subfigure) in both accuracy and cost when the constraint is applied: the yellow region (\texttt{``with'' constraint}) consistently lies closer to the bottom-right corner, representing the more desirable adversarial outcome where the system is simultaneously less accurate and more expensive to run.

From the left subfigure, we also observe that the degradation is not only stronger \texttt{with the constraint}, but more importantly, the performance remains stable when transitioning from phase-1 (attacking $f^1$ with an accuracy-focused objective) to phase-2 (attacking $g^1$ with a cost-focused objective). The increase highlighted by the solid colored lines is notably smaller, indicating that our constraint effectively preserves the adversarial effect achieved in earlier stages rather than allowing phase-2 optimization to undo it.
This figure shows the effect of increasing the total suffix length (equivalently, the number of optimization rounds $r$), plotted along the x-axis. While larger suffixes generally strengthen the attack, we observe diminishing returns as the length grows. For efficiency and consistency with prior work, we adopt the default setting of a 20-token suffix, which already produces strong adversarial effects; subsequent results could be further amplified with larger suffixes at the cost of additional optimization time.

Moreover, as shown in Figure~\ref{fig:tradeoff_comparison1}--\ref{fig:tradeoff_comparison2}, comparing our default setting with two alternatives—\texttt{``SLM only''} (attacking only the small prediction model) and \texttt{``DM only''} (attacking only the decision module)—demonstrates that our alternating joint-attack strategy (see \sectionautorefname~\ref{subsec:alternating} and Algorithm~\ref{alg:attack_framework}) is generally more effective at degrading both accuracy and inference cost. Notably, only with our alternating optimization technique can the overall cascade performance be pushed to extreme failure regimes that are unattainable when attacking either component in isolation. This is expected: when only one component is attacked, the untouched component may partially compensate for or correct the induced errors, limiting the overall degradation. The results therefore validate the necessity and effectiveness of our proposed joint attack design, which explicitly leverages the interdependence between prediction stages and decision modules to destabilize the cascade holistically.

\vspace{1cm}

\begin{table*}[!thbp]
\centering
\scriptsize
\aboverulesep=0ex
\belowrulesep=0ex
\setlength{\tabcolsep}{2.8pt}
\renewcommand{\arraystretch}{1.15}
\resizebox{\textwidth}{!}{
\begin{tabular}{
>{\centering\arraybackslash}m{1.7cm}
>{\centering\arraybackslash}m{1.45cm}
>{\centering\arraybackslash}m{1.25cm}
>{\centering\arraybackslash}m{1.75cm}
>{\centering\arraybackslash}m{2.0cm}
>{\centering\arraybackslash}m{2.0cm}
>{\centering\arraybackslash}m{2.0cm}
>{\centering\arraybackslash}m{2.0cm}
>{\centering\arraybackslash}m{2.0cm}
}
\toprule
\textbf{Cascade system} & \textbf{Method} & \textbf{Setting} & \textbf{Metric} & \textbf{Headline} & \textbf{Overruling} & \textbf{IMDB} & \textbf{AGNews} & \textbf{WebQuestions} \\
\midrule

\multirow{11}{*}{\parbox[c][11.0cm][c]{1.7cm}{\centering\textbf{GPT2-medium}\\\textbf{+}\\\textbf{BERT-base}\\\textbf{+}\\\textbf{Phi-3-medium-4k-instruct}}}
& \multicolumn{2}{c}{\textbf{Initial}}
& \makecell[c]{Performance\\Token cost\\Time cost (s)\\Passrate (\%)}
& \makecell[c]{$78.00_{\pm 0.00}$\\$13.43_{\pm 0.00}$\\$4.80_{\pm 0.06}$\\$55.00_{\pm 0.00}$}
& \makecell[c]{$71.00_{\pm 0.00}$\\$13.47_{\pm 0.00}$\\$3.95_{\pm 0.05}$\\$54.00_{\pm 0.00}$}
& \makecell[c]{$82.00_{\pm 0.00}$\\$14.11_{\pm 0.00}$\\$3.94_{\pm 0.04}$\\$59.00_{\pm 0.00}$}
& \makecell[c]{$78.00_{\pm 0.00}$\\$17.38_{\pm 0.00}$\\$4.73_{\pm 0.05}$\\$62.50_{\pm 0.00}$}
& \makecell[c]{$61.50_{\pm 0.00}$\\$9.67_{\pm 0.00}$\\$3.74_{\pm 0.04}$\\$63.00_{\pm 0.00}$}
\\
\cmidrule(lr){2-9}

& \multicolumn{2}{c}{\textbf{LLM w/o cascade}}
& \makecell[c]{Performance\\Token cost\\Time cost (s)\\Passrate (\%)}
& \makecell[c]{$76.00_{\pm 0.00}$\\$25.23_{\pm 0.00}$\\$7.12_{\pm 0.07}$\\$100.00_{\pm 0.00}$}
& \makecell[c]{$72.00_{\pm 0.00}$\\$21.42_{\pm 0.00}$\\$6.67_{\pm 0.07}$\\$100.00_{\pm 0.00}$}
& \makecell[c]{$82.50_{\pm 0.00}$\\$19.47_{\pm 0.00}$\\$5.94_{\pm 0.06}$\\$100.00_{\pm 0.00}$}
& \makecell[c]{$78.50_{\pm 0.00}$\\$26.71_{\pm 0.00}$\\$7.14_{\pm 0.07}$\\$100.00_{\pm 0.00}$}
& \makecell[c]{$62.00_{\pm 0.00}$\\$15.14_{\pm 0.00}$\\$5.75_{\pm 0.06}$\\$100.00_{\pm 0.00}$}
\\
\cmidrule(lr){2-9}

& \multicolumn{2}{c}{\textbf{Random noise}}
& \makecell[c]{Performance\\Token cost\\Time cost (s)\\Passrate (\%)}
& \makecell[c]{$69.50_{\pm 0.50}$\\$16.83_{\pm 0.34}$\\$5.41_{\pm 0.06}$\\$69.00_{\pm 0.58}$}
& \makecell[c]{$65.50_{\pm 0.50}$\\$16.02_{\pm 0.12}$\\$4.83_{\pm 0.08}$\\$62.50_{\pm 1.00}$}
& \makecell[c]{$74.50_{\pm 0.50}$\\$15.26_{\pm 0.11}$\\$4.89_{\pm 0.05}$\\$74.00_{\pm 0.50}$}
& \makecell[c]{$69.50_{\pm 0.50}$\\$21.43_{\pm 0.37}$\\$6.17_{\pm 0.06}$\\$75.50_{\pm 1.00}$}
& \makecell[c]{$53.50_{\pm 0.50}$\\$11.47_{\pm 0.09}$\\$4.37_{\pm 0.05}$\\$68.00_{\pm 0.50}$}
\\
\cmidrule(lr){2-9}

& \multirow{2}{*}{\parbox[c][1.5cm][c]{1.45cm}{\centering\textbf{Single-target}\\\textbf{attack}}}
& SLM only
& \makecell[c]{Performance\\Token cost\\Time cost (s)\\Passrate (\%)}
& \makecell[c]{$69.00_{\pm 1.00}$\\$20.21_{\pm 0.15}$\\$5.64_{\pm 0.06}$\\$72.00_{\pm 0.50}$}
& \makecell[c]{$66.50_{\pm 0.50}$\\$17.51_{\pm 0.13}$\\$5.61_{\pm 0.06}$\\$74.00_{\pm 2.00}$}
& \makecell[c]{$72.00_{\pm 1.50}$\\$16.78_{\pm 0.12}$\\$4.50_{\pm 0.05}$\\$71.50_{\pm 1.00}$}
& \makecell[c]{$68.00_{\pm 1.50}$\\$20.09_{\pm 0.14}$\\$5.73_{\pm 0.06}$\\$70.00_{\pm 0.50}$}
& \makecell[c]{$57.00_{\pm 0.50}$\\$12.97_{\pm 0.10}$\\$4.84_{\pm 0.05}$\\$78.00_{\pm 2.00}$}
\\
\cmidrule(lr){3-9}

&
& DM only
& \makecell[c]{Performance\\Token cost\\Time cost (s)\\Passrate (\%)}
& \makecell[c]{$46.50_{\pm 0.50}$\\$15.17_{\pm 0.12}$\\$4.17_{\pm 0.05}$\\$45.00_{\pm 1.50}$}
& \makecell[c]{$48.50_{\pm 0.50}$\\$12.59_{\pm 0.16}$\\$3.19_{\pm 0.04}$\\$47.00_{\pm 1.00}$}
& \makecell[c]{$63.00_{\pm 2.00}$\\$11.74_{\pm 0.09}$\\$3.51_{\pm 0.04}$\\$51.50_{\pm 0.50}$}
& \makecell[c]{$56.50_{\pm 1.00}$\\$15.17_{\pm 0.12}$\\$4.38_{\pm 0.05}$\\$55.00_{\pm 0.50}$}
& \makecell[c]{$41.00_{\pm 2.50}$\\$6.20_{\pm 0.07}$\\$2.82_{\pm 0.03}$\\$49.50_{\pm 0.50}$}
\\
\cmidrule(lr){2-9}

& \multirow{6}{*}{\parbox[c][4.85cm][c]{1.45cm}{\centering\textbf{Joint-target}\\\textbf{attack}}}
& P=0
& \makecell[c]{Performance\\Token cost\\Time cost (s)\\Passrate (\%)}
& \makecell[c]{$23.00_{\pm 2.50}$\\$7.26_{\pm 0.08}$\\$2.47_{\pm 0.03}$\\$21.00_{\pm 1.50}$}
& \makecell[c]{$21.00_{\pm 0.29}$\\$4.78_{\pm 0.06}$\\$2.04_{\pm 0.03}$\\$17.50_{\pm 0.29}$}
& \makecell[c]{$24.00_{\pm 1.00}$\\$2.97_{\pm 0.04}$\\$1.49_{\pm 0.02}$\\$26.50_{\pm 1.50}$}
& \makecell[c]{$19.50_{\pm 0.50}$\\$5.78_{\pm 0.07}$\\$2.25_{\pm 0.03}$\\$18.00_{\pm 1.50}$}
& \makecell[c]{$16.50_{\pm 0.50}$\\$1.78_{\pm 0.03}$\\$1.71_{\pm 0.02}$\\$22.50_{\pm 1.00}$}
\\
\cmidrule(lr){3-9}

& & P=0.2
& \makecell[c]{Performance\\Token cost\\Time cost (s)\\Passrate (\%)}
& \makecell[c]{$41.00_{\pm 0.50}$\\$14.62_{\pm 0.12}$\\$3.30_{\pm 0.04}$\\$37.50_{\pm 1.00}$}
& \makecell[c]{$37.00_{\pm 1.50}$\\$7.94_{\pm 0.08}$\\$2.86_{\pm 0.03}$\\$29.50_{\pm 0.50}$}
& \makecell[c]{$38.00_{\pm 1.50}$\\$6.64_{\pm 0.07}$\\$2.39_{\pm 0.03}$\\$38.50_{\pm 1.00}$}
& \makecell[c]{$37.50_{\pm 1.00}$\\$9.64_{\pm 0.09}$\\$3.15_{\pm 0.04}$\\$31.50_{\pm 2.50}$}
& \makecell[c]{$30.50_{\pm 1.50}$\\$3.94_{\pm 0.05}$\\$2.41_{\pm 0.03}$\\$41.00_{\pm 0.50}$}
\\
\cmidrule(lr){3-9}

& & P=0.4
& \makecell[c]{Performance\\Token cost\\Time cost (s)\\Passrate (\%)}
& \makecell[c]{$43.50_{\pm 1.00}$\\$15.99_{\pm 0.13}$\\$4.04_{\pm 0.05}$\\$46.00_{\pm 3.00}$}
& \makecell[c]{$45.50_{\pm 1.50}$\\$12.57_{\pm 0.10}$\\$3.21_{\pm 0.04}$\\$47.00_{\pm 2.50}$}
& \makecell[c]{$47.50_{\pm 1.50}$\\$9.17_{\pm 0.08}$\\$3.11_{\pm 0.04}$\\$46.00_{\pm 1.00}$}
& \makecell[c]{$44.50_{\pm 2.50}$\\$13.85_{\pm 0.11}$\\$4.01_{\pm 0.05}$\\$47.00_{\pm 1.00}$}
& \makecell[c]{$36.00_{\pm 0.29}$\\$6.21_{\pm 0.07}$\\$2.84_{\pm 0.03}$\\$49.50_{\pm 2.00}$}
\\
\cmidrule(lr){3-9}

& & P=0.6
& \makecell[c]{Performance\\Token cost\\Time cost (s)\\Passrate (\%)}
& \makecell[c]{$50.00_{\pm 0.50}$\\$18.34_{\pm 0.14}$\\$4.78_{\pm 0.05}$\\$62.00_{\pm 2.00}$}
& \makecell[c]{$57.00_{\pm 1.50}$\\$14.66_{\pm 0.32}$\\$4.46_{\pm 0.05}$\\$59.50_{\pm 0.50}$}
& \makecell[c]{$58.00_{\pm 0.50}$\\$12.61_{\pm 0.27}$\\$3.86_{\pm 0.04}$\\$57.00_{\pm 0.50}$}
& \makecell[c]{$53.50_{\pm 1.50}$\\$17.24_{\pm 0.23}$\\$4.76_{\pm 0.05}$\\$61.00_{\pm 1.00}$}
& \makecell[c]{$44.00_{\pm 1.00}$\\$8.02_{\pm 0.08}$\\$3.16_{\pm 0.04}$\\$55.00_{\pm 2.50}$}
\\
\cmidrule(lr){3-9}

& & P=0.8
& \makecell[c]{Performance\\Token cost\\Time cost (s)\\Passrate (\%)}
& \makecell[c]{$56.50_{\pm 1.00}$\\$20.09_{\pm 0.15}$\\$5.59_{\pm 0.06}$\\$71.00_{\pm 1.00}$}
& \makecell[c]{$61.50_{\pm 0.50}$\\$16.94_{\pm 0.43}$\\$5.27_{\pm 0.06}$\\$70.00_{\pm 0.50}$}
& \makecell[c]{$64.50_{\pm 1.00}$\\$15.83_{\pm 0.22}$\\$4.41_{\pm 0.05}$\\$70.50_{\pm 0.50}$}
& \makecell[c]{$61.50_{\pm 1.50}$\\$20.97_{\pm 0.34}$\\$5.84_{\pm 0.06}$\\$72.50_{\pm 1.50}$}
& \makecell[c]{$50.00_{\pm 1.00}$\\$11.18_{\pm 0.09}$\\$4.28_{\pm 0.05}$\\$67.50_{\pm 2.00}$}
\\
\cmidrule(lr){3-9}

& & P=1
& \makecell[c]{Performance\\Token cost\\Time cost (s)\\Passrate (\%)}
& \makecell[c]{$63.00_{\pm 1.50}$\\$22.45_{\pm 0.16}$\\$6.56_{\pm 0.07}$\\$84.50_{\pm 2.00}$}
& \makecell[c]{$67.50_{\pm 1.00}$\\$19.46_{\pm 0.14}$\\$6.10_{\pm 0.06}$\\$85.50_{\pm 0.50}$}
& \makecell[c]{$75.50_{\pm 2.00}$\\$17.81_{\pm 0.13}$\\$5.37_{\pm 0.06}$\\$86.00_{\pm 0.50}$}
& \makecell[c]{$70.50_{\pm 0.00}$\\$24.15_{\pm 0.17}$\\$6.59_{\pm 0.07}$\\$85.50_{\pm 1.00}$}
& \makecell[c]{$56.00_{\pm 1.00}$\\$14.47_{\pm 0.11}$\\$5.04_{\pm 0.05}$\\$85.50_{\pm 1.50}$}
\\

\bottomrule
\end{tabular}
}
\vspace{16pt}
\caption{\textbf{Results on the GPT-2+BERT+Phi3 cascade system.}}
\label{tab:gpt2_bert_phi3_results}
\par
\raggedright
\footnotesize
\end{table*}

\begin{table*}[!thbp]
\centering
\scriptsize
\aboverulesep=0ex
\belowrulesep=0ex
\setlength{\tabcolsep}{2.8pt}
\renewcommand{\arraystretch}{1.15}

\resizebox{\textwidth}{!}{
\begin{tabular}{
>{\centering\arraybackslash}m{2.0cm}
>{\centering\arraybackslash}m{1.45cm}
>{\centering\arraybackslash}m{1.25cm}
>{\centering\arraybackslash}m{1.85cm}
>{\centering\arraybackslash}m{2.0cm}
>{\centering\arraybackslash}m{2.0cm}
>{\centering\arraybackslash}m{2.0cm}
>{\centering\arraybackslash}m{2.0cm}
>{\centering\arraybackslash}m{2.0cm}
}
\toprule
\textbf{Cascade system} & \textbf{Method} & \textbf{Setting} & \textbf{Metric} & \textbf{AGNews} & \textbf{CommonGen} & \textbf{SQuAD2.0} & \textbf{SVAMP} & \textbf{WildJailbreak} \\
\midrule

\multirow{11}{*}{\parbox[c][11.0cm][c]{2.0cm}{\centering\textbf{Qwen2.5-3B}\\\textbf{+}\\\textbf{DistilBERT-base}\\\textbf{+}\\\textbf{Phi-3.5-MoE-16$\times$3.8B}}}
& \multicolumn{2}{c}{\textbf{Initial}}
& \makecell[c]{Performance\\Token cost\\Time cost (s)\\Passrate (\%)}
& \makecell[c]{$91.50_{\pm 0.00}$\\$63.57_{\pm 0.00}$\\$4.79_{\pm 0.06}$\\$48.00_{\pm 0.00}$}
& \makecell[c]{$93.00_{\pm 0.00}$\\$36.84_{\pm 0.00}$\\$2.02_{\pm 0.03}$\\$37.00_{\pm 0.00}$}
& \makecell[c]{$84.00_{\pm 0.00}$\\$102.41_{\pm 0.00}$\\$4.61_{\pm 0.06}$\\$42.00_{\pm 0.00}$}
& \makecell[c]{$75.00_{\pm 0.00}$\\$54.17_{\pm 0.00}$\\$6.09_{\pm 0.07}$\\$57.50_{\pm 0.00}$}
& \makecell[c]{$97.00_{\pm 0.00}$\\$24.80_{\pm 0.00}$\\$2.39_{\pm 0.03}$\\$67.00_{\pm 0.00}$}
\\
\cmidrule(lr){2-9}

& \multicolumn{2}{c}{\textbf{LLM w/o cascade}}
& \makecell[c]{Performance\\Token cost\\Time cost (s)\\Passrate (\%)}
& \makecell[c]{$94.00_{\pm 0.00}$\\$112.58_{\pm 0.00}$\\$8.12_{\pm 0.09}$\\$100.00_{\pm 0.00}$}
& \makecell[c]{$96.00_{\pm 0.00}$\\$96.14_{\pm 0.00}$\\$4.41_{\pm 0.06}$\\$100.00_{\pm 0.00}$}
& \makecell[c]{$86.50_{\pm 0.00}$\\$277.24_{\pm 0.00}$\\$8.49_{\pm 0.07}$\\$100.00_{\pm 0.00}$}
& \makecell[c]{$76.50_{\pm 0.00}$\\$92.71_{\pm 0.00}$\\$9.02_{\pm 0.04}$\\$100.00_{\pm 0.00}$}
& \makecell[c]{$98.50_{\pm 0.00}$\\$40.04_{\pm 0.00}$\\$3.49_{\pm 0.04}$\\$100.00_{\pm 0.00}$}
\\
\cmidrule(lr){2-9}

& \multicolumn{2}{c}{\textbf{Random noise}}
& \makecell[c]{Performance\\Token cost\\Time cost (s)\\Passrate (\%)}
& \makecell[c]{$78.50_{\pm 0.50}$\\$99.78_{\pm 3.47}$\\$7.07_{\pm 0.08}$\\$80.50_{\pm 1.00}$}
& \makecell[c]{$86.50_{\pm 1.50}$\\$52.42_{\pm 2.17}$\\$2.63_{\pm 0.04}$\\$50.50_{\pm 0.50}$}
& \makecell[c]{$68.50_{\pm 1.00}$\\$215.18_{\pm 4.49}$\\$7.30_{\pm 0.08}$\\$77.50_{\pm 1.00}$}
& \makecell[c]{$67.50_{\pm 0.50}$\\$69.55_{\pm 2.16}$\\$7.04_{\pm 0.08}$\\$71.50_{\pm 1.00}$}
& \makecell[c]{$84.50_{\pm 2.00}$\\$28.97_{\pm 0.68}$\\$2.54_{\pm 0.03}$\\$77.50_{\pm 1.00}$}
\\
\cmidrule(lr){2-9}

& \multirow{2}{*}{\parbox[c][1.8cm][c]{1.45cm}{\centering\textbf{Single-target}\\\textbf{attack}}}
& SLM only
& \makecell[c]{Performance\\Token cost\\Time cost (s)\\Passrate (\%)}
& \makecell[c]{$74.00_{\pm 0.50}$\\$96.41_{\pm 4.24}$\\$6.87_{\pm 0.07}$\\$74.00_{\pm 1.00}$}
& \makecell[c]{$84.00_{\pm 1.50}$\\$79.63_{\pm 3.74}$\\$3.84_{\pm 0.05}$\\$79.50_{\pm 1.50}$}
& \makecell[c]{$70.00_{\pm 1.00}$\\$242.57_{\pm 6.39}$\\$8.02_{\pm 0.14}$\\$87.00_{\pm 2.00}$}
& \makecell[c]{$61.00_{\pm 2.00}$\\$67.47_{\pm 2.31}$\\$6.83_{\pm 0.27}$\\$68.00_{\pm 0.00}$}
& \makecell[c]{$81.00_{\pm 1.50}$\\$29.63_{\pm 1.74}$\\$2.67_{\pm 0.03}$\\$78.50_{\pm 2.00}$}
\\
\cmidrule(lr){3-9}

&
& DM only
& \makecell[c]{Performance\\Token cost\\Time cost (s)\\Passrate (\%)}
& \makecell[c]{$69.50_{\pm 0.50}$\\$78.60_{\pm 4.16}$\\$6.01_{\pm 0.06}$\\$54.00_{\pm 1.50}$}
& \makecell[c]{$80.50_{\pm 1.00}$\\$65.81_{\pm 3.71}$\\$3.26_{\pm 0.04}$\\$66.50_{\pm 1.00}$}
& \makecell[c]{$53.50_{\pm 1.50}$\\$168.22_{\pm 8.32}$\\$5.87_{\pm 0.06}$\\$60.50_{\pm 1.00}$}
& \makecell[c]{$54.50_{\pm 1.00}$\\$49.86_{\pm 2.86}$\\$5.29_{\pm 0.06}$\\$50.50_{\pm 1.50}$}
& \makecell[c]{$66.00_{\pm 1.50}$\\$19.27_{\pm 1.04}$\\$2.16_{\pm 0.03}$\\$53.00_{\pm 1.50}$}
\\
\cmidrule(lr){2-9}

& \multirow{6}{*}{\parbox[c][4.45cm][c]{1.45cm}{\centering\textbf{Joint-target}\\\textbf{attack}}}
& P=0
& \makecell[c]{Performance\\Token cost\\Time cost (s)\\Passrate (\%)}
& \makecell[c]{$25.50_{\pm 0.50}$\\$21.44_{\pm 2.03}$\\$2.58_{\pm 0.03}$\\$21.00_{\pm 0.50}$}
& \makecell[c]{$52.50_{\pm 1.00}$\\$21.89_{\pm 1.79}$\\$1.27_{\pm 0.02}$\\$19.00_{\pm 0.50}$}
& \makecell[c]{$16.00_{\pm 1.00}$\\$61.27_{\pm 4.02}$\\$3.28_{\pm 0.04}$\\$26.00_{\pm 0.00}$}
& \makecell[c]{$25.50_{\pm 1.50}$\\$26.49_{\pm 1.23}$\\$2.94_{\pm 0.03}$\\$23.00_{\pm 0.50}$}
& \makecell[c]{$18.00_{\pm 0.50}$\\$6.43_{\pm 0.84}$\\$1.11_{\pm 0.07}$\\$22.50_{\pm 1.00}$}
\\
\cmidrule(lr){3-9}

& & P=0.2
& \makecell[c]{Performance\\Token cost\\Time cost (s)\\Passrate (\%)}
& \makecell[c]{$40.50_{\pm 1.50}$\\$39.84_{\pm 3.94}$\\$3.61_{\pm 0.04}$\\$34.50_{\pm 1.00}$}
& \makecell[c]{$60.00_{\pm 2.50}$\\$34.52_{\pm 3.06}$\\$1.96_{\pm 0.03}$\\$36.00_{\pm 0.50}$}
& \makecell[c]{$29.50_{\pm 1.50}$\\$94.11_{\pm 6.45}$\\$4.34_{\pm 0.05}$\\$39.50_{\pm 0.50}$}
& \makecell[c]{$37.50_{\pm 0.00}$\\$39.62_{\pm 2.92}$\\$4.33_{\pm 0.05}$\\$40.00_{\pm 1.00}$}
& \makecell[c]{$38.50_{\pm 1.00}$\\$13.70_{\pm 0.89}$\\$1.69_{\pm 0.02}$\\$38.50_{\pm 0.50}$}
\\
\cmidrule(lr){3-9}

& & P=0.4
& \makecell[c]{Performance\\Token cost\\Time cost (s)\\Passrate (\%)}
& \makecell[c]{$49.50_{\pm 2.00}$\\$60.17_{\pm 5.71}$\\$4.42_{\pm 0.05}$\\$44.50_{\pm 1.50}$}
& \makecell[c]{$67.50_{\pm 0.50}$\\$48.77_{\pm 4.68}$\\$2.48_{\pm 0.03}$\\$48.50_{\pm 0.50}$}
& \makecell[c]{$41.00_{\pm 0.00}$\\$141.49_{\pm 9.27}$\\$5.22_{\pm 0.06}$\\$51.50_{\pm 0.50}$}
& \makecell[c]{$44.50_{\pm 1.00}$\\$48.17_{\pm 2.65}$\\$5.07_{\pm 0.06}$\\$49.50_{\pm 1.50}$}
& \makecell[c]{$47.50_{\pm 0.50}$\\$19.89_{\pm 1.34}$\\$2.08_{\pm 0.03}$\\$54.50_{\pm 0.50}$}
\\
\cmidrule(lr){3-9}

& & P=0.6
& \makecell[c]{Performance\\Token cost\\Time cost (s)\\Passrate (\%)}
& \makecell[c]{$61.00_{\pm 0.50}$\\$82.29_{\pm 5.07}$\\$5.74_{\pm 0.06}$\\$59.50_{\pm 1.50}$}
& \makecell[c]{$75.00_{\pm 0.00}$\\$60.58_{\pm 4.11}$\\$3.03_{\pm 0.04}$\\$60.50_{\pm 0.00}$}
& \makecell[c]{$49.50_{\pm 0.50}$\\$172.32_{\pm 12.42}$\\$6.07_{\pm 0.07}$\\$62.00_{\pm 0.50}$}
& \makecell[c]{$53.00_{\pm 0.50}$\\$59.73_{\pm 6.29}$\\$6.31_{\pm 0.07}$\\$61.00_{\pm 2.00}$}
& \makecell[c]{$58.50_{\pm 1.00}$\\$25.12_{\pm 3.52}$\\$2.44_{\pm 0.03}$\\$67.50_{\pm 1.00}$}
\\
\cmidrule(lr){3-9}

& & P=0.8
& \makecell[c]{Performance\\Token cost\\Time cost (s)\\Passrate (\%)}
& \makecell[c]{$69.50_{\pm 1.00}$\\$93.07_{\pm 7.22}$\\$6.59_{\pm 0.07}$\\$69.50_{\pm 0.50}$}
& \makecell[c]{$80.50_{\pm 0.50}$\\$71.27_{\pm 8.95}$\\$3.52_{\pm 0.05}$\\$70.00_{\pm 0.50}$}
& \makecell[c]{$57.50_{\pm 0.50}$\\$214.35_{\pm 10.83}$\\$7.24_{\pm 0.08}$\\$77.50_{\pm 1.00}$}
& \makecell[c]{$61.50_{\pm 0.50}$\\$70.54_{\pm 7.95}$\\$7.11_{\pm 0.08}$\\$72.50_{\pm 0.50}$}
& \makecell[c]{$73.00_{\pm 1.50}$\\$30.98_{\pm 3.44}$\\$2.71_{\pm 0.03}$\\$79.00_{\pm 1.00}$}
\\
\cmidrule(lr){3-9}

& & P=1
& \makecell[c]{Performance\\Token cost\\Time cost (s)\\Passrate (\%)}
& \makecell[c]{$75.50_{\pm 0.50}$\\$104.37_{\pm 8.36}$\\$7.41_{\pm 0.08}$\\$83.50_{\pm 1.00}$}
& \makecell[c]{$85.00_{\pm 0.00}$\\$87.59_{\pm 7.11}$\\$4.22_{\pm 0.05}$\\$88.00_{\pm 0.50}$}
& \makecell[c]{$66.00_{\pm 1.50}$\\$254.87_{\pm 15.54}$\\$8.18_{\pm 0.09}$\\$91.00_{\pm 0.00}$}
& \makecell[c]{$69.00_{\pm 1.00}$\\$84.21_{\pm 7.36}$\\$8.24_{\pm 0.09}$\\$87.00_{\pm 1.50}$}
& \makecell[c]{$86.50_{\pm 2.00}$\\$37.44_{\pm 3.55}$\\$3.18_{\pm 0.04}$\\$91.50_{\pm 0.00}$}
\\

\bottomrule
\end{tabular}
}
\par
\vspace{12pt}
\raggedright
\caption{\textbf{Results on the Qwen2.5+DistilBERT+Phi-3.5-MoE cascade system.}}
\label{tab:qwen_distilbert_phi35_results}
\end{table*}

\begin{table*}[!thbp]
\centering
\scriptsize
\aboverulesep=0ex
\belowrulesep=0ex
\setlength{\tabcolsep}{2.8pt}
\renewcommand{\arraystretch}{1.20}
\resizebox{\textwidth}{!}{
\begin{tabular}{
>{\centering\arraybackslash}m{2.45cm}
>{\centering\arraybackslash}m{1.45cm}
>{\centering\arraybackslash}m{1.25cm}
>{\centering\arraybackslash}m{2.05cm}
>{\centering\arraybackslash}m{2.30cm}
>{\centering\arraybackslash}m{2.30cm}
>{\centering\arraybackslash}m{2.30cm}
}
\toprule
\textbf{Cascade system} & \textbf{Method} & \textbf{Setting} & \textbf{Metric} & \textbf{AGNews} & \textbf{SQuAD2.0} & \textbf{WildJailbreak} \\
\midrule

\multirow{11}{*}{\parbox[c][13.2cm][c]{2.45cm}{\centering\textbf{Qwen2.5}\\\textbf{+}\\\textbf{BERT-base}\\\textbf{+}\\\textbf{Mistral-7B}\\\textbf{+}\\\textbf{BERT-base}\\\textbf{+}\\\textbf{Phi-3.5-MoE}}}
& \multicolumn{2}{c}{\textbf{Initial}}
& \makecell[c]{Performance\\Token cost\\Time cost (s)\\Passrate$_1$ (\%)\\Passrate$_2$ (\%)}
& \makecell[c]{$89.50_{\pm 0.00}$\\$52.31_{\pm 0.00}$\\$4.13_{\pm 0.04}$\\$55.0_{\pm 0.00}$\\$29.0_{\pm 0.00}$}
& \makecell[c]{$81.00_{\pm 0.00}$\\$110.31_{\pm 0.00}$\\$4.97_{\pm 0.06}$\\$51.5_{\pm 0.00}$\\$24.0_{\pm 0.00}$}
& \makecell[c]{$98.00_{\pm 0.00}$\\$20.73_{\pm 0.00}$\\$2.11_{\pm 0.03}$\\$58.0_{\pm 0.00}$\\$34.0_{\pm 0.00}$}
\\
\cmidrule(lr){2-7}

& \multicolumn{2}{c}{\textbf{LLM w/o cascade}}
& \makecell[c]{Performance\\Token cost\\Time cost (s)\\Passrate$_1$ (\%)\\Passrate$_2$ (\%)}
& \makecell[c]{$94.00_{\pm 0.00}$\\$112.58_{\pm 0.00}$\\$8.12_{\pm 0.07}$\\$100.0_{\pm 0.00}$\\$100.0_{\pm 0.00}$}
& \makecell[c]{$86.50_{\pm 0.00}$\\$277.24_{\pm 0.00}$\\$8.49_{\pm 0.09}$\\$100.0_{\pm 0.00}$\\$100.0_{\pm 0.00}$}
& \makecell[c]{$98.50_{\pm 0.00}$\\$40.04_{\pm 0.00}$\\$3.49_{\pm 0.04}$\\$100.0_{\pm 0.00}$\\$100.0_{\pm 0.00}$}
\\
\cmidrule(lr){2-7}

& \multicolumn{2}{c}{\textbf{Random noise}}
& \makecell[c]{Performance\\Token cost\\Time cost (s)\\Passrate$_1$ (\%)\\Passrate$_2$ (\%)}
& \makecell[c]{$72.50_{\pm 0.50}$\\$64.14_{\pm 2.64}$\\$5.07_{\pm 0.06}$\\$57.5_{\pm 1.00}$\\$17.5_{\pm 0.50}$}
& \makecell[c]{$69.00_{\pm 1.50}$\\$141.89_{\pm 4.21}$\\$7.04_{\pm 0.07}$\\$52.0_{\pm 2.00}$\\$24.5_{\pm 0.50}$}
& \makecell[c]{$82.00_{\pm 0.50}$\\$23.85_{\pm 1.55}$\\$2.24_{\pm 0.03}$\\$61.5_{\pm 0.50}$\\$11.5_{\pm 1.50}$}
\\
\cmidrule(lr){2-7}

& \multirow{2}{*}{\parbox[c][2.2cm][c]{1.45cm}{\centering\textbf{Single-target}\\\textbf{attack}}}
& SLM only
& \makecell[c]{Performance\\Token cost\\Time cost (s)\\Passrate$_1$ (\%)\\Passrate$_2$ (\%)}
& \makecell[c]{$79.00_{\pm 0.50}$\\$92.18_{\pm 4.87}$\\$6.91_{\pm 0.17}$\\$71.0_{\pm 1.00}$\\$52.0_{\pm 0.50}$}
& \makecell[c]{$71.00_{\pm 0.50}$\\$227.38_{\pm 6.52}$\\$8.27_{\pm 0.28}$\\$69.5_{\pm 2.00}$\\$48.5_{\pm 1.00}$}
& \makecell[c]{$79.00_{\pm 0.00}$\\$30.12_{\pm 3.17}$\\$2.55_{\pm 0.13}$\\$61.5_{\pm 1.50}$\\$37.5_{\pm 0.50}$}
\\
\cmidrule(lr){3-7}

&
& DM only
& \makecell[c]{Performance\\Token cost\\Time cost (s)\\Passrate$_1$ (\%)\\Passrate$_2$ (\%)}
& \makecell[c]{$64.50_{\pm 1.00}$\\$67.85_{\pm 4.28}$\\$5.58_{\pm 0.16}$\\$50.0_{\pm 1.00}$\\$26.0_{\pm 0.50}$}
& \makecell[c]{$56.00_{\pm 0.00}$\\$161.08_{\pm 7.16}$\\$5.79_{\pm 0.17}$\\$54.5_{\pm 1.00}$\\$20.5_{\pm 0.50}$}
& \makecell[c]{$64.00_{\pm 1.00}$\\$17.80_{\pm 1.45}$\\$1.94_{\pm 0.04}$\\$46.5_{\pm 2.00}$\\$14.0_{\pm 1.00}$}
\\
\cmidrule(lr){2-7}

& \multirow{6}{*}{\parbox[c][6.6cm][c]{1.45cm}{\centering\textbf{Joint-target}\\\textbf{attack}}}
& P=0
& \makecell[c]{Performance\\Token cost\\Time cost (s)\\Passrate$_1$ (\%)\\Passrate$_2$ (\%)}
& \makecell[c]{$21.50_{\pm 0.50}$\\$19.40_{\pm 1.46}$\\$2.27_{\pm 0.09}$\\$24.0_{\pm 0.50}$\\$8.0_{\pm 1.00}$}
& \makecell[c]{$12.50_{\pm 1.00}$\\$75.11_{\pm 5.42}$\\$3.53_{\pm 0.21}$\\$26.5_{\pm 0.50}$\\$10.5_{\pm 0.50}$}
& \makecell[c]{$31.00_{\pm 1.50}$\\$7.91_{\pm 0.73}$\\$1.22_{\pm 0.14}$\\$28.5_{\pm 1.00}$\\$16.0_{\pm 0.50}$}
\\
\cmidrule(lr){3-7}

& & P=0.2
& \makecell[c]{Performance\\Token cost\\Time cost (s)\\Passrate$_1$ (\%)\\Passrate$_2$ (\%)}
& \makecell[c]{$35.50_{\pm 1.00}$\\$28.97_{\pm 2.44}$\\$3.18_{\pm 0.31}$\\$39.5_{\pm 2.50}$\\$17.5_{\pm 0.50}$}
& \makecell[c]{$30.00_{\pm 1.50}$\\$109.40_{\pm 11.26}$\\$4.31_{\pm 0.72}$\\$36.0_{\pm 1.00}$\\$14.5_{\pm 1.00}$}
& \makecell[c]{$41.50_{\pm 0.50}$\\$13.80_{\pm 1.57}$\\$1.68_{\pm 0.07}$\\$43.5_{\pm 2.00}$\\$20.5_{\pm 1.50}$}
\\
\cmidrule(lr){3-7}

& & P=0.4
& \makecell[c]{Performance\\Token cost\\Time cost (s)\\Passrate$_1$ (\%)\\Passrate$_2$ (\%)}
& \makecell[c]{$42.00_{\pm 0.50}$\\$39.77_{\pm 3.32}$\\$3.96_{\pm 0.24}$\\$50.5_{\pm 1.00}$\\$22.5_{\pm 0.50}$}
& \makecell[c]{$36.50_{\pm 0.50}$\\$138.02_{\pm 12.19}$\\$5.02_{\pm 0.28}$\\$49.5_{\pm 1.00}$\\$27.5_{\pm 2.00}$}
& \makecell[c]{$50.00_{\pm 0.00}$\\$20.94_{\pm 2.03}$\\$1.94_{\pm 0.03}$\\$54.0_{\pm 0.50}$\\$30.5_{\pm 1.50}$}
\\
\cmidrule(lr){3-7}

& & P=0.6
& \makecell[c]{Performance\\Token cost\\Time cost (s)\\Passrate$_1$ (\%)\\Passrate$_2$ (\%)}
& \makecell[c]{$50.00_{\pm 1.00}$\\$50.41_{\pm 4.43}$\\$4.41_{\pm 0.05}$\\$62.5_{\pm 1.00}$\\$25.5_{\pm 0.50}$}
& \makecell[c]{$44.50_{\pm 1.50}$\\$168.87_{\pm 14.72}$\\$5.94_{\pm 0.06}$\\$58.5_{\pm 1.50}$\\$24.0_{\pm 0.50}$}
& \makecell[c]{$61.50_{\pm 0.50}$\\$26.55_{\pm 2.47}$\\$2.44_{\pm 0.03}$\\$66.0_{\pm 2.00}$\\$27.0_{\pm 1.50}$}
\\
\cmidrule(lr){3-7}

& & P=0.8
& \makecell[c]{Performance\\Token cost\\Time cost (s)\\Passrate$_1$ (\%)\\Passrate$_2$ (\%)}
& \makecell[c]{$59.50_{\pm 1.00}$\\$64.58_{\pm 4.89}$\\$5.16_{\pm 0.06}$\\$70.5_{\pm 2.50}$\\$30.0_{\pm 0.50}$}
& \makecell[c]{$52.00_{\pm 1.00}$\\$192.47_{\pm 18.56}$\\$6.85_{\pm 0.07}$\\$67.5_{\pm 1.00}$\\$37.5_{\pm 1.00}$}
& \makecell[c]{$74.50_{\pm 1.50}$\\$32.04_{\pm 3.01}$\\$2.62_{\pm 0.07}$\\$74.0_{\pm 0.50}$\\$33.5_{\pm 1.50}$}
\\
\cmidrule(lr){3-7}

& & P=1
& \makecell[c]{Performance\\Token cost\\Time cost (s)\\Passrate$_1$ (\%)\\Passrate$_2$ (\%)}
& \makecell[c]{$66.50_{\pm 1.00}$\\$77.50_{\pm 3.25}$\\$5.79_{\pm 0.07}$\\$81.0_{\pm 1.50}$\\$36.0_{\pm 0.50}$}
& \makecell[c]{$59.50_{\pm 0.50}$\\$207.02_{\pm 19.17}$\\$7.79_{\pm 0.08}$\\$83.5_{\pm 1.00}$\\$42.0_{\pm 2.00}$}
& \makecell[c]{$88.50_{\pm 1.00}$\\$39.17_{\pm 2.66}$\\$3.04_{\pm 0.04}$\\$88.5_{\pm 0.50}$\\$37.0_{\pm 0.00}$}
\\
\bottomrule
\end{tabular}
}
\par
\vspace{12pt}
\raggedright
\caption{\textbf{Results for attacking the first Decision maker in the Qwen2.5 + BERT-base + Mistral-7B + BERT-base + Phi-3.5-MoE cascade system.}}
\label{tab:qwen_mistral_phi35_block1_results}
\end{table*}

\begin{table*}[!thbp]
\centering
\scriptsize
\aboverulesep=0ex
\belowrulesep=0ex
\setlength{\tabcolsep}{2.8pt}
\renewcommand{\arraystretch}{1.20}
\resizebox{\textwidth}{!}{
\begin{tabular}{
>{\centering\arraybackslash}m{2.45cm}
>{\centering\arraybackslash}m{1.45cm}
>{\centering\arraybackslash}m{1.25cm}
>{\centering\arraybackslash}m{2.05cm}
>{\centering\arraybackslash}m{2.30cm}
>{\centering\arraybackslash}m{2.30cm}
>{\centering\arraybackslash}m{2.30cm}
}
\toprule
\textbf{Cascade system} & \textbf{Method} & \textbf{Setting} & \textbf{Metric} & \textbf{AGNews} & \textbf{SQuAD2.0} & \textbf{WildJailbreak} \\
\midrule

\multirow{11}{*}{\parbox[c][13.2cm][c]{2.45cm}{\centering\textbf{Qwen2.5}\\\textbf{+}\\\textbf{BERT-base}\\\textbf{+}\\\textbf{Mistral-7B}\\\textbf{+}\\\textbf{BERT-base}\\\textbf{+}\\\textbf{Phi-3.5-MoE}}}
& \multicolumn{2}{c}{\textbf{Initial}}
& \makecell[c]{Performance\\Token cost\\Time cost (s)\\Passrate$_1$ (\%)\\Passrate$_2$ (\%)}
& \makecell[c]{$89.50_{\pm 0.00}$\\$52.31_{\pm 0.00}$\\$4.13_{\pm 0.04}$\\$55.0_{\pm 0.00}$\\$29.0_{\pm 0.00}$}
& \makecell[c]{$81.00_{\pm 0.50}$\\$110.31_{\pm 1.20}$\\$4.97_{\pm 0.06}$\\$51.5_{\pm 0.50}$\\$24.0_{\pm 0.00}$}
& \makecell[c]{$98.00_{\pm 0.50}$\\$20.73_{\pm 0.35}$\\$2.11_{\pm 0.03}$\\$58.0_{\pm 0.00}$\\$34.0_{\pm 0.50}$}
\\
\cmidrule(lr){2-7}

& \multicolumn{2}{c}{\textbf{LLM w/o cascade}}
& \makecell[c]{Performance\\Token cost\\Time cost (s)\\Passrate$_1$ (\%)\\Passrate$_2$ (\%)}
& \makecell[c]{$94.00_{\pm 0.00}$\\$112.58_{\pm 0.00}$\\$8.12_{\pm 0.09}$\\$100.0_{\pm 0.00}$\\$100.0_{\pm 0.00}$}
& \makecell[c]{$86.50_{\pm 0.00}$\\$277.24_{\pm 0.00}$\\$8.49_{\pm 0.07}$\\$100.0_{\pm 0.00}$\\$100.0_{\pm 0.00}$}
& \makecell[c]{$98.50_{\pm 0.00}$\\$40.04_{\pm 0.00}$\\$3.49_{\pm 0.04}$\\$100.0_{\pm 0.00}$\\$100.0_{\pm 0.00}$}
\\
\cmidrule(lr){2-7}

& \multicolumn{2}{c}{\textbf{Random noise}}
& \makecell[c]{Performance\\Token cost\\Time cost (s)\\Passrate$_1$ (\%)\\Passrate$_2$ (\%)}
& \makecell[c]{$72.50_{\pm 0.50}$\\$64.14_{\pm 2.89}$\\$5.07_{\pm 0.06}$\\$57.5_{\pm 1.00}$\\$17.5_{\pm 1.00}$}
& \makecell[c]{$69.00_{\pm 1.00}$\\$141.89_{\pm 7.42}$\\$7.04_{\pm 0.07}$\\$52.0_{\pm 2.00}$\\$24.5_{\pm 0.00}$}
& \makecell[c]{$82.00_{\pm 0.50}$\\$23.85_{\pm 1.45}$\\$2.24_{\pm 0.03}$\\$61.5_{\pm 1.00}$\\$11.5_{\pm 0.50}$}
\\
\cmidrule(lr){2-7}

& \multirow{2}{*}{\parbox[c][2.2cm][c]{1.45cm}{\centering\textbf{Single-target}\\\textbf{attack}}}
& SLM only
& \makecell[c]{Performance\\Token cost\\Time cost (s)\\Passrate$_1$ (\%)\\Passrate$_2$ (\%)}
& \makecell[c]{$76.00_{\pm 0.50}$\\$91.41_{\pm 7.26}$\\$6.83_{\pm 0.07}$\\$66.0_{\pm 1.00}$\\$45.5_{\pm 0.50}$}
& \makecell[c]{$68.00_{\pm 1.00}$\\$206.40_{\pm 14.25}$\\$7.64_{\pm 0.08}$\\$60.5_{\pm 0.50}$\\$42.0_{\pm 1.00}$}
& \makecell[c]{$76.00_{\pm 1.00}$\\$28.31_{\pm 2.17}$\\$2.41_{\pm 0.03}$\\$59.0_{\pm 1.00}$\\$41.5_{\pm 1.50}$}
\\
\cmidrule(lr){3-7}

&
& DM only
& \makecell[c]{Performance\\Token cost\\Time cost (s)\\Passrate$_1$ (\%)\\Passrate$_2$ (\%)}
& \makecell[c]{$70.50_{\pm 0.50}$\\$72.46_{\pm 6.41}$\\$5.74_{\pm 0.06}$\\$58.5_{\pm 0.50}$\\$39.0_{\pm 0.50}$}
& \makecell[c]{$60.00_{\pm 1.50}$\\$174.98_{\pm 12.84}$\\$6.41_{\pm 0.07}$\\$54.0_{\pm 1.00}$\\$38.5_{\pm 1.00}$}
& \makecell[c]{$83.00_{\pm 0.50}$\\$29.82_{\pm 3.58}$\\$2.78_{\pm 0.03}$\\$60.5_{\pm 0.50}$\\$46.0_{\pm 1.00}$}
\\
\cmidrule(lr){2-7}

& \multirow{6}{*}{\parbox[c][6.6cm][c]{1.45cm}{\centering\textbf{Joint-target}\\\textbf{attack}}}
& P=0
& \makecell[c]{Performance\\Token cost\\Time cost (s)\\Passrate$_1$ (\%)\\Passrate$_2$ (\%)}
& \makecell[c]{$34.50_{\pm 1.50}$\\$37.19_{\pm 2.43}$\\$3.45_{\pm 0.04}$\\$62.0_{\pm 1.00}$\\$7.0_{\pm 0.50}$}
& \makecell[c]{$28.50_{\pm 0.50}$\\$90.89_{\pm 8.76}$\\$4.37_{\pm 0.05}$\\$54.5_{\pm 0.50}$\\$8.5_{\pm 1.00}$}
& \makecell[c]{$58.00_{\pm 0.00}$\\$15.82_{\pm 1.58}$\\$1.69_{\pm 0.02}$\\$58.0_{\pm 1.50}$\\$6.5_{\pm 0.50}$}
\\
\cmidrule(lr){3-7}

& & P=0.2
& \makecell[c]{Performance\\Token cost\\Time cost (s)\\Passrate$_1$ (\%)\\Passrate$_2$ (\%)}
& \makecell[c]{$42.00_{\pm 0.00}$\\$49.88_{\pm 4.24}$\\$4.02_{\pm 0.05}$\\$60.5_{\pm 0.50}$\\$14.5_{\pm 1.00}$}
& \makecell[c]{$36.00_{\pm 0.00}$\\$131.33_{\pm 11.23}$\\$5.05_{\pm 0.06}$\\$52.0_{\pm 1.00}$\\$15.0_{\pm 0.50}$}
& \makecell[c]{$62.50_{\pm 0.50}$\\$19.17_{\pm 1.14}$\\$2.01_{\pm 0.03}$\\$59.0_{\pm 1.00}$\\$13.5_{\pm 1.50}$}
\\
\cmidrule(lr){3-7}

& & P=0.4
& \makecell[c]{Performance\\Token cost\\Time cost (s)\\Passrate$_1$ (\%)\\Passrate$_2$ (\%)}
& \makecell[c]{$50.00_{\pm 0.00}$\\$60.57_{\pm 4.89}$\\$4.84_{\pm 0.05}$\\$56.0_{\pm 1.50}$\\$22.0_{\pm 0.50}$}
& \makecell[c]{$43.50_{\pm 0.50}$\\$158.54_{\pm 13.64}$\\$5.89_{\pm 0.06}$\\$54.0_{\pm 0.50}$\\$24.0_{\pm 1.00}$}
& \makecell[c]{$69.00_{\pm 0.50}$\\$24.23_{\pm 3.17}$\\$2.35_{\pm 0.03}$\\$60.5_{\pm 0.50}$\\$19.5_{\pm 1.00}$}
\\
\cmidrule(lr){3-7}

& & P=0.6
& \makecell[c]{Performance\\Token cost\\Time cost (s)\\Passrate$_1$ (\%)\\Passrate$_2$ (\%)}
& \makecell[c]{$59.50_{\pm 1.50}$\\$77.31_{\pm 7.46}$\\$5.61_{\pm 0.06}$\\$58.0_{\pm 1.00}$\\$30.5_{\pm 1.50}$}
& \makecell[c]{$52.50_{\pm 1.00}$\\$191.74_{\pm 16.46}$\\$6.72_{\pm 0.07}$\\$55.0_{\pm 1.50}$\\$32.0_{\pm 0.50}$}
& \makecell[c]{$75.50_{\pm 0.50}$\\$29.41_{\pm 3.54}$\\$2.72_{\pm 0.03}$\\$57.5_{\pm 1.00}$\\$27.5_{\pm 1.00}$}
\\
\cmidrule(lr){3-7}

& & P=0.8
& \makecell[c]{Performance\\Token cost\\Time cost (s)\\Passrate$_1$ (\%)\\Passrate$_2$ (\%)}
& \makecell[c]{$68.00_{\pm 0.00}$\\$89.74_{\pm 7.27}$\\$6.42_{\pm 0.07}$\\$57.0_{\pm 0.50}$\\$39.0_{\pm 1.00}$}
& \makecell[c]{$63.00_{\pm 1.00}$\\$216.30_{\pm 15.37}$\\$7.62_{\pm 0.08}$\\$59.5_{\pm 1.00}$\\$41.5_{\pm 1.50}$}
& \makecell[c]{$81.00_{\pm 1.50}$\\$32.77_{\pm 3.79}$\\$3.01_{\pm 0.04}$\\$49.0_{\pm 1.50}$\\$36.0_{\pm 1.00}$}
\\
\cmidrule(lr){3-7}

& & P=1
& \makecell[c]{Performance\\Token cost\\Time cost (s)\\Passrate$_1$ (\%)\\Passrate$_2$ (\%)}
& \makecell[c]{$77.50_{\pm 0.50}$\\$105.49_{\pm 6.42}$\\$7.39_{\pm 0.08}$\\$56.0_{\pm 1.00}$\\$47.5_{\pm 1.50}$}
& \makecell[c]{$70.50_{\pm 1.50}$\\$242.17_{\pm 19.53}$\\$8.27_{\pm 0.09}$\\$61.5_{\pm 0.50}$\\$50.5_{\pm 1.00}$}
& \makecell[c]{$92.50_{\pm 0.50}$\\$38.42_{\pm 4.49}$\\$3.37_{\pm 0.04}$\\$54.0_{\pm 1.00}$\\$46.0_{\pm 2.00}$}
\\

\bottomrule
\end{tabular}
}
\par
\vspace{12pt}
\raggedright
\caption{\textbf{Results for attacking the second Decision maker in the Qwen2.5 + BERT-base + Mistral-7B + BERT-base + Phi-3.5-MoE cascade system.}}
\label{tab:qwen_mistral_phi35_block2_results}
\end{table*}


\section{Algorithm Pseudocode}
The algorithm Pseudocode (using prediction accuracy as the task metric for demonstration) that summarizes Section~\ref{sec:method} is provided in Algorithm~\ref{alg:attack_framework}. 

\begin{algorithm*}[!h]
\caption{Attack Algorithm}
\label{alg:attack_framework}
\KwIn{Input $\vx$, label $\vy$, cascade $\Phi(\cdot;\mathcal{F},\mathcal{G})$, suffix $\delta$ with bounded length (reflecting $N_\epsilon$), rounds $R$, update iterations $T$, target pass-rate $p$, target cascade system's layer index $i$}
\KwOut{Adversarial example $\vx'$}
\BlankLine
 Set the $\delta^{(f)}_{0,T}$, $\delta^{(g)}_{0,T}$ to be empty string, $\delta=\delta^{(f)}_{0,T} \,\|\, \delta^{(g)}_{0,T}$ \;
\For{$r \gets 1$ \KwTo $R$}{
  \tcp{\color{blue}{Prediction-targeted phase: update $\delta^{(f)}_{r,t}$ to decrease the early-stage prediction models' task performance.}}
  Initialize $\delta^{(f)}_{r,0}$ randomly\;
  \For{$t \gets 1$ \KwTo $T$}{
  $\delta^{(f)}_{r,t}
\;\leftarrow\;
\mathrm{Update} \;\big(\delta^{(f)}_{r,t-1};\, \mathcal{L}_{\mathrm{acc}}(\delta^{(f)}_{r,t-1}, \delta)\big)$  where 
$\mathcal{L}_{\mathrm{acc}}(\delta^{(f)}_{r,t-1}, \delta)\big)
= \Loss\;\big(f^i(\vx\,\|\, \delta \,\|\, \delta^{(f)}_{r,t-1}),\,\vy\big)
$\;
  }
  Update the final suffix $\delta=\delta \,\|\, \delta^{(f)}_{r,T}$\;
  Compute intermediate predictions $\hat{\vy}^i_r \gets f^i(\vx \,\|\, \delta)$\;
  \tcp{\color{blue}{Decision-targeted phase: update $\delta^{(g)}_{r,t}$ to control passrate (inference cost)}}
  Initialize $\delta^{(g)}_{r,0}$ randomly\;
  \For{$t \gets 1$ \KwTo $T$}{
    Generate candidate set $\mathcal{C}_{r,t} \subset N_\epsilon(\delta^{(g)}_{r,t-1})$\;
    Sample $b\in \{0,1\}$ from Bernoulli distribution $\{p,1-p\}$ \;
    \If{$\hat{\vy}^i_r=\vy$ \textbf{\textrm{or}} $b=1$}{ 
\scalebox{0.9}{$
\delta^{(g)}_{r,t}
\leftarrow
\mathrm{Update}\;\big(\delta^{(g)}_{r,t-1};\,
\mathcal{L}_{\mathrm{cost}}(\vx \,\|\, \delta \,\|\, \delta^{(g)}_{r,t-1})\big)
$\, where \, $
\mathcal{L}_{\mathrm{cost}}(\vx \,\|\, \delta \,\|\, \delta^{(g)}_{r,t-1})
= g^i(\vx \,\|\,  \delta,\delta^{(g)}_{r,t-1}, \, f^i(\vx \,\|\, \delta \,\|\, \delta^{(g)}_{r,t-1}))
$}\;
    Select $\delta^{(g)}$ maximizing $\mathcal{L}_{\mathrm{cost}}(\vx \,\|\,  \delta \,\|\, \delta^{(g)})$ under the priorities listed in \sectionautorefname~\ref{subsubsec:constraint_preserve}:\
     }
    \Else{ 
     $\delta^{(g)}_{r,t}
\;\leftarrow\;
\mathrm{Update} \;\big(\delta^{(g)}_{r,t-1};\, - \mathcal{L}_{\mathrm{cost}}(\vx \,\|\, \delta \,\|\, \delta^{(g)}_{r,t-1})\big)$ \;
    Select $\delta^{(g)}$ minimizing $\mathcal{L}_{\mathrm{cost}}(\vx \,\|\, \delta \,\|\, \delta^{(g)})$ under the priorities listed in \sectionautorefname~\ref{subsubsec:constraint_preserve}:\
    }
     \begin{enumerate}[label=(\roman*), leftmargin=*]
    \item[(i)] Prefer candidates that keep the intermediate prediction unchanged:
    $$f^i(\vx \,\|\, \delta \,\|\, \delta^{(g)}) = \hat{\vy}^i_r $$
    \item[(ii)] If none satisfy (i), prefer candidates that maintain previously incorrect predictions as still incorrect:
    \[
    f^i(\vx \,\|\, \delta \,\|\, \delta^{(g)}) \neq \vy 
    \]
    \item[(iii)] If neither (i) nor (ii) can be satisfied, drop the constraint and minimize \(\mathcal{L}_{\mathrm{cost}}\) freely.
    \end{enumerate}

  }
  Update the final suffix $\delta=\delta \,\|\, \delta^{(g)}_{r,T}$\;
}
\Return $\vx' = \vx \,\|\, \delta$\;
\end{algorithm*}

\end{document}